\documentclass[aps,prx,twocolumn,superscriptaddress,nofootinbib,notitlepage,longbibliography]{revtex4-1}
\usepackage{graphicx}
\usepackage{amsmath}
 \usepackage{amstext}
\usepackage{amssymb}
\usepackage{enumitem}   
\usepackage{fancybox}
\usepackage{graphicx}
\usepackage{amsmath}
\usepackage{amstext}
\usepackage{amssymb}
\usepackage{amsfonts}
\usepackage{longtable,booktabs}
\usepackage{hyperref}\usepackage{url}
\usepackage{subfigure}%
\usepackage{dsfont}
\usepackage{booktabs}
\usepackage{amsbsy}
\usepackage{dcolumn} 
\usepackage{amsthm}
\usepackage{bm}
\usepackage{esint}
\usepackage{multirow}
\usepackage{hyperref}
\usepackage{ulem}
\hypersetup{
    colorlinks, bookmarks=true,breaklinks=true,linkcolor=magenta, citecolor=blue, linktocpage=true, urlcolor=black
}
\usepackage{cleveref}
  \usepackage{extarrows}
\usepackage{mathrsfs}
\usepackage{amsfonts}
\usepackage{amsbsy}
\usepackage{dcolumn}
\usepackage{bm}
\usepackage{multirow}
\usepackage{color}
\def\Z{\mathbb{Z}}

\usepackage{datetime}
\usepackage{comment}
\usepackage[super]{nth}
 \usepackage{varwidth}
\usepackage{epstopdf}
\usepackage{float}
 \usepackage{amssymb}
\usepackage[T1]{fontenc} 
\usepackage{amsmath}
\usepackage{graphicx}
\usepackage{float}
\newcommand{\tabincell}[2]{\begin{tabular}{@{}#1@{}}#2\end{tabular}}

\newcommand{\bpm}{\begin{pmatrix}}
\newcommand{\epm}{\end{pmatrix}}
 \def\Z{\mathbb{Z}}

\begin{document}
 
 \title{Fractionalizing Global Symmetry on Looplike Topological Excitations}
 
 \author{Shang-Qiang Ning}
\altaffiliation{Present address: Department of Physics, The Chinese University of Hong Kong, Shatin, New Territories, Hong Kong, China}
\affiliation{Department of Physics, The University of Hong Kong, Pokfulam Road, Hong Kong, China}
\affiliation{Institute for Advanced Study, Tsinghua University, Beijing, 100084, China}
\author{Zheng-Xin Liu}
\affiliation{Department of Physics, Renmin University of China, Beijing, 100872, China}
\author{Peng Ye}
 \email{Corresponding author: yepeng5@mail.sysu.edu.cn}
\affiliation{School of Physics, State Key Laboratory of Optoelectronic Materials and Technologies, and Guangdong Provincial Key Laboratory of Magnetoelectric Physics and Devices, Sun Yat-sen University, Guangzhou, 510275,
China}   
\affiliation{Department of Physics and Institute for Condensed Matter Theory, University of Illinois at Urbana-Champaign, Illinois 61801, USA}
 
    \begin{abstract} 
Symmetry fractionalization on topological excitations is one of the most remarkable quantum phenomena in topological orders with symmetry, i.e., symmetry-enriched topological phases. While much progress has been theoretically and experimentally made in two dimensions (2D), the understanding on symmetry fractionalization in 3D is far from complete. A long-standing challenge is to understand symmetry fractionalization on looplike topological excitations which  are spatially extended objects. In this work, we construct a powerful topological-field-theoretical framework approach for 3D topological orders, which leads to a systematic characterization and classification of symmetry fractionalization.   For systems with  Abelian gauge groups ($G_g$) and  Abelian symmetry groups ($G_s$),  we successfully establish equivalence classes that lead to a finite number of patterns of symmetry fractionalization, although there are notoriously infinite number of Lagrangian-descriptions of the system.    Based on this, we  compute topologically distinct types of fractional symmetry charges carried by particles. Then, for each type, we compute topologically distinct statistical phases of braiding processes among loop excitations and external symmetry fluxes. As a result, we are able to unambiguously list all physical observables for each pattern of symmetry fractionalization. We present detailed calculations on many concrete examples. As an example, we  find  that the symmetry fractionalization in an untwisted $\mathbb{Z}_2\times \mathbb{Z}_2$ topological order with $\mathbb{Z}_2$ symmetry is classified by $ (\mathbb{Z}_2)^6\oplus  (\mathbb{Z}_2)^2\oplus (\mathbb{Z}_2)^2\oplus (\mathbb{Z}_2)^2$. If the topological order is twisted, the classification reduces to $(\mathbb{Z}_2)^6$ in which particle excitations always carry integer charge.      Inspired by the field-theoretical analysis, we find that our classification of symmetry fractionalization with twist $\omega$ can be formally organized into an algebraic formalism:    $    \bigoplus_{\nu_i} \mathcal H^4 ( G_g\leftthreetimes_{\nu_i}G_s, U (1)) /\Gamma_\omega ({\nu_i}) \,,
  $ where anomaly-free symmetry fractionalization on particles $\nu_i \in \mathcal{H}_{non}^2 (G_s, G_g)$.
  Despite the lack of detailed dependence of  $\Gamma_\omega (\nu_i)$ on $\omega$ and $\nu_i$, the present   field-theoretical approach allows us to efficiently calculate and understand  $\Gamma_\omega (\nu_i)$. Several future directions are present at the end of this paper.\end{abstract} 
\date{{\color{blue}{\today}}}
\maketitle

\tableofcontents

\section{Introduction}\label{section_introduction}
 
    \subsection{Research background}\label{subsection_research_background}
Symmetry fractionalization (SF)\footnote{Some frequently used abbreviations are collected in Appendix~\ref{appendix_abbr_list}.} is  a remarkable quantum phenomenon arising from the interplay of global symmetry and topology in strongly correlated systems \cite{set_review_chen16,wen_quantum_order_2002,WenRMP,Barkeshli722}. For example, in the  superconductivity  theory of doped cuprates \cite{WenRMP}, electrons are fractionalized into holons and spinons carrying charge and spin respectively, which behave  differently from electrons.  
Especially, the fractional quantum Hall systems \cite{RevModPhys.89.041004} as typical topological phases of matter, support fractional charge excitations called anyons, which has been experimentally observed \cite{goldman1995resonant, de1998direct, reznikov1999observation, PhysRevLett.79.2526, martin2004localization}. 
  On the other hand, in   quantum spin liquids \cite{Savary_2016, ZhouY_RMP,MF9400}, the spinon excitations  carry half-spin quantum numbers,  in contrast to   spin-integer magnon   excitations in magnetically ordered materials.    Such kind of symmetric topological orders is often referred to as ``symmetry-enriched topological phases (SET)'' \cite{local_unitary_transf_quantum_entanglement_chen_gu_wen_2010_prb}.   The patterns of symmetry fractionalization in SET phases, as illustrated in above examples, are important for both theoretical and experimental sides. Intuitively, the existence of symmetry fractionalization on particles can be encoded, indirectly, by the  Aharonov-Bohm (AB) phase $e^{iq\phi}$ of braiding a particle carrying fractional charge around a $U(1)$ symmetry flux. If $q$ takes integer, the resulting phase is said to be trivial and can be used to establish equivalence relations for  fractional parts of $q$.  This AB phase counting motivates us to transform the problem of SF into the problem of braiding statistics of underlying topological excitations in the presence of external symmetry fluxes. In fact, we will use this \textit{mixed} braiding statistics involving both topological excitations and external symmetry fluxes to efficiently study SF in the present paper.

One of  the key tasks in  the study of SET  phases is to understand how global symmetry   acts on  topological excitations  in a self-consistent fashion.  In an anomaly-free SET, global symmetry transformations on the whole spectrum are compatible with full data of   topological order. Otherwise, the constructed SET is said to be anomalous, which might be realizable  on the boundary of some higher dimensional topological phases including symmetry-protected topological phases (SPTs) \cite{chenguliuwenscience,chenguliuwenprb}.  For example, the surface of bosonic topological insulators [i.e., 3D bosonic SPT with $U(1)$ and time-reversal symmetry] supports a $\Z_2$ topological order  with unbroken  symmetry, such that both $e$ anyon  and $m$ anyon carry half-quantized electric charge. Such a symmetric 2D topological order can  exist   as a surface state \cite{corbodism3,bti1,bti2,bti3,bti4,bti5,bti6,bti7}.

      Since  1D gapped phases do not admit topological order  \cite{PhysRevB.83.035107}, SET phases only  exist in two or higher dimensions.   2D SETs and the associated SF have been studied from various theoretical approaches  \cite{set_review_chen16, wen_quantum_order_2002, PhysRevB.74.174423, PhysRevLett.103.196803,  lu_ashvin_set_2016_prb, hung_wen_set_2013_prb, PhysRevB.87.195103, Barkeshli2014, Ran_SET_2013_prb, Bombin2010,Teo2015,Modular_Extension_LT_17, Lan2017}. From group-theoretical point of view, one of the key points in 2D SET phases is that anyonic excitations can carry \textit{projective representations} of the symmetry group of the system. The product of two projective representations may result in a linear representation, which is consistent with the following two facts (i) anyons must be created in pairs    and (ii) the whole many-body wavefunction must transform linearly under symmetry transformations. Furthermore, one needs to consider the situations that symmetry may permute different types of anyons and that some patterns of symmetry fractionalization may have anomaly. Sometimes, the permutation invariance among anyons is called ``anyonic / topological symmetry'' in the literature.       
        To elaborate,   symmetry enrichment of   2D topological orders involves three aspects:       \begin{enumerate}[label=\textbf{(\roman*)}]
\item permutation of superselection sectors under  symmetry transformations; 
\item symmetry fractionalization  of topological excitations; and 
\item stacking with nontrivial SPT layers. 
\end{enumerate}
In more sophisticated situations,  nontrivial permutation among different topological excitations would make it hard to define   projective representation of symmetry group.   A  systematic framework called ``$G^{\times}$ tensor category'' for 2D bosonic SET is established \cite{Barkeshli2014}, which is related to the mathematical framework  in Ref.~\cite{etingof2010fusion}. The authors in Refs.~\cite{Modular_Extension_LT_17, Lan2017} proposed  a general framework for 2D fermionic SETs, which is the so-called modular extension of unitary fusion category. It  is equivalent to the ``$G^{\times}$ tensor category'' when   2D bosonic SETs are considered. Very recently, frameworks of physical characterization and classification for {2D} fermionic SETs are establised in Ref.~\cite{Maissam2021a,Aasen2021,Bulmash2021a,Bulmash2021b,Ning2021a}. For more about 2D SF, we suggest the review article~\cite{set_review_chen16}.

In 3D or higher dimensions, quantum statistics of particles is less interesting compared to 2D: anyons do not exist \cite{Leinaas1977, YSWu84a} and mutual statistics among particles is trivial. Nevertheless, topological excitations with spatially extended shapes emerge in the low-energy physics. As a result,  braiding data become far more fruitful and complicated  \cite{wang_levin1, string5, 2016arXiv161209298P, ye17b, PhysRevResearch.3.023132, Zhang:2021ycl,jian_qi_14}.  In the long-wavelength limit, such spatially extended objects   geometrically  form  closed manifolds\footnote{As a side note, non-manifold-like excitations \cite{PhysRevB.101.245134,Li:2021umc} are found in a class of exactly solvable models that exhibit exotic fracton-topological order.}, such as loops\footnote{ In this paper, we  use  ``loop excitations'' rather than ``string excitations'' although the latter   are also popular in the literature. Furthermore,   ``particle excitations'' and ``loop excitations'' are also frequently simplified as ``particles'' and ``loops'' unless otherwise specified.}, membranes,  $\cdots$. For example,  3D toric code model is the simplest toy model exhibiting 3D topological order, where topological excitations include a particle carrying $\mathbb{Z}_2$ gauge charge and a loop carrying $\mathbb{Z}_2$ gauge flux. 
  In the literature, 
  some concrete examples of 3D SETs have been studied  \cite{xucenke_2013_3dset, bti6,maciejkoFTI,maciejko2015, swingle2011, ye16a,YW13a, Ye:2017aa, sahoo_sirota_cho_teo_2017_fti}. 
For instance, 3D fractional topological insulators in both bosonic and fermionic systems  \cite{maciejkoFTI, maciejko_model, maciejko2015, swingle2011, ye16a, YW13a,Ye:2017aa, sahoo_sirota_cho_teo_2017_fti, cho_teo_fradkin_2017, levin_lattice, swingle_fti_2012} are typical  3D SETs in which time-reversal symmetry plays a critical role and the axion angle is quantized at fractional but time-reversal invariant value.  One generalization of the anyonic symmetry to 3D SET phases was discussed in Ref.~\cite{ye16a}, which are dubbed charge-loop excitation symmetry.  Anomalous SETs in 3D were considered in Ref.~\cite{2016arXiv161008645Y, 2019RyoheiarXiv} and surface theory of 3D SETs was partially studied in Ref.~\cite{bti6,Ye:2017aa}.     In Refs.~\cite{chen_hermele_3dset2016,2015arXiv151102563C},  the authors relate  symmetric properties of loop excitations to properties of specific lower-dimensional topological systems.   Ref.~\cite{WenPRB16}  constructed  many local bosonic models in  3D which can realize various types of simplest topological orders with time-reversal symmetry. 
In Ref.~\cite{ye16_set}, the authors initiate  the field-theoretical line of thinking to study SET physics, especially emphasizing the critical role of symmetry-enriched gauge theories.      
Very recently, two formal frameworks based on higher-form symmetry and spectral sequences in Refs.~\cite{Hsin2020JHEP,Wang2021LHS} are proposed to classify  symmetry enriched gauge theories and SET phases.

Despite the above progresses, the theoretical study of 3D SET phases is far from complete and satisfactory, as compared with the aforementioned triumph in 2D SET phases. In   higher-dimensional topological orders, one may ask how to systematically classify SF on   spatially extended topological excitations. And, what are   relevant physical observables responsible for such kind of SF phenomena?  For example, one may wonder rigorously how many distinct phases are there when $\Z_2$ topological order is enriched by $\Z_2$ symmetry? And, how can we distinguish them via physical observables of either realistic experiments or thought experiments? And more seriously, what is meaning of SF on $\Z_2$ gauge fluxes (i.e., $\Z_2$ loop excitations)? Motivated by these questions, in the present work,    based on previous efforts, we attempt to complete the construction of a systematic field-theoretical description of 3D SET phases for the purpose of  physical  characterization and systematic classification of patterns of symmetry fractionalization in 3D topological orders. Specially, attentions will  be paid to loop excitations, which are entirely absent in lower dimensions.  Inspired by the AB phase counting picture mentioned at the very beginning of this paper, we expect braiding statistics in the presence of external symmetry fluxes will be a very important way to characterize SF on loops.


          \begin{table*}[t] 
    \centering
     \caption{\textbf{Classification of   symmetry fractionalization}.  We use    gauge group $G_g$ and ``twist'' to determine a topological order.  $n=1,2,\cdots$.  $\Z_N$ topological order is always untwisted, so we denote the twist as $-$.    We use  a pair of integers  (coefficients of two linearly independent twisted topological terms) to label a twist. If both integers vanish, the term is untwisted.  The classification is expressed  as $\mathcal{C}_1\oplus \mathcal{C}_2 \oplus\cdots $, where $i=1,2,...$ of $\mathcal{C}_i$ label the $i$-th type of SFP (symmetry fractionalization on particles).  Within a given type $i$, $\mathcal{C}_i$ corresponds to the classification of SFL (symmetry fractionalization on loops). $\oplus$ denotes a   summation over topologically distinct types of SFP.  We organize SFP types such that the less ``fractionalized'' SFP comes first. That is, $\mathcal{C}_1$ always denotes the SFL classification when SFP is trivial. The symbol ``$ (\Z_n)^m$'' is defined as $(\Z_n)^m\equiv\Z_n\times \Z_n\cdots \Z_n$, where $m$ denotes the number of $\Z_n$'s repeated in the direct product.  The symbol ``$k (\Z_n)^m$'' is defined as: $ k (\Z_n)^m\equiv(\Z_n)^m \oplus  (\Z_n)^m \oplus\cdots\oplus  (\Z_n)^m$, where $k$ denotes the number of   $ (\Z_{n})^m$'s repeated in the summation. For example, when gauge group $G_g=\Z_2$ and symmetry group $G_s= \Z_{2^n}$, the    classification is given by $\Z_2\times \Z_2\oplus \Z_1$, \textit{i.e.}, $2^2+1=5$ distinct patterns of symmetry fractionalization and thus distinct SET phases. If  $e$ particle  (\textit{i.e.}, the particle with unit $\Z_2$ gauge charge) carries integer symmetry charge, the classification of SFL on the unit $\Z_2$ gauge flux (i.e., elementary loop excitation denoted as $\Sigma$) is $\Z_2\times \Z_2\equiv (\Z_2)^2$,  (\textit{i.e.}, 4 different classes that form a cyclic group); if $e$ particle carries one-half symmetry charge, the classification result is $\Z_1$ (\textit{i.e.}, only the trivial one). For $G_g=\Z_2\times \Z_2$, there are four different twisted terms labeled by $(0,0)$, $(2,0)$, $(0,2)$, and $(2,2)$. The corresponding topological orders are  denoted by $\mathsf{TO}_1$, $\mathsf{TO}_2$, $\mathsf{TO}_3$, and $\mathsf{TO}_4$. But the difference between $\mathsf{TO}_2$ and $\mathsf{TO}_3$ is trivial as they are changed to each other upon switching two $\Z_2$ gauge subgroups. For this reason, in this table, we only preserve $(2,0)$. Each SET phase is characterized by a set of physical observables, which  are shown in Table~\ref{table_z4z4sfl},~\ref{table_z2z2_z2_sfl}, and~\ref{table_z2_z2z2_sfl}.}
    \begin{tabular*}{\textwidth}{@{\extracolsep{\fill}}cccc}

   \hline
   
   \hline
   
    \hline 
      \textbf{Gauge group $G_g$}   & \textbf{Twist} &\textbf{Symmetry group $G_s$} & \textbf{Classification} 
   \\ \hline

  \hline  
    $\Z_2$ & --& $\Z_{2n+1}$ & \small{${\Z_1}$}\\ 
  $\Z_2$  & -- &$\Z_{2^n}$ & \small{$ { (\Z_2)^2}  \oplus  \Z_1$} \\ 
 $\Z_2$  & -- &$\Z_{2^n} \times \Z_{2^n}$ &\small{
$ (\Z_2)^6\oplus  \Z_{2} \oplus \Z_{2} \oplus \Z_{2} $  }\\ 
 $\Z_4$  &-- & $\Z_{4^n}$ & \small{$  (\Z_4)^2\oplus  \Z_2\oplus \Z_1 \oplus \Z_1  $} \\ 
      $\Z_2\times \Z_2$ &  (0, 0) & $\Z_{2^n}$ & \small{\tabincell{c}{ $ (\Z_2)^6\oplus  (\Z_2)^2 \oplus   (\Z_2)^2 \oplus  (\Z_2)^2$}}\\ 
 $\Z_2\times \Z_2$ &  (2, 0) & $\Z_{2^n}$ & \small{$ (\Z_2)^6$} \\ 
      $\Z_2\times \Z_2$ &  (2, 2) & $\Z_{2^n}$ & \small{$ (\Z_2)^6$} \\

  \hline
  
  \hline
  
  \hline

    \end{tabular*}
   
    \label{znset}
\end{table*} 

 \begin{table*}[t] 
    \centering\caption{\textbf{$\Z_4$ topological order with  $\Z_4$ symmetry.} In this case, $G_g=G_s=\Z_4$. In the column of SFP, $e$ denotes the particle excitation that carries unit gauge charge of $G_g$ gauge group. The symbols ``$e0$'', ``$eC$'', ``$eQ_{\pm}$'' mean  that $e$ carries integer, half-integer, $\pm \frac{1}{4}$ symmetry charge of $\Z_4$ symmetry group, respectively. For each given SFP, we can characterize SFL via a set of   M3L statistical phases (denoted by $\theta_{...}$) that are uniquely determined by linearly independent integers with proper periods (\textit{i.e.}, $p_1,p_2,\cdots$).    
  $\Sigma$ denotes the elementary loop excitation that carries unit $G_g$ gauge flux. $\sigma$ denotes the unit quantized  $G_s$ symmetry flux. $\theta_{ab;c}$ denotes an M3L (Mixed 3-Loop)statistical phase of the M3L  braiding process: loop-like object $a$ (either loop excitation or symmetry flux) is braided around $b$ both of which are   linked to $c$.   If $a$ and $b$ are identical, the braiding process, e.g., $\theta_{a;c}$ is the exchanging phase upon exchanging two identical loop-like objects denoted as $a$, both of which are linked to $c$.    The periods indicate the minimal trivial values of the   M3L statistical phases. Two  values of a given M3L statistical phase are said to be equivalent if they differ by a period.       
 The classification   is given by $(\Z_4)^2\oplus \Z_2\oplus 2\Z_1$ which is also summarized in Table~\ref{znset}. The exposition of technical details of this table is  present in Sec.~\ref{SET:ZN}.  
 }
    \begin{tabular}{ccccccccc}
 \hline
    
    \hline
    
    \hline
  \multicolumn{1}{c}{ \multirow{2}{*}{\quad \quad \textbf{\quad SFP \quad }\quad \quad}} &\multicolumn{7}{c}{ \textbf{Characterizing SFL by  M3L   statistical phases}} & \multicolumn{1}{c}{ \multirow{2}{*}{
 \quad \quad \textbf{Classification}\quad \quad} }\\
  &\multicolumn{1}{c}{} &  \multicolumn{1}{c}{\quad $\theta_{\Sigma,\sigma;\sigma}$ \quad \quad} &  \multicolumn{1}{c}{\quad$\theta_{\sigma,\sigma;\Sigma}$ \quad\quad } & \multicolumn{1}{c}{\quad $\theta_{\sigma;\Sigma}$ \quad\quad}  & \multicolumn{1}{c}{ \quad $\theta_{\sigma,\Sigma;\Sigma}$ \quad\quad }  &  \multicolumn{1}{c}{\quad $\theta_{\Sigma,\Sigma;\sigma}$ \quad\quad }  & \multicolumn{1}{c}{ \quad $\theta_{\Sigma;\sigma}$ \quad\quad }  &\multicolumn{1}{c}{} \\
   \hline
   
   \hline
{ \multirow{2}{*}{ $e0$}} & \multicolumn{1}{c}{ value: }& \multicolumn{1}{c}{${3\pi p_1}/{8}$} &${\pi p_1}/{4}$ & ${\pi p_1}/{8}$& ${3\pi p_2}/{8}$ & ${\pi p_2}/{4}$&${\pi p_2}/{8}$&  \multicolumn{1}{c}{\multirow{2}{*}{ $p_1 \in \Z_4$, $p_2 \in \Z_4$}}\\
  & \multicolumn{1}{c}{periods:} & \multicolumn{1}{c}{${\pi}/{2}$} &${\pi}/{2}$ &  ${\pi}/{2} $& ${\pi}/{2}$ &${\pi}/{2}$  &${\pi}/{2}$ & \multicolumn{1}{c}{}\\
   \\

    \multirow{2}{*}{ $eC$} &\multicolumn{1}{c}{value: } & \multicolumn{1}{c}{${\pi p_3}/{8}$} &$0$ & ${\pi p_3}/{8}$& $0$& $0$&${\pi p_3}/{4}$&  \multicolumn{1}{c}{\multirow{2}{*}{$p_3\in \Z_2$}}\\
&\multicolumn{1}{c}{periods:}&  \multicolumn{1}{c}{${\pi}/{4}$} &${\pi}/{4}$ &${\pi}/{4}$ &${\pi}/{4}$ &${\pi}/{2}$ & ${\pi}/{2} $ &  \multicolumn{1}{c}{\multirow{2}{*}{}}\\
\\

 \multirow{2}{*}{ $eQ_{+}$ or $eQ_{-}$ } &\multicolumn{1}{c}{value: } & {0} &0& 0& 0& 0&0 &  \multicolumn{1}{c}{\multirow{2}{*}{$\Z_1$}}\\
  &\multicolumn{1}{c}{periods:} & \multicolumn{1}{c}{${\pi}/{8}$} &${\pi}/{8}$ &${\pi}/{8}$& ${\pi}/{8}$ &${\pi}/{2}$  &${\pi}/{2}$ &  \multicolumn{1}{c}{}\\
     
   \hline
    
    \hline
    
    \hline
\end{tabular}
  \label{table_z4z4sfl}
\end{table*}

\begin{table*}[t] 
    \centering
    \caption{\textbf{$\Z_2\times \Z_2$ topological orders with  $\Z_2$ symmetry.}  In this case, $G_g=\Z_{N_1}\times\Z_{N_2}$ and $G_s=\Z_2$ with $N_1=N_2=2$.  In the column of SFP, $e_i$ denotes the particle excitation that carries unit gauge charge of $\Z_{N_i}$ gauge subgroup. The symbols ``$e_10$'' and ``$e_1C$'' mean that $e_1$ carries integer and half-integer symmetry charge of $\Z_2$ symmetry group, respectively. For each given SFP, we can characterize SFL via a set of   M3L  statistical phases (denoted by $\theta_{...}$) that are uniquely determined by linearly independent $\Z_2$ integers (\textit{i.e.}, $p_1,p_2,\cdots,p_{12}$).    
  $\Sigma^{i}$  denotes an elementary loop excitation that  carries   unit gauge flux of $\Z_{N_i}$ gauge subgroup.  We   note that there are four   topological orders with the same gauge group $\Z_2\times\Z_2$ (two of them are actually identical; see the caption of Table~\ref{znset}). The last column specifies  topological orders that can, in an anomaly-free way, realize the SFP and SFL in the same row.   We also note that for the first row, i.e., $e_10e_20$, the $(\Z_2)^6$ classification for SFL can be realized by  all $\Z_2\times \Z_2$ topological orders while the other three rows can only be realized by  $\mathsf{TO}_1$, \textit{i.e.}, untwisted topological order.  Therefore, as shown in Table~\ref{znset}, the classification for untwisted $\mathsf{TO}_1$ is $(\Z_2)^6\oplus 3(\Z_2)^2$ while for twisted one, it is $(\Z_2)^6$ only. The exposition of technical details of this table is  present in Sec.~\ref{SET:ZNZN}. For saving space, this table does not include  $\theta_{\Sigma^1,\Sigma^1;\sigma}, \theta_{\Sigma^2,\Sigma^2;\sigma}, \theta_{\sigma,\sigma;\Sigma^1}, \theta_{\sigma,\sigma;\Sigma^2}$, all of which are vanishing up to proper periods.  
 }

    \resizebox{\textwidth}{!}{%
    \begin{tabular}{ccccccccccccccccc}
    \hline
    
    \hline
    
    \hline
   \multirow{2}{*}{  \textbf{ SFP  }} & \multicolumn{12}{c}{ \textbf{Characterizing SFL by   M3L   statistical phases}} & \multicolumn{1}{c}{ \multirow{2}{*}{ \textbf{Classification}}}& \multicolumn{1}{c}{ \multirow{2}{*}{ $\mathsf{TO}$ }}\\
   & \multicolumn{1}{c}{} &  \multicolumn{1}{c}{  $\theta_{\Sigma^1,\sigma;\sigma}$  }    & \multicolumn{1}{c}{  $\theta_{\sigma;\Sigma^1}$ }  &  \multicolumn{1}{c}{  $\theta_{\sigma,\Sigma^1;\Sigma^1}$  }  &  \multicolumn{1}{c}{ $\theta_{\Sigma^1;\sigma}$  }  &     \multicolumn{1}{c}{  $\theta_{\Sigma^2,\sigma;\sigma}$  }  & \multicolumn{1}{c}{  $\theta_{\sigma;\Sigma^2}$ }  &  \multicolumn{1}{c}{  $\theta_{\sigma,\Sigma^2;\Sigma^2}$  } &    \multicolumn{1}{c}{  $\theta_{\Sigma^2;\sigma}$  } &   \multicolumn{1}{c}{  $\theta_{\sigma,\Sigma^2;\Sigma^1}$  } &   \multicolumn{1}{c}{  $\theta_{\Sigma^2,\Sigma^1;\sigma}$  }  &   \multicolumn{1}{c}{  $\theta_{\sigma,\Sigma^1;\Sigma^2}$  } &  \multicolumn{1}{c}{} &\multicolumn{1}{c}{} \\
      \hline
      
      \hline
     \multirow{2}{*}{  $e_10e_20$} & \multicolumn{1}{c}{value: }& \multicolumn{1}{c}{$\frac{p_{ 1} \pi}{2}$}   &{ $\frac{p_1\pi}{2}$}&{$\frac{p_2 \pi}{2}$}&{$\frac{p_2 \pi}{2}$}&{ $\frac{p_3\pi}{2}$} &{ $\frac{p_3\pi}{2}$}&{ $\frac{p_4\pi}{2}$} &{ $\frac{p_4\pi}{2}$}&{ $\frac{p_5\pi}{2}$}&{ $\frac{p_6\pi}{2}$}&\multicolumn{1}{c}{ { $\frac{(p_5+p_6)\pi}{2}$}}&\multirow{2}{*}{$(p_1,\cdots,p_6)\in (\Z_2)^6$}  &\multicolumn{1}{c}{\multirow{2}{*}{all $\mathsf{TO}_i$}   }\\
   &\multicolumn{1}{c}{ periods:}& \multicolumn{1}{c}{$\pi$} & {$\pi$}& {$\pi$} & {$\pi$}& {$\pi$} & {$\pi$}& {$\pi$} & {$\pi$}& {$\pi$}& {$\pi$}& \multicolumn{1}{c}{$\pi$}& &\multicolumn{1}{c}{}\\
\\
   \multirow{2}{*}{  $e_1Ce_20$} & \multicolumn{1}{c}{value:   }& \multicolumn{1}{c}{ 0} 
    &{ 0}&{ 0} &{ 0}&{$\frac{p_7 \pi}{4}$}  & {$\frac{p_{7} \pi}{4}$} &{ 0} &{ $\frac{p_8\pi}{2}$}&{ 0}&{ $\frac{p_7\pi}{2}$}&\multicolumn{1}{c}{ 0}&\multirow{2}{*}{$(p_7,p_8)\in(\Z_2)^2$} &\multicolumn{1}{c}{\multirow{2}{*}{$\mathsf{TO}_1$}}\\
   &\multicolumn{1}{c}{ periods:} & \multicolumn{1}{c}{$\frac{\pi}{2}$}  & {$\frac{\pi}{2}$}& {$\frac{\pi}{2}$} & {${\pi}$}& {$\frac{\pi}{2}$} & {$\frac{\pi}{2}$}& {$\frac{\pi}{2}$} & {${\pi}$}& {$\frac{\pi}{2}$}& {${\pi}$}& \multicolumn{1}{c}{$\frac{\pi}{2}$}& &\multicolumn{1}{c}{}\\
\\
     \multirow{2}{*}{  $e_10e_2C$} &  \multicolumn{1}{c}{ value: }& \multicolumn{1}{c}{ $\frac{p_{ 9} \pi}{4}$}  &{$\frac{p_{ 9} \pi}{4}$}&{ 0} &$\frac{p_{10}\pi}{2}$&{0}  & {0} &{ 0} &{ 0}&{ 0}&{ $\frac{p_9\pi}{2}$}&\multicolumn{1}{c}{ 0}&\multirow{2}{*}{$(p_9,p_{10})\in (\Z_2)^2$} &\multicolumn{1}{c}{\multirow{2}{*}{$\mathsf{TO}_1$}}\\
 &\multicolumn{1}{c}{ periods:} & \multicolumn{1}{c}{$\frac{\pi}{2}$}  & {$\frac{\pi}{2}$}& {$\frac{\pi}{2}$}  & {${\pi}$}& {$\frac{\pi}{2}$} & {$\frac{\pi}{2}$}& {$\frac{\pi}{2}$} & {${\pi}$}& {$\frac{\pi}{2}$}& {${\pi}$}& \multicolumn{1}{c}{$\frac{\pi}{2}$}& &\multicolumn{1}{c}{}\\
\\
     \multirow{2}{*}{  $e_1Ce_2C$} &   \multicolumn{1}{c}{  value: }& \multicolumn{1}{c}{$\frac{p_{11} \pi}{4}$}  &\multicolumn{1}{c}{$\frac{p_{11} \pi}{4}$}&{ 0} &{ $\frac{p_{12} \pi}{2}$}&{$\frac{p_{ 11} \pi}{4}$}  & {{$\frac{p_{11} \pi}{4}$}} &{ 0} &{ $\frac{(p_{11}+p_{12}) \pi}{2}$}&{ $0$}&{ $\frac{p_{11} \pi}{2}$}&\multicolumn{1}{c}{ $0$}&\multirow{2}{*}{$(p_{11},p_{12})\in (\Z_2)^2$} &\multicolumn{1}{c}{\multirow{2}{*}{$\mathsf{TO}_1$}}\\
 & \multicolumn{1}{c}{ periods:} & \multicolumn{1}{c}{$\frac{\pi}{2}$} & {$\frac{\pi}{2}$}& {$\frac{\pi}{2}$} & {${\pi}$}& {$\frac{\pi}{2}$} & {$\frac{\pi}{2}$}& {$\frac{\pi}{2}$} & {${\pi}$}& {$\frac{\pi}{2}$}& {${\pi}$}& \multicolumn{1}{c}{$\frac{\pi}{2}$} & &\multicolumn{1}{c}{}\\
\hline

\hline

\hline
\end{tabular}
}
\label{table_z2z2_z2_sfl}
\end{table*}
 
\begin{table*}[] 
    \centering
    \caption{\textbf{$\Z_2$ topological order with  $\Z_2\times \Z_2$ symmetry.}  In this case, $G_g=\Z_2$ and $G_s=\Z_{K_1}\times\Z_{K_2}$ with $K_1=K_2=2$.   The symbols ``$e0$'' and ``$eC$'' mean that $e$ carries integer and half-integer symmetry charge of one of the two $\Z_2$ symmetry groups, respectively. $\sigma^i$ ($i=1,2$) denotes a unit symmetry flux of $\Z_{K_i}$ symmetry subgroup. $\sigma^{12}$ is a composite of $\sigma^1$ and $\sigma^2$ as a result of fusion process. For each given SFP, we can characterize SFL via a set of   M3L statistical phases (denoted by $\theta_{...}$) that are uniquely determined by linearly independent $\Z_2$ integers (\textit{i.e.}, $p_1,p_2,\cdots,p_9$).  
    The exposition of technical details of this table is  present in Sec.~\ref{sec_example_znzkzk}.  
 }

    \resizebox{\textwidth}{!}{%
    \begin{tabular}{cccccccccccccccccc}
    \hline
    
    \hline
    
    \hline
   \multirow{2}{*}{ \textbf{ SFP  } } & \multicolumn{14}{c}{ \textbf{Characterizing SFL by    M3L   statistical phases}} & \multicolumn{1}{c}{ \multirow{2}{*}{ \textbf{Classification}}}\\
   & \multicolumn{1}{c}{} &  \multicolumn{1}{c}{  $\theta_{\sigma^1,\Sigma;\Sigma}$  }    & \multicolumn{1}{c}{  $\theta_{\Sigma;\sigma^1}$ }  &  \multicolumn{1}{c}{  $\theta_{\Sigma,\sigma^1;\sigma^1}$  }  &  \multicolumn{1}{c}{ $\theta_{\sigma^1;\Sigma}$  }  &     \multicolumn{1}{c}{  $\theta_{\sigma^2,\Sigma;\Sigma}$  }  & \multicolumn{1}{c}{  $\theta_{\Sigma;\sigma^2}$ }  &  \multicolumn{1}{c}{  $\theta_{\Sigma,\sigma^2;\sigma^2}$  } &    \multicolumn{1}{c}{  $\theta_{\sigma^2;\Sigma}$  } &   \multicolumn{1}{c}{  $\theta_{\Sigma,\sigma^2;\sigma^1}$  } &   \multicolumn{1}{c}{  $\theta_{\sigma^2,\sigma^1;\Sigma}$  }  &   \multicolumn{1}{c}{  $\theta_{\Sigma,\sigma^1;\sigma^2}$  } & \multicolumn{1}{c}{  $\theta_{\Sigma,\sigma^{12};\sigma^1}$  }&\multicolumn{1}{c}{  $\theta_{\Sigma,\sigma^{12};\sigma^2}$  }  &  \multicolumn{1}{c}{} \\
      \hline
      
      \hline
     \multirow{2}{*}{  $e00$} & \multicolumn{1}{c}{value: }& \multicolumn{1}{c}{$\frac{p_{ 1} \pi}{2}$}   &{ $\frac{p_1\pi}{2}$}&{$\frac{p_2 \pi}{2}$}&{$\frac{p_2 \pi}{2}$}&{ $\frac{p_3\pi}{2}$} &{ $\frac{p_3\pi}{2}$}&{ $\frac{p_4\pi}{2}$} &{ $\frac{p_4\pi}{2}$}&{ $\frac{p_5\pi}{2}$}&{ $\frac{(p_5+p_6)\pi}{2}$}&\multicolumn{1}{c}{ { $\frac{p_6\pi}{2}$}}&  \multicolumn{1}{c}{ $\frac{(p_2+p_5)\pi}{2}$}& \multicolumn{1}{c}{ $\frac{(p_2+p_6)\pi}{2}$}   &\multirow{2}{*}{$(p_1,\cdots,p_6)\in (\Z_2)^6$}  \\
   &\multicolumn{1}{c}{ periods:}& \multicolumn{1}{c}{$\pi$} & {$\pi$}& {$\pi$} & {$\pi$}& {$\pi$} & {$\pi$}& {$\pi$} & {$\pi$}& {$\pi$}& {$\pi$}& \multicolumn{1}{c}{$\pi$}& \multicolumn{1}{c}{$\pi$}& \multicolumn{1}{c}{$\pi$} &\\
\\
   \multirow{2}{*}{  $eC0$} & \multicolumn{1}{c}{ value:  }& \multicolumn{1}{c}{ 0} 
    &{ 0}&{ 0} &{ 0}&{0}  & {0} &{ 0} &{ 0}&{ $\frac{p_7 \pi}{2}$}&{ 0}&\multicolumn{1}{c}{ 0}&\multicolumn{1}{c}{ 0}&\multicolumn{1}{c}{ 0}&\multirow{2}{*}{$p_7\in \Z_2$} \\
   &\multicolumn{1}{c}{ periods:} & \multicolumn{1}{c}{$\frac{\pi}{2}$}  & {${\pi}$}& {$\frac{\pi}{2}$} & {$\frac{\pi}{2}$}& {${\pi}$} & {${\pi}$}& {${\pi}$} & {${\pi}$}& {${\pi}$}& {$\frac{\pi}{2}$}& \multicolumn{1}{c}{$\frac{\pi}{2}$}& \multicolumn{1}{c}{$\frac{\pi}{2}$}&\multicolumn{1}{c}{$\frac{\pi}{2}$} &\multicolumn{1}{c}{}&\\
\\
     \multirow{2}{*}{  $e0C$} & \multicolumn{1}{c}{ value:  }& \multicolumn{1}{c}{ 0} 
    &{ 0}&{ 0} &{ 0}&{0}  & {0} &{ 0} &{ 0}&{ 0}&{ 0}&\multicolumn{1}{c}{ $\frac{p_8 \pi}{2}$}&\multicolumn{1}{c}{ 0} & \multicolumn{1}{c}{ 0} &\multirow{2}{*}{$p_8\in \Z_2$} \\
   &\multicolumn{1}{c}{ periods:} & \multicolumn{1}{c}{${\pi}$}  & {$\pi$}& {${\pi}$} & {${\pi}$}& {$\frac{\pi}{2}$} & {${\pi}$}& {$\frac{\pi}{2}$} & {$\frac{\pi}{2}$}& {$\frac{\pi}{2}$}& {$\frac{\pi}{2}$}& \multicolumn{1}{c}{${\pi}$}& \multicolumn{1}{c}{$\frac{\pi}{2}$} &\multicolumn{1}{c}{$\frac{\pi}{2}$} &\multicolumn{1}{c}{}&\\
\\
       \multirow{2}{*}{  $eCC$} & \multicolumn{1}{c}{  value: }& \multicolumn{1}{c}{ 0} 
    &{ 0}&{ 0} &{ 0}&{0}  & {0} &{ 0} &{ 0}&{ 0}&{ 0}&\multicolumn{1}{c}{ 0}&\multicolumn{1}{c}{ $\frac{p_9 \pi}{2}$}&\multicolumn{1}{c}{ $\frac{p_9 \pi}{2}$} &\multirow{2}{*}{$p_9\in \Z_2$} \\
   &\multicolumn{1}{c}{ periods:} & \multicolumn{1}{c}{$\frac{\pi}{2}$}  & {${\pi}$}& {$\frac{\pi}{2}$} & {$\frac{\pi}{2}$}& {$\frac{\pi}{2}$} & {${\pi}$}& {$\frac{\pi}{2}$} & {$\frac{\pi}{2}$}& {$\frac{\pi}{2}$}& {$\frac{\pi}{2}$}& \multicolumn{1}{c}{$\frac{\pi}{2}$}& \multicolumn{1}{c}{${\pi}$}&\multicolumn{1}{c}{${\pi}$} &\multicolumn{1}{c}{}&\\
\hline

\hline

\hline
\end{tabular}
}
\label{table_z2_z2z2_sfl}
\end{table*}


  \subsection{Synopsis and summary of key findings}
\label{section_key_findings_intro}
In this paper, we consider systems with discrete global symmetry and Abelian topological order described by topological quantum field theories (TQFTs) \cite{sarma_08_TQC}, where both the symmetry group $G_s$ and the gauge group $G_g$ are Abelian.  We develop a step-by-step way to compute SFP (symmetry fractionalization on particles) and  SFL (symmetry fractionalization on loops), which eventually renders  a classification of SF patterns. The classification is complete within the present field-theoretical framework.

      The classification is illustrated in Table~\ref{znset} via several typical examples. Each 3D topological order      is labeled by two input data: a gauge group $G_g$ and   a  cocycle $\omega\in \mathcal{H}^4(G_g, U(1))$\footnote{Some technical description of such bosonic 3D topological order can be found in the literature see, \textit{e.g.}, Refs.~\cite{Lan_wen_3DTO_AB_prx,PhysRevB.97.085147}. Appendix~\ref{sec_prel_1}  also provides  useful information. But if taking Borromean rings braiding into account, there are more types of 3D topological order \cite{PhysRevResearch.3.023132,ye17b} which will not be considered in this work.}. Each cocycle, in our field-theoretical formalism, is uniquely represented by a twisted topological term (i.e., twist in Table~\ref{znset}).  
   It should be noted that the trivial element in the classification means `\textit{SET with trivial SF} or  \textit{SF-trivial  SET}'.           The SF-trivial SET for a $G_g$ topological order with a twist $\omega$ and a symmetry $G_s$ can be viewed as stacking a   $G_s$ SPT on top of a $G_g$ topological order with a twist $\omega$.  In contrast, an SET with \textit{trivial symmetry enrichment} is a concept with much stronger constraints, where   the following aspects are required to be simultaneously trivial:  permutation among excitations,   SFP, SFL and stacked SPTs. Therefore,   the classification we obtain in   Table~\ref{znset} is the classification of patterns of symmetry fractionalization, which is an important building block of the final SET classification.

Patterns of symmetry fractionalization   in an SET phase are composed by the patterns of SFP and SFL. These data, as classified in Table~\ref{znset},  are physically explained in   Tables~\ref{table_z4z4sfl}, \ref{table_z2z2_z2_sfl} and \ref{table_z2_z2z2_sfl} for various representative combinations of $G_s$ and $G_g$.   To characterize SFP, one can braid particle excitations around externally inserted symmetry fluxes and collect all inequivalent braiding phases. The resulting braiding phases in turn reflect how symmetry charge is fractionalized on particle excitations with nontrivial gauge charges. Among all types of SFP, we  also carefully  exclude anomalous types. The patterns of anomaly-free SFP are shown in  the first columns in Tables~\ref{table_z4z4sfl}, \ref{table_z2z2_z2_sfl} and \ref{table_z2_z2z2_sfl}.  To characterize SFL that is much more complex than SFP, we compute  a set of physical observables, which we dub  `mixed three loop braiding' (M3L) statistical phases. We study the quantization rules and periods of all M3L statistical phases, which are utilized to characterize and classify SFL when SFP is specified.  The patterns of SFL for any given SFP are listed in the main body of  Tables~\ref{table_z4z4sfl}, \ref{table_z2z2_z2_sfl} and \ref{table_z2_z2z2_sfl}, which are physical observables of symmetry fractionalization.

  In summary,  the key findings of the present work can be summarized into the following aspects:  
      \begin{enumerate}[label=\textbf{(\roman*)}]
 \item  A systematic field-theoretical framework   is finally established for SF patterns of {SET phases with} Abelian gauge group{s} and Abelian symmetry group{s}.   Several technical difficulties  are solved.   Especially, among   infinite number of Lagrangians, we figure out  equivalence relations which give rise to a finite number of equivalence classes;    
 
 \item   The characterization of SFP and especially SFL is provided. Physical observables are constructed to   unambiguously label SETs  with distinct SF patterns; 
 
 \item  We find many exotic SET phases with nontrivial SFP and SFL. For example, we find an SET phase with $\Z_4$ symmetric $\Z_4$ topological order where  both SFP and SFL are nontrivial, i.e., gauge charge  carries one half $\Z_4$ symmetry charge and nontrivial M3L statistics $\theta_{\sigma;\Sigma}=\pi/8$, as shown in Table~\ref{table_z4z4sfl}.  We also  find many exotic SET phases with the $\Z_2$ symmetric twisted $\Z_2\times \Z_2$ topological orders, in which  there are $(\Z_2)^6$ classes with different types of SFL even though  their particle excitations carry integer charges, as shown  in Table~\ref{table_z2z2_z2_sfl}.    For  $\Z_2\times \Z_2$ symmetric $\Z_2$ topological order, there is an exotic  SET  phase where gauge charge carries nontrivial SFP (either $eC0, e0C$ or $eCC$) and the SFL are characterized by M3L statistics involving both two symmetry flux defects and flux loop excitation, as shown in Table~\ref{table_z2_z2z2_sfl}; 
 
 \item    Within the present field-theoretical framework, a complete classification of SF is naturally obtained, some of which are collected in Table~\ref{znset};   
 
  \item   Inspired by the present field-theoretical framework, the underlying algebraic structure of SF patterns is   briefly discussed, which leads to several stimulating questions.

 \end{enumerate}

\subsection{Technical aspects and outline}

 Below we  elaborate the key technical steps of our calculations. The low-energy effective field theory of topological order is TQFT \cite{sarma_08_TQC}. Therefore, what we have  explored is essentially the interplay of TQFT and global symmetry. More specifically, if a topological order is described by TQFT of some gauge group, the existence of anomaly-free SETs necessarily enforces stringent   conditions such that gauge transformations   and   symmetry transformations are mutually compatible.  In TQFT, to detect symmetry fractionalization, one can externally insert symmetry fluxes such that braiding statistical phases among topological excitations (both particles and loops) and symmetry fluxes can be used to describe the interplay of symmetry and topological order. Then, by computing   all braiding data via TQFT, one can characterize and classify SF patterns of SET phases.

   In this paper,  we first exhaust all types of SFP via field-theoretical calculation on braiding of particle and symmetry-flux, which can be done as long as gauge group $G_g$ and symmetry group $G_s$ are given. This calculation is conducted in Sec.~\ref{section_main_preli}. Several technical difficulties, e.g., treatment on infinite number of possible ``charge matrices'', are successfully solved. We note that, some of SFP types found in this section are essentially anomalous when $G_s$-topological order is twisted.

  Then, for a given  type of SFP, we exhaust all types of SFL via field-theoretical calculation on M3L statistical phases, as shown in Sec.~\ref{section_main_sfloop}.   M3L statistical phases are closely related to the so-called three-loop braiding statistical phases where ``three loops'' are really three loop excitations \cite{wang_levin1}. But, in an M3L process, ``three loops'' means a  mixture of loop excitations and external symmetry fluxes. The latter are geometrically loop-like too. Even though it might be conceptually  straightforward  to shift from three-loop braiding to the M3L braiding,  it is very challenging to formulate them in a precise and significant manner. For example, how many   M3L statistical phases should we take into account for a certain symmetry, given a 3D topological order? What are the   equivalence  relations among M3L statistical phases affected by both gauge group and symmetry group? What are the quantization rules, linear dependence, and periods of these M3L statistical phases?

   To address these questions, we realize that these M3L statistical phases are closely related to a set of invariants, which we dub \textit{M3L invariants} as shown in Table~\ref{table_three_M3L_example_list} of Sec.~\ref{section_main_sfloop}. The M3L invariants are expressed in the spacetime manifolds, from which we can determine  the M3L statistical phases, similar to the fact that the Hopf link invariants in $(2+1)$D spacetime can determine mutual statistics of anyons in 2D topological orders. All these M3L invariants can be derived from TQFTs with certain twisted topological terms, which provides  a platform for systematic study of    M3L invariants and   M3L statistical phases. We expect the present framework we adopt is universal enough to capture the most general   M3L statistical phases for the cases we consider in this paper. We also believe this framework can be generalized to study more complicated cases.

The remainder of this paper is organized as follows:
 
       \begin{enumerate}[label=\textbf{(\roman*)}]

   \item  In Sec.~\ref{section_main_preli}, we discuss symmetry fractionalization on particles (SFP).     In Sec.~\ref{sec_adding_symmetry}, we review TQFT description of 3D topological order with symmetry, where several relevant terminologies are introduced. Appendix~\ref{sec_prel_1} is relevant to this subsection.  In Sec.~\ref{SF}, we study SFP from the field-theoretical viewpoint, where we especially discuss the reduction of charge matrices.  Appendix~\ref{subsection_syfra_particle_general} is relevant to this subsection.  And, several frequently used abbreviations and concepts are collected in Appendix~\ref{appendix_abbr_list}.

  \item  In Sec.~\ref{section_main_sfloop}, we systematically discuss symmetry fractionalization on loops (SFL). In Sec.~\ref{sec_twisted_quantized_period_coe}, we study the quantization and periods of coefficients of twisted terms, which will be useful in the discussion of SFL.  Appendix~\ref{appendix_SEG_quantization} is relevant to this subsection.    In Sec.~\ref{section_mmt_general_discussion}, we introduce M3L braiding, M3L invariants, and M3L statistical phases. Appendix~\ref{footnote_multiloopinv} is relevant to this subsection.  In   Sec.~\ref{mmt_general_discussion_more_General}, we present the general steps for computing characterization and classification   of SFL. In  Sec.~\ref{obstruction}, anomalous SETs are discussed. In  Sec.~\ref{classification_principle}, we discuss the classification from the algebraic viewpoint. Appendix~\ref{section_gauging_loops} is relevant to this subsection.

 \item   Starting from Sec.~\ref{SET:ZN}, we present three typical   examples to demonstrate how to classify and characterize SET phases. In Sec.~\ref{SET:ZN}, we study symmetry fractionalization in untwisted topological orders, by taking $\Z_4$ topological order enriched by $\Z_4$  symmetry as an example. Appendix~\ref{sec_app_znzk} is relevant to this section.  
In Sec.~\ref{SET:ZNZN}, we study symmetry fractionalization in  topological orders that can be either untwisted or twisted. We take $\Z_{2} \times \Z_{2}$  topological orders enriched by $\Z_2$ symmetry as an example. Appendices~\ref{appendix_substituting} and \ref{sec_app_znznzk}  are relevant to this section. Sec.~\ref{sec_example_znzkzk} is devoted to topological orders with multi $\Z_{K}$ symmetry subgroups where interplay of two symmetry subgroups play a critical role. We take $\Z_2$ topological order enriched by $\Z_2\times \Z_2$ symmetry as an example. Appendix~\ref{sec_app_znzkzk} is relevant to this section.

\item    We conclude the paper in Sec.~\ref{section_summary_outlook} with several future directions.   
       
 \end{enumerate}
 
  \section{Symmetry fractionalization on particles}
 \label{section_main_preli}
 
 Here we  warm up with some preliminary steps, including how to implement the global symmetry, and how to classify  SFP.  In order to detect SF, we turn on the external probe fields  and minimally couple them to the conserved currents written in terms of gauge fields. The coupling coefficients are  integer-valued and form   a so-called \text{charge matrix}.   In this paper, we only consider the Abelian  SFP:  particles   (gauge charges)  carry  one-dimensional irreducible projective representations of  the symmetry group. Higher-dimensional projective representations carried by gauge charges are beyond the scope of the present work and will be left for future study.

\subsection{Topological actions and implementation of global symmetry}
\label{sec_adding_symmetry}

The 3D     topological orders we consider here are bosonic and Abelian,  which   can be described by topological   $BF$  gauge theories \cite{PhysRevLett.114.031601,wang_levin1,string5,string10,YeGu2015, ye17b, 2016arXiv161008645Y,3loop_ryu, 2016arXiv161209298P,Tiwari:2016aa, string6, bti2, Kapustin2014, PhysRevB.95.035131,PhysRevResearch.3.023132, PhysRevB.97.085147,QRW2019prb,PhysRevB.99.205120}  whose action take the form of \footnote{We neglect the wedge product symbol $\wedge$ for simplifying formulas. It should also be noted that, if explicitly writing spacetime indices, there is an additional $\frac{1}{2}$ prefactor for each $b^ida^i$ since $b^i$ is of 2-form, which results in $\frac{N_i}{4\pi}$. }
\begin{align}
S=  \int \frac{N_i}{2\pi} b^i d a^i+\int \frac{q_{ijk}}{4\pi^2}  a^i a^j d a^k\, 
\label{eq:action_of_pure_gauge_3lb}
 \end{align}
 with gauge group $G_g=\prod_{i=1}^n \Z_{N_i}$ where repeated indices are summed implicitly. 
 The first term is a set of $BF$ terms that  describe   charge-loop braiding processes. The second term, which we denote  by $S_{int}^0$, can be regarded as interaction between gauge fields. $S_{int}^0$ is a set of  twisted terms that describe    three-loop braiding processes \cite{wang_levin1}.    One can also add some  trivial layers corresponding to level-1 gauge fields\footnote{ {Definition of ``level-$1$ gauge field''}. In this paper,   gauge fields are specially called ``level-$1$'' gauge fields if they appear in level-$1$   $BF$ terms. Each BF term is called ``a layer''. Those $BF$ terms with level-$1$ are said to be trivial layers.\label{footnote_level1}} (\textit{i.e.}, $N_i=1$ in action (\ref{eq:action_of_pure_gauge_3lb})) to the system, which    does not affect the topological order but may potentially induce important effect on  SFL. (One can refer to Appendix~\ref{sec_prel_1} for more detailed review on topological $BF$ gauge theories.) Then, $i$ can take $1,..., n+n'$ as we have taken into account adding $n'$ trivial-layers ($N_i=1$ for $i=n+1,\cdots,n+n'$). 

The  global symmetry group to be considered   is $G_s=\prod_{k=1}^m\Z_{K_k}$. Below we encode the information of global symmetry into the field theoretical formalism. In the action (\ref{eq:action_of_pure_gauge_3lb}),  we can define  a $1$-form    conserved current $J^i$ for each $i$ as $*J^i=\frac{1}{2\pi}db^i\,$,  where $*$ denotes the Hodge dual operation. Symmetry charges are then carried by those  conserved currents.
 As a result, each conserved current (labeled by $i$) minimally couples to the probe field $A^i$, where the probe field $A^i$  \cite{bti6} is used to probe global symmetry $Z_{K_i}$.    Generally, one    conserved current can carry different symmetry charges, therefore the general form of the minimal coupling term is
\begin{align} 
S_{\text{c}}\!= Q_{ij} \!\!\int \!\!J^j * A^{i}=\frac{Q_{ij}}{2\pi}\int A^{i} d b^j\,,\label{eq:coupling}
\end{align}
where the summation over repeated indices is implicit. Since $i$ can take $1, 2, \cdots, m$ and $j$ can take $1, 2, \cdots, n+n'$, $Q$ is a $m\times (n+n')$ integer matrix, called \textit{charge matrix}, and its element $Q_{ij}$ could be any integer now. 
{The probe field} $A^i$ takes values continuously in $U(1)$ group but  are in fact  ``higgsed''   to $\mathbb Z_{K_i}$ such that the symmetry flux piercing  any one-dimensional closed path is quantized as $\frac{2\pi p}{K_i}$ with $p\in \Z_{K_i}$, that is,
\begin{equation}
\frac{K_i}{2\pi}\int_{\mathcal{M}^1}A^{i}\in\Z\, \label{eq:holo}
\end{equation} 
 for $\Z_{K_i}$  symmetry subgroup where $\mathcal{M}^1$ denotes a closed worldline. A direct consequence of such a quantized symmetry flux (\ref{eq:holo}) is that 
the symmetry charge $Q_{ij}=K_i$ is topologically equivalent to $Q_{ij}=0$.

%

%

      \subsection{Equivalence relations and reduction of charge matrix} \label{SF}

Below, we aim to present how to characterize and classify    SFP.   We   use the fractionalized symmetry charge to characterize    SFP and further consider the equivalence relations among    fractionalized symmetry charges to find out the minimal symmetry charge that is trivial.   We assume that there is no permutation of topological sectors under symmetry transformations. We will first illustrate the basic idea   and then discuss the main results for the general cases with detailed derivations in Appendix~\ref{subsection_syfra_particle_general}.

To illustrate, we discuss a simple example:  $\Z_4$ gauge theory enriched by $\Z_{16}$ symmetry:
 \begin{equation}
 S_1= \frac{4}{2\pi} \int b  da + \frac{Q}{2\pi} \int A  db +  q_g \int a  *j\,, 
 \end{equation}
where $j^{\mu}$ is the excitation current formed by particles that carry $q_g$ unit gauge charge of the gauge field $a$. { If  $q_g=1$, these particle excitations carry unit gauge charge. In general, $q_g$ can take arbitrary integers. The gauge group we consider here is $\Z_4$, which means $q_g=4$ and $q_g=0$ are topologically equivalent. Therefore, $q_g$ can take four inequivalent integers: $0,1,2,3$.  Keeping the additivity property of Abelian gauge charge in mind, it is sufficient and necessary to study SFP  in which particle excitations\footnote{Such particle excitations are called ``elementary particle excitations'' and denoted by $e_i$ for $\Z_{N_i}$ gauge subgroup. Analogously, ``elementary loop excitations'' are loop excitations that carry unit gauge flux, denoted by   $\Sigma_i$ for $\Z_{N_i}$ gauge subgroup. }  carry unit gauge charge, \textit{i.e.}, $q_g=1$ (denoted by $e$).  }Upon integrating out $b$, we obtain   $a=-\frac{Q}{4} A$ under a proper gauge, then: 
\begin{equation}
S_\text{eff}=-\frac{Q}{4} \int A  *j\,.
\label{z4z4:fractionalization}
\end{equation}
From this effective action, we find that  the particle current is effectively  coupled to the external probe field with coefficient $-Q/4$, which indicates that  $e$ carries $-Q/4$ symmetry charge of symmetry group $\Z_4$.  
We note that here we do not have any requirement on the integral $Q$ \textit{i.e.}, it might take arbitrary integers, so it seems that there are infinite choices of   symmetry charges, either fractional or integral, carried by $e$. 

To figure out equivalence relations among them, 
 by noting that $Q$ is the coefficient of minimal coupling in $S_1$, $Q$ is equivalent to $Q+K$, \textit{i.e.},
 \begin{align}
 Q\sim Q+K
 \label{eqn:symmetry_charge_equivalence1}
 \end{align}
  with $K=16$ for the symmetry group $\Z_{16}$, as discussed at the end of Sec.~\ref{sec_adding_symmetry}. 
 On the other hand, there may be other equivalence relations among symmetry charges. One may perform the integral-by-part and obtain  ``$\frac{1}{2\pi} \int  Q b dA$'', which indicates that the external $\Z_4$ symmetry flux is also charged in the gauge group $\Z_{4}$. Therefore, the coefficient $Q$ should be equivalent to $Q+N$, \textit{i.e.},
 \begin{align}
 Q\sim Q+N
 \label{eqn:symmetry_charge_equivalence2}
 \end{align}with $N=4$ for SFP.  
 Intuitively, we can also understand this equivalence relation of $Q$ by considering the attachment of a trivial particle (\textit{i.e.}, particles with gauge charge $0\text{ mod }4$ in $\Z_4$ gauge group) to an elementary particle excitation $e$ that carries unit gauge charge.   This attachment process  does not alter topological properties, \textit{i.e.}, braiding data of $e$ particle, so that after attachment, the composite of $e$ and   trivial particles is topologically   equivalent to $e$. Nevertheless, the attachment of a trivial particle may shift the symmetry charge $Q/N$ carried by $e$ to $Q/N+1$ as trivial particles carry integer charge. As a result, we obtain the equivalence relation (\ref{eqn:symmetry_charge_equivalence2}).

By taking the two equivalence relations (\ref{eqn:symmetry_charge_equivalence1}) and (\ref{eqn:symmetry_charge_equivalence2}) into account, we have\footnote{$\gcd$ stands for greatest common divisor.} $$Q\sim Q+\gcd(K,N).$$ Thus, $Q/4$ is equivalent to $Q/4+1$, which eventually leads to   four types of SFP: $0$, $1/4$, $1/2$, and $3/4$. 
 Among these four types, the first one is trivial, which actually means that the symmetry charge is not  fractionalized and any integral charges are equivalent to zero.

The above analysis can be easily generalized to more general and complicated situations. 
 For a general charge matrix as (\ref{eq:coupling}), it turns out that its element $Q_{ij}$ have the following  equivalence relations (see Appendix~\ref{subsection_syfra_particle_general} for detailed derivations)
 \begin{align}
 Q_{ij}\sim Q_{ij} +\gcd(K_i, N_j).\label{equation_charge_matrix_equivalent_SFP_purpose}
 \end{align}
 Accordingly, if $N_j=1$, \textit{i.e.}, it is a trivial layer, $Q_{ij}\sim 0$, which means there is no nontrivial SFP. 
 We can further define the so-called \text{reduced charge matrix} $\mathcal{Q}$, whose elements are defined as:
  $ \mathcal{Q}_{ij} =Q_{ij} \text{ mod }   \gcd(K_i,\! N_j)  \,,
   1\le i \le m  \,,   1\le j\le n\,;$   $ \mathcal{Q}_{ij}= 0  \,,\text{ otherwise} $. 
  Therefore, the reduced charge matrix is one-to-one correspondence to the Abelian SFP. In fact,
we can  use the \textit{SFP matrix} $\mathcal{C}_{ij}$ to denote  the corresponding $\Z_{K_i}$ symmetry charge   carried by $\Z_{N_j}$ elementary particle excitations denoted as $e_i$ (\textit{i.e.}, particle with only  unit $\Z_{N_j}$ gauge charge)\footnote{We note that the period of $\mathcal{C}_{ij}$ can be fractionalized as long as $\gcd(K_i,N_j)<N_j$. Consequently, when $ {Q}_{ij}/N_j$ is fractional, it doesn't mean that the elementary particle excitation $e_j$ really carries fractional charge of $\Z_{K_i}$ symmetry group. It is possible that the fractional number may be potentially removed by proper periods. In the following discussion, by ``fractional charge'', we really mean that  the fractionalization exists against any periodic shift. In other words, we'd better define fractional charge in its ``1st Brillouin zone''.}: 
\begin{align}
\text{SFP matrix: }\,\,\,\mathcal{C}_{ij}=\frac{ {Q}_{ij}}{N_j}\text{ mod }\frac{\text{gcd}(K_i,N_j)}{N_j}\,.\label{eq_mathcalc}
\end{align}
Therefore, it has $\text{gcd}(K_i,N_j)$ inequivalent values,  indicating that there are $\text{gcd}(K_i,N_j)$ types of SFP\footnote{For $j=n+1,\cdots, n+n'$, we have trivial $BF$ levels, \textit{i.e.}, $N_j=1$, therefore, $\mathcal{C}_{ij}$ is always $0\text{ mod }1$, \textit{i.e.}, non-fractionalized. As such, for the purpose of computing SFP, it is sufficient to consider reduced charge matrices   of a reduced size $m\times n$. \label{footnote_reduced_column}}. 
In total, we have $N_v$ different types of SFP labeled by $\nu_i$ whose $i=1,2,\cdots, N_v$ with
 \begin{align}
N_v=\prod_{i=1}^m \prod_{j=1}^n \text{gcd} (K_i, N_j)\,.\label{eq_trivial_SFL_nu}
\end{align}
We note that $\nu_1$ denotes the trivial SFP in which symmetry charge is integral.

We now apply the above results to three  examples with untwisted topological orders (\textit{i.e.}, $S^0_{int}=0$): 
       \begin{enumerate}[label=\textbf{(\roman*)}]
\item   \textit{$\Z_N$ topological order and $\Z_K$ symmetry.} There are $\text{gcd} (N,K)$    types of SFP in $\Z_N$ topological order with $\Z_K$ symmetry, which can be respectively labeled by   symmetry charges that are carried by  $e$: $ \frac{1}{N},\frac{2}{N},\cdots,\frac{\text{gcd} (N,K)}{N} \text{ mod }\frac{\text{gcd} (N,K)}{N}$. These different choices of symmetry charges  exhaust all  types of SFP. Each type corresponds to a projective representation, \textit{i.e.}, an element in the second cohomology:  $\mathcal H^2 (\Z_K, \Z_N)=\Z_{\text{gcd} (N,K)}$.

\item   \textit{$\Z_{N_1}\times \Z_{N_2}$ \textit{untwisted} topological order and $\Z_K$ symmetry.}
In such untwisted topological order, there are two kinds of elementary particle excitations, denoted by $e_1$ and $e_2$ respectively. Each elementary particle $e_i$ can carry $\text{gcd} (N_i, K)$ different kinds of $\Z_K$ symmetry charge and in total there are $\text{gcd} (N_1, K) \times \text{gcd} (N_2, K)$  types of SFP. Each type corresponds to a projective representation, \textit{i.e.}, an element in the second cohomology $\mathcal H^2 (\Z_K, \Z_{N_1}\times \Z_{N_2})=\Z_{\text{gcd} (N_1,K)} \times \Z_{\text{gcd} (N_2,K)}$. 

\item   \textit{$\Z_{N}$ topological order and $\Z_{K_1}\times \Z_{K_2}$ symmetry.} Denote $ (s_1, s_2)$ as the symmetry charge  carried by the elementary particle excitation, where $s_i=\frac{k_i}{N}$ and $k_i=1,2,\cdots, \text{gcd} (N,K_i)$. As a result, there are in total $\text{gcd} (N,K_1)\times \text{gcd} (N,K_2)$   types of SFP, which exhausts all     one-dimensional projective representations carried by particle excitations. SFP with non-Abelian representation  cannot be realized by the present field-theoretical framework. 

  \end{enumerate}
  
   One caveat is:  {some types among the  $N_v$ types may be potentially anomalous, which must be removed from the final classification.}    The key reason why so far we are unable to identify anomalous SF is that twisted terms in (\ref{eq:action_of_pure_gauge_3lb}) do not enter the calculation of SFP  (see  Appendix~\ref{subsection_syfra_particle_general}). In other words, the above classification  is always valid for untwisted topological orders but potentially contains anomalous types for twisted topological orders that are determined by twisted gauge theories in Eq.~(\ref{eq:action_of_pure_gauge_3lb}).   For twisted gauge theories with nontrivial twisted terms (\textit{i.e.}, all gauge fields are not from trivial layers), collected as $S^0_{int}$, the reduced charge matrices $\mathcal{Q}$ are not always simultaneously compatible  with   gauge invariance and global symmetry.  We will discuss the impact of $S^0_{int}$    in   Sections~\ref{obstruction} and also in the concrete example in Sec.~\ref{SET:ZNZN}.

 \section{Symmetry fractionalization on loop excitations}\label{section_main_sfloop}

   In this section, we present how to characterize  and classify SFL, \textit{i.e.}, symmetry-fractionalization on loop excitations. 
  We use the  mixed three loop (M3L) braiding statistics to characterize SFL. Different from the (intrinsic) three-loop  braidings where all loops are really loop excitations of bulk topological order,  {participants of each M3L braiding process simultaneously consist of    loop excitations of bulk topological order and quantized symmetry fluxes of external gauge (i.e., probe) fields.} 
  Generally speaking, the M3L  braiding statistical phase is defined to be the phase accumulated in  one of the two types of processes: (1)  one loop excitation or symmetry flux defect is braided fully around another loop excitation or symmetry flux defect which are both linked to the third loop excitation or symmetry flux defect, (2)  two identical loop excitations or  symmetry flux defects are exchanged which are both linked to the third loop excitation or symmetry flux defect.
 The total number $\mathcal{N}$ of loop excitations and symmetry fluxes participating   M3L braiding processes is 3. More generally, there are  mixed four loop braiding statistics with $\mathcal{N}=4$, which however is of non-Abelian nature. In this paper we only consider simplest cases: M3L, which is sufficient for all examples we will compute.
  We begin with the discussion of coefficients of twisted topological terms that is closely related to the M3L braiding staitistics.

 \subsection{Coefficients of twisted topological terms affected by global symmetry}\label{sec_twisted_quantized_period_coe}

 Here we discuss  the quantization levels and periods of coefficients of twisted topological terms in the absence or presence of general charge matrices. We will see that global symmetry will drastically change the quantization levels and {the} periods of {the} coefficients.  We  list   {some} useful formulas here. Detailed derivation  can be found in 
Appendix~\ref{appendix_SEG_quantization}.  In the absence of any global symmetry, as  mentioned in Appendix~\ref{sec_prel_1},  the coefficients  $q_{ijk}$  of twisted terms are quantized and periodic: 
\begin{align}
q_{ijk}=  r \frac{ N_iN_j }{N_{ij}}, \,r\in \Z_{N_{ijk}}  
\label{eqn_eqi_rela^2}\,,
\end{align}
where the symbol $N_{ij\cdots}$ is introduced to denote the greatest common divisor of $N_i, N_j,\cdots$, \textit{i.e.}, $N_{ij\cdots}=\text{gcd}(N_i,N_j,\cdots)$.
 
  In the presence of symmetry,    additional constraints should be taken into account, which often changes  the quantization levels  and  periods  of the twisted-term coefficients. Here we mainly consider two typical cases with general charge matrices. See Appendix~\ref{appendix_SEG_quantization} for the detailed derivation of the two cases and more general cases. 
   We  consider  a gauge theory  with gauge group   $\prod_{i=1}^{n}\Z_{N_i} $ and symmetry group   $\prod_{i=1}^m\Z_{K_i}$ in the presence of a twisted topological term {involving the gauge fields $a^1,a^2,a^3$}. The action is given by:  
\begin{align}
S=&S_0+S_c\,, \label{action:review2_main}
 \\
S_0=&\sum_{i=1}^{n+n'} \int \frac{N_i}{2\pi} a^i   db^i +\frac{q_{123}}{4\pi^2} a^1 a^2 da^3  \,, \\
S_c=&\sum_{i=1}^m\sum_{j=1}^{n+n'} \int \frac{Q_{ij}}{2\pi} A^i  db^j\,,
\end{align}
where the $BF$ level for trivial layers is unit:
 $N_i=1
  $ for $i=n+1,\cdots,n+n'$. 
The coefficient $q_{123}$ is quantized and periodically identified as
\begin{align}
q_{123}=k  M_{123}\,, k\in\Z_{\Gamma_{123}}\,,
\end{align}
where the integers $M_{123}$ and $\Gamma_{123}$ are defined as:
\small
\begin{align}
\!\!\!&M_{123}=\text{lcm}\bigg[N_1,N_2,\frac{N_1K_1}{\gcd(\hat Q_{11}, N_1K_1)},\frac{N_2K_1}{\gcd(\hat Q_{12},N_2K_1)},\nonumber\\
&\qquad\qquad\qquad\frac{N_1K_2}{\gcd(\hat Q_{21}, N_1K_2)},\frac{N_2K_2}{\gcd(\hat Q_{22},N_2K_2)},  
\cdots,\nonumber \\
&\qquad\qquad\qquad \frac{N_1K_m}{\gcd(\hat Q_{m1}, N_1K_m)},\frac{N_2K_m}{\gcd(\hat Q_{m2},N_2K_m)}\bigg]\,,
\label{eqn:twisted_coff_general_1} 
\end{align}
and
\begin{align}
&\Gamma_{123}=\gcd\bigg\{\text{lcm}\bigg[\frac{N_1N_2K_j}{\gcd(N_1N_2K_j,  \hat Q_{j2}M_{123})},\nonumber \\ &\qquad\qquad\qquad\,\,\frac{N_1N_2K_i K_j}{\gcd({N_1N_2K_i K_j}, \hat Q_{i1} \hat Q_{j2} M_{123})}, \forall i,j=1,...,m\bigg], \nonumber \\
&\qquad \qquad\, \,\text{lcm}\bigg[\frac{N_1N_2K_j}{\gcd(N_1N_2K_j,  \hat Q_{j1}M_{123})},\nonumber \\ &\qquad\qquad\qquad\,\,
\frac{N_1N_2K_i K_j}{\gcd({N_1N_2K_i K_j}, \hat Q_{i1} \hat Q_{j2} M_{123})}, \forall i,j=1,...,m\bigg],\nonumber \\
&\qquad \qquad\, \,\text{lcm}\bigg[\frac{N_1N_3K_j}{\gcd(N_1N_3K_j,  \hat Q_{j1}M_{123})},\,\frac{N_2N_3K_j}{\gcd(N_2N_3K_j,  \hat Q_{j2}M_{123})},\nonumber \\
&\qquad \qquad\quad\, \,\,\,\,\,\,\,\frac{N_1N_3K_i K_j}{\gcd({N_1N_3K_i K_j}, \hat Q_{i3} \hat Q_{j1} M_{123})}, \nonumber \\ &\qquad\qquad\quad\,\,\,\,\,\,\,\frac{N_2N_3K_i K_j}{\gcd({N_2N_3K_i K_j}, \hat Q_{i3} \hat Q_{j2} M_{123})}, \forall i,j=1,...,m\bigg]
\bigg\}.\!\!\!\!\!
\label{eqn:twisted_coff_general_2}
\end{align}
Here,  the symbol ``lcm'' and ``gcd'' stand for the least common multiple and greatest common divisor respectively. 
 
 There is a very important  notation $\hat{Q}_{ij}$ defined as:
  \begin{align}
&  \hat Q_{ij}=Q_{ij}\,,\,\text{if  $Q_{ij}\neq 0$}\,;\, \hat Q_{ij}=K_i\,,\,\text{if $Q_{ij}=0$}\,.\label{eqn_definition_hat_Q_matrix}
\end{align}  
  Using this definition, the expressions of $M_{123}$ and $\Gamma_{123}$ can be conveniently applied to both $Q_{ij}=0$ and $Q_{ij}\neq 0$, which largely simplifies the remaining calculation of this paper. 

  
Now we consider another type of twisted term: $a^1a^2da^2$.
 We  consider  a gauge theory  with gauge group   $\prod_{i=1}^{n}\Z_{N_i} $ and symmetry group   $\prod_{i=1}^m\Z_{K_i}$ and  the following  action 
\begin{eqnarray}
S&=&S_0+S_c \,,\label{action:review2_main}
 \\
S_0&=&\sum_{i=1}^{n+n'} \int \frac{N_i}{2\pi} a^i   db^i +\frac{q_{122}}{4\pi^2} a^1 a^2 da^2  \,, \\
S_c&=&\sum_{i=1}^m\sum_{j=1}^{n+n'} \int \frac{Q_{ij}}{2\pi} A^i  db^j\,.  
\end{eqnarray}
The coefficient $q_{122}$ is quantized and periodically identified as
\begin{align}
q_{122}=k’  M_{122}, k'\in\Z_{\Gamma_{122}}\,,
\end{align}
where the two integers $M_{122}$ and $\Gamma_{122}$ are defined respectively by: 
\small
\begin{align}
\!\!\!&M_{122}=\text{lcm}\bigg[N_1,N_2,\frac{N_1K_1}{\gcd(\hat Q_{11}, N_1K_1)},\frac{N_2K_1}{\gcd(\hat Q_{12},N_2K_1)},\nonumber\\
&\qquad\qquad\qquad\frac{N_1K_2}{\gcd(\hat Q_{21}, N_1K_2)},\frac{N_2K_2}{\gcd(\hat Q_{22},N_2K_2)},  
\cdots,\nonumber \\
&\qquad\qquad\qquad \frac{N_1K_m}{\gcd(\hat Q_{m1}, N_1K_m)},\frac{N_2K_m}{\gcd(\hat Q_{m2},N_2K_m)}\bigg]\,,
\label{eqn:twisted_coff_general_11} 
\end{align}
and 
\begin{align}
&\Gamma_{122}=\gcd\bigg\{\text{lcm}\bigg[\frac{N_1N_2K_j}{\gcd(N_1N_2K_j,  \hat Q_{j2}M_{122})},\nonumber \\ &\qquad\qquad\qquad\,\,\frac{N_1N_2K_i K_j}{\gcd({N_1N_2K_i K_j}, \hat Q_{i1} \hat Q_{j2} M_{122})}, \forall i,j=1,...,m\bigg], \nonumber \\
&\qquad \qquad\, \,\text{lcm}\bigg[\frac{N_1N_2K_j}{\gcd(N_1N_2K_j,  \hat Q_{j1}M_{122})},\nonumber \\ &\qquad\qquad\,\,
\frac{N_1N_2K_i K_j}{\gcd({N_1N_2K_i K_j}, \hat Q_{i1} \hat Q_{j2} M_{122})}, \forall i,j=1,...,m\bigg]\bigg\}
\label{eqn:twisted_coff_general_3}\,.\!\!\!\!\nonumber \\
\end{align}
From the above two actions with different twisted terms ($a^1a^2da^3$ and $a^1a^2da^2$), we reach the following useful conclusions:
      \begin{enumerate}[label=\textbf{(\roman*)}]
\item For the purpose of concreteness, the above formulas are written by explicitly choosing the first three layers (1st, 2nd, 3rd) that forms $a^1a^2da^3$ and $a^1a^2da^2$. It is natural to generalize these formulas to  three \textit{arbitrarily} selected   layers as long as we carefully replace layer indices in order.  
\item The coefficients $q_{123}$ and $q_{122}$ follow the same quantization rule, i.e.,  $M_{123}=M_{122}$. 
So, the general minimal quantized value is in fact fully determined by information  of 1st and 2nd layers. We introduce a new symbol $M_{12}$:
\begin{align}  
M_{12}\equiv M_{123}=M_{122}\,.
\end{align}

 \item  In contrast to $M_{122}$ and $M_{123}$ which are equivalent to each other,  the expressions of the two periods, i.e., $\Gamma_{122}$ and $\Gamma_{123}$ are not the same, so they should be computed separately. 

\item   When the twisted term is $a^1a^2da^3$,  $\Gamma_{123}$   is fully determined by $N_1$, $N_2$, $N_3$, $Q_{i1}$,  $Q_{i2}$, and $Q_{i3}$ with $i=1,2,\cdots, m$.   When the twisted term is $a^1a^2da^2$,  $\Gamma_{122}$   is fully determined by $N_1$, $N_2$, $Q_{i1}$, and $Q_{i2}$  with $i=1,2,\cdots, m$. Therefore, we reach a general result that $\Gamma_{122}$  and $\Gamma_{123}$
 are fully determined by the information of gauge fields (i.e.,   levels and coupling matrix elements) that appear in the twisted term $aada$.

\item If $a^2a^1da^1$ is considered,  it is clear that $M_{211}=M_{122}$ and $\Gamma_{211}=\Gamma_{122}$.  So,  it is unnecessary to compute the coefficient of $a^2a^1da^1$ if the coefficient of $a^1a^2da^2$ has been computed.

\item If symmetry group $G_s=\Z_K$, then the above formulas can be largely simplified by directly setting  $i=j=1=m, K_i=K_j=K$.

\end{enumerate}

\subsection{M3L invariants, M3L statistical phases, and two origins of periods}
\label{section_mmt_general_discussion}

Here we will show, via an example, how to obtain the  inequivalent quantized values of  M3L statistical phases that are originated from the same  M3L invariant. More explicitly, we first need to determine the quantization rules  of M3L invariants, which can  determine the quantization of the corresponding M3L statistical phases. Then, we need to determine the minimal periods of the corresponding M3L statistical phases.

To illustrate,  we  take   {one $\Z_4$ symmetric}  $\Z_4$ topological order  as an example{, with the following action under proper gauge  choice:
\begin{align}
 S= &\frac{4}{2\pi} \int b^1   da^1 + \frac{1}{2\pi}\int b^2   da^2 + \frac{t}{4\pi^2} \int a^1  a^2 da^2
 \nonumber\\
 &+\frac{Q_{12}}{2\pi} \int A   db^2 +  \int a^i   *j^i +\int b^1   * \Sigma^1 \,,
 \label{eqn_action_z4_z4}
 \end{align}
where $2$-form $\Sigma^1$ represents the world-sheet of the loop excitation current and $\Sigma^2$ is omitted since it is a  trivial loop excitation. Both $b^2$ and $a^2$ are level-$1$ gauge fields (see footnote \ref{footnote_level1}). In this action, $S^0_{int}$ doesn't exist since  $\Z_4$ topological order is always untwisted. But $S^1_{int}=\frac{q_{122}}{4\pi^2} \int a^1  a^2 da^2$  is included in the action (see Eq.~\ref{equation_twisted_term_split} for definition of $S_{int}^1$).   
{For simplifying notation, we temporarily use $t$ to replace ``$q_{122}$'' in this subsection.
 }

Upon integrating out $b^1$ and $b^2$, we get $a^1=-\frac{\pi}{2}{d^{-1}(* \Sigma^1)}$ and $a^2= -Q_{12}A=-{\pi Q_{12}\over2}d^{-1}(*\sigma)$. Here $d^{-1}$ formally denotes the inverse of differential operator $d$\footnote{While the formal notation $d^{-1}$ works well, a more mathematically rigorous derivation can be achieved by using Seifert surface and $\delta$-function of manifolds, see, e.g., Refs.~\cite{Zhang:2021ycl,2016arXiv161209298P,PhysRevResearch.3.023132,bti6,zee2010quantum}. }, and the $2$-form tensor $\sigma=*\frac{2}{\pi} dA$ represents the world-sheet of the symmetry flux-line. In the expression of $\sigma$, we choose the  coefficient to be $2/\pi$    such that $\sigma$ counts {the} number of $\Z_4$ unit flux lines, by noting that $dA=\frac{2\pi}{4}\times \Z$. Formally, $A$ can be resolved as: $A=\frac{\pi}{2} {d^{-1}(*\sigma)} $.  Then we obtain 
\begin{align}
S_{\text{eff}}=&-\frac{\pi}{2} \int( * j^1 ) d^{-1} {(* \Sigma^1)}-Q_{12}\int A  *j^2 \nonumber\\
&+\frac{Q_{12}^2 t\pi}{32} \int  d^{-1} (*\Sigma^1)\wedge  d^{-1} (*\sigma)  \wedge (*\sigma)\,.  \label{eq_mml_first}
\end{align}

The first term in Eq.~(\ref{eq_mml_first}) is a  Hopf term and  leads to the particle-loop braiding process. Such a process   involves  one  particle  with unit gauge charge  and  one loop excitation with unit gauge flux.  

 The second term  in Eq.~(\ref{eq_mml_first}) indicates  the particle (denoted by $e_2$) with unit gauge charge of the level-$1$ gauge field  $a^2$  carries integral symmetry charge,  as expected. As a side, this term can be rewritten as $-\frac{\pi Q_{12}}{2}\int(* j^2)d^{-1} *\sigma $. If $Q_{12}=1$, this Hopf term leads to a $\frac{\pi}{2}$ phase once $e_2$ is braided around $\sigma$.

The third term in Eq.~(\ref{eq_mml_first}) describes M3L braiding processes, which contributes to three M3L statistical phases (see Appendix~\ref{footnote_multiloopinv}): $\theta_{\sigma, \sigma; \Sigma^1}=\frac{Q_{12}^2 t \pi }{16}$, $\theta_{\sigma; \Sigma^1}=\frac{Q_{12}^2t \pi }{32}$ and  $\theta_{ \Sigma^1,\sigma; \sigma}=-\frac{Q_{12}^2 t \pi }{32}$.  Here, $\theta_{\sigma, \sigma; \Sigma^1}$ is the statistical phase of braiding one  symmetry flux  around  another  symmetry flux,     both of which are    linked with a loop excitation $\Sigma^1$, and $\theta_{ \Sigma^1,\sigma; \sigma}$  is the statistical phase of   braiding    one symmetry flux    around $\Sigma^1$, both of which are    linked with   another symmetry flux.  $\theta_{\sigma; \Sigma^1}$ is the exchange statistical phase of two ‘identical’ symmetry fluxes both of which are   linked to   $\Sigma^1$.

 Next we investigate  all   inequivalent values (quantization and periods) of these M3L statistical phases. We take  $\theta_{\sigma, \sigma; \Sigma^1}$ and $\theta_{\sigma; \Sigma^1}$ as      examples.   
 By applying Eqs.~(\ref{eqn:twisted_coff_general_11}) and (\ref{eqn:twisted_coff_general_3}),  we can obtain the quantization for $t$\footnote{Since $Q_{11}=0$ in Eq.~(\ref{eqn_action_z4_z4}), we set $\hat{Q}_{11}=K=4$. $N_1=4$, $N_2=1$. $Q_{12}=0,1,2,3$, which corresponds to $\hat{Q}_{12}=4,1,2,3$.}: 
\begin{align}
t=4k\,, \label{equation_syfra_values_q}
\end{align}
where $k$ is a periodic integer:  
\begin{align}
k\in \Z_{{4}/{\gcd (4, \hat{Q}_{12})}}.\label{eqn:loops_period}
\end{align} 
 From Eq.~(\ref{equation_syfra_values_q}), the quantized values of  $\theta_{\sigma, \sigma; \Sigma^1}$  and $\theta_{\sigma; \Sigma^1}$ take $\frac{Q_{12}^2  \pi}{4} \cdot k$ and $\frac{Q_{12}^2  \pi}{8} \cdot k$  with integral $k$, respectively. Therefore, when $Q_{12}=1$,  they  can take the minimal (\textit{i.e.}, most general) quantized value, \textit{i.e.}, $\theta_{\sigma, \sigma; \Sigma^1}=\frac{\pi k}{4}$,  $\theta_{\sigma; \Sigma^1}=\frac{\pi k}{8}$. 
 
 Next, we determine the minimal period of the two M3L statistical phases:
      \begin{enumerate}[label=\textbf{(\roman*)}]
\item  First, 
we note that Eq.~(\ref{eqn:loops_period}) provides a period of $t$. If $Q_{12}=1$, {hence} $\hat{Q}_{12}=1$, the period of $t$ is $16$. This period is required by gauge invariance and symmetry invariance. Thus, the statistical phase $\theta_{\sigma, \sigma; \Sigma^1}=\frac{Q_{12}^2t \pi }{16}$ is defined modulo $\pi$ and $\theta_{\sigma; \Sigma^1}=\frac{Q_{12}^2 t \pi }{32}$ is defined modulo $\frac{\pi}{2}$.

\item  Second, we  consider the periods from attaching particles to one dimensional objects, as we want to quotient out the phase ambiguity caused by attaching  {particles}.  More concretely, the trivial particle $e_2$ can be attached onto  {the} symmetry fluxes. As a result,  attaching an $e_2$ particle to one of the symmetry flux     will shift $\theta_{\sigma, \sigma; \Sigma^1}$  by $\frac{\pi}{2}$ due to the AB phase between $e_2$ and the $\Z_4$ symmetry flux $\sigma$. (See the texts below Eq.~(\ref{eq_mml_first})). Therefore, following Bezout's lemma, the minimal period of $\theta_{\sigma, \sigma; \Sigma^1}$ is determined by the greatest common divisor of the above two periods: $\pi$ and $\frac{\pi}{2}$, that is: $\pi\times\text{gcd}(1,\frac{1}{2})=\frac{\pi}{2}$. Likewise, in the ``exchange'' phase $\theta_{\sigma; \Sigma^1}$, in order to keep both $\sigma$'s  identical to each other,  {we attach one $e_2$ particle to each of the $\sigma$ loops}. Then, the exchange phase $\theta_{\sigma; \Sigma^1}$ is shifted by $\frac{1}{2}\times \frac{\pi}{2}+\frac{1}{2}\times \frac{\pi}{2}= \frac{\pi}{2}$. Thus,  the minimal period of the exchange phase is $\pi\times\text{gcd}(\frac{1}{2},\frac{1}{2})=\frac{\pi}{2}$.  

 \end{enumerate}

 In summary, the inequivalent quantized values of  $\theta_{\sigma; \Sigma^1}$ are $0, \frac{\pi }{8}, \frac{\pi }{4}, \frac{3\pi }{8}\text{ mod } \frac{\pi}{2}$ and those of $\theta_{\sigma, \sigma; \Sigma^1}$ are $0, \frac{\pi }{4} \text{ mod }\frac{\pi}{2}$.  
In a similar fashion, we can obtain the inequivalent values of $\theta_{ \Sigma^1,\sigma; \sigma}=0, \frac{\pi }{8}, \frac{\pi}{4}, \frac{3\pi }{8}\text{ mod }\frac{\pi}{2}$.  If $Q_{12}$ takes other values, these M3L statistical phases are not beyond those with $Q_{12}=1$. Also, if $n'>1$, the M3L statistical phases are not beyond those with $n'=1$, $Q_{11}=0$ and $Q_{12}=1$. Nevertheless, the M3L statistical phases may be drastically altered if $Q_{11}\neq 0$, which we will discuss in Sec.~\ref{SET:ZN}.

 \subsection{General procedure for computing SFL}
\label{mmt_general_discussion_more_General}
In the following, we shall   discuss, on a general ground, how to obtain   SFL for a given  SFP. Several important terminologies will also be introduced.  
As mentioned, the M3L statistical phases can  be used to characterize   SFL,  which are determined by the related M3L invariants. So we first enumerate all different possible M3L invariants.  Then we discuss how to determine classification (\textit{i.e.}, equivalence relations) among the M3L statistical phases.

\subsubsection{Enumerating all independent M3L invariants and M3L statistical   phases}\label{section_M3Linvariant_statistical_phases_relation}
 
In contrast to SFP, twisted terms play a critical role in SFL. Twisted terms contain two pieces as shown in Eq.~(\ref{equation_twisted_term_split}) in Appendix~\ref{subsection_syfra_particle_general}.  If we start with the action (\ref{couplingaction}) with gauge group $G_g$ and symmetry $G_s$, by integrating out all dynamical gauge fields, i.e., all $b^i$'s and $a^i$'s in Eq.~(\ref{equation_twisted_term_split}),  a quantity $\mathcal{I}$ shown below appears as an effective action that involves three loop-like objects in which   loop excitations and symmetry fluxes must coexist:
\begin{align}
\mathcal{I}=2\pi \mathcal{I}_{ \mathsf{ijk}}\int_{M^4}  (*d^{-1} \Omega_{\mathsf{ i}}) (*d^{-1} \Omega_{\mathsf{ j}})  (* \Omega_{\mathsf{ k}})\,,
\label{MML_ivariant}
\end{align}
where $2\pi \mathcal{I}_{\mathsf{ijk}}$ is a real-valued coefficient. {We should note that, the notations $\Omega_{\mathsf{i}},\Omega_{\mathsf{j}},\cdots$ denote  either unit symmetry fluxes  $\sigma$ or elementary loop excitations $\Sigma$. Therefore, the subscripts ``$\mathsf{i,j,k}$''   are intentionally  written in  fonts different from ``$i,j,k$''. }

 From now on, we also use the coefficient  $ \mathcal{I}_{\mathsf{ijk}}$ to represent the invariant given by (\ref{MML_ivariant}).
The  invariant $ \mathcal{I}_{\mathsf{ijk}}$ can be used to describe the following   braiding processes:
      \begin{enumerate}[label=\textbf{(\roman*)}]
\item A 
statistical phase  ``$2\pi  \mathcal{I}_{\mathsf{ijk}}$'' (denoted as $\theta_{\Omega_{\mathsf{ k}}, \Omega_{\mathsf{ j}};\Omega_{\mathsf{ i}}}$)  is accumulated once $\Omega_{\mathsf{ k}}$ is braided around $\Omega_{\mathsf{ j}}$ both of which are   simultaneously linked to $\Omega_{\mathsf{ i}}$. 

\item  A statistical phase  ``$-2\pi \mathcal{I}_{\mathsf{ijk}}$'' (denoted as $
\theta_{\Omega_{\mathsf{ k}}, \Omega_{\mathsf{ i}};\Omega_{\mathsf{ j}}}$) is accumulated once  $\Omega_{\mathsf{ k}}$ is braided around $\Omega_{\mathsf{ i}}$ both of which are  simultaneously linked to $\Omega_{\mathsf{ j}}$.
\item  In general, $\Omega_{\mathsf{ i}}$, $\Omega_{\mathsf{ j}}$ and $\Omega_{\mathsf{ k}}$ are different from each other.  If $
\Omega_{\mathsf{ j}}$ and $\Omega_{\mathsf{ k}}$ are identical, the statistical phase $\theta_{\Omega_{\mathsf{ k}}, \Omega_{\mathsf{k}};\Omega_{\mathsf{ i}}}=4\pi  \mathcal{I}_{\mathsf{ikk}} $. If the process only involves the exchange of two  identical  $\Omega_{\mathsf{ k}}$'s   both of which are simultaneously  linked to $\Omega_{\mathsf{ i}}$,  we can have a statistical phase $\theta_{\Omega_{\mathsf{ k}};\Omega_{\mathsf{ i}}}=2\pi  \mathcal{I}_{\mathsf{ikk}}$.  
 \end{enumerate}
 
Now we discuss how to  enumerate all different M3L invariants for  a  gauge group $G_g$ and a  symmetry group $G_s$. 
From the definition of $ \mathcal{I}_{\mathsf{ijk}}$, the following relation holds: 
\begin{align}
 \mathcal{I}_{\mathsf{ijk}}=- \mathcal{I}_{\mathsf{jik}}\, \label{equation_I_skewsymmetry}
 \end{align}
  which means that exchanging the positions of the first two loop objects does not provide a new invariant.  
Suppose symmetry group $G_s=\prod_i^m \Z_{K_i}$ and gauge group $G_g=\prod_j^n \Z_{N_j}$. Thus,  there are $m$ topologically distinct unit  symmetry fluxes  denoted as  $\sigma^1, \sigma^2,\sigma^3,\cdots$ respectively from symmetry subgroups $\Z_{K_1},\Z_{K_2},\Z_{K_3},\cdots$. There are  $n$ topologically distinct loop excitations  denoted as $\Sigma^1, \Sigma^2,\Sigma^3,\cdots$ respectively from gauge subgroups $\Z_{N_1},\Z_{N_2},\Z_{N_3},\cdots$. 

\textbf{Definition of universal notation:}  For the latter convenience, we introduce the following notation system. We use Greek letters $\alpha,\beta,\gamma,\cdots$ as the subscript of $\mathcal{I}_{\cdots}$  if the corresponding loop-like objects are unit quantized symmetry fluxes $\sigma^1, \sigma^2,\sigma^3,\cdots$ of symmetry subgroups $\Z_{K_1},\Z_{K_2},\cdots$ respectively. We use Latin letters $a,b,c,\cdots$ as the subscript of $\mathcal{I}_{\cdots}$ if the corresponding loop-like objects are elementary loop excitations $\Sigma^1,\Sigma^2,\Sigma^3,\cdots$  of gauge subgroups $\Z_{N_1},\Z_{N_2},\cdots$ respectively. In this notation system, we can distinguish two sets of ``$1,2,3,\cdots$'' and thus unambiguously  express M3L invariants,  as shown in   Table~\ref{table_three_M3L_example_list}.

 For the purpose of enumerating all the independent types of M3L invariants, 
we follow the two rules: 
      \begin{enumerate}[label=\textbf{(\roman*)}]
\item Enumerate all   kinds of mixture of loop excitations and symmetry fluxes. Among three loop-like objects, if  {two of them are} loop excitations, they can be identical (e.g., both are $\Sigma^1$). Likewise, if there are two symmetry fluxes, they can also be identical (e.g., both are $\sigma^1$).
\item  List all  M3L invariants for each choice of combination. Apply Eq.~(\ref{equation_I_skewsymmetry})  to eliminate redundancy. 
 \end{enumerate}
 \begin{table*}[t]
    \centering  \caption{List of M3L invariants when gauge group and symmetry group are given.  Once the quantized values and periods of M3L invariants (i.e., $\mathcal{I}_{...}$) are computed,  the quantized values of M3L statistical phases (i.e., $\theta_{...}$) can be fixed from the fourth column. In addition to the periods of M3L invariants, the  periods of M3L statistical phases are also affected by `particle attachment'.  The final periods are fixed by  both factors. More details are available in Sec.~\ref{mmt_general_discussion_more_General}.}
    \begin{tabular}{cccc}
    \hline
    
    \hline
  \text{ }  \text{ } \text{ } \text{ } \text{ }$G_g$ \text{ } \text{ } \text{ } \text{ }\text{ }&  \text{ } \text{ } \text{ } \text{ }   $ G_s $ \text{ } \text{ } \text{ } \text{ }& \text{ } \text{ } \text{ } \text{ }   \text{ } \text{ } \text{ } \text{ }  M3L invariants\text{ } \text{ } \text{ } \text{ }  \text{ } \text{ }\text{ } \text{ }&M3L statistical phases\\
\\            \hline \\
     $\Z_N$  & $\Z_K$  & \begin{minipage}{2.2in}\centering  $\mathcal{I}_{a\alpha\alpha}$ \\$\mathcal{I}_{\alpha aa}$ \end{minipage}& \begin{minipage}{2.2in}\centering $ \theta_{\sigma;\Sigma}= -\theta_{\Sigma, \sigma;\sigma}=2\pi \mathcal{I}_{a\alpha \alpha}$ \\$\theta_{\sigma, \sigma;\Sigma}=4\pi \mathcal{I}_{a \alpha \alpha}$ \\ $ \theta_{\Sigma;\sigma}=-\theta_{\sigma, \Sigma;\Sigma}=2\pi \mathcal{I}_{\alpha aa}$ \\$ \theta_{\Sigma,\Sigma;\sigma}=4\pi\mathcal{I}_{\alpha aa}$      \end{minipage}  
 \\ \\\hline\\
  $\Z_{N_1}\times \Z_{N_2}$  & $\Z_K$  & 
  \begin{minipage}{2.2in}\centering   $\mathcal{I}_{a\alpha\alpha}$ \\ $\mathcal{I}_{b\alpha\alpha}$\\ $\mathcal{I}_{\alpha aa}$\\ $\mathcal{I}_{\alpha bb}$\\ $\mathcal{I}_{\alpha ab}$\\$\mathcal{I}_{\alpha ba}$\\$\mathcal{I}_{ab \alpha}$ \end{minipage}& 
  \begin{minipage}{2.2in}
  \centering  $\theta_{\sigma; \Sigma^1}=-\theta_{\Sigma^1, \sigma; \sigma}=2\pi \mathcal{I}_{a\alpha\alpha}$\\
  $\theta_{\sigma,\sigma; \Sigma^1}=4\pi \mathcal{I}_{a\alpha\alpha}$\\
 $\theta_{\Sigma^1; \sigma}=-\theta_{\sigma, \Sigma^1; \Sigma^1}=2\pi \mathcal{I}_{\alpha aa}$ \\
 $\theta_{\Sigma^1,\Sigma^1; \sigma}=4\pi \mathcal{I}_{\alpha aa}$ \\
  $\theta_{\sigma; \Sigma^2}$ $=-\theta_{\Sigma^2, \sigma; \sigma}=2\pi \mathcal{I}_{b\alpha\alpha}$ \\  
  $\theta_{\sigma,\sigma; \Sigma^2}$ $=4\pi \mathcal{I}_{b\alpha\alpha}$ \\
  $\theta_{\Sigma^2; \sigma}=-\theta_{\sigma, \Sigma^2; \Sigma^2}=2\pi \mathcal{I}_{\alpha bb}$\\
   $\theta_{\Sigma^2,\Sigma^2; \sigma}=4\pi \mathcal{I}_{\alpha bb}$\\
$\theta_{\sigma, \Sigma^2; \Sigma^1}=2\pi (\mathcal{I}_{ ab\alpha}-\mathcal{I}_{\alpha ab})$  \\$\theta_{\Sigma^2, \Sigma^1; \sigma}$$=2\pi (\mathcal{I}_{\alpha ab}+\mathcal{I}_{\alpha ba})$ \\
 $\theta_{\sigma, \Sigma^1; \Sigma^2}=-2\pi (\mathcal{I}_{\alpha ba}+\mathcal{I}_{ ab\alpha})$     \end{minipage}  
\\ \\\hline \\
  $\Z_{N} $  &$\Z_{K_1}\times\Z_{K_2}$    &\begin{minipage}{2.2in}\centering   $\mathcal{I}_{\alpha aa}$  \\$\mathcal{I}_{\beta aa}$  \\$\mathcal{I}_{a\alpha\alpha}$ \\ $\mathcal{I}_{a\beta\beta}$  \\$\mathcal{I}_{ \alpha a \beta}$ \\ $\mathcal{I}_{   \beta   a \alpha}$\\ $\mathcal{I}_{\alpha \beta a}$ \end{minipage}& \begin{minipage}{2.2in} $ \theta_{\Sigma;\sigma^1}=-\theta_{\sigma^1,\Sigma;\Sigma}=2\pi \mathcal{I}_{\alpha aa}$\\
  $ \theta_{\Sigma,\Sigma;\sigma^1}=4\pi \mathcal{I}_{\alpha aa}$\\
   $ \theta_{\Sigma;\sigma^2}=-\theta_{\sigma^2,\Sigma;\Sigma}=2\pi \mathcal{I}_{\beta aa}$\\
  $ \theta_{\Sigma,\Sigma;\sigma^2}=4\pi \mathcal{I}_{\beta aa}$\\  
   $ \theta_{\sigma^1;\Sigma}=-\theta_{\Sigma,\sigma^1;\Sigma}=2\pi \mathcal{I}_{a\alpha\alpha}$\\
  $ \theta_{\sigma^1,\sigma^;\Sigma}=4\pi \mathcal{I}_{a\alpha\alpha}$\\
  $ \theta_{\sigma^2;\Sigma}=-\theta_{\Sigma,\sigma^2;\Sigma}=2\pi \mathcal{I}_{a\beta\beta}$\\
  $ \theta_{\sigma^2,\sigma^;\Sigma}=4\pi \mathcal{I}_{a\beta\beta}$\\
  $ \theta_{\Sigma,\sigma^2;\sigma^1}=2\pi (\mathcal{I}_{\alpha \beta a}+\mathcal{I}_{  \alpha a \beta})$\\
   $  \theta_{\sigma^2,\sigma^1;\Sigma}=2\pi (\mathcal{I}_{a\alpha \beta }-\mathcal{I}_{ \beta  a\alpha})$\\
   $  \theta_{\sigma^1,\Sigma;\sigma^2}=2\pi (\mathcal{I}_{  \beta a\alpha}-\mathcal{I}_{  \alpha \beta a})$\\ 
\end{minipage}  
 \\ \\
  \hline
  
  \hline
  \end{tabular}
    \label{table_three_M3L_example_list}
\end{table*}

To make the above general statements more concrete, we list three examples in Table~\ref{table_three_M3L_example_list}. 
For instance,   $G_g=\Z_N$ and $G_s=\Z_K$. In this case, there is only one kind of elementary loop excitations denoted by $\Sigma$ and one kind of unit symmetry fluxes denoted by $\sigma$. There are two independent M3L invariants denoted by $\mathcal{I}_{a\alpha\alpha}$ and $\mathcal{I}_{\alpha aa}$, whose effective actions are respectively written as $2\pi \mathcal{I}_{a\alpha\alpha} \int (d^{-1} *\Sigma) (d^{-1}* \sigma)  (* \sigma)\,$ and $2\pi \mathcal{I}_{\alpha aa} \int (d^{-1} *\sigma) (d^{-1} *\Sigma)  (* \Sigma)\,$.
 From the M3L invariants, we can obtain the following M3L statistical phases:  $ \theta_{\sigma;\Sigma}= -\theta_{\Sigma, \sigma;\sigma}=2\pi \mathcal{I}_{a\alpha \alpha}$, $\theta_{\sigma, \sigma;\Sigma}=4\pi \mathcal{I}_{a \alpha \alpha}$, $ \theta_{\Sigma;\sigma}=-\theta_{\sigma, \Sigma;\Sigma}=2\pi \mathcal{I}_{\alpha aa}$, $ \theta_{\Sigma,\Sigma;\sigma}=4\pi\mathcal{I}_{\alpha aa}$. The formulas in Table~\ref{table_three_M3L_example_list} will be used in Sec.~\ref{SET:ZN}, \ref{SET:ZNZN}, and \ref{sec_example_znzkzk}.

\subsubsection{Quantization rules and periods of M3L   statistical phases}

Next, we discuss how M3L invariants  as shown in Table~\ref{table_three_M3L_example_list}, are quantized.  From  the derivation   in Sec.~\ref{section_mmt_general_discussion},  we see that  a nonvanishing M3L invariant exists only when  twisted terms\footnote{Since twisted terms contain two different pieces   $S_{int}=S^{0}_{int}+S^{1}_{int}$ as  shown in Eq.~(\ref{equation_twisted_term_split}), even though the topological order is untwisted, i.e., $S^0_{int}=0$, one can still  add trivial layers and consider $S^1_{int}$. The coefficients in $S^1_{int}$ are periodically equivalent to zero if symmetry is not considered. But when symmetry is considered, they have nontrivial values that cannot be periodically identified as zero.} are present. In other words,  M3L invariants numerically depend on  the coefficients of twisted terms $q_{...}$, charge matrix elements $Q_{ij}$, $\{N_i\}$, and $\{K_i\}$.

In practice, a specific twisted term can contribute to many different M3L invariants. Meanwhile,  multiple twisted terms can contribute to the same M3L invariant. Therefore, in general, an M3L invariant is quantitatively determined by coefficients of twisted terms.  Since the coefficients are quantized and periodic, M3L invariants are also quantized and periodic. 
  Once we know how M3L invariants are quantized, we can straightforwardly deduce M3L statistical phases via the fourth column of Table~\ref{table_three_M3L_example_list}.

Once quantization is obtained, the next step is to determine the \textit{minimal} periods of   M3L statistical phases.  The minimal period of each M3L statistical phase is the greatest common divisor of all possible periods.
In this paper, we consider the following two   sources of periodic identification for M3L statistical phases:
      \begin{enumerate}[label=\textbf{(\roman*)}]
\item First, we note that each M3L statistical phases can be expressed in terms   of M3L invariants. So, the periodicity of M3L invariants provides a period for each M3L statistical phase. 

\item Second,  in  M3L braiding processes, one can always attach particle excitations onto either $\Omega_{\mathsf{j}}$ or $\Omega_{\mathsf{k}}$ or both. This attachment leaves flux contents unaltered but may potentially shift the final statistical  phases by an AB phase. As a result, we obtain another period.
\end{enumerate}
Finally, we can characterize and classify SFL via M3L statistical phases for each SFP.

   \subsection{Anomalous  symmetry fractionalization}
\label{obstruction}

  {In the following  we show that some types of SFP labeled by $\nu_i$ may be inconsistent with the twisted topological terms, which is said to be anomalous.   For this reason, we must carefully remove the anomalous patterns in our classification of SF.} 

Let us go back to the derivation of SFP in Sec.~\ref{SF} and  Appendix~\ref{subsection_syfra_particle_general}. In the   derivation of SFP, we did not include the twisted terms $S_{int}$.   Hence, it seems that for a specific  (Abelian, finite) gauge group,  no matter  whether the topological orders are twisted or not, they can always admit  the same   SFP. However,   {this is not always the case}. The reason lies in the fact that the minimal coupling with external probe field(s) (\textit{i.e.},  the charge matrix) may substantially affect the quantization and periods  of the twisted terms which  in turn determine whether a topological order is twisted or not. We will see that some types of SFP cannot  consistently coexist with a  topological order  with a specific twist, which means such an SFP is obstructed in certain twisted topological order.

For example, we consider the simplest twisted topological order, that is $\Z_2\times \Z_2$ topological order and $\Z_{2K}$ symmetry with the  action:
\begin{align}
S= \frac{2}{2\pi} \int \sum_{i=1}^2 b^i   da^i  + \frac{q}{4\pi^2} \int a^1  a^2 da^2 
 +\frac{1}{2\pi} \int A db^1
\end{align}
where $q=2$. 
Following the method above, we can know that here $e_1$ carries half integer symmetry charge while $e_2$ carries integer symmetry charge.
Using the results (\ref{eqn:twisted_coff_general_1}) and (\ref{eqn:twisted_coff_general_3}), the allowed coefficients of $q$ are $0, 4K$ mod $8K$, which are all equivalent to zero if  {the symmetry is broken}. In other words, the anomaly-free symmetric twisted $\Z_2\times \Z_2$ topological order cannot have such SFP  (corresponding to a symmetry assginment denoted by the charge martix $Q= (Q_{11},Q_{12})= (1,0)$).
On the other hand, if a twisted $\Z_2\times \Z_2$ topological order has such a SFP, then it has to be anomalous. The existence of this kind of anomaly was first pointed out in Ref.~\cite{2016arXiv161008645Y} in a very specific  example, but a systematic treatment was lacking there. In the present work, we are able to   characterize and classify symmetry fractionalization in a systematic and powerful field-theoretical approach, in which the above anomaly can be safely removed from the final classification.

In Sec.~\ref{SET:ZNZN}, we will concretely analyze the symmetry fractionalization in $\Z_2\times \Z_2$ topological orders with $\Z_2$ symmetry. When the bulk topological order is twisted, we find in Sec.~\ref{section_concrete_example_of_anomaly_z2z2z2} that nontrivial SFP of $\Z_2$ symmetry is impossible in twisted $\Z_2\times\Z_2$ topological order.

       \subsection{Preliminary observations on algebraic structure and procedure of gauging}

     \label{classification_principle}

     \label{general}
 In the field of topological phases of matter, it is usually a hallmark of significant theoretical progress  when an algebraic theory (i.e., abstract mathematics) of a given class of topological phases of matter is established. For example,  group cohomology theory~\cite{chenguliuwenscience,chenguliuwenprb} is the algebraic theory of the classification of bosonic SPTs. As mentioned in Sec.~\ref{subsection_research_background}, a  systematic framework called ``$G^{\times}$ tensor category'' for 2D bosonic SET is established \cite{Barkeshli2014}, which is related to the mathematical framework  in Ref.~\cite{etingof2010fusion}.      While  a pure algebraic theory of symmetry fractionalization in 3D is beyond the scope of the present field-theoretical approach, it may be still  inspiring to present some interesting observations.

     For  a given  SFP $\nu_i$ and a fixed topological order $\omega$, 
      the classification of SFL can be denoted as $\mathcal{C}_\omega({\nu_i})$, where $\omega\in \mathcal H^4 (G_g,U (1))$}   labels the underlying {Dijkgraaf-Witten type} topological order {with gauge group $G_g$}. Then total classification of SF in 3D SET  phases can be  enumerated  as: 
$
      \Sigma_\omega=\bigoplus_{i=1}^{N_\nu} \mathcal{C}_\omega (\nu_i)\,,
 $  
 where
$N_\nu=\prod_{i=1}^n \prod_{j=1}^m \gcd (K_j, N_i)$ (see Sec.~\ref{SF} and Appendix~\ref{subsection_syfra_particle_general}), and the symbol $\oplus$ denotes a   summation over all  SFP types.   From Sec.~\ref{obstruction}, it is possible that for certain $w$, SFP $\nu_i $ is anomalous. If this happens,  the corresponding $\mathcal{C}_\omega(\nu_i)$ is replaced by the empty set.   {In our field theoretical approach, we have embedded the symmetry group into a continuous group $U(1)\otimes\cdots\otimes U(1)$, therefore all the}   particles carry one-dimensional  irreducible representations of the symmetry group, which   {can be} either linear or projective.   {In other words, all the patterns of  SFP we have considered are Abelian.} But more generally, particles can   carry higher dimensional irreducible projective representations of  {the symmetry group},  {belonging to} a general element $\nu_i \in \mathcal H^2 (G_s, G_g)$.\footnote{We assume $G_g$ to be Abelian.}
 In addition, the types of SFL are characterized by M3L statistical phases, which are Abelian phases. This means we only consider Abelian symmetry fractionalization on both particles and loops in the present field-theoretical framework.

Physically we can gauge the symmetry group $G_s$ and obtain an enlarged gauge group \begin{align}
G_g^*\equiv G_g\leftthreetimes_{\nu_i}G_s
\end{align} which depends on the choice of  ${\nu_i}$ that is non-anomalous. 
 Noticing that $\nu_i \in \mathcal H^2 (G_s, G_g)$ essentially defines a central extension of  $G_s$ by $G_g$\footnote{More formally, for an element $\nu_i \in \mathcal H^2 (G_s, G_g)$ with Abelian $G_g$, we have a  short exact  sequence: 
 $ 0 \longrightarrow G_g \longrightarrow G_g^* \longrightarrow G_s \longrightarrow 0\,  $  which determines an extended group $G_g^*$. } and that the projective representations of $G_s$ are linear representations of the extended group $G_g^*$,
 the SFP can be physically understood as `gauging' the symmetry group which results in a larger gauge group $G_g^*$.   
For the (non-anomalous) Abelian $\nu_i$, $G_g^*$ can be obtained by  the procedures  in Appendix~\ref{section_gauging_loops}. The algorithm presented in the appendix provides an efficient route to   $G_g^*$. {The SFL, on the other hand, are closely related to (but not the same with) the (twisted) gauge theory after the gauging.}
 
 {With} this observation, we \textit{conjecture}   that
 \begin{align}\mathcal{C}_{w}(\nu_i) \cong \mathcal H^4 (G_g\leftthreetimes_{\nu_i}G_s, U (1)) / \Gamma_{\omega} (\nu_i)\,
 \end{align} for non-anomalous $\nu_i$, where $\Gamma_{\omega} (\nu_i)$  stands  for an `equivalence relation' in the sense that both of the topological order and the  SF patterns are equivalent.  Therefore, the classification of symmetry fractionalization in 3D SET for a bosonic topological order $\omega$ with  gauge group $G_g$ and  symmetry $G_s$  {reads:}   
 \begin{align}  
  \Sigma_\omega\cong\bigoplus_{\nu_i \in H_{non}^2 (G_s, G_g)} \mathcal H^4 ( G_g\leftthreetimes_{\nu_i}G_s, U (1)) /\Gamma_\omega ({\nu_i}) \,,
 \end{align} where we have collected all non-anomalous $\nu_i$ as $H_{non}^2 (G_s, G_g)$.
  The abstract dependence of $\Gamma_\omega (\nu_i)$ on $\omega$ and $\nu_i$ is complicated.
 Despite the lack of abstract formalism, our present down-to-earth  field-theoretical analysis provides some insights in the form of $\Gamma_\omega (\nu_i)$.  {In the following we analyze the form of $\Gamma_\omega (\nu_i)$ via some examples.}

For the $\Z_2$ topological order labeled by $w_0$ enriched by $\Z_2\times \Z_2$ symmetry, we consider two cases. \textbf{(i)} SFP is trivial (i.e., $e00$, denoted by $\nu_0$), the classification of SFL is $(\Z_2)^6$, while for the trivial SFP, $G_g^*=(\Z_2)^3$ and  $\mathcal{H}^4((\Z_2)^3, U(1))=(\Z_2)^8$. Then we can see that  $\Gamma_{\omega_0}(\nu_0)=(\Z_2)^2$, such that $\mathcal{H}^4((\Z_2)^3, U(1))/\Gamma_{\omega_0}(\nu_0)=(\Z_2)^6$. In this case, the $\Gamma_{\omega_0}(\nu_0)$ has physical meaning that they are the stacking $\Z_2\times \Z_2$ SPT. \textbf{(ii)} SFP is $eC0$ (denoted by $\nu_1$), the group $G_g^*=\Z_2\times \Z_4$, which implies $\mathcal{H}^4(G_g^*, U(1))=(\Z_2)^2$, and the classification of SFL is $\Z_2$, then we can see that $\Gamma_{\omega_0}(\nu_1)=\Z_2$. The physical meaning of this $\Gamma_{\omega_0}(\nu_1)$ is interesting and from our calculation in Sec.~\ref{section_Z2_Z2Z2_computing_SFL_for_SFP_eC0_e0C}, it is related to the trivalization of M3L statistical phases due to the presence of fractionalized symmetry charge carried by gauge charge. 

Another interesting example is untwisted $\Z_2\times \Z_2$ topological order (denoted as $ w_0$) enriched by $\Z_2$ symmetry. As for trivial SFP $e_10e_20$ (denoted by $\nu_0$), the group $G_g^*$ is $(\Z_2)^3$ which implies that $\mathcal{H}^4(G_g^*,U(1))=(\Z_2)^8$, and the classification of SFL now is $(\Z_2)^6$, then we can see that $\Gamma_{ w_0}( \nu_0)=(\Z_2)^2$. The physical meaning of this $\Gamma_{ w_0}( \nu_0)$ is that it is related to the classification of $\Z_2\times \Z_2$ gauge theory.

 \section{Symmetry fractionalization in untwisted topological orders: taking $\Z_4$ topological order enriched by $\Z_4$  symmetry as an example}
  \label{SET:ZN}
  
   From this section to Sec.~\ref{sec_example_znzkzk}, we present three  typical  examples, \textit{i.e.}, $\Z_4$ topological order enriched by $\Z_4$  symmetry, $\Z_{2} \times \Z_{2}$  topological order enriched by $\Z_2$ symmetry and $\Z_2$ topological order enriched by $\Z_2\times \Z_2$ symmetry.  More examples can be found in appendices.

  In the present section, we present how to {obtain} the classification  for  $\Z_4$ topological order enriched by $\Z_4$ symmetry.  First, we   derive classification of SFP. Then,  for each SFP class, we derive SFL classification. 

 \subsection{Computation of SFP}

 \label{sfgg1}
According to Eq.~(\ref{eq_mathcalc}),   there are four different  SFP types,  \textit{i.e.},  $\mathcal{C}=0, \frac{1}{2}, \pm \frac{1}{4}$ fractionalized symmetry charge carried by the elementary particles, which we {denote as} $e0, eC, eQ_{\pm}$ respectively  (see Table~\ref{SFGgZ4Z4}). Correspondingly, the reduced charge matrices are only one dimensional and  $\mathcal{Q}=\mathcal{Q}_{11}=0, \,2, \,\pm 1$. Therefore, SFP is classified by $\Z_4$, which is consistent to   $\mathcal{H}^2 (\Z_4, \Z_4)=\Z_4$. Upon gauging, the resulting gauge groups are $\Z_4\times \Z_4$, $\Z_2\times \Z_8$, $\Z_{16}$ and $\Z_{16}$ respectively. 
 \begin{table}[h]
    \centering  \caption{Four  SFP types  for $G_g=\Z_4$ and $G_s=\Z_4$.   $\mathcal{C}$ denotes the SFP matrix defined in Eq.~(\ref{eq_mathcalc}). $\hat{Q}$ is defined in Eq.~(\ref{eqn_definition_hat_Q_matrix}). $G_g^*$ is the gauge group after gauging the symmetry.}
    \begin{tabular}{ccccc}
    \hline
    
    \hline
  \text{ }  \text{ } \text{ } SFP  \text{ } \text{ }\text{ }&$Q_{11}$&  \text{ } \text{ }  \quad\quad  $\mathcal{C}$ \quad\quad \text{ } \text{ }&   $\hat{Q}_{11}$  & \text{ } \text{ } \text{ }    $G_g^*$ \text{ }\text{ } \text{ } \text{ } \\
            \hline
     $e0$  &$0$& $0\text{ mod }1$  & 4& $\Z_{4} \times \Z_{4}$ \\
      $eC$&$2$    &   ${1}/{2}\text{ mod }1$  & $2$& $\Z_{2}\times \Z_8$   \\
      $eQ_+$ &$1$   & ${1}/{4}\text{ mod }1$  & $ 1$&  $\Z_{16}$     \\
      $eQ_-$  &$3$&$-{1}/{4}\text{ mod }1$   & $3$ & $\Z_{16}$ \\
  \hline
  
  \hline
  \end{tabular}
    \label{SFGgZ4Z4}
\end{table}
 \subsection{Minimal model for computing  SFL}\label{section_generality_SFL_action}

We note that, in deriving SFP, twisted terms and trivial layers are not considered. But in deriving SFL, it is crucial to incorporate twisted terms and trivial layers in order to exhaust all possible inequivalent types of SFL. Below, we shall show that, for $\Z_N$ topological order enriched by $\Z_K$ symmetry, among infinite ways of expressing action, it is sufficient to consider a minimal model given by  Eq.~(\ref{eq_main_z4z4_SFL_action0}). For the present example, $N=K=4$.

In general, we can add arbitrarily many but finite trivial layers with level-$1$ (see footnote~\ref{footnote_level1}), labeled by  $i=2,...,1+n'$, which could couple to external $\Z_N$ probe field $A$ by $Q_{1i}$. We  can also add arbitrarily many possible twisted terms that involve one or two   gauge fields of level-$1$, which does not alter bulk topological order as explained below Eq.~(\ref{equation_gamma_gg}). For the purpose of obtaining M3L invariants, we   only need to consider  the twisted terms that at least contain the level-$4$ gauge field $a^1$.   
Thus, we can  divide   twisted terms into two sets:
      \begin{enumerate}[label=\textbf{(\roman*)}]
\item For those involving two different gauge fields  i.e., $a^1$ and $a^i$, such as twisted terms $a^ia^1da^1$ and $a^1a^ida^i$, we collect them as $S_{int}^{1,1}=\frac{1}{(2\pi)^2}\int \sum_{i=2}^{1+n'} q_{i11} a^ia^1da^1+q_{1ii} a^1a^ida^i\,.$
\item For those involving three different   gauge fields, i.e., $a^1$, $a^i$ and $a^j$ with $j\neq i$, such as  twisted terms $a^1a^ida^j$ and $a^1a^jda^i$, we collect them as
 $
S_{int}^{1,2}=\frac{1}{(2\pi)^2}\int \sum_{i<j}^{1+n'} q_{1ij} a^1a^ida^j+q_{1ji} a^1a^jda^i+q_{ij1}a^ia^jda^1\,.$
 \end{enumerate}
As a result, we consider the  action below ($N=K=4$):
\begin{align}
S=&S_0+S_{int}^{1,1}+S_{int}^{1,2}+S_c+S_{sr} \,,\\
 S_0= & \frac{1}{2\pi} \int N b^1   da^1 + \sum_{i=2}^{1+n'} b^i   da^i   \,,\\
 S_c=&\frac{1}{2\pi}\int  Q_{11}  A   db^1 +\sum_{i=2}^{1+n'}Q_{1i}A   db^i  \,,  \\
 S_{sr}=&   \int a^1   *j^1 +  \sum_{i=2}^{1+n'}a^i   *j^i+ b^1   * \Sigma \,,
 \label{}
 \end{align}
where $\Sigma$ is the elementary loop excitation of $\Z_N$ gauge field.

Integrating out $b^1$ and all $b^i$ gauge fields, we obtain
\begin{align}
&a^1=-\frac{2\pi }{N} d^{-1}* \Sigma-\frac{2\pi Q_{11}}{NK} d^{-1}*\sigma\,,\\
&a^i=-\frac{2\pi Q_{1i}}{K} d^{-1}*\sigma\,,
\label{Eqnmotion}
\end{align}
where  $A=\frac{2\pi}{K}d^{-1}*\sigma$. Upon substituting  them   into   twisted terms, we can obtain the aforementioned M3L invariants. More explicitly, from $q_{i11} a^ia^1da^1$, we obtain 
\begin{align}
\frac{q_{i11}}{4\pi^2} a^ia^1da^1&=-\frac{2\pi q_{i11} Q_{1i}}{N^2K} (d^{-1}*\sigma)(d^{-1}*\Sigma)(*\Sigma)\nonumber \\
&\,\,\,\,\,\,\,+\frac{2\pi q_{i11}Q_{11}Q_{1i}}{(NK)^2} (d^{-1}*\Sigma)(d^{-1}*\sigma)(*\sigma)\,,
\end{align}
where we have used the fact that the integration  $\int (d^{-1}*\sigma)(d^{-1}*\sigma)(*\Sigma)$ is equal to zero since for any one-form $f$ and two form $g$ we always have $\int f\wedge f \wedge g=0$. 

Similarly, from $q_{1ii}a^1a^ida^i$, we obtain
\begin{align}
\frac{q_{1ii}}{4\pi^2} a^1a^ida^i&=-\frac{2\pi q_{1ii}Q_{1i}^2}{NK^2}(d^{-1}*\Sigma)(d^{-1}*\sigma)(*\sigma) \,.
\end{align}
And from $q_{1ij}a^1a^ida^j$, we obtain
\begin{align}
\frac{q_{1ij}}{4\pi^2} a^1a^ida^j=-\frac{2\pi q_{1ij}Q_{1i}Q_{1j}}{NK^2}(d^{-1}*\Sigma)(d^{-1}*\sigma)(*\sigma).
\end{align}
(For $q_{1ji}a^1a^jda^i$, the calculation is very similar to $q_{1ij}a^1a^ida^j$, just by exchanging $i$ and $j$.) However, from $q_{ij1}a^ia^jda^1$, there is no contribution to M3L invariant since $\int (d^{-1}*\sigma)(d^{-1}*\sigma)(*\Sigma)$ is vanishing. Therefore, the M3L invariants from all these twisted terms are collected together as follows
\begin{align}
\!\!\!\!\!\!\!\mathcal{I}_{a\alpha \alpha }\!=\!&\sum_{i=2}^{1+n'}  \!q_{i11}\frac{ Q_{11}Q_{1i}}{(NK)^2}\!-\! \sum_{i=2}^{1+n'} \!q_{1ii}\!\frac{Q_{1i}^2}{NK^2}  \!  - \!\sum_{\substack{i\neq j}}^{1+n'} q_{1ij}\frac{  Q_{1i}Q_{1j}}{NK^2}
  \label{eqn_simplified_Ia_a_alpha11},\!\!\\
 \!\!\!\!\!\!\!\!\mathcal{I}_{\alpha aa}\!=\!&-\!\sum_{i=2}^{1+n'}\!q_{i11}\!\frac{   Q_{1i}}{N^2K}\label{eqn_simplified_Ialpha_a_a11}\,,\!\!
\end{align} 
where all $i$ and $j$ take values in $[2,1+n']$.   
Furthermore,  each summation corresponds to a special way of selecting three layers for constituting twisted topological terms. Within each summation in the above two formulas, all summed terms share the same  periods and quantization rules of the coefficients. More concretely, in the summation ``$\sum_{i=2}^{1+n'} \frac{ q_{i11}Q_{11}Q_{1i}}{(NK)^2}=\frac{  Q_{11} }{(NK)^2}\sum_{i=2}^{1+n'}    q_{i11} Q_{1i} $'', $q_{i11} Q_{1i}$ is a function of $Q_{11},N,K,Q_{1i}$, which is fully determined by the 1st and $i$th layers according to Sec.~\ref{sec_twisted_quantized_period_coe}. 
So, as a minimal model, it is enough to consider just one typical term in each summation, e.g., nonzero $q_{211}$ while all other $q_{i11}$'s can be turned off. Likewise, we can further simplify $\mathcal{I}_{a\alpha \alpha } $, which results in ($N=K=4$ in the present example):
\begin{align}
\mathcal{I}_{a\alpha \alpha } 
=& \,q_{211}\frac{ Q_{11}Q_{12} }{(NK)^2}-  \frac{q_{122}Q^2_{12}}{NK^2}    -   \frac{ q_{123}Q_{12}Q_{13}}{NK^2}\label{eqn_simplified_Ia_a_alpha123}\,,\\\mathcal{I}_{\alpha aa}=&-\frac{ q_{211}Q_{12}}{N^2K}\label{eqn_simplified_Ialpha_a_a123}\,.
\end{align} 
In other words, for characterizing SFL, it is sufficient to consider nontrivial $q_{211}$, $q_{122}$ and $q_{123}$, which also means $n'=2$.

  Therefore, without loss of generality, we obtain the following \textit{minimal model}: 
\begin{subequations}
\label{eq_main_z4z4_SFL_action0}
\begin{align}
S=&S_0+S_{int}+S_c+S_{sr}\label{eq_main_z4z4_SFL_action}\,,\\
 S_0= & \frac{1}{2\pi} \int N b^1   da^1 +  b^2   da^2 + b^3   da^3  \,,\\
 S_{int}=& \frac{1}{4\pi^2}  \int q_{211}  a^2  a^1da^1+q_{122}  a^1  a^2da^2+q_{123}  a^1  a^2da^3
 \,, \\
 S_c=&\frac{1}{2\pi}\int  Q_{11}  A   db^1 +   Q_{12}A  db^2+ Q_{13} A   db^3   \,, \\
 S_{sr}=&   \int a^1   *j^1 +  a^2   *j^2+a^3*j^3+ b^1   * \Sigma \,,
 \end{align}
 \end{subequations}
where $b^2,a^2, b^3, a^3$ are from trivial layers, and are called   level-$1$ gauge fields (footnote~\ref{footnote_level1}).

To quantitatively determine the quantization of M3L invariants, one can apply formulas in Sec.~\ref{sec_twisted_quantized_period_coe}. Before calculation, here is one more subtle thing. According to Eq.~(\ref{equation_charge_matrix_equivalent_SFP_purpose}), it is enough to restrict all charge matrix elements $Q_{ij}$ (here $i$ and $j$ are arbitrary layer indices) to the domain $[0,1,\cdots,\gcd(K_i,N_j)-1]$ for the purpose of SFP classification. However, for SFL, we need to extend the domain to a larger one, i.e., 
\begin{align}
Q_{ij}\in [0,1,\cdots,K_i-1]\label{equation_q_domain_SFL}
\end{align}  as it is potentially possible that $Q_{ij}$ and $Q_{ij}+\gcd(K_i,N_j)$ lead to the same SFP but completely distinct SFL. But it is unnecessary to go beyond Eq.~(\ref{equation_q_domain_SFL}). By   noting that $\int A_i$ is quantized at $2\pi/K_i$,   shifting $Q_{ij}$ by $K_i$ only generates $2\pi$ quantized terms in Eq.~(\ref{eq_main_z4z4_SFL_action}),    which doesn't change the partition function of the system.  
Back to the present example, there is only one $K_i$, i.e., $K_i=K=4$. Therefore, $Q_{11},Q_{12},Q_{13}\in\{0,1,2,3\}$, which also uniquely determines $\hat{Q}_{ij}$ via the definition in Eq.~(\ref{eqn_definition_hat_Q_matrix}).

Despite the above fact,  we perform all calculations in Sec.~\ref{SET:ZN} and Sec.~\ref{SET:ZNZN} by $Q_{ij}=1$ for all trivial layers for the sake of simplifying calculation. For   concrete examples we study in these two sections, this special choice   gives us     correct classification. We will study the most general choices of $Q_{ij}$ of trivial layers in Sec.~\ref{sec_example_znzkzk} and appendices. For this reason, below we set $Q_{12}=Q_{13}=1$, which corresponds to $\hat{Q}_{12}=\hat{Q}_{13}=1$. The two M3L invariants reduce to:
 \begin{align}
\mathcal{I}_{a\alpha \alpha } 
=& \,q_{211}\frac{ Q_{11}  }{(NK)^2}-  \frac{q_{122} }{NK^2}    -   \frac{ q_{123} }{NK^2}\label{eqn_simplified_Ia_a_alpha}\,,\\
\mathcal{I}_{\alpha aa}=&-\frac{ q_{211} }{N^2K}\label{eqn_simplified_Ialpha_a_a}\,.
\end{align} 
By setting $m=1, N_1=4, N_2=1, K_1=4$  in Eq.~(\ref{eqn:twisted_coff_general_11}),   we have the following quantization rules:
\begin{align}
q_{211}  =k   M \,, q_{122} =\bar k   M   \,,q_{123}&=\widetilde k  M\label{eqn_example_z4z4_3} \,,
\end{align}
where $k,\bar k$ and $\widetilde k$ are three independent and arbitrary integers, and the minimally quantized value $M$ is given by: 
\begin{align}
M=\frac{16}{\gcd(\hat Q_{11},16)}\,.
\label{eqn_example_z4z4_4} 
\end{align}
 Then, from  Eq.~(\ref{eqn:twisted_coff_general_3}), we can obtain the period of $k,\bar k$:  
\begin{align}
\Gamma=4\,, \text{ i.e., } k\in\Z_4\,, \bar k\in\Z_4.
\label{eqn_example_z4z4_5} 
\end{align}
And from  Eq.~(\ref{eqn:twisted_coff_general_2}), we can obtain the period of $\widetilde k$:
\begin{align}
\widetilde \Gamma=4\,, \text{ i.e., } \widetilde{k}\in\Z_4.
\label{eqn_example_z4z4_6} 
\end{align}

 \subsection{Computation of  SFL for each SFP}\label{section_computation_SFL_for_SFPZ4Z4}

  Below we begin to compute SFL when SFP is given.    The results are summarized in Table \ref{table_z4z4sfl}.

 \subsubsection{Computing SFL when SFP is $e0$ }

First we consider SFL when SFP is $e0$. In this SFP,  {$Q_{11}=0$ (hence $\hat Q_{11}=K=4$), thus   the  elementary particle excitations  carry integer $\Z_4$ symmetry charge}.
From (\ref{eqn_example_z4z4_4}), we have:
\begin{align}
M=4\,.
\end{align}
 Then from  
 from  Eq.~(\ref{eqn_example_z4z4_3}), we can obtain the two M3L invariants   
 $\mathcal{I}_{\alpha aa}$  and $\mathcal{I}_{a\alpha\alpha}$ defined in Eqs.~(\ref{eqn_simplified_Ialpha_a_a}) and (\ref{eqn_simplified_Ia_a_alpha}): 
\begin{align}
\mathcal{I}_{a \alpha \alpha}&=-\frac{   \bar k+\widetilde k}{16}\,,\,\,\,\mathcal{I}_{\alpha aa}=-\frac{ k}{16}\,.
\end{align}
 
By noting that the three integers, i.e., $k$, $\bar k$, and $\widetilde k$, are arbitrary and mutually independent, 
 we may conclude that both M3L invariants are independently quantized at $\frac{1}{16}$. Then, we can obtain quantization rules for all M3L statistical phases listed in  Table~\ref{table_three_M3L_example_list} in Sec.~\ref{mmt_general_discussion_more_General}:  
 $ \theta_{\sigma;\Sigma}=-\theta_{\Sigma, \sigma;\sigma}=\frac{\pi p_1}{8}\,,
 \theta_{\sigma, \sigma;\Sigma}=\frac{\pi  p_1}{4}\,,
 \theta_{\Sigma;\sigma}= -\theta_{\sigma, \Sigma;\Sigma}=\frac{\pi p_2}{8}\,,
 \theta_{\Sigma,\Sigma;\sigma}=\frac{\pi p_2 }{4}$, 
 where $p_1=-(\bar k +\widetilde k)$ and $p_2 =-k$.   
In summary, the six M3L statistical phases are fully controlled by two independent integers and have a very natural period $2\pi$ since the actual physical observables have a form like $e^{i\theta_{...}}$. Nevertheless, this $2\pi$ period is not  the minimal one. The latter is what we shall look for below.

The first period origin comes from Eqs.~(\ref{eqn_example_z4z4_5}) and (\ref{eqn_example_z4z4_6}) which state that $k$, $\bar k$, and $\widetilde{k}$ are periodically identified.  Due to elementary number theory, the integers $p_1$ and $p_2$ also have exactly the same periods, i.e., $p_1\in\Z_4$ and $p_2\in\Z_4$.  Therefore, M3L statistical phases have the following periods:
\begin{align}
 \theta_{\sigma;\Sigma}=&\frac{\pi p_1 }{8} \text{ mod } \frac{\pi}{2}\,,\label{eqn_1337}\\
\theta_{\Sigma, \sigma;\sigma}=&-\frac{\pi p_1 }{8} \text{ mod } \frac{\pi}{2}\,,\\
 \theta_{\sigma, \sigma;\Sigma}= &\frac{\pi p_1}{4} \text{ mod }  \pi\,,\label{eqn_1339}\\ 
 \theta_{\Sigma;\sigma}= &\frac{\pi p_2 }{8} \text{ mod } \frac{\pi}{2}\,,\label{eqn_1341}\\ 
 \theta_{\sigma, \Sigma;\Sigma}=&-\frac{\pi p_2 }{8} \text{ mod } \frac{\pi}{2}\,,\\ 
 \theta_{\Sigma,\Sigma;\sigma}=&\frac{\pi p_2 }{4} \text{ mod } {\pi}\,. \label{eqn_1342}
\end{align}

The second period origin comes from particle attachment.  For each M3L braiding process, there are two loop-like objects that are both linked to another ``base loop'' and are braided around each other. We may always attach  a particle onto braided loop-like objects in the M3L process, then there is a possible phase ambiguity due to particle-loop braiding. This phase ambiguity  provides a new periodic identification for each M3L statistical phase. Then, the final minimal period of each M3L statistical phase is the greatest common divisor of all aforementioned periods, including all kinds of attachment and the period due to periodic $k,\bar k, \widetilde k$. Below, we examine them one by one.

For the M3L statistical phases  $\theta_{\Sigma,\sigma;\sigma}$ and $\theta_{\sigma,\Sigma;\Sigma}$, the elementary loop excitation $\Sigma$ and the unit symmetry flux $\sigma$ are braided around each other, both of which are linked to a base loop. The base loop for   $\theta_{\Sigma,\sigma;\sigma}$ is another $\sigma$ and the base loop for $\theta_{\sigma,\Sigma;\Sigma}$ is another $\Sigma$. Therefore, the elementary particle excitation $e$ can be attached onto $\sigma$ during the M3L braiding process. Since the elementary particle excitation $e$ carries unit $\Z_4$ gauge charge (i.e., $\mathcal{C}=0\text{ mod }1$ in Table~\ref{SFGgZ4Z4}),  there is a  $\frac{\pi}{2}$ phase shift due to the AB phase $\frac{2\pi}{4}=\frac{\pi}{2}$ from the particle-loop braiding between the elementary loop excitation $\Sigma$ and the elementary particle excitation $e$.   As a result, the final period is still   $\frac{\pi}{2}$. That is to say, the above attachment consideration doesn't change anything in the present example.  Finally, we end up with the   result listed in Table~\ref{table_z4z4sfl}:
\begin{align}
\theta_{\Sigma,\sigma;\sigma}=&-\frac{\pi p_1}{8} \, \text{mod }\frac{\pi}{2}=\frac{3\pi p_1}{8} \text{ mod }\frac{\pi}{2}\,,\label{eqn_1final}\\
\theta_{\sigma,\Sigma;\Sigma}=&-\frac{\pi p_2}{8} \, \text{mod }\frac{\pi}{2}=\frac{3\pi p_2}{8} \text{ mod }\frac{\pi}{2}\,,\label{eqn_2final}
\end{align}  
where we have removed minus sign  by means of the periodic shift.
 
  For $\theta_{\sigma,\sigma;\Sigma}$, an $e$  can be attached onto one of   $\sigma$'s in the M3L braiding process. A   $\frac{\pi}{2}$ phase shift   is generated, which is smaller than the period $\pi$ in Eq.~(\ref{eqn_1339}).  Therefore, the final period should be $\text{gcd}(\frac{\pi}{2},\pi)=\frac{\pi}{2}$, and we end up with the result listed in Table~\ref{table_z4z4sfl}:
   \begin{align}
 \theta_{\sigma, \sigma;\Sigma}= &\frac{\pi p_1}{4} \text{ mod }  \frac{\pi}{2}\,.\label{eqn_1339_final} 
\end{align}  

 For $\theta_{\Sigma,\Sigma;\sigma}$, an  $e$ can be attached onto one of $\Sigma$'s in the M3L braiding process.  A   $\frac{\pi}{2}$ phase shift   is generated since there is a $1\times\frac{2\pi}{N}=\frac{2\pi}{4}=\frac{\pi}{2}$ phase once braiding   $e$ around $\Sigma$. This new period is smaller than the period $\pi$ in Eq.~(\ref{eqn_1342}).  Therefore, the final period should be $\text{gcd}(\frac{\pi}{2},\pi)=\frac{\pi}{2}$, and we end up with the result listed in Table~\ref{table_z4z4sfl}:
   \begin{align}
 \theta_{\Sigma,\Sigma;\sigma}=&\frac{\pi p_2 }{4} \text{ mod } {\frac{\pi}{2}}\,. \label{eqn_1342_final}
\end{align}

   For $\theta_{\sigma;\Sigma}$,  in the exchanging braiding of two identical $\sigma$'s, a   $\frac{1}{2} \frac{\pi}{2}+\frac{1}{2}  \frac{\pi}{2}=  \frac{\pi}{2}$ phase shift is caused by attaching  one $e$ to  each $\sigma$  so that both $\sigma$'s still keep identical.  Therefore, the result in Eq.~(\ref{eqn_1337}) is   unchanged, and we rewrite the same formula:
   \begin{align}
 \theta_{\sigma;\Sigma}=&\frac{\pi p_1 }{8} \text{ mod } \frac{\pi}{2}\,\label{eqn_3final}
   \end{align}
   which is listed  in Table~\ref{table_z4z4sfl}.

   For $\theta_{\Sigma;\sigma}$,  in the exchanging braiding of two identical $\Sigma$'s, a   $\frac{1}{2} \frac{\pi}{2}+\frac{1}{2}  \frac{\pi}{2}=  \frac{\pi}{2}$ phase shift is caused by attaching  one $e$ to  each $\Sigma$  so that both $\Sigma$'s still keep identical.  Therefore, the result in Eq.~(\ref{eqn_1341}) is   unchanged, and we rewrite the result again:
   \begin{align}
 \theta_{\Sigma;\sigma}= &\frac{\pi p_2 }{8} \text{ mod } \frac{\pi}{2}\,\label{eqn_1341_final}
   \end{align}
   which is listed  in Table~\ref{table_z4z4sfl}.
   
        Eqs.~(\ref{eqn_1final}), (\ref{eqn_2final}), (\ref{eqn_1339_final}), (\ref{eqn_1342_final}),   (\ref{eqn_3final}), and (\ref{eqn_1341_final}) together form a complete set of physical observables that classify and characterize SFL when SFP is $e0$. The classification of SFL can be formally written as a cyclic group $\Z_4\times \Z_4$, which indicates that there are in total $16$ kinds of topologically distinct SFL classes. We can use the above six M3L braiding statistical phases to distinguish   $16$ classes, which is exhausted in Table~\ref{Table_list_of_all_phases_e0Z4Z4}.

    \begin{table*}[t]
    \centering  \caption{Six M3L statistical phases [Eqs.~(\ref{eqn_1final}), (\ref{eqn_2final}), (\ref{eqn_1339_final}), (\ref{eqn_1342_final}),   (\ref{eqn_3final}), and (\ref{eqn_1341_final})] when SFP is $e0$ in $\Z_4$ topological order enriched by $\Z_4$ global symmetry. As all phases have the same period $\frac{\pi}{2}$, we restrict all phases into   ``the first Brillouin zone'': $[0,\frac{\pi}{2})$ by properly adding or subtracting multiple of $\frac{\pi}{2}$. For SFP $e0$, there are in total $16$ topologically distinct SFL classes each of which is uniquely labeled by a pair of integers $(p_1,p_2)$ with $p_1\in\Z_4$ and $p_2\in\Z_4$. }
    \begin{tabular}{ccccccccccccccccccc}
    \hline
    
    \hline
  $\begin{pmatrix}p_1\\p_2\end{pmatrix}$ &$\begin{pmatrix}0\\0\end{pmatrix}$ &$\begin{pmatrix}0\\1\end{pmatrix}$ &$\begin{pmatrix}0\\2\end{pmatrix}$ &$\begin{pmatrix}0\\3\end{pmatrix}$ &$\begin{pmatrix}1\\0\end{pmatrix}$ &$\begin{pmatrix}1\\1\end{pmatrix}$ &$\begin{pmatrix}1\\2\end{pmatrix}$ &$\begin{pmatrix}1\\3\end{pmatrix}$ &$\begin{pmatrix}2\\0\end{pmatrix}$ &$\begin{pmatrix}2\\1\end{pmatrix}$ &$\begin{pmatrix}2\\2\end{pmatrix}$ &$\begin{pmatrix}2\\3\end{pmatrix}$ &$\begin{pmatrix}3\\0\end{pmatrix}$ &$\begin{pmatrix}3\\1\end{pmatrix}$ &$\begin{pmatrix}3\\2\end{pmatrix}$ &$\begin{pmatrix}3\\3\end{pmatrix}$  \\ 
              \hline
      $\theta_{\Sigma,\sigma;\sigma}$  &$0$&$0$&$0$&$0$&$\frac{3\pi}{8}$&$\frac{3\pi}{8}$&$\frac{3\pi}{8}$&$\frac{3\pi}{8}$&$\frac{\pi}{4}$&$\frac{\pi}{4}$&$\frac{\pi}{4}$&$\frac{\pi}{4}$&$\frac{\pi}{8}$&$\frac{\pi}{8}$&$\frac{\pi}{8}$&$\frac{\pi}{8}$ \\ \\
      $\theta_{\sigma,\sigma;\Sigma}$  &$0$&$0$&$0$&$0$&$\frac{\pi}{4}$&$\frac{\pi}{4}$&$\frac{\pi}{4}$&$\frac{\pi}{4}$&$0$&$0$&$0$&$0$&$\frac{\pi}{4}$&$\frac{\pi}{4}$&$\frac{\pi}{4}$&$\frac{\pi}{4}$\\ \\
     $\theta_{\sigma;\Sigma}$               &$0$&$0$&$0$&$0$ &$\frac{\pi}{8}$ &$\frac{\pi}{8}$ &$\frac{\pi}{8}$ &$\frac{\pi}{8}$ &$\frac{\pi}{4}$&$\frac{\pi}{4}$&$\frac{\pi}{4}$&$\frac{\pi}{4}$&$\frac{3\pi}{8}$&$\frac{3\pi}{8}$&$\frac{3\pi}{8}$&$\frac{3\pi}{8}$ \\ \\
      $\theta_{\sigma,\Sigma;\Sigma}$  &$0$&$\frac{3\pi}{8}$&$\frac{\pi}{4}$&$\frac{\pi}{8}$&$0$&$\frac{3\pi}{8}$&$\frac{\pi}{4}$&$\frac{\pi}{8}$&$0$&$\frac{3\pi}{8}$&$\frac{\pi}{4}$&$\frac{\pi}{8}$&$0$&$\frac{3\pi}{8}$&$\frac{\pi}{4}$&$\frac{\pi}{8}$\\ \\
     $\theta_{\Sigma,\Sigma;\sigma}$      &$0$&$\frac{\pi}{4}$&$0$&$\frac{\pi}{4}$  &$0$&$\frac{\pi}{4}$&$0$&$\frac{\pi}{4}$  &$0$&$\frac{\pi}{4}$&$0$&$\frac{\pi}{4}$  &$0$&$\frac{\pi}{4}$&$0$&$\frac{\pi}{4}$\\  \\
      $\theta_{\Sigma;\sigma}$   &$0$&$\frac{\pi}{8}$&$\frac{\pi}{4}$&$\frac{3\pi}{8}$&$0$&$\frac{\pi}{8}$&$\frac{\pi}{4}$&$\frac{3\pi}{8}$&$0$&$\frac{\pi}{8}$&$\frac{\pi}{4}$&$\frac{3\pi}{8}$&$0$&$\frac{\pi}{8}$&$\frac{\pi}{4}$&$\frac{3\pi}{8}$\\
   
  \hline
  
  \hline
  \end{tabular}
    \label{Table_list_of_all_phases_e0Z4Z4}
\end{table*}

   \subsubsection{Computing SFL when SFP is $eC$ }

 In the SFP denoted as $eC$, $Q_{11}=2$, \textit{i.e.}, $\hat Q_{11}=2$, thus  the elementary particle excitation $e$    carries half integer $\Z_4$ symmetry charge, i.e., $\mathcal{C}=\frac{1}{2}$ in Table~\ref{SFGgZ4Z4}.  
 From (\ref{eqn_example_z4z4_4}), we have:
\begin{align}
M=8\,.
\end{align}
 Then from  
 Eq.~(\ref{eqn_example_z4z4_3}), we can obtain the two M3L invariants   
 $\mathcal{I}_{\alpha aa}$  and $\mathcal{I}_{a\alpha\alpha}$ defined in Eqs.~(\ref{eqn_simplified_Ialpha_a_a}) and (\ref{eqn_simplified_Ia_a_alpha}): 
\begin{align}
\mathcal{I}_{a \alpha \alpha}=-\frac{ k }{16}-\frac{   \bar k+\widetilde k}{8}\,,\,\,\,\mathcal{I}_{\alpha aa}&=-\frac{ k}{8}\,.
\end{align}
  It is convenient to introduce two integers $p_3$ and $p_4$ via $p_3=k$ and $p_4=-(\bar k+\widetilde k)$. Then, these two integers are mutually independent and have the same period: $p_3\in\Z_4$ and $p_4\in\Z_4$ according to Eqs.~(\ref{eqn_example_z4z4_5}) and (\ref{eqn_example_z4z4_6}). Thus, $p_3+2p_4$ also has the same period: $p_3+2p_4\in\Z_4$ according to the elementary number theory. The six M3L statistical phases can be expressed as:
  \begin{align}
 \theta_{\sigma;\Sigma}=&\frac{\pi p_3 }{8} +\frac{\pi p_4}{4}\text{ mod } \frac{\pi}{2}\,,\label{eqn_133711}\\
\theta_{\Sigma, \sigma;\sigma}=&-\frac{\pi p_3 }{8} -\frac{\pi p_4}{4}\text{ mod } \frac{\pi}{2}\,,\\
 \theta_{\sigma, \sigma;\Sigma}= &\frac{\pi p_3 }{4} +\frac{\pi p_4}{2}\text{ mod } {\pi},\label{eqn_133911}\\ 
 \theta_{\Sigma;\sigma}= &-\frac{\pi p_3}{4} \text{ mod } {\pi}\,,\label{eqn_134111}\\ 
 \theta_{\sigma, \Sigma;\Sigma}=&\frac{\pi p_3}{4} \text{ mod } {\pi}\,,\\ 
 \theta_{\Sigma,\Sigma;\sigma}=&-\frac{\pi p_3 }{2} \text{ mod } {2\pi}\,. \label{eqn_134211}
\end{align}
 Next, we will consider the effect of particle attachment. 
 
   For both $\theta_{\Sigma,\sigma;\sigma}$ and  $\theta_{\sigma,\Sigma;\Sigma}$, one can consider the attachment of $e$ onto braided $\Sigma$. Since the latter is braided around $\sigma$, $e$ carrying $\frac{1}{2}$ symmetry charge is simultaneously braided around the unit $\Z_4$ symmetry flux $\sigma$, rendering  a $\frac{1}{2}\times\frac{2\pi}{4}=\frac{\pi}{4}$ phase shift. This $\frac{\pi}{4}$ phase shift changes the periods of the two M3L statistical phases  to $\frac{\pi}{4}$, leading to the result listed in Table~\ref{table_z4z4sfl}:
\begin{align}
\theta_{\Sigma, \sigma;\sigma}=&-\frac{\pi p_3 }{8} \, (\text{mod } \frac{\pi}{4})=\frac{\pi p_3 }{8} \text{ mod } \frac{\pi}{4}\,,\label{eqn_1final1111}\\
 \theta_{\sigma, \Sigma;\Sigma}=&0\text{ mod }\frac{\pi}{4}\,.\label{eqn_2final1111}
\end{align}  
In Eq.~(\ref{eqn_1final1111}), $p_4$-term is exactly multiple of the minimal period $\pi/4$, and is removed. Likewise,  in Eq.~(\ref{eqn_2final1111}), $p_3$-term is exactly multiple of the minimal period  $\pi/4$, and is removed.
 
  For $\theta_{\sigma,\sigma;\Sigma}$, an $e$ can be attached onto one of   $\sigma$'s in the M3L braiding process. A   $\frac{\pi}{4}$ phase shift   is generated, which is smaller than the period $\pi$ in Eq.~(\ref{eqn_133911}).  Therefore, the final period should be $\text{gcd}(\frac{\pi}{4},\pi)=\frac{\pi}{4}$, and we end up with the result listed in Table~\ref{table_z4z4sfl}:
   \begin{align}
 \theta_{\sigma, \sigma;\Sigma}= &0 \text{ mod }  \frac{\pi}{4}\,.\label{eqn_1339_final11111} 
\end{align}

 For $\theta_{\Sigma,\Sigma;\sigma}$, an $e$ can be attached onto one of $\Sigma$'s in the M3L braiding process.  A   $\frac{\pi}{2}$ phase shift   is generated, which is smaller than the period $\pi$ in Eq.~(\ref{eqn_134211}).  Therefore, the final period should be $\text{gcd}(\frac{\pi}{2},\pi)=\frac{\pi}{2}$, and we end up with the result listed in Table~\ref{table_z4z4sfl}:
   \begin{align}
 \theta_{\Sigma,\Sigma;\sigma}=&0 \text{ mod } {\frac{\pi}{2}}\,. \label{eqn_1342_final11111}
\end{align}

   For $\theta_{\sigma;\Sigma}$,  in the exchanging braiding of two identical $\sigma$'s, a   $\frac{1}{2} \frac{\pi}{4}+\frac{1}{2}  \frac{\pi}{4}=  \frac{\pi}{4}$ phase shift is caused by attaching  one $e$ onto  each $\sigma$  so that both $\sigma$'s still keep identical.  Therefore,  we have:
   \begin{align}
 \theta_{\sigma;\Sigma}=&\frac{\pi p_3 }{8} \text{ mod } \frac{\pi}{4}\,\label{eqn_3final1111}
   \end{align}
   which is listed  in Table~\ref{table_z4z4sfl}.

   For $\theta_{\Sigma;\sigma}$,  in the exchanging braiding of two identical $\Sigma$'s, a   $\frac{1}{2} \frac{\pi}{2}+\frac{1}{2}  \frac{\pi}{2}=  \frac{\pi}{2}$ phase shift is caused by attaching  one $e$ to  each $\Sigma$  so that both $\Sigma$'s still keep identical.  Therefore, the result in Eq.~(\ref{eqn_134111}) is   unchanged, and we rewrite the result again:
   \begin{align}
 \theta_{\Sigma;\sigma}= &-\frac{\pi p_3}{4} \,\text{mod } \frac{\pi}{2}=\frac{\pi p_3}{4} \text{ mod } \frac{\pi}{2}\, \label{eqn_1341_final1111}
   \end{align}
   which is listed  in Table~\ref{table_z4z4sfl}.
    
        Eqs.~(\ref{eqn_1final1111}), (\ref{eqn_2final1111}), (\ref{eqn_1339_final11111}), (\ref{eqn_1342_final11111}),   (\ref{eqn_3final1111}), and (\ref{eqn_1341_final1111}) together form a complete set of physical observables that classify and characterize SFL when SFP is $eC$. It is apparent that $p_4$ disappears in these formulas. The remaining parameter $p_3$ is periodically identified as $p_3\sim p_3+4$. However, upon carefully tracking  how the six M3L statistical phases change with respect to $p_3$, we find that $p_3$ has a smaller period: $p_3\sim p_3+2$, i.e., $p_3\in\Z_2$. Thus, the classification of SFL is given by $\Z_2$.  We can use the above six M3L braiding statistical phases to distinguish   the two SFL classes, which is exhausted in Table~\ref{Table_list_of_all_phases_eCZ4Z4}.
  \begin{table}[t]
    \centering  \caption{Six M3L statistical phases [Eqs.~(\ref{eqn_1final1111}), (\ref{eqn_2final1111}), (\ref{eqn_1339_final11111}), (\ref{eqn_1342_final11111}),   (\ref{eqn_3final1111}), and (\ref{eqn_1341_final1111})] when SFP is $eC$ in $\Z_4$ topological order enriched by $\Z_4$ global symmetry.  We restrict all phases into   their own ``the first Brillouin zone''. For SFP $eC$, there are in total $2$ topologically distinct SFL classes each of which is uniquely labeled by an integer $p_3$ with $p_3\in\Z_2$. }
    \begin{tabular}{ccccccccccccccccccc}
    \hline
    
    \hline
   $p_3$   &
      $\theta_{\Sigma,\sigma;\sigma}$  &
      $\theta_{\sigma,\sigma;\Sigma}$   & 
     $\theta_{\sigma;\Sigma}$            &  
      $\theta_{\sigma,\Sigma;\Sigma}$   &
     $\theta_{\Sigma,\Sigma;\sigma}$     &  
      $\theta_{\Sigma;\sigma}$   \\\hline
      $0$&$0$&$0$&$0$&$0$&$0$&$0$\\
      $1$&$\frac{\pi}{8}$&$0$&$\frac{\pi}{8}$&$0$&$0$&$\frac{\pi}{4}$ \\
   
  \hline
  
  \hline
  \end{tabular}
    \label{Table_list_of_all_phases_eCZ4Z4}
\end{table}

  \subsubsection{Computing SFL when SFP is either $eQ_{+}$ or $eQ_{-}$ }

  For these cases, $Q_{11}=1$ or 3 for $eQ_{\pm}$, so $e$ can  carry  (plus or minus) one fourth  $\Z_4$ symmetry charge. From (\ref{eqn_example_z4z4_4}), we have $M=16$, and then from  Eq.~(\ref{eqn_example_z4z4_3}), we have
  \begin{align}
\mathcal{I}_{\alpha aa}&=-\frac{ k }{4}\,,\,\,\,\,\mathcal{I}_{a \alpha \alpha}=\frac{ k Q_{11} }{16}-\frac{    \bar k +\widetilde k }{4}\,.
\end{align}
 As such, the  quantized values of the
 statistical phases  $  \theta_{\sigma, \Sigma;\Sigma}=-\theta_{ \Sigma;\sigma}={\pi k }/{2}$, $\theta_{ \Sigma,\Sigma;\sigma}=-\pi k $ and $ \theta_{\sigma;\Sigma}=-\theta_{\Sigma, \sigma;\sigma}={\pi  (k-4\bar k -4\widetilde k )  }/{8}$ and $\theta_{\sigma, \sigma;\Sigma}={\pi  (k-4\bar k -4\widetilde k)  }/{4}$.  

Next we consider particle attachment.  In the present SFP, $e$ carries $\pm\frac{1}{4}$ symmetry charge, which leads to a $\frac{1}{4}\times\frac{2\pi}{4}=\frac{\pi}{8}$ phase shift if braiding $e$ around a unit $\Z_4$ symmetry flux $\sigma$ in a M3L braiding process.   Again, there is a $1\times\frac{2\pi}{4}=\frac{\pi}{2}$ phase shift if braiding $e$ around an elementary loop excitation $\Sigma$ in a M3L braiding process. By carefully taking these two kinds of phase shift into account, following the similar discussion, we obtain the result listed in Table~\ref{table_z4z4sfl}. The result indicates that if SFP is either $eQ_+$ or $eQ_-$, there is no nontrivial SFL. Formally, we use $\Z_1$ to denote the classification of SFL when SFP is either  $eQ_+$ or $eQ_-$. 

 In summary, the total classification of $\Z_4$ topological order enriched by $\Z_4$ global symmetry is given by: $(\Z_4\times \Z_4)\oplus \Z_2\oplus \Z_1\oplus\Z_1$, as also shown in Table~\ref{table_z4z4sfl}. When SFP is $e0$,  $e$ carries integer symmetry charge and SFL is classified by $\Z_4\times\Z_4$. When SFP is $eC$, $e$ carries half-integer symmetry charge and SFL is classified by $\Z_2$. When SFP is either $eQ_+$ or $eQ_-$, $e$ carries either $1/4$ or $-1/4$ symmetry charge and SFL is always trivial (denoted as $\Z_1$).
In Appendix~\ref{sec_app_znzk}, we present details for  general $\Z_N$ topological orders enriched by $\Z_K$ symmetry.   All results are collected in Table~\ref{znset}.

\section{Symmetry fractionalization in twisted topological orders: taking $\Z_{2} \times \Z_{2}$  topological orders enriched by $\Z_2$ symmetry as an example}
\label{SET:ZNZN}

In this section, we study $\Z_{2} \times \Z_{2}$  topological orders enriched by $\Z_2$ symmetry. In contrast to the previous section where there is only one $\Z_2$ topological order,  $\Z_{2} \times \Z_{2}$  topological orders include one untwisted topological order and three twisted topological orders (two of them are identical to each other upon group automorphism). To distinguish them, it is necessary to calculate three-loop braiding statistical phases (called ``intrinsic three-loop'' or ``I3L'' in the following texts) among three elementary loop excitations. After imposing symmetry, we need to collect I3L and M3L statistical phases together, from which we can read symmetry fractionalization and its connection to a specific  $\Z_2\times \Z_2$ topological order.

This section is organized as follows. In Sec.~\ref{section_z2z2_z2_I3L}, we compute I3L statistical phases  (Table~\ref{Table_Z2Z2bulk_topo_order_LIST}) in order to identify topological orders with the same gauge group $G_g=\Z_2\times\Z_2$.   In Sec.~\ref{section_computing_sfgg2}, we compute SFP and collect results into Table~\ref{tab_SFGg2}. In Sec.~\ref{SF_loop_z2z2_z2}, we derive the minimal model [Eq.~(\ref{eqn_action_z2z2_z2})] that is served as the starting point for the computation of SFL. 
In Sec.~\ref{section_concrete_example_of_anomaly_z2z2z2}, we demonstrate the quantum anomaly in $\Z_{2} \times \Z_{2}$  topological orders enriched by $\Z_2$ symmetry.  In Sec.~\ref{section_Z2Z2_Z2_computing_SFL_for_SFP}, we concretely compute SFL of each SFP, which renders the final classification of symmetry fractionalization of $\Z_{2} \times \Z_{2}$  topological orders enriched by $\Z_2$ symmetry.

\subsection{I3L statistical phases for  $\Z_{2} \times \Z_{2}$   topological orders}\label{section_z2z2_z2_I3L}
  
 In contrast  to 3D $\Z_N$ topological order, $\Z_{N_1}\times \Z_{N_2}$ topological order can have many phases even in the absence of symmetry. In terms of twisted gauge theory language, when the gauge group is $\Z_{N_1}\times \Z_{N_2}$, there can be  nontrivial twisted terms in Eq.~(\ref{eq:action_of_pure_gauge_3lb}): 
 $ S= \int \!\!\frac{1}{2\pi}  \sum_{i=1}^2  N_i b^i da^i 
\!+ \! \frac{q_{122}}{4\pi^2}  a^1 a^2 da^2 \!+\!\frac{{q}_{211}}{4\pi^2} a^2a^1 da^1\,, $ 
where $q_{122}=k \frac{N_1N_2}{N_{12}},  q_{211}=\bar k \frac{N_1N_2}{N_{12}}, $ with $k,\bar k\in \Z_{N_{12}}$ ($N_{12}$ is the greatest common divisor of $N_1$ and $N_2$).

   Physically,   three-loop braiding statistics among nontrivial elementary loop excitations, i.e., $\Sigma^1$ and $\Sigma^2$ can be used to distinguish different topological orders with the same gauge group $G_g=\Z_{N_1}\times \Z_{N_2}$. More specifically, I3L statistical phases can be obtained from the following effective action of $\Sigma^1$ and $\Sigma^2$ ($N_1=N_2=N$ is assumed):
   \begin{align}
S_{TL}= &\int \frac{- 2\pi q_{122}}{N^3} (d^{-1}*\Sigma^1)(d^{-1}*\Sigma^2)(*\Sigma^2)\nonumber\\
 &+  
\frac{-2\pi q_{211}}{N^3} (d^{-1}*\Sigma^2)(d^{-1}*\Sigma^1)(*\Sigma^1)\,.\label{eqn_I3L}
\end{align} 
The two terms in $S_{TL}$ of Eq.~(\ref{eqn_I3L}) lead to the following I3L statistical phases ($N=2$):   
   \begin{align}
&   \theta_{\Sigma^2;\Sigma^1}=-\theta_{\Sigma^1,\Sigma^2;\Sigma^2}=\frac{\pi q_{122}}{4}\,,\theta_{\Sigma^2,\Sigma^2;\Sigma^1}=\frac{\pi q_{122}}{2} \,,\label{eqn_three_loop_braiding_z2z2_line1}
\\
&\theta_{\Sigma^1;\Sigma^2} =-\theta_{\Sigma^1,\Sigma^2;\Sigma^1}= \frac{\pi q_{211}}{4}\,,\theta_{\Sigma^1,\Sigma^1;\Sigma^2}=\frac{\pi q_{211}}{2} \,.\label{eqn_three_loop_braiding_z2z2_line2}
\end{align}     
 In the absence of symmetry, the coefficients $q_{122}$ and $q_{211}$ are quantized and periodic in the following way: 
 \begin{align}
 q_{122}=2k ,q_{211}=2\bar k\label{eqn_without_symmetry_z2z2_qq}
 \end{align}
  with $k,\bar k\in \Z_2$ by applying the universal formula (\ref{eqn_eqi_rela^2}), which leads to:
   \begin{align}
&   \theta_{\Sigma^2;\Sigma^1}= \frac{\pi k}{2}\text{  mod }\pi\,,  \, \theta_{\Sigma^1,\Sigma^2;\Sigma^2}=\frac{\pi k}{2}\text{  mod }\pi\,, \nonumber\\
&\theta_{\Sigma^2,\Sigma^2;\Sigma^1}= \pi k \text{  mod }2\pi\,,\theta_{\Sigma^1;\Sigma^2}  = \frac{\pi \bar k }{2}\text{  mod }\pi\,, \nonumber\\
&\theta_{\Sigma^1,\Sigma^2;\Sigma^1}= \frac{\pi \bar k }{2}\text{  mod }\pi\,, \,\theta_{\Sigma^1,\Sigma^1;\Sigma^2}= \pi \bar k\text{  mod }2\pi \,.\nonumber
\end{align} 
   We should also consider the effect of attaching particles. In the present topological order, there is a $\pi$ phase  shift if we braid $e_i$ (i=1,2) around $\Sigma^i$(i=1,2).  As a result, we obtain the following new set of phases:
   \begin{align}
&   \theta_{\Sigma^2;\Sigma^1}= \frac{\pi k}{2}\text{  mod }\pi\,,  \label{eqn_pure_z2z2_topo_order_3loopdata1}\\  &\theta_{\Sigma^1,\Sigma^2;\Sigma^2}=\frac{\pi k}{2}\text{  mod }\pi\,, \label{eqn_pure_z2z2_topo_order_3loopdata2}\\
&\theta_{\Sigma^2,\Sigma^2;\Sigma^1}= 0 \text{  mod }\pi\,,
\label{eqn_pure_z2z2_topo_order_3loopdata3}\\
&\theta_{\Sigma^1;\Sigma^2}  = \frac{\pi \bar k }{2}\text{  mod }\pi\,,\label{eqn_pure_z2z2_topo_order_3loopdata4}\\
&  \theta_{\Sigma^1,\Sigma^2;\Sigma^1}= \frac{\pi \bar k }{2}\text{  mod }\pi\,,\label{eqn_pure_z2z2_topo_order_3loopdata5}\\
&\theta_{\Sigma^1,\Sigma^1;\Sigma^2}= 0  \text{  mod }\pi \,.\label{eqn_pure_z2z2_topo_order_3loopdata6}
\end{align}

Thus, by using these six braiding statistical phases to characterize the bulk topological order, one can realize that there are totally four    distinct $\Z_2\times \Z_2$ topological orders denoted as $\mathsf{TO_1}$, $\mathsf{TO_2}$, $\mathsf{TO_3}$, $\mathsf{TO_4}$\footnote{In fact, $\mathsf{TO_2}$ and $\mathsf{TO_3}$ are the same topological order since they just relate each other by relabelling fluxes (mathematically, they are related by a group automorphism action). }.  We list their physical properties in Table~\ref{Table_Z2Z2bulk_topo_order_LIST}. 
 \begin{table*}[t]
    \centering  \caption{I3L statistical phases  [Eqs.~(\ref{eqn_pure_z2z2_topo_order_3loopdata1}),~(\ref{eqn_pure_z2z2_topo_order_3loopdata2}),~(\ref{eqn_pure_z2z2_topo_order_3loopdata3}),~(\ref{eqn_pure_z2z2_topo_order_3loopdata4}),~(\ref{eqn_pure_z2z2_topo_order_3loopdata5}), and~(\ref{eqn_pure_z2z2_topo_order_3loopdata6})]  of $\Z_2\times\Z_2$ topological orders.  $\mathsf{TO_2}$ and $\mathsf{TO_3}$ are identical to each other upon group automorphism.}
    \begin{tabular}{ccccccccccccccccccc}
    \hline
    
    \hline
     Topological Orders    &\quad\quad
      $\theta_{\Sigma^2;\Sigma^1}$  \quad\quad&\quad\quad
      $\theta_{\Sigma^1,\Sigma^2;\Sigma^2}$   \quad\quad& \quad\quad
     $\theta_{\Sigma^2,\Sigma^2;\Sigma^1}$           \quad\quad & \quad\quad 
      $\theta_{\Sigma^1;\Sigma^2}$ \quad\quad  &\quad\quad
     $\theta_{\Sigma^2,\Sigma^1;\Sigma^1}$    \quad\quad &  \quad\quad
      $\theta_{\Sigma^1,\Sigma^1;\Sigma^2}$  \quad\quad \\\hline
      $\mathsf{TO_1}$&$0\text{ mod } \pi$ &$0\text{ mod } \pi$ &$0\text{ mod } \pi$ &$0\text{ mod } \pi$ &$0\text{ mod } \pi$ &$0\text{ mod } \pi$ \\
       $\mathsf{TO_2}$&$\frac{\pi}{2}\text{ mod } \pi$ &$\frac{\pi}{2}\text{ mod } \pi$ &$0\text{ mod } \pi$ &$0\text{ mod } \pi$ &$0\text{ mod } \pi$ &$0\text{ mod } \pi$ \\
        $\mathsf{TO_3}$&$0\text{ mod } \pi$ &$0\text{ mod } \pi$ &$0\text{ mod } \pi$ &$\frac{\pi}{2}\text{ mod } \pi$ &$\frac{\pi}{2}\text{ mod } \pi$ &$0\text{ mod } \pi$ \\
         $\mathsf{TO_4}$& $\frac{\pi}{2}\text{ mod } \pi$ &$\frac{\pi}{2}\text{ mod } \pi$ &$0\text{ mod } \pi$ &$\frac{\pi}{2}\text{ mod } \pi$ &$\frac{\pi}{2}\text{ mod } \pi$ &$0\text{ mod } \pi$  \\ 
   
  \hline
  
  \hline
  \end{tabular}
    \label{Table_Z2Z2bulk_topo_order_LIST}
\end{table*}
This table will be useful for identify topological orders after symmetry is imposed.

%
 \subsection{Computation of SFP}
\label{section_computing_sfgg2}
 
 \begin{table*}[t] \caption{Four  SFP types for $G_g=\Z_2\times \Z_2$ topological order and $G_s=\Z_2$ symmetry.  $\mathcal{C}$ denotes the SFP matrix defined in Eq.~(\ref{eq_mathcalc}). $\mathcal{C}_{11}$ and $\mathcal{C}_{12}$ are  the fractionalized symmetry charges carried by the elementary particle excitation $e_1$ and $e_2$ respectively.  $\hat{Q}$ is defined in Eq.~(\ref{eqn_definition_hat_Q_matrix}). $G_g^*$ is the gauge group after gauging the symmetry.      }
    \label{tab_SFGg2}
    \centering
    \begin{tabular}{ccccc}
    \hline
     
    \hline
  \text{ }  SFP  \text{ }&\quad\quad\quad$(Q_{11},Q_{12})$\quad\quad\quad& \quad \quad\quad\quad  $(\mathcal{C}_{11},\mathcal{C}_{12})$\quad\quad \quad \quad&   $ (\hat{Q}_{11},\hat{Q}_{12})$ &\quad\quad \quad\quad$G_g^*$\quad\quad \quad\quad \\
            \hline
     $e_10e_20$  &$(0,0)$&    $(0\text{ mod }1, 0\text{ mod }1)$ & (2,2)& $\Z_{2}  \times \Z_{2} \times \Z_2$\\
     $e_1Ce_20$ &$(1,0)$  &    $(\frac{1}{2}\text{ mod }1, 0\text{ mod }1)$ &  ($1,2$) & $\Z_{4} \times \Z_{2} $\\
     $e_10e_2C$   &$(0,1)$&    $(0\text{ mod }1, \frac{1}{2}\text{ mod }1)$  & ($2, 1$) &$\Z_{2} \times \Z_{4} $ \\
     $e_1Ce_2C$   &$(1,1)$&    $ (\frac{1}{2}\text{ mod }1, \frac{1}{2}\text{ mod }1)$ &($1,1$)  &$\Z_{4} \times \Z_{2}$\\
  \hline

  \hline
  \end{tabular}
   
\end{table*}

 From  (\ref{eq_mathcalc}), depending on whether $e_1 (e_2)$ carries integer or half integer symmetry charge, there are four different SFP, denoted as $e_10 (C)e_20 (C)$ (see Table~\ref{tab_SFGg2}).
 The corresponding   charge matrices are $ (\hat{Q}_{11}, \hat{Q}_{12})= (2,2),  (2,1), (1,2), (1,1)$.   Upon gauging, the resulting gauge groups  are $(\Z_2)^3$, $\Z_2\times \Z_4$, $\Z_4\times \Z_2$, and   $\Z_2\times \Z_4$.

 \subsection{Minimal model for computing SFL}
  \label{SF_loop_z2z2_z2}

  Here we will derive the minimal model, to be given in Eq.~(\ref{eqn_action_z2z2_z2}), which is  sufficient  for the purpose of classification of SFL. We start from the most general action and then derive the most general form of different M3L invariants, through which we can show the action  (\ref{eqn_action_z2z2_z2}) is sufficient enough for classification of SFL. The derivations for the minimal model apply to  $\Z_N\times \Z_N$ toplogical order enriched by $\Z_K$ symmetry with general $N$ and $K$. For the present example, $N=K=2$.

We first consider different types of twisted terms that could be related to M3L invariants, which are classified in four classes ($\nu=1,2; i,j=3,4,\cdots, 2+n'$;  $n'$ is the total number of trivial layers):
      \begin{enumerate}[label=\textbf{(\roman*)}]
\item Twisted terms with  gauge fields from  two nontrivial layers.  
 $S_{int}^0=\frac{1}{(2\pi)^2} \int q_{122} a^1a^2da^2 + q_{211} a^2a^1da^1$. 
These two terms determine what topological order belongs to;
\item Twisted terms with  gauge fields from  one nontrivial layer and one trivial layer.   
$
S_{int}^{1,1}=\frac{1}{(2\pi)^2} \int \sum_{i=3}^{2+n'} q_{\nu ii} a^\nu a^ida^i + q_{i\nu\nu} a^ia^\nu da^\nu$; 
 \item Twisted terms with  gauge fields from one nontrivial layer and two trivial layers.  
 $S_{int}^{1,2}=\frac{1}{(2\pi)^2} \int \sum_{j>i\ge 3}^{2+n'} q_{\nu ij} a^\nu a^ida^j + q_{\nu ji} a^\nu a^j da^i  + q_{ij\nu} a^i a^j da^\nu$; 
\item Twisted terms with gauge fields from two nontrivial layers and one trivial layers.  
$ S_{int}^{1,3}=\frac{1}{(2\pi)^2} \int \sum_{i= 3}^{2+n'} q_{12i} a^1 a^2da^i + q_{i12} a^i a^1 da^2 + q_{i21} a^i a^2 da^1. $
\end{enumerate}
Then we consider the most general action 
\begin{align}
S=&S_0+S_{int}^0+S_{int}^{1,1}+S_{int}^{1,2}+S_{int}^{1,3}+S_c+S_{sr}\,, \\
 S_0= & \frac{1}{2\pi} \int N \sum_{\nu=1,2} b^\nu   da^\nu + \sum_{i=3}^{2+n'} b^i   da^i \,,  \\
 S_c=&\frac{1}{2\pi}\int  \sum_{\nu=1,2}Q_{1\nu}  A   db^\nu +\sum_{i=3}^{2+n'}Q_{1i}A   db^i   \,, \\
 S_{sr}=&   \int \sum_{\nu=1,2}(a^\nu   *j^\nu + b^\nu  * \Sigma^\nu)  +\sum_{i=3}^{2+n'}a^i   *j^i \,.
 \label{}
 \end{align}
  Integrating out  $b^{1,2}$ and $b^i$, we obtain ($\nu=1,2$) 
  \begin{align}
\! \!\!\! a^\nu&\!\!=\!\!-\frac{2\pi}{N}d^{-1}*\Sigma^{\nu}-\frac{2\pi Q_{1\nu}}{NK} d^{-1}*\sigma\,,\,
  a^i\!\!=\!\!-\frac{2\pi Q_{1i}}{K} d^{-1}*\sigma,\!\label{eqn_formula_of_substituting3}
  \end{align}
  where   $N=K=2$ and $A=\frac{2\pi}{K}d^{-1}*\sigma$. Upon substituting  them   into the twisted terms, we can obtain the aforementioned M3L invariants (see Appendix~\ref{appendix_substituting}).   The resulting effective action is given by:
 $S_\text{eff}= S_{AB}+S_{TL}+S_{M3L}
 $, where $S_{AB}$ collects the  terms describing the braiding between  $e_\nu$ and $\Sigma^\nu$ ($\nu=1,2$).  $S_{TL}$ is given by Eq.~(\ref{eqn_I3L}), which collects two terms involving three-loop braiding statistics among elementary loop excitations $\Sigma^1$ and $\Sigma^2$, which is used to  characterize topological order without symmetry. And $S_{M3L}$ collects all M3L braiding processes:  
\begin{align}
\!& S_{M3L}\!=\!2\pi \int  \mathcal{I}_{ab\alpha} (d^{-1}*\Sigma^1)(d^{-1}*\Sigma^2)(*\sigma) \nonumber\\
 \!&\!+ \!\mathcal{I}_{\alpha ab} (d^{-1}*\sigma)(d^{-1}*\Sigma^1)(*\Sigma^2)\!+\!
\mathcal{I}_{\alpha ba} (d^{-1}*\sigma)(d^{-1}*\Sigma^2)(*\Sigma^1)
\nonumber\\
 \!&\!+\!\mathcal{I}_{a\alpha \alpha}(d^{-1}*\Sigma^1)(d^{-1}*\sigma)(*\sigma)
\!+ \!\mathcal{I}_{b\alpha \alpha}  (d^{-1}*\Sigma^2)(d^{-1}*\sigma)(*\sigma)
\nonumber\\
 \!&\!+\! \mathcal{I}_{\alpha aa} (d^{-1}*\sigma)(d^{-1}*\Sigma^1)(*\Sigma^1)\!+\!\mathcal{I}_{\alpha bb} (d^{-1}*\sigma)(d^{-1}*\Sigma^2)(*\Sigma^2)\,.\label{eqn_main_z2z2_z2_m3laction}
\end{align} The seven coefficients are M3L invariants. For those involving two elementary loop excitations and one symmetry flux, M3L invariants are given by:
\begin{align}
\!&\mathcal{I}_{\alpha aa}\!\!=\! \!-\frac{ q_{211}Q_{12}}{N^3K}\!\!-\!\!\sum_{i=3}^{2+n'}\!\!\frac{ q_{i11}Q_{1i}}{N^2K}, 
\mathcal{I}_{\alpha bb}\!\!=\!\!-\frac{ q_{122}Q_{11}}{N^3K}\!-\!\sum_{i=3}^{2+n'}\!\!\frac{ q_{i22}Q_{1i}}{N^2K},\nonumber\\
\!&\mathcal{I}_{\alpha ab}\!\!=\!\! \frac{ q_{122}Q_{12}}{N^3K}\!-\!\sum_{i=3}^{2+n'}\!\!\frac{ q_{i12}Q_{1i}}{N^2K} , 
\mathcal{I}_{\alpha ba}\!\!=\!\!\frac{ q_{211}Q_{11}}{N^3K}\!-\!\sum_{i=3}^{2+n'}\!\!\frac{ q_{i21}Q_{1i}}{N^2K} ,\nonumber\\
\!&\mathcal{I}_{ab\alpha} \!\!=\!\! -\frac{ q_{122}Q_{12}}{N^3K}+\frac{ q_{211}Q_{11}}{N^3K}\!-\!\sum_{i=3}^{2+n'}\!\!\frac{ q_{12i}Q_{1i}}{N^2K} .\nonumber
\end{align}
For those involving one elementary loop excitation and two symmetry fluxes, M3L invariants are given by:
\begin{align}
&\mathcal{I}_{a\alpha \alpha}=\nonumber\\
& \frac{- q_{122}Q_{12}^2}{N^3K^2}+\frac{ q_{211}Q_{11}Q_{12}}{N^3K^2} +\sum_{i=3}^{2+n'}\frac{ q_{i11}Q_{1i}Q_{11}}{N^2K^2}-\sum_{i=3}^{2+n'}\frac{ q_{1ii} Q^2_{1i}}{NK^2}\nonumber \\
&-\sum_{i\neq j }^{2+n'}\frac{ q_{1ij}Q_{1i} Q_{1j}}{NK^2} +\sum_{i=3}^{2+n'}\frac{ q_{i12}Q_{1i}Q_{12}}{N^2K^2}-\sum_{i=3}^{2+n'}\frac{ q_{12i}Q_{12}Q_{1i}}{N^2K^2}\,,\nonumber\\
& \mathcal{I}_{b\alpha \alpha}=
 \nonumber\\
& \frac{ q_{122}Q_{11}Q_{12}}{N^3K^2}-\frac{ q_{211}Q_{11}^2}{N^3K^2}+\sum_{i=3}^{2+n'}\frac{ q_{i22}Q_{1i}Q_{12}}{N^2K^2}-\sum_{i=3}^{2+n'} \frac{ q_{2ii}Q^2_{1i} }{N K^2}\nonumber \\
&-\sum_{i\neq j }^{2+n'}\frac{ q_{2ij}Q_{1i}Q_{1j}}{N K^2} +\sum_{i=3}^{2+n'}\frac{ q_{i21}Q_{1i}Q_{11}}{N^2K^2}+\sum_{i=3}^{2+n'}\frac{q_{12i}Q_{11}Q_{1i}}{N^2K^2} \,.\nonumber
\end{align}
Following the same analysis as Sec.~\ref{section_generality_SFL_action}, it is sufficient to consider $n'=2$ and preserve one typical term in each summation. Then, we have the following simplified formulas:
\begin{subequations}
\label{eqn_main_z2z2_z2_m3l_typical}
  \begin{align}
\!&\mathcal{I}_{\alpha aa}\!\!=\! \!-\frac{ q_{211}Q_{12}}{N^3K}\!-\!\frac{ q_{311}Q_{13}}{N^2K}, \\
&\mathcal{I}_{\alpha bb}\!\!=\!\!-\frac{ q_{122}Q_{11}}{N^3K}\!-\! \!\!\frac{ q_{322}Q_{13}}{N^2K},\\
\!&\mathcal{I}_{\alpha ab}\!\!=\!\! \frac{ q_{122}Q_{12}}{N^3K}\!-\!  \frac{ q_{312}Q_{13}}{N^2K} , \\
&\mathcal{I}_{\alpha ba}\!\!=\!\!\frac{ q_{211}Q_{11}}{N^3K}\!-\!  \frac{ q_{321}Q_{13}}{N^2K} ,\\
\!&\mathcal{I}_{ab\alpha} \!\!=\!\! -\frac{ q_{122}Q_{12}}{N^3K}+\frac{ q_{211}Q_{11}}{N^3K}\!-\!\frac{ q_{123}Q_{13}}{N^2K} ,\\
&\mathcal{I}_{a\alpha \alpha}=\nonumber\\
& \frac{- q_{122}Q_{12}^2}{N^3K^2}+\frac{ q_{211}Q_{11}Q_{12}}{N^3K^2} +\frac{ q_{311}Q_{13}Q_{11}}{N^2K^2}- \frac{ q_{133} Q^2_{13}}{NK^2}\nonumber \\
&- \frac{ q_{134}Q_{13} Q_{14}}{NK^2} + \frac{ q_{312}Q_{13}Q_{12}}{N^2K^2}- \frac{ q_{123}Q_{12}Q_{13}}{N^2K^2}\,,\\
& \mathcal{I}_{b\alpha \alpha}=
 \nonumber\\
& \frac{ q_{122}Q_{11}Q_{12}}{N^3K^2}-\frac{ q_{211}Q_{11}^2}{N^3K^2}+ \frac{ q_{322}Q_{13}Q_{12}}{N^2K^2}- \frac{ q_{233}Q^2_{13} }{N K^2}\nonumber \\
&- \frac{ q_{234}Q_{13}Q_{14}}{N K^2} + \frac{ q_{321}Q_{13}Q_{11}}{N^2K^2}+ \frac{q_{123}Q_{11}Q_{13}}{N^2K^2} \,.
\end{align}
\end{subequations}
  Therefore, without loss of generality, we obtain the following minimal model ($N=2$): 
  \begin{subequations}
  \label{eqn_action_z2z2_z20}
  \begin{align}
&S=S_0+S_{int}+S_c+S_{sr}\label{eqn_action_z2z2_z2}\,,\\
 &S_0=  \frac{1}{2\pi}\int \sum_{v=1}^2 N b^v   da^v+ b^3   da^3 +b^4   da^4 \,, \\
  &S_c=\frac{1}{2\pi}\int  \sum_{v=1}^2 Q_{1v}  A   db^v +  Q_{13}A   db^3+Q_{14}A   db^4 \,,   \\
 &S_{sr}=   \int \sum_{v=1}^2 (a^v   *j^v +b^v   * \Sigma^v)+ a^3*j^3+ a^4   *j^4 \,,  \\
 &S_{int}= \frac{1}{4\pi^2}  \int  q_{122}  a^1  a^2da^2 +q_{211}  a^2  a^1da^1  \nonumber \\
&+ \sum_{v=1}^2 q_{3vv}a^3  a^vda^v+q_{v33}  a^v  a^3da^3+q_{v34}  a^v  a^3da^4 + \nonumber \\
&\quad \qquad q_{123}a^1a^2da^3+q_{312}a^3a^1da^2+q_{321}a^3a^2da^1\,.
 \end{align}
 \end{subequations}

By further setting $Q_{ij}=1$ for all trivial layers, all M3L invariants reduce to:
\begin{subequations}
\label{eqn_main_z2z2_z2_invariant_expr}
\begin{align}
\mathcal{I}_{\alpha aa}=& -\frac{ q_{211}Q_{12}}{N^3K}- \frac{ q_{311}}{N^2K}\,, \\
\mathcal{I}_{\alpha bb}=&-\frac{ q_{122}Q_{11}}{N^3K}- \frac{ q_{322} }{N^2K}\,,\\
\mathcal{I}_{\alpha ab}=& \frac{ q_{122}Q_{12}}{N^3K}- \frac{ q_{312} }{N^2K}\,,\\
\mathcal{I}_{\alpha ba}=&\frac{ q_{211}Q_{11}}{N^3K}- \frac{ q_{321} }{N^2K}\,,\\
\mathcal{I}_{ab\alpha} =&-\frac{ q_{122}Q_{12}}{N^3K}+\frac{ q_{211}Q_{11}}{N^3K}- \frac{ q_{123} }{N^2K}\,,\nonumber\\
\mathcal{I}_{a\alpha \alpha}=& \frac{- q_{122}Q_{12}^2}{N^3K^2}+\frac{ q_{211}Q_{11}Q_{12}}{N^3K^2} + \frac{ q_{311} Q_{11}}{N^2K^2}- \frac{ q_{133}}{NK^2}\nonumber \\
&- \frac{ q_{134} }{NK^2}+ \frac{ q_{312} Q_{12}}{N^2K^2}- \frac{ q_{123}Q_{12} }{N^2K^2} \,, \\
\mathcal{I}_{b\alpha \alpha}= &\frac{ q_{122}Q_{11}Q_{12}}{N^3K^2}-\frac{q_{211}Q_{11}^2}{N^3K^2}+ \frac{ q_{322}Q_{12}}{N^2K^2} -\frac{ q_{233} }{N K^2}\nonumber \\
& - \frac{ q_{234} }{N K^2}+ \frac{ q_{321}Q_{11}}{N^2K^2}+ \frac{q_{123}Q_{11} }{N^2K^2} \,.
\end{align}
\end{subequations}
The coefficients are quantized as:
$ q_{133}=k_1 M_{13} \,,\,q_{134}=\bar k_1 M_{13}\,,\,q_{311}=k_2 M_{13}\,, 
q_{233}=k_3 M_{23} \,,\,q_{234}=\bar k_3 M_{23} \,,\,q_{322}=k_4 M_{23} \,,\,
q_{312}=k_5 M_{13} \,,\,q_{321}=k_6 M_{23}  
\,,\,
q_{123}=k_7 M \,,\,q_{122}=k M \,,\,q_{211}=\bar k M 
$,  
where $k_1,\bar k_1, k_2,k_3,\bar k_3,k_4,k_5,k_6,k_7,k$ and $\bar k$ are arbitrary integers. $M$, $M_{13}$ and $M_{23}$ are given by:  
\begin{align}
M=\text{lcm}\left[\frac{4}{\gcd(\hat Q_{11}, 4)},\frac{4}{\gcd(\hat Q_{12}, 4)}\right],
\label{eqn_z2z2_z2_00_qntz_1_main}
\end{align}
  and 
 \begin{align}
 M_{13}&=\frac{4}{\gcd(\hat Q_{11},4)}\,, \, M_{23}=\frac{4}{\gcd(\hat Q_{12},4)}\,.\label{eqn_z2z2_z2_quant_mul2}
 \end{align}
Furthermore, one can check one by one that the periods  of all integers i.e., $k_1,\bar k_1, k_2,k_3,\bar k_3, k_4, k_5, k_6$ and $k_7$ are given by
\begin{align}
\Gamma=2\, 
\label{eqn_z2z2_z2_period1}
\end{align}
regardless of the choices of $\hat{Q}_{11}$ and $\hat{Q}_{12}$. $k$ and $\bar k$ also have the same period: 
\begin{align}
& \Gamma=2  \text{ for } e_10e_20,e_1Ce_20, e_10e_2C\,, \label{eqn_z2z2_z2_period2}\\
 &\Gamma=4 \text{ for } e_1Ce_2C.\label{eqn_z2z2_z2_period3}
  \end{align}
  
According to Table~\ref{table_three_M3L_example_list}, we can write totally $15$ M3L statistical phases as ($N=2; K=2$):
 \begin{align}
&  \theta_{\sigma; \Sigma^1}=-\theta_{\Sigma^1, \sigma; \sigma}=\frac{1}{2}\theta_{\sigma,\sigma; \Sigma^1}\nonumber\\
=&\frac{- 2\pi  kMQ_{12}^2}{N^3K^2}+\frac{ 2\pi  \bar k M Q_{11}Q_{12}}{N^3K^2} + \frac{ 2\pi k_2 M_{13} Q_{11}}{N^2K^2}- \frac{ 2\pi k_1 M_{13}}{NK^2}\nonumber \\
&- \frac{2 \pi  \bar k_1 M_{13}  }{NK^2}+ \frac{ 2\pi k_5 M_{13} Q_{12}}{N^2K^2}- \frac{ 2\pi k_7 M Q_{12} }{N^2K^2}\,,  \label{equation_thetaz2z2_z2_1}\\
 &\theta_{\Sigma^1; \sigma}=-\theta_{\sigma, \Sigma^1; \Sigma^1}\nonumber\\
 =&\frac{1}{2}\theta_{\Sigma^1,\Sigma^1; \sigma} =-\frac{ 2\pi\bar k M Q_{12}}{N^3K}- \frac{2\pi k_2M_{13}}{N^2K}\,,\label{equation_thetaz2z2_z2_2}\\
  &\theta_{\sigma; \Sigma^2} =-\theta_{\Sigma^2, \sigma; \sigma}= \frac{1}{2} \theta_{\sigma,\sigma; \Sigma^2}\nonumber
  \\
  =&\frac{ 2\pi     k M Q_{11}Q_{12}}{N^3K^2}-\frac{ 2\pi \bar k M Q_{11}^2}{N^3K^2}+ \frac{ 2\pi  k_4 M_{23} Q_{12}}{N^2K^2} -\frac{ 2\pi  k_3 M_{23}}{N K^2}\nonumber \\
& - \frac{  2\pi \bar k_3 M_{23} }{N K^2}+ \frac{ 2\pi  k_6 M_{23} Q_{11}}{N^2K^2}+ \frac{ 2\pi k_7 M Q_{11} }{N^2K^2}\,,\label{equation_thetaz2z2_z2_3}\\  
  &\theta_{\Sigma^2; \sigma}=-\theta_{\sigma, \Sigma^2; \Sigma^2}= \frac{1}{2} \theta_{\Sigma^2,\Sigma^2; \sigma}\nonumber\\
  =& -\frac{ 2\pi kM  Q_{11}}{N^3K}- \frac{ 2\pi  k_4 M_{23} }{N^2K}\,,\label{equation_thetaz2z2_z2_4}\\
&\theta_{\sigma, \Sigma^2; \Sigma^1}=  -\frac{4\pi k M Q_{12}}{N^3K}+\frac{  2\pi \bar k M Q_{11}}{N^3K}- \frac{  2\pi k_7 M }{N^2K}+ \frac{  2\pi k_5 M_{13} }{N^2K} \,, \nonumber\\ 
&\theta_{\Sigma^2, \Sigma^1; \sigma}\nonumber\\
=&\frac{ 2\pi  k M Q_{12}}{N^3K}- \frac{2\pi  k_5 M_{13} }{N^2K}+  \frac{ 2\pi  \bar k M Q_{11}}{N^3K}- \frac{ 2\pi k_6 M_{23}  }{N^2K}\,,\label{equation_thetaz2z2_z2_5}\\
 &\theta_{\sigma, \Sigma^1; \Sigma^2}\nonumber\\
 =&  -\frac{ 4\pi  \bar k M Q_{11}}{N^3K}+ \frac{ 2\pi k_6 M_{23} }{N^2K}  +\frac{2\pi  k M Q_{12}}{N^3K}+ \frac{2\pi  k_7 M  }{N^2K}\,.\label{equation_thetaz2z2_z2_6}
\end{align}

 \subsection{Fractional $\Z_2$ symmetry charge in  twisted $\Z_2\times\Z_2$ topological orders must be anomalous}\label{section_concrete_example_of_anomaly_z2z2z2}
 From Table~\ref{table_z2z2_z2_sfl}, we see that among all   $\Z_2\times\Z_2$ topological orders, which are denoted by $\mathsf{TO_i}$ with $i=1,2,3,4$, only untwisted one denoted as $\mathsf{TO_1}$ can support nontrivial SFP. For all twisted ones,   SFP  must be trivial, denoted by $e_10e_20$. This is the first concrete example that demonstrates the quantum anomaly introduced in Sec.~\ref{obstruction}. In other words, nontrivial SFP, i.e., fractional charge carried by elementary particle excitations, must be anomalous in twisted $\Z_2\times \Z_2$ topological order with $\Z_2$ symmetry.

   To be more specific,    in the presence of symmetry,  $q_{122}$ and $q_{211}$ are quantized as
$ q_{122}=k M$, $q_{211}=\bar k M$ where $M$ is given by Eq.~(\ref{eqn_z2z2_z2_00_qntz_1_main}).  There are four independent combinations for     $\hat{Q}_{11}$  and $\hat{Q}_{12}$, as shown in Table~\ref{tab_SFGg2}, which respectively correspond  to four different types of SFP. 

If we consider the trivial SFP denoted as $e_10e_20$,  i.e,  $\hat{Q}_{11}=\hat{Q}_{12}=2$, Eq.~(\ref{eqn_z2z2_z2_00_qntz_1_main}) gives $M=2$. Thus, $q_{122}=2k$, $q_{211}=2\bar k$,   $k\in\Z_2$,  and $\bar k\in\Z_2$, which are exactly the same as Eq.~(\ref{eqn_without_symmetry_z2z2_qq}). Thus all four topological orders in Table~\ref{Table_Z2Z2bulk_topo_order_LIST} are realizable when SFP is trivial.   In other words, all topological orders do support trivial SFP, as shown in Table~\ref{table_z2z2_z2_sfl}.

But if we consider any kind of nontrivial SFP,  either $\hat{Q}_{11}$ or $\hat{Q}_{12}$ equals to $1$. Then, Eq.~(\ref{eqn_z2z2_z2_00_qntz_1_main}) leads to $M=4$, i.e., $q_{122}$ and $q_{211}$ must be multiple of $4$ in order to respect global symmetry.  As a result, from Eqs.~(\ref{eqn_three_loop_braiding_z2z2_line1}) and (\ref{eqn_three_loop_braiding_z2z2_line2}),  we can verify that the values of these six phases   always correspond to  $\mathsf{TO_1}$ in Table~\ref{Table_Z2Z2bulk_topo_order_LIST}.

Therefore, it is impossible to have nontrivial SFP (i.e., $e_1Ce_20$, $e_10e_2C$, and $e_1Ce_2C$) in $\Z_2\times\Z_2$ twisted topological orders (i.e., $\mathsf{TO_2}$, $\mathsf{TO_3}$, and $\mathsf{TO_4}$) with $\Z_2$ symmetry. Alternatively speaking, fractional $\Z_2$ symmetry charge carried by elementary particle excitations in twisted $\Z_2\times\Z_2$ topological orders   must be anomalous. As discussed in Sec.~\ref{obstruction}, such an anomaly cannot be identified if twisted terms are not taken into account. 
In Sec.~\ref{section_Z2Z2_Z2_computing_SFL_for_SFP}, we will compute SFL for each SFP in Table~\ref{section_computing_sfgg2}, and will clearly see that nontrivial SFP can only be realized when topological order is untwisted for $G_g=\Z_2\times \Z_2$ and $G_s=\Z_2$.

  \subsection{Computing SFL for each SFP}\label{section_Z2Z2_Z2_computing_SFL_for_SFP}

  \subsubsection{Computing SFL when SFP is $e_10e_20$}\label{section_Z2Z2_Z2_computing_SFL_for_SFP_e0e0}

In this SFP, ${Q}_{11}={Q}_{12}=0$, $\hat{Q}_{11}=\hat{Q}_{12}=K=2$, $M=M_{23}=M_{13}=2$. We  can write totally $15$ M3L statistical phases as:
 \begin{align}
& \theta_{\sigma; \Sigma^1}=-\theta_{\Sigma^1, \sigma; \sigma}= \frac{\pi}{2}(-k_1-\bar k_1)  \text{ mod }\pi\,,\nonumber\\
  &\theta_{\sigma,\sigma; \Sigma^1}= \pi(-k_1-\bar k_1)  \text{ mod }2\pi\,,\nonumber\\
 &\theta_{\Sigma^1; \sigma}=-\theta_{\sigma, \Sigma^1; \Sigma^1}= -\frac{ \pi}{2}k_2 \text{ mod }\pi \,,\nonumber\\
 &\theta_{\Sigma^1,\Sigma^1; \sigma}=-\pi k_2  \text{ mod }2\pi \nonumber\,,\\
  &\theta_{\sigma; \Sigma^2} =-\theta_{\Sigma^2, \sigma; \sigma}=  \frac{\pi}{2}(-k_3-\bar k_3)   \text{ mod }\pi\nonumber\,,\\  
  &\theta_{\sigma,\sigma; \Sigma^2} = \pi(-k_3-\bar k_3)  \text{ mod }2\pi\nonumber\,,\\
  &\theta_{\Sigma^2; \sigma}=-\theta_{\sigma, \Sigma^2; \Sigma^2}=  -\frac{ \pi}{2}k_4 \text{ mod }\pi\nonumber\,,\\
   &\theta_{\Sigma^2,\Sigma^2; \sigma}= -{ \pi}k_4  \text{ mod }2\pi\nonumber\,,\\
&\theta_{\sigma, \Sigma^2; \Sigma^1}=  \frac{\pi}{2}(-k_7+k_5) \text{ mod }\pi \,, \nonumber\\ 
&\theta_{\Sigma^2, \Sigma^1; \sigma}=\frac{\pi}{2}(-k_5-k_6) \text{ mod }\pi\,,\nonumber\\
 &\theta_{\sigma, \Sigma^1; \Sigma^2}=\frac{\pi}{2}(k_6+k_7) \text{ mod }\pi\nonumber\,,
\end{align}
where the periods are due to   Eqs.~(\ref{eqn_z2z2_z2_period1}) and (\ref{eqn_z2z2_z2_period2}). All kinds of    particle attachment   provide  a minimal $\pi$ period:
      \begin{enumerate}[label=\textbf{(\roman*)}]
\item Braiding $e_1$ around $\sigma$ gives a $\pi$ phase shift. 
\item Braiding $e_1$ around $\Sigma^1$ gives a ${\pi}$ phase shift. 
\item Braiding $e_2$ around $\sigma$ gives a $\pi$ phase shift. 
\item Braiding $e_2$ around $\Sigma^2$ gives a $\pi$ phase shift. 
\end{enumerate}
So   $\theta_{\sigma,\sigma; \Sigma^1}$, $\theta_{\Sigma^1,\Sigma^1; \sigma}$, $\theta_{\sigma,\sigma; \Sigma^2}$, and $\theta_{\Sigma^2,\Sigma^2; \sigma}$ vanish up to $\pi$. 

In the remaining $11$ M3L statistical phases,    one   can further simplify the above $11$ M3L statistical phases by introducing six \textit{linearly independent} integers $p_1,\cdots, p_6\in\Z_2$:
  \begin{align}
& \theta_{\sigma; \Sigma^1}= \theta_{\Sigma^1, \sigma; \sigma}= \frac{\pi}{2}p_1 \text{ mod }\pi \label{eqn_theta_1}\,, \\
  &\theta_{\Sigma^1; \sigma}= \theta_{\sigma, \Sigma^1; \Sigma^1}=\frac{\pi}{2}p_2 \text{ mod }\pi  \label{eqn_theta_2}\,,\\
   &\theta_{\sigma; \Sigma^2} = \theta_{\Sigma^2, \sigma; \sigma}= \frac{\pi}{2}p_3 \text{ mod }\pi    \label{eqn_theta_3}\,,\\  
   &\theta_{\Sigma^2; \sigma}= \theta_{\sigma, \Sigma^2; \Sigma^2}=  \frac{\pi}{2}p_4 \text{ mod }\pi\label{eqn_theta_4}\,,\\
 &\theta_{\sigma, \Sigma^2; \Sigma^1}= \frac{\pi}{2}p_5 \text{ mod }\pi\label{eqn_theta_5}  \,,\\ 
&\theta_{\Sigma^2, \Sigma^1; \sigma}=\frac{\pi}{2}p_6 \text{ mod }\pi\label{eqn_theta_6}\,,\\
 &\theta_{\sigma, \Sigma^1; \Sigma^2}=\frac{\pi}{2}( p_5+p_6) \text{   mod   }\pi\, \label{eqn_theta_7}
\end{align}
as shown in Table~\ref{table_z2z2_z2_sfl}.. Since  $q_{122}=Mk=2k$ and $q_{211}=M\bar k=2\bar k$, we find that  all four    topological orders in Table~\ref{Table_Z2Z2bulk_topo_order_LIST} can be exhausted. Therefore, the present SFP can be realized by all four topological orders. This result is consistent to the discussion in Sec.~\ref{section_concrete_example_of_anomaly_z2z2z2}.       In summary, when SFP is $e_10e_20$, SFL is classified by $(\Z_2)^6$ for all $\Z_2\times\Z_2$ topological orders, regardless of twists.   
  \subsubsection{Computing SFL when SFP is either $e_1Ce_20$ or $e_10e_2C$}\label{section_z2z2_z2_sub_e1ce20}

In the following, we first consider SFL when SFP is $e_1Ce_20$, then $e_10e_2C$ can be obtained by switching indices properly.

In the SFP $e_1Ce_20$, $Q_{11}=\hat{Q}_{11}=1, Q_{12}=0, \hat{Q}_{12}=2$, $M=4, M_{13}=4, M_{23}=2$.   We  can write totally $15$ M3L statistical phases as:
 \begin{align}
& \theta_{\sigma; \Sigma^1}=
-\theta_{\Sigma^1, \sigma; \sigma}= \frac{\pi}{2} k_2    \text{ mod }\pi\,,\nonumber\\
  &\theta_{\sigma,\sigma; \Sigma^1}=  \pi k_2    \text{ mod }2\pi \,,\nonumber\\
 &\theta_{\Sigma^1; \sigma}=-\theta_{\sigma, \Sigma^1; \Sigma^1}=    -\pi k_2 \text{ mod }2\pi\,,\nonumber \\
 &\theta_{\Sigma^1,\Sigma^1; \sigma}=  0 \text{ mod }2\pi\,,\nonumber \\
  &\theta_{\sigma; \Sigma^2} =-\theta_{\Sigma^2, \sigma; \sigma}= \frac{\pi}{4}(k_6-\bar k)  \text{ mod }\frac{\pi}{2}\nonumber\\  
  &\theta_{\sigma,\sigma; \Sigma^2} =\frac{\pi}{2}(k_6-\bar k)  \text{ mod }{\pi}\,,\nonumber\\
  &\theta_{\Sigma^2; \sigma}=-\theta_{\sigma, \Sigma^2; \Sigma^2}=  \frac{ \pi}{2}(-k-k_4) \text{ mod }\pi\,,\nonumber\\
   &\theta_{\Sigma^2,\Sigma^2; \sigma}= { \pi}(-k-k_4) \text{ mod }2\pi\,,\nonumber\\
&\theta_{\sigma, \Sigma^2; \Sigma^1}=  \frac{\pi}{2} \bar k  \text{ mod }\pi  \,,\nonumber\\ 
&\theta_{\Sigma^2, \Sigma^1; \sigma}=\frac{\pi}{2}( \bar k- k_6) \text{ mod }\pi\,,\nonumber\\
 &\theta_{\sigma, \Sigma^1; \Sigma^2}=\frac{\pi}{2} k_6  \text{ mod }\pi\,.\nonumber
\end{align} where the periods are due to   Eqs.~(\ref{eqn_z2z2_z2_period1}) and (\ref{eqn_z2z2_z2_period2}). 

The elementary particle excitations $e_1$ and $e_2$ carry respectively $1/2$ and $1$ symmetry charges. So, there are following different kinds of particle attachment, which provides various kinds of phase shift. 
      \begin{enumerate}[label=\textbf{(\roman*)}]
\item Braiding $e_1$ around $\sigma$ gives a $\frac{\pi}{2}$ phase shift. 
\item Braiding $e_1$ around $\Sigma^1$ gives a ${\pi}$ phase shift. 
\item Braiding $e_2$ around $\sigma$ gives a $\pi$ phase shift. 
\item Braiding $e_2$ around $\Sigma^2$ gives a $\pi$ phase shift. 
\end{enumerate}
By properly attaching these particles into M3L braiding processes, we reach the following results with new minimal periods: 
 \begin{align}
& \theta_{\sigma; \Sigma^1} =  0 \text{ mod }\frac{\pi}{2}\,,\, \theta_{\Sigma^1, \sigma; \sigma}= 0 \text{ mod }\frac{\pi}{2}\,,\nonumber\\
  &\theta_{\sigma,\sigma; \Sigma^1}= 0 \text{ mod }\frac{\pi}{2} \,,\,  \theta_{\Sigma^1; \sigma}=      0 \text{ mod }{\pi} \,, \nonumber\\
 & \theta_{\sigma, \Sigma^1; \Sigma^1}=  0 \text{ mod }\frac{\pi}{2} \,,\, \theta_{\Sigma^1,\Sigma^1; \sigma}=  0 \text{ mod }\pi \,,\nonumber\\
  &\theta_{\sigma; \Sigma^2}  =  \theta_{\Sigma^2, \sigma; \sigma}= \frac{\pi}{4}(k_6-\bar k)  \text{ mod }\frac{\pi}{2}\,,\nonumber\\  
  &\theta_{\sigma,\sigma; \Sigma^2} = 0  \text{ mod }\frac{\pi}{2}\,,\, \theta_{\Sigma^2; \sigma}=   \frac{ \pi}{2}(-k-k_4) \text{ mod }\pi\,,\nonumber\\
   &  \theta_{\sigma, \Sigma^2; \Sigma^2}=  0 \text{ mod }\frac{\pi}{2}\,,\, \theta_{\Sigma^2,\Sigma^2; \sigma}= 0 \text{ mod }\pi\,,\nonumber\\
&  \theta_{\sigma, \Sigma^2; \Sigma^1}=  0  \text{ mod }\frac{\pi}{2} \,,\, \theta_{\Sigma^2, \Sigma^1; \sigma}=\frac{\pi}{2}( \bar k- k_6) \text{ mod }\pi\,,\nonumber\\
 &\theta_{\sigma, \Sigma^1; \Sigma^2}= 0 \text{ mod }\frac{\pi}{2}\,.\nonumber
\end{align}
As a result, there are only four nonvanishing phases parameterized by two independent $\Z_2$ integers, $p_7=k_6-\bar k$ and $p_8=k_4+k$.
\begin{align}
  &\theta_{\sigma; \Sigma^2}  = \frac{\pi}{4}p_7 \text{ mod }\frac{\pi}{2}\,, \,  \theta_{\Sigma^2, \sigma; \sigma}= \frac{\pi}{4}p_7   \text{ mod }\frac{\pi}{2}\,,\\  
  &\theta_{\Sigma^2; \sigma}=   \frac{ \pi}{2}p_8\text{ mod }\pi\,, \,\theta_{\Sigma^2, \Sigma^1; \sigma}=\frac{\pi}{2} p_7 \text{ mod }\pi\,.
  \end{align}
In this new parametrization, I3L statistical phases and M3L statistical phases are uncorrelated. Since  $q_{122}=Mk=4k$ and $q_{211}=M\bar k=4\bar k$, we find that only the untwisted topological order denoted as $\mathsf{TO_1}$ is possible.  Therefore, the present SFP can be realized only by $\mathsf{TO}_1$. This result is consistent to the discussion in Sec.~\ref{section_concrete_example_of_anomaly_z2z2z2}.

In summary, in SFP $e_1Ce_20$, the classification of SFL is given by $\Z_2\times\Z_2$, which can only be realized in $\mathsf{TO_1}$ as shown in Table~\ref{table_z2z2_z2_sfl}.  The case of SFP $e_10e_2C$ can be analyzed in the same way, also resulting in $\Z_2\times\Z_2$ classification labeled by two $\Z_2$ integers $p_{9}$ and $p_{10}$ in Table~\ref{table_z2z2_z2_sfl}. 
  
   \subsubsection{Computing SFL when SFP is $e_1Ce_2C$}

In the SFP $e_1Ce_2C$, $Q_{11}=\hat{Q}_{11}=1, Q_{12}=\hat{Q}_{12}=1$, $M=4, M_{13}=4, M_{23}=4$.  We  can write totally $15$ M3L statistical phases as:
 \begin{align}
& \theta_{\sigma; \Sigma^1}=-\theta_{\Sigma^1, \sigma; \sigma}= \frac{\pi}{4}(-k+\bar k)+ \frac{\pi}{2}(k_2-k_1-\bar k_1+k_5-k_7 )  \text{ mod }\pi\,,\nonumber\\
  &\theta_{\sigma,\sigma; \Sigma^1}=  \frac{\pi}{2}(-k+\bar k)+  {\pi} (k_2-k_1-\bar k_1+k_5-k_7 )  \text{ mod }2\pi,\nonumber\\
 &\theta_{\Sigma^1; \sigma}=-\theta_{\sigma, \Sigma^1; \Sigma^1}= -\frac{ \pi}{2}\bar k-\pi k_2  \text{ mod }2\pi \,,\nonumber\\
 &\theta_{\Sigma^1,\Sigma^1; \sigma}=-  \pi \bar k \text{ mod }2\pi \nonumber\,,\\
  &\theta_{\sigma; \Sigma^2} =-\theta_{\Sigma^2, \sigma; \sigma}=   \frac{\pi}{4}(k-\bar k)+\frac{\pi}{2}( k_4+k_6+k_7)   \text{ mod }\pi\nonumber\,,\\  
  &\theta_{\sigma,\sigma; \Sigma^2} =   \frac{\pi}{2}(k-\bar k)+{\pi}( k_4+k_6+k_7)   \text{ mod }2\pi\nonumber\,,\\  
  &\theta_{\Sigma^2; \sigma}=-\theta_{\sigma, \Sigma^2; \Sigma^2}=  -\frac{ \pi}{2}k-\pi k_4  \text{ mod }2\pi\nonumber\,,\\
   &\theta_{\Sigma^2,\Sigma^2; \sigma}= -\pi k  \text{ mod }2\pi\nonumber\,,\\
&\theta_{\sigma, \Sigma^2; \Sigma^1}=  -\pi k +\frac{\pi}{2}\bar k-\pi k_7+\pi k_5  \text{ mod }2\pi \,, \nonumber\\ 
&\theta_{\Sigma^2, \Sigma^1; \sigma}=\frac{\pi}{2}(k+\bar k)-\pi (k_5+ k_6 ) \text{ mod }2\pi\,,\nonumber\\
 &\theta_{\sigma, \Sigma^1; \Sigma^2}=-\pi \bar k +\frac{\pi}{2}k+\pi k_6+\pi k_7  \text{ mod }2\pi \nonumber\,,
\end{align}
where the periods are due to   Eqs.~(\ref{eqn_z2z2_z2_period1}) and (\ref{eqn_z2z2_z2_period3}). 

The elementary particle excitations $e_1$ and $e_2$ carry respectively $1/2$ and $1$ symmetry charges. So, there are following different kinds of particle attachment, which provides various kinds of phase shift. 
\begin{enumerate}
\item Braiding $e_1$ around $\sigma$ gives a $\frac{\pi}{2}$ phase shift. 
\item Braiding $e_1$ around $\Sigma^1$ gives a ${\pi}$ phase shift. 
\item Braiding $e_2$ around $\sigma$ gives a $\frac{\pi}{2} $ phase shift. 
\item Braiding $e_2$ around $\Sigma^2$ gives a $\pi$ phase shift. 
\end{enumerate}

By properly attaching these particles into M3L braiding processes, we reach the following results with new minimal periods: 
   \begin{align}
& \theta_{\sigma; \Sigma^1}=\frac{\pi}{4}(k +\bar k  )  \text{ mod }\frac{\pi}{2}\,,\,\theta_{\Sigma^1, \sigma; \sigma}= \frac{\pi}{4}( k+\bar k  )  \text{ mod }\frac{\pi}{2}\,,\nonumber\\
  &\theta_{\sigma,\sigma; \Sigma^1}= 0 \text{ mod }{\frac{\pi}{2}}\,, \,\theta_{\Sigma^1; \sigma}=   \frac{\pi}{2}   \bar k  \text{ mod }\pi\,,\nonumber \\
 & \theta_{\sigma, \Sigma^1; \Sigma^1}=   0     \text{ mod }\frac{\pi}{2}\,,\,\theta_{\Sigma^1,\Sigma^1; \sigma}=0 \text{ mod }\pi\,,\nonumber \\
  &\theta_{\sigma; \Sigma^2} = \frac{\pi}{4}(k+\bar k)  \text{ mod }\frac{\pi}{2}\,,\, \theta_{\Sigma^2, \sigma; \sigma}= \frac{\pi}{4}(k+\bar k)  \text{ mod }\frac{\pi}{2}\,,\nonumber\\  
  &\theta_{\sigma,\sigma; \Sigma^2} = 0  \text{ mod }{\frac{\pi}{2}}\,,\,\theta_{\Sigma^2; \sigma}=   \frac{ \pi}{2}k \text{ mod }\pi\,,\nonumber\\
  & \theta_{\sigma, \Sigma^2; \Sigma^2}= 0 \text{ mod }\frac{\pi}{2}\,,\, \theta_{\Sigma^2,\Sigma^2; \sigma}= 0 \text{ mod }\pi\,,\nonumber\\
&\theta_{\sigma, \Sigma^2; \Sigma^1}= 0 \text{ mod }\frac{\pi}{2} \,, \,\theta_{\Sigma^2, \Sigma^1; \sigma}=\frac{\pi}{2}( k+\bar k ) \text{ mod }\pi\,,\nonumber\\
 &\theta_{\sigma, \Sigma^1; \Sigma^2}= 0\text{ mod }\frac{\pi}{2}\,.\nonumber
\end{align}
By introducing two new $\Z_2$ integers via $p_{11}=k+\bar k$ and $p_{12}=\bar k$, we arrive at the results listed in Table~\ref{table_z2z2_z2_sfl}. Following the same analysis in Sec.~\ref{section_z2z2_z2_sub_e1ce20}, only $\mathsf{TO_1}$ can realize $e_1Ce_2C$.

In summary, the total classification of the untwisted $\Z_2\times\Z_2$ topological order enriched by $\Z_2$ global symmetry is given by $(\Z_2)^6\oplus (\Z_2)^2\oplus(\Z_2)^2\oplus(\Z_2)^2$. For any twisted topological order, the classification is given by $(\Z_2)^6$ in which only trivial SFP denoted as $e_10e_20$ is realizable. Thus, nontrivial SFP (i.e., fractional $\Z_2$ symmetry charge) in twisted $\Z_2\times\Z_2$ topological orders must be anomalous.

       \section{Symmetry fractionalization in topological orders with multi $\Z_{K}$ symmetry subgroups: taking $\Z_2$ topological order enriched by $\Z_{2}\times \Z_2$ symmetry as an example} 
       \label{sec_example_znzkzk}
In the previous two sections, we have studied SET examples where symmetry group is simply $\Z_K$ and the unit external symmetry flux is denoted as $\sigma$. In this section, we consider  the $\Z_2$ topological order enriched by $\Z_{2}\times \Z_2$ symmetry, in which two symmetry subgroups coexist. Consequently, we must introduce two distinct symmetry fluxes denoted as $\sigma^1$ and $\sigma^2$, in order to detect symmetry fractionalization on particles and loops. Physically, one may expect that the final classification have two typical parts. The first  part includes SET phases in which   $\Z_2$ symmetry subgroups are independently fractionalized or non-fractionalized.  The second  part includes SET phases in which both symmetry subgroups are ``intertwined'' together such that it is impossible to regard the fractionalization patterns as independent combinations of data from two symmetry subgroups. Indeed, our calculation shows that there are M3L braiding processes in which both $\sigma^1$ and $\sigma^2$ are involved. Furthermore,  we should consider M3L braiding processes in which external symmetry flux (denoted as $\sigma^{12}$) is a composite of $\sigma^1$ and $\sigma^2$. These M3L braiding processes are used to detect the second part of classification, which unveils nontrivial interplay of two symmetry subgroups.  The results are summarized in Table \ref{table_z2_z2z2_sfl}.

This section is organized as follows. In Sec.~\ref{section_z2_z2z2_computing_SFP_table},   we compute SFP and collect results into Table~\ref{SFGg3}. In Sec.~\ref{eqn_action_z2z2_z2_minimal_model}, we derive the minimal model [Eq.~(\ref{eq_main_z2_z2z2_SFL_action})] that serves as the starting point for the computation of SFL. 
  In Sec.~\ref{sec_loop_SF_z2_z2z2_cases}, we concretely compute SFL of each SFP, which renders the final classification of symmetry fractionalization of $\Z_{2}$  topological order enriched by $\Z_2\times \Z_2$ symmetry.  

\subsection{Computation of SFP}\label{section_z2_z2z2_computing_SFP_table}
There are four different types of SFP. From (\ref{eq_mathcalc}), the elementary gauge charge, denoted as $e$, can carry one-half or integer  charge of the two $\Z_{2}$ symmetry subgroups.  We denote these four types as $e00$ and $e0C$, $eC0$ and $eCC$, respectively. Upon gauging, the four SFP types correspond to  gauged groups $\Z_2\times \Z_2 \times \Z_2$, $\Z_2\times \Z_4$, $\Z_4\times \Z_2$ and $\Z_2\times \Z_4$ respectively (see Table~\ref{SFGg3}). 
 
\begin{table*}[t]
   \caption{The four  SFP types for $G_g=\Z_2$ topological order with $G_s=\Z_2\times \Z_2$ symmetry.  $\mathcal{C}$ denotes the SFP matrix defined in Eq.~(\ref{eq_mathcalc}). $\mathcal{C}_{11}$ and $\mathcal{C}_{21}$ are  the fractionalized symmetry charges carried by $e$, which are respectively in the first and second $\Z_2$   subgroups of $G_s$.  $\hat{Q}$ is defined in Eq.~(\ref{eqn_definition_hat_Q_matrix}). $G_g^*$ is the gauge group after gauging the symmetry. }
    \label{SFGg3}
    \centering
    \begin{tabular}{ccccc}
    \hline
     
    \hline
  \text{ }  SFP  \text{ }&\quad\quad$(Q_{11},Q_{21})$\quad\quad&    \text{ }\text{ }  \text{ } \text{ }\text{ }\text{ }\text{ }  $\mathcal{C}$\text{ }\text{ }\text{ }\text{ }\text{ } \text{ }\text{ }\text{ } &  \quad$(\hat{Q}_{11},\hat{Q}_{21})$\quad& \quad\quad\quad$G_g^*$\quad\quad\quad \text{ }\\
            \hline
     $e00$ &$(0,0)$ &    $(1, 1)^T$ & $(2,2)$& $\Z_{2}  \times \Z_{2}\times \Z_2$\\
     $eC0$ &$(1,0)$ &    $(1/2, 1)^T$ &  $(1,2)$ & $\Z_{4} \times \Z_{2}$\\
     $e0C$  &$(0,1)$&    $(1, 1/2)^T$  & $(2, 1)$ &$\Z_{2} \times \Z_{4}$ \\
     $eCC$  &$(1,1)$&     $(1/2, 1/2)^T$ &$(1,1)$  &$\Z_{4} \times \Z_{2}$\\
  \hline

  \hline
  \end{tabular}
  
\end{table*} 
 \subsection{Minimal model for computing SFL}\label{eqn_action_z2z2_z2_minimal_model}
 
  Here we show that the action (\ref{eq_main_z2_z2z2_SFL_action}) we consider in the example is sufficient  for the purpose of classification of SFL. We start from the most general action and then derive the most general form of different M3L invariants, through which we can show the action  (\ref{eq_main_z2_z2z2_SFL_action}) is sufficient enough for classification of SFL. Our derivations apply to general $\Z_N$ and $\Z_K\times \Z_K$ symmetry. For the present example, $N=K=2$.
   
   We first consider different types of twisted terms that could be related to M3L invariants, which are classified in two classes:
      \begin{enumerate}[label=\textbf{(\roman*)}]
\item Twisted terms with  gauge fields from  one nontrivial layer and one trivial layer.  
 $
S_{int}^{1,1}=\frac{1}{(2\pi)^2} \int \sum_{i=2}^{1+n'} q_{1 ii} a^1 a^ida^i + q_{i11} a^ia^1 da^1$. 
\item Twisted terms with  gauge fields from  one nontrivial layer and two trivial layers.  
 $
S_{int}^{1,2}=\frac{1}{(2\pi)^2} \int \sum_{j>i\ge 2}^{1+n'} q_{1 ij} a^1 a^ida^j + q_{1 ji} a^1 a^j da^i+ q_{ij1} a^i a^j da^1$. 
\end{enumerate}
Then we consider the following general   action:
\begin{align}
&S=S_0+S_{int}^{1,1}+S_{int}^{1,2}+S_c+S_{sr}\nonumber\\
 &S_0=  \frac{1}{2\pi}\int  N b^1   da^1+ \sum_{i=2}^{1+n'}b^i   da^i \nonumber \\
  &S_c=\frac{1}{2\pi}\int  \sum_{k=1}^2 (Q_{k1}  A^k   db^1 +\sum_{i=2}^{1+n'}Q_{ki}A^k   db^i )\nonumber \\
 &S_{sr}=   \int   a^1   *j^1 +b^1   * \Sigma+ \sum_{i=2}^{1+n'}a^i*j^i\,, \nonumber\\
 &S_{int}= \frac{1}{4\pi^2}  \int \sum_{i=2}^{1+n'} q_{i11}a^i  a^1da^1+q_{1ii}  a^1  a^ida^i+\nonumber \\
&\quad \qquad \qquad \quad\, \sum_{j>i\ge 2}^{1+n'} q_{1ij}  a^1  a^ida^j+q_{1ji}  a^1  a^jda^i +q_{ij1}a^ia^jda^1\nonumber
 \end{align}
 where we have ignored the loop excitations corresponding to the level-1 gauge fields since they are trivial.
Integrating out  $b^{1}$ and $b^i$, we obtain 
 $  a^1 =-\frac{2\pi}{N}d^{-1}*\Sigma-\frac{2\pi Q_{11}}{NK} d^{-1}*\sigma^1-\frac{2\pi Q_{21}}{NK}d^{-1}*\sigma^2$ and 
  $a^i =-\frac{2\pi Q_{1i}}{K} d^{-1}*\sigma^1-\frac{2\pi Q_{2i}}{K} d^{-1}*\sigma^2$ and $A^i=\frac{2\pi}{K}d^{-1}*\sigma^i$. Upon substituting  them   into the twisted terms, we can obtain the  M3L invariants:
 \begin{align}
&\mathcal{I}_{\alpha aa}=-\sum_{i=2}^{1+n'}\frac{q_{i11}Q_{1i}}{N^2K}\nonumber\\
&\mathcal{I}_{\beta aa}=-\sum_{i=2}^{1+n'}\frac{q_{i11}Q_{2i}}{N^2K}\nonumber
\\
&\mathcal{I}_{a\alpha\alpha}=\sum_{i=2}^{1+n'}[\frac{q_{i11} Q_{1i}Q_{11}}{(NK)^2}-\frac{q_{1ii}Q^2_{1i} }{NK^2}]-\sum_{i\ne j}\frac{q_{1ij}Q_{1i}Q_{1j}}{NK^2}\nonumber\\
&\mathcal{I}_{a\beta\beta}=\sum_{i=2}^{1+n'}[\frac{q_{i11} Q_{2i}Q_{21}}{(NK)^2}-\frac{q_{1ii}Q^2_{2i}}{NK^2}]-\sum_{i\ne j}\frac{q_{1ij}Q_{2i}Q_{2j}}{NK^2}\nonumber\\
&\mathcal{I}_{\alpha a \beta}=\sum_{i=2}^{1+n'}[-\frac{q_{i11}Q_{1i}Q_{21}}{(NK)^2}+\frac{q_{1ii}Q_{1i}Q_{2i}}{NK^2}]+\sum_{i\ne j}\frac{q_{1ij}Q_{1i}Q_{2j}}{NK^2}\nonumber\\
&\mathcal{I}_{\beta a \alpha}=\sum_{i=2}^{1+n'}[-\frac{q_{i11}Q_{2i}Q_{11}}{(NK)^2}+\frac{q_{1ii}Q_{2i}Q_{1i}}{NK^2}]+\sum_{i\neq j}\frac{q_{1ij}Q_{2i}Q_{1j}}{NK^2} \nonumber\\
&\mathcal{I}_{\alpha \beta a}=\sum_{i=2}^{1+n'}\frac{q_{i11}(Q_{2i}Q_{11}-Q_{1i}Q_{21})}{(NK)^2} \text{+} \sum_{i\neq j}\frac{q_{ij1}(\text{$Q_{1j}Q_{2i}$$-$$Q_{1i}Q_{2j}$)}}{NK^2}\nonumber
\end{align}
Following the same analysis as previous sections,  we preserve one term per summation:
\begin{subequations}
\label{eq_main_invariants_expr_znzkzk}
\begin{align}
&\mathcal{I}_{\alpha aa}=-\frac{q_{211}Q_{12}}{NK^2}\\
&\mathcal{I}_{\beta aa}=-\frac{q_{211}Q_{22}}{NK^2}\\
&\mathcal{I}_{a\alpha\alpha}=\frac{q_{211} Q_{12}Q_{11}}{(NK)^2}-\frac{q_{122}Q_{12}Q_{12}}{NK^2}-\frac{q_{123}Q_{12}Q_{13}}{NK^2}\label{eq_main_znzkzk_invriant_3}\\
&\mathcal{I}_{a\beta\beta}=\frac{q_{211} Q_{22}Q_{21}}{(NK)^2}-\frac{q_{122}Q_{22}Q_{22}}{NK^2}-\frac{q_{123}Q_{22}Q_{23}}{NK^2}\label{eq_main_znzkzk_invriant_4}\\
&\mathcal{I}_{\alpha a \beta}=-\frac{q_{211}Q_{12}Q_{21}}{(NK)^2}+\frac{q_{122}Q_{12}Q_{22}}{NK^2}+\frac{q_{123}Q_{12}Q_{23}}{NK^2}\label{eq_main_znzkzk_invriant_5}\\
&\mathcal{I}_{\beta a \alpha}=-\frac{q_{211}Q_{22}Q_{11}}{(NK)^2}+\frac{q_{122}Q_{22}Q_{12}}{NK^2}+\frac{q_{123}Q_{22}Q_{13}}{NK^2} \label{eq_main_znzkzk_invriant_6}\\
&\mathcal{I}_{\alpha \beta a}=\frac{q_{211}(Q_{22}Q_{11}-Q_{12}Q_{21})}{(NK)^2}  +\frac{q_{231}(Q_{13}Q_{22}-Q_{12}Q_{23})}{NK^2}\label{eq_main_znzkzk_invriant_7}
\end{align}
\end{subequations}
These M3L invariants are generated by the following minimal action:
\begin{subequations}
\label{eq_main_z2_z2z2_SFL_action}
\begin{align}
&S=S_0+S_{int}+S_c+S_{sr}
 \\
 &S_0=  \frac{1}{2\pi}\int  N b^1   da^1+ b^2   da^2 +b^3   da^3  \\
  &S_c=\frac{1}{2\pi}\int  \sum_{i=1}^2 Q_{i1}  A^i   db^1 +Q_{i2}A^i   db^2+Q_{i3} A^i   db^3    \\
 &S_{sr}=   \int   a^1   *j^1 +b^1   * \Sigma+ a^2*j^2+ a^3   *j^3 \,,  \\
 &S_{int}= \frac{1}{4\pi^2}  \int q_{211}a^2 a^1da^1+q_{122}  a^1  a^2da^2+  \\
&\quad \qquad \qquad \quad\, q_{123}  a^1  a^2da^3 +q_{231}a^2a^3da^1.
 \end{align}
 \end{subequations}
 Note that if $Q_{21}=Q_{22}=Q_{23}=0$, the second $\Z_{2}$ symmetry subgroup is not coupled to the system so that the system looks like only have  first $\Z_{2}$ symmetry.  
Similarly, if $Q_{11}=Q_{12}=Q_{13}=0$, the first $\Z_{2}$ symmetry subgroup is not coupled to the system so that the system looks like only have second $\Z_{2}$ symmetry.
  
From (\ref{eqn:twisted_coff_general_1}) and (\ref{eqn:twisted_coff_general_11}), we have 
\begin{align} 
q_{211}=kM_{12}\,,\\
 q_{122}=\bar k M_{12}\,,\\  
 q_{123}=\widetilde k M_{12}\,,\\
 q_{231}= k' M_{23}
 \end{align} where $k,\bar k,\widetilde k$, $k'$ are arbitrary integers  and
\begin{align}
M_{12}=&\text{lcm}[2,\frac{4}{\gcd(4,\hat Q_{11})}, \frac{ 2}{\gcd( 2, \hat Q_{12})}, \nonumber \\
&\qquad   \frac{4}{\gcd(4,\hat Q_{21})}, \frac{ 2}{\gcd( 2,\hat Q_{22})}]\label{eqn_main_znzk1zk2_multiple_1}\\
M_{23}=&\text{lcm}[\frac{ 2}{\gcd( 2,\hat Q_{12})}, \frac{ 2}{\gcd( 2, \hat Q_{13})}, \nonumber \\
&\qquad   \frac{ 2}{\gcd( 2,\hat Q_{22})},\frac{  2}{\gcd( 2,\hat Q_{23})}].
\label{eqn_main_znzk1zk2_multiple_2}
\end{align}
Plugging these results into the above M3L invariants, we can easily obtain quantization rules of the corresponding M3L statistical phases.  The periods of $k, \bar k, \widetilde k$, $k'$  will be discussed case by case below.

  \subsection{Computing SFL for each SFP}\label{sec_loop_SF_z2_z2z2_cases}
Below we discuss the SFL for different SFP respectively.

  \subsubsection{Computing SFL when SFP is $e00$}\label{section_Z2_Z2Z2_computing_SFL_for_SFP_e00}

We first discuss  the case with SFP being $e00$, namely the gauge charge $e$ carries  integer symmetry charge of both the two $\Z_2$ subgroups. In this case, $Q_{11}=0$, $Q_{21}=0$ and then $\hat Q_{11}=2, \hat Q_{21}=2$. Then (\ref{eqn_main_znzk1zk2_multiple_1}) and (\ref{eqn_main_znzk1zk2_multiple_2}) reduce to
\begin{align}
M_{12}
  =&2\,,
\label{eqn_main_znzk1zk2_invariant_1aa}\\
M_{23}=&\text{lcm}[\frac{ 2}{\gcd( 2,\hat Q_{12})}, \frac{ 2}{\gcd( 2, \hat Q_{13})}, \nonumber \\
&\qquad   \frac{ 2}{\gcd( 2,\hat Q_{22})},\frac{  2}{\gcd( 2,\hat Q_{23})}].
\label{eqn_main_znzk1zk2_invariant_2aa}
\end{align}
Therefore, the above invariants become
\begin{subequations}
\label{eq_main_invariants_expr_znzkzk_e00}
\begin{align}
&\mathcal{I}_{\alpha aa}=-\frac{kQ_{12}}{4}\label{eq_main_znzkzk_invriant_1__e00}\\
&\mathcal{I}_{\beta aa}=-\frac{kQ_{22}}{4}\label{eq_main_znzkzk_invriant_2__e00}\\
&\mathcal{I}_{a\alpha\alpha}=-\frac{\bar k Q_{12}Q_{12}}{4}-\frac{\widetilde k Q_{12}Q_{13}}{4}\label{eq_main_znzkzk_invriant_3__e00}\\
&\mathcal{I}_{a\beta\beta}=-\frac{\bar k Q_{22}Q_{22}}{4}-\frac{\widetilde k Q_{22}Q_{23}}{4}\label{eq_main_znzkzk_invriant_4__e00}\\
&\mathcal{I}_{\alpha a \beta}=\frac{\bar k Q_{12}Q_{22}}{4}+\frac{\widetilde k Q_{12}Q_{23}}{4}\label{eq_main_znzkzk_invriant_5__e00}\\
&\mathcal{I}_{\beta a \alpha}=\frac{\bar k Q_{22}Q_{12}}{4}+\frac{\widetilde k Q_{22}Q_{13}}{4} \label{eq_main_znzkzk_invriant_6__e00}\\
&\mathcal{I}_{\alpha \beta a}=\frac{k' M_{23}(Q_{13}Q_{22}-Q_{12}Q_{23})}{8}\label{eq_main_znzkzk_invriant_7__e00}
\end{align}
\end{subequations}

We first discuss the  invariant $\mathcal{I}_{\alpha aa}$. 
If $Q_{12}$ is zero, then $\mathcal{I}_{\alpha aa}$ in (\ref{eq_main_znzkzk_invriant_1__e00}) vanishes. So we assume $Q_{12}$ is nonzero when discuss the  invariant. Under this assumption, $\hat Q_{12}=Q_{12}$.  
 According to Table~\ref{table_three_M3L_example_list}, the M3L statistical phases take
\begin{align}
&\theta_{\Sigma;\sigma^1}=\frac{-2\pi  k Q_{12} }{4}\label{eq_main_z2zkzk_e00_i11_phase1}\\
&  \theta_{\sigma^1,\Sigma;\Sigma}=\frac{2\pi  k  Q_{12}}{4}\label{eq_main_z2zkzk_e00_i11_phase2}
\end{align} 
When $Q_{12}=1$, these two statistical phases take the most general quantized values, that is, $\theta_{\Sigma;\sigma^1}=-\frac{2\pi k}{4}$ and  $\theta_{\sigma^1,\Sigma;\Sigma}=\frac{2\pi k}{4}$.

Now we discuss the periods of the two statistical phases. First of all, we consider those from attaching particles.  Since $e$ carries only integer symmetry charge, then by attaching particle, the statistical phases $\theta_{\Sigma;\sigma^1}$ can shift by $\pi$, and  $\theta_{\Sigma,\sigma^1;\Sigma}$ can also shift by $\pi$. Secondly, we can also consider the periods from those of twisted coefficients. However,  the periods from twisted coefficients are always multiple of the ones from attaching particles, as we show below.

From (\ref{eqn:twisted_coff_general_3}),  we set $m=2,N_1=2,N_2=1$ and $M_{12}$ is given by (\ref{eqn_main_znzk1zk2_invariant_1aa}),    the period of $k$   is given by
\begin{align}
\Gamma=&\text{lcm}[\frac{2}{\gcd(2,  Q_{12})}, \frac{2}{\gcd(2,  Q_{22})}].
\label{eq_main_z2zkzk_e00_i11_p1}
\end{align}
Plugging (\ref{eq_main_z2zkzk_e00_i11_p1}) into (\ref{eq_main_z2zkzk_e00_i11_phase1}) and (\ref{eq_main_z2zkzk_e00_i11_phase2}), then periods of these two statistical phases $\theta_{\Sigma;\sigma^1}$ and $\theta_{\Sigma,\sigma^1;\Sigma}$ due to $\Gamma$  are
\begin{align}
\frac{2\pi  Q_{12}}{4}&\text{lcm}[\frac{2}{\gcd(2, Q_{12})}, \frac{2}{\gcd(2, Q_{22})}]\,.
\end{align}
It is easy to verify that, regardless of $Q_{12}$ and $Q_{22}$, this period is always multiple of $\pi$ that is the period from particle attachment. 
 Therefore, the statistical phase $\theta_{\Sigma;\sigma^1}$ and $\theta_{\sigma^1,\Sigma;\Sigma}$ can take  two inequivalent values: $\frac{\pi p_1}{2}$ with $p_1\in \Z_2$. In summary, these statistical phases contribute to one $\Z_2$ classification for SFL.

Similarly, the invariant $\mathcal{I}_{\beta aa}$ can also be discussed and the M3L statistical phases are quantized to be
{\small{
\begin{align}
&\theta_{\Sigma;\sigma^2}=\frac{-2\pi  k  Q_{22}}{4}\\
&  \theta_{\sigma^2,\Sigma;\Sigma}=\frac{2\pi  k Q_{22} }{4}
\end{align}}}
and they can take the most general values
\begin{align}
&\theta_{\Sigma;\sigma^2}=-\frac{2\pi k}{4}\\
&\theta_{\Sigma,\sigma^2;\Sigma}=\frac{2\pi k}{4}
\end{align}
whose periods are both  $\pi$.

We note that even though the two invariants  $\mathcal{I}_{\alpha aa}$  and $\mathcal{I}_{\beta aa}$ can come from the same twisted term $q_{211}$, however depending on different $Q_{12}$ and $Q_{22}$, they can appear independently by choosing different values of $Q_{12}$ and $Q_{22}$ so that they independently contribute to the classification.
Therefore,  the statistical phases $\theta_{\Sigma;\sigma^2}=\theta_{\sigma^2,\Sigma;\Sigma}=\frac{\pi p_3}{2}$ with $p_3\in \Z_2$ can also contribute to another $\Z_2$ classification for SFL.

Now we consider the invariant $\mathcal{I}_{a\alpha \alpha}$. 
  We  assume $Q_{12}$ to be nonzero, otherwise both parts vanish.    
  According to Table~\ref{table_three_M3L_example_list}, the  M3L statistical phases $\theta_{\sigma^1;\Sigma}$ and $\theta_{\Sigma, \sigma^1;\sigma^1}$ from  $\mathcal{I}_{a\alpha \alpha}$ are quantized to 
\begin{align}
 &\theta_{\sigma^1;\Sigma}=-\frac{2\pi (\bar k Q_{12}+ \widetilde k Q_{13} ) Q_{12} }{4}\label{eqn_main_znzkzk_e00_5}\\
&\theta_{\Sigma,\sigma^1;\sigma^1}=\frac{2\pi (\bar k Q_{12}+ \widetilde k Q_{13} ) Q_{12} }{4}\label{eqn_main_znzkzk_e00_6}
\end{align}
When $Q_{12}=Q_{13}=1$, the statistical phases  $\theta_{\sigma^1;\Sigma}$ and $\theta_{\Sigma,\sigma^1;\sigma^1} $ can take the most general values, i.e.,
\begin{align}
&\theta_{\sigma^1;\Sigma}=-\frac{2\pi (\bar k + \widetilde k  )  }{4} \label{eqn_main_znzkzk_e00_6_0}\\
&\theta_{\Sigma,\sigma^1;\sigma^1}=\frac{2\pi (\bar k + \widetilde k  )  }{4}\label{eqn_main_znzkzk_e00_6_1}
\end{align}
with $\bar k$ and $\widetilde k$ being integral. 

The periods of these statistical phases can come from attaching particles and periods of $\bar k$ and $\widetilde k$.  From attaching particles, both $\theta_{\sigma^1;\Sigma}$ and $\theta_{\Sigma,\sigma^1;\sigma^1}$ can shift by $\pi$. On the other hand, the periods of the two statistical phases from the ones of $\bar k$ and $\widetilde k$ are always multiple of that from attaching particles, \textit{i.e.},  $\pi$, as follows.

The period of $\bar k$ is  given by (\ref{eq_main_z2zkzk_e00_i11_p1}). From (\ref{eqn:twisted_coff_general_2}), we set $m=2, N_1=2,N_2=N_3=1, K_1=K_2=2$ and $M_{12}$  is given by (\ref{eqn_main_znzk1zk2_invariant_1aa}), we obtain the period of $\widetilde k$:
\begin{align}
\widetilde \Gamma
=\gcd\{&\text{lcm}[\frac{2}{\gcd(2,  Q_{12} )}, \frac{2}{\gcd(2, Q_{22} )}],\nonumber \\
&\text{lcm}[\frac{2}{\gcd(2,  \hat Q_{13})}, \frac{2}{\gcd(2,  \hat Q_{23})}] \}
\label{eq_main_z2zkzk_e00_1ij_p1}
\end{align}
From (\ref{eqn_main_znzkzk_e00_5}) and (\ref{eqn_main_znzkzk_e00_6}), we see that $\Gamma  Q_{12}$ and $\widetilde \Gamma Q_{13} Q_{12}$ must be even, so the periods of $\theta_{\sigma^1;\Sigma}$ and $\theta_{\Sigma,\sigma^1;\sigma^1}$ from  those of $\bar k$ and $\widetilde k$ must be multiple of $\pi$.

Therefore, the statistical phases  ($\theta_{\sigma^1;\Sigma}$, $\theta_{\Sigma,\sigma^1;\sigma^1}$) can take two inequivalent values, that is, ($\frac{\pi p_2}{2}$, $-\frac{\pi p_2}{2}$) with $p_2\in \Z_2$, which contribute to one $\Z_2$ classification for SFL. 

Similarly to $\mathcal{I}_{a\alpha \alpha}$, two statistical phases  ($\theta_{\sigma^2;\Sigma}$, $\theta_{\Sigma,\sigma^2;\sigma^2}$) determined by the invariant $\mathcal{I}_{a\beta \beta}$ also take two inequivalent values,  that is, ($\frac{\pi p_4}{2}$, $-\frac{\pi p_4}{2}$) with $p_4\in \Z_2$, which contributes to another $\Z_2$ classification for SFL. 

We also note that even though the two invariants $\mathcal{I}_{a\alpha \alpha}$ and $\mathcal{I}_{a\beta \beta}$ can come from the same twisted term $q_{122}$ and $q_{123}$, however depending on different $Q_{12}$ and $Q_{22}$, they can appear independently so that they independently contribute to the classification.

Now we discuss the invariant $\mathcal{I}_{\alpha a \beta}$ which now beomes (\ref{eq_main_znzkzk_invriant_5__e00}). 
 Then the M3L statistical phases $\theta_{\Sigma, \sigma^2;\sigma^1}$ and $\theta_{\sigma^2,\sigma^1;\Sigma}$ from the invariant  $\mathcal{I}_{\alpha  a\beta}$ are quantized to 
\begin{align}
 &\theta_{\Sigma,\sigma^2;\sigma^1}=\frac{2\pi (\bar k Q_{22}+ \widetilde k Q_{23} ) Q_{12} }{4}\label{eqn_main_znzkzk_e00_9}\\
&\theta_{\sigma^2,\sigma^1;\Sigma}=-\frac{2\pi (\bar k Q_{22}+ \widetilde k Q_{23} ) Q_{12} }{4}\label{eqn_main_znzkzk_e00_10}
\end{align}
When $Q_{12}=Q_{13}=1$ and $Q_{22}=Q_{23}=1$, the statistical phases  $\theta_{\sigma^1;\Sigma}$ and $\theta_{\Sigma,\sigma^1;\sigma^1} $ can take the most general values, i.e.,
\begin{align}
&\theta_{\Sigma,\sigma^2;\sigma^1}=\frac{2\pi (\bar k + \widetilde k  )  }{4}\label{eqn_main_znzkzk_e00_10_1}\\
&\theta_{\sigma^2,\sigma^1;\Sigma}=-\frac{2\pi (\bar k + \widetilde k  )  }{4}\label{eqn_main_znzkzk_e00_10_2}
\end{align}
with $\bar k$ and $\widetilde k$ being integral.  

Now we consider the periods of  $\theta_{\Sigma,\sigma^2;\sigma^1}$ and $\theta_{\sigma^2,\sigma^1;\Sigma} $. The periods of these statistical phases can come from attaching particles and periods of $\bar k$ and $\widetilde k$.  From attaching particles, both $\theta_{\Sigma,\sigma^2;\sigma^1}$ and $\theta_{\sigma^2,\sigma^1;\Sigma}$ can shift by $\pi$. On the other hand, the periods of the two statistical phases from the ones of $ \bar k$ and $\widetilde k$ are always multiple of that from attaching particles, \textit{i.e.},  $\pi$. 
 The periods of $\bar k$ and $\widetilde k$ are also   $\Gamma$ and $\widetilde \Gamma$, given in (\ref{eq_main_z2zkzk_e00_i11_p1})  and (\ref{eq_main_z2zkzk_e00_1ij_p1}). It is easy to see that  $\Gamma Q_{12}$ and $\widetilde \Gamma Q_{23} Q_{12}$ must be even, so the periods of $\theta_{\Sigma,\sigma^2;\sigma^1}$ and $\theta_{\sigma^2,\sigma^1;\Sigma}$  from those of $\bar k$ and $\widetilde k$ must be multiple of $\pi$.

Therefore, the statistical phases can take two inequivalent values, \textit{i.e.}, $(\frac{\pi  p_5}{2}, -\frac{\pi  p_5}{2})$  with $p_5\in \Z_2$, which contribute to one $\Z_2$ classification for SFL.

Similarly, we can also discuss the invariant $\mathcal{I}_{\beta a \alpha }$. We note that one can observe that simply by exchanging the two $\Z_2$  symmetry subgroups, the above discussions can be carried over for the case of  $\mathcal{I}_{\beta a \alpha }$. Therefore, we can straightforwardly obtain the conclusion: from $\mathcal{I}_{\beta a \alpha }$, the statistical phases $\theta_{\Sigma,\sigma^1;\sigma^2}$ and $\theta_{\sigma^2,\sigma^1;\Sigma}$ can take two inequivalent values that is, $(\frac{\pi p_6}{2}, -\frac{\pi p_6}{2})$  with $p_6\in \Z_2$, which contributes to one $\Z_2$ classification for SFL. We note that the statistical phase $\theta_{\sigma^2,\sigma^1;\Sigma}$ can also originate from the invariant  $\mathcal{I}_{\alpha a \beta }$(the total contribution is then $\frac{\pi (p_5+p_6)}{2}$ in Table ~\ref{table_z2_z2z2_sfl}).

We note that  the four invariants (\ref{eq_main_znzkzk_invriant_3__e00})-(\ref{eq_main_znzkzk_invriant_6__e00}) can come from the same twisted terms $q_{122}$ and $q_{123}$,  however the corresponding statistical phases can take different values independently by choosing different $\bar k$, $\widetilde k$ and also $Q_{12}, Q_{22}, Q_{13}, Q_{23}$. For example, for $\mathcal{I}_{a\alpha \alpha}$, one can choose  $\widetilde k=0$ and $Q_{12}=1$ and other elements of charge matrix are zero and $\bar k$ take different values in $\Z_2$; for $\mathcal{I}_{a\beta \beta}$, one can choose  $\widetilde k=0$ and $Q_{22}=1$ and other elements of charge matrix are zero and $\bar k$ take different values in $\Z_2$;   for   $\mathcal{I}_{\alpha a\beta}$, one can choose $\bar k=0, Q_{12}=Q_{23}=1$ and other elements of charge matrix are zero with $\widetilde k$ take different values in $\Z_2$; for   $\mathcal{I}_{\beta a\alpha}$, one can choose $\bar k=0, Q_{22}=Q_{13}=1$ and other elements of charge matrix are zero with $\widetilde k$ take different values in $\Z_2$. 
Therefore these four invariants can contribute   independently to the classification.

Finally, we come to discuss the invariant $\mathcal{I}_{\alpha \beta a}$. For convenience, we  can divide the expression in (\ref{eq_main_znzkzk_invriant_7__e00}) into two parts, \textit{i.e.},
\begin{align} 
&\mathcal{I}_{\alpha \beta a}^{(1)}=\frac{k' M_{23}Q_{13}Q_{22}}{8}\\
&\mathcal{I}_{\alpha \beta a}^{(2)}=\frac{-k'M_{23}Q_{12}Q_{23}}{8}.
\end{align}
 For convenience, we define $p_{23}$ and $\bar p_{23}$ through
\begin{align}
&\hat Q_{12} M_{23}=2 p_{23}  \label{eq_main_znzkzk_aux_5} \\
&\hat Q_{22} M_{23}=2 \bar p_{23}.  \label{eq_main_znzkzk_aux_6}
\end{align}
We first discuss  $\mathcal{I}_{\alpha \beta a}^{(1)}$. If $Q_{22}=0$ or $Q_{13}=0$, it vanishes. So we focus on the case with nonzero $Q_{22}$ and nonzero $Q_{13}$. Then it can be simplified to
$\mathcal{I}_{\alpha \beta a}^{(1)}=\frac{k' Q_{13} \bar p_{23}}{4}$,
which determines the statistical phases
\begin{align}
&\theta_{\Sigma, \sigma^1;\sigma^2}^{(1)}=-\frac{2\pi k’ Q_{13} \bar p_{23} }{4}\label{eqn_main_znzkzk_e00_11}\\
&\theta_{\Sigma, \sigma^2;\sigma^1}^{(1)}=\frac{2\pi k' Q_{13} \bar p_{23} }{4}\label{eqn_main_znzkzk_e00_12}
\end{align}
When $Q_{22}=1$, then $\bar p_{23}=1$. Further if $Q_{13}=1$, the $\theta_{\Sigma, \sigma^1;\sigma^2}^{(1)}$ and $\theta_{\Sigma, \sigma^2;\sigma^1}^{(1)}$ can take the most general quantized values, that is,
$\theta_{\Sigma, \sigma^1;\sigma^2}^{(1)}=-\frac{\pi k’ }{2}
$, 
$\theta_{\Sigma, \sigma^2;\sigma^1}^{(1)}=\frac{\pi k' }{2}.
$

Similarly, we disucss the second part of the invariant $\mathcal{I}_{\alpha \beta a}^{(2)}$.  If $Q_{12}=0$ or $Q_{23}=0$, it vanishes. So we focus on the case with nonzero $Q_{12}$ and nonzero $Q_{23}$. Then it can be simplified to
$\mathcal{I}_{\alpha \beta a}^{(2)}=-\frac{k' Q_{23} p_{23}}{4},
$
which determines the statistical phases
\begin{align}
&\theta_{\Sigma, \sigma^1;\sigma^2}^{(2)}=\frac{2\pi k’ Q_{23} p_{23} }{4}\label{eqn_main_znzkzk_e00_15}\\
&\theta_{\Sigma, \sigma^2;\sigma^1}^{(2)}=-\frac{2\pi k' Q_{23} p_{23} }{4}\label{eqn_main_znzkzk_e00_16}
\end{align}
When $Q_{12}=1$, then $p_{23}=1$. Further if $Q_{23}=1$, the $\theta_{\Sigma, \sigma^1;\sigma^2}^{(2)}$ and $\theta_{\Sigma, \sigma^2;\sigma^1}^{(2)}$ can take the most general quantized values, that is,
$\theta_{\Sigma, \sigma^1;\sigma^2}^{(2)}=\frac{\pi k’ }{2}$,
$\theta_{\Sigma, \sigma^2;\sigma^1}^{(2)}=-\frac{\pi k' }{2}$.

From  the above discussion, 
  we see the most general quantized values of the two statistical phases from $\mathcal{I}_{\alpha \beta a}$ are
\begin{align}
&\theta_{\Sigma, \sigma^1;\sigma^2}=\frac{\pi k'}{2}\label{eqn_main_znzkzk_e00_27}\\
&\theta_{\Sigma, \sigma^2;\sigma^1}=-\frac{\pi k' }{2}.\label{eqn_main_znzkzk_e00_28}
\end{align}
with $k'$ being integral.

We recall that these two statistical phases $\theta_{\Sigma, \sigma^1;\sigma^2}$ and $\theta_{\Sigma, \sigma^2;\sigma^1}$ can also detemined by the invariants $\mathcal{I}_{\beta a \alpha}$ and $\mathcal{I}_{\alpha a \beta}$, respectively. As compared to (\ref{eqn_main_znzkzk_e00_15}) and (\ref{eqn_main_znzkzk_e00_10_1}),  the values of these two statistical phases from $\mathcal{I}_{\alpha \beta a}$ are not beyond those in (\ref{eqn_main_znzkzk_e00_15}) and (\ref{eqn_main_znzkzk_e00_10_1}).  And any pattern of the two statistical phases in (\ref{eqn_main_znzkzk_e00_27}) and (\ref{eqn_main_znzkzk_e00_28}) are not independent since they can be viewed as combination of those from invariants $\mathcal{I}_{\beta a \alpha}$ and $\mathcal{I}_{\alpha a \beta}$. While they do not contribute to new pattern of SFL,  $\mathcal{I}_{\alpha \beta a}$ might contribute to smaller periods of   these statistical phases.  However as we will show below it does not contribute any smaller period of these statistical phases, that is, they are always multiple of $\pi$.

As we have discussed the periods of $\theta_{\Sigma, \sigma^1;\sigma^2}$ and $\theta_{\Sigma, \sigma^2;\sigma^1}$ from attaching particles above, here we only need to focus on the ones from twisted coefficients. 

We first discuss the period of $k'$.  From (\ref{eqn:twisted_coff_general_2}),   we set $m=2$, $N_1=N_2=1$ and $N_3=2$ and $K_1=K_2=2$, and $M_{23}$ is given by (\ref{eqn_main_znzk1zk2_invariant_2aa}), and we obtain the period of $k'$:
\begin{align}
\Gamma'=\gcd\{ \text{lcm}[&\frac{2}{\gcd(2, p_{23} \hat Q_{13})}, \frac{2}{\gcd(2, p_{23} \hat Q_{23})}, \nonumber \\
&\frac{2}{\gcd(2, \bar p_{23} \hat Q_{13})}, \frac{2}{\gcd(2, \bar p_{23} \hat Q_{23})}], \nonumber \\
\text{lcm}[&\frac{2}{\gcd(2, p_{23} )},\frac{2}{\gcd(2, \bar p_{23} )}, \nonumber \\
&\frac{2}{\gcd(2, p_{32})},\frac{2}{\gcd(2, \bar p_{32} )}
]\}
\end{align}
It is easy to see that $\Gamma' Q_{13} \bar p_{23}$ is even. So from (\ref{eqn_main_znzkzk_e00_11}), (\ref{eqn_main_znzkzk_e00_12}), (\ref{eqn_main_znzkzk_e00_15}), and (\ref{eqn_main_znzkzk_e00_16}),  we see that the periods of  $\theta_{\Sigma, \sigma^1;\sigma^2}$ and $\theta_{\Sigma, \sigma^2;\sigma^1}$ are both multiple of $\pi$.

 In fact, we can also consider other M3L statistical phases, such as $\theta_{\Sigma,\sigma^{12};\sigma^i}$ with $i=1,2$, involving $\sigma^{12}$ which can be viewed as fusion of $\sigma^1$ and $\sigma^2$, having  the unit symmetry fluxes of both the two $\Z_2$ symmetries. However, in this case, they are not independent. For example, the quantization of $\theta_{\Sigma,\sigma^{12};\sigma^i}$ are determined by $\theta_{\Sigma,\sigma^1;\sigma^i}$ and $\theta_{\Sigma,\sigma^2;\sigma^i}$, i.e.,
\begin{align}
\theta_{\Sigma,\sigma^{12};\sigma^i}=\theta_{\Sigma,\sigma^1;\sigma^i}+\theta_{\Sigma,\sigma^2;\sigma^i}.
\label{eqn_main_z2_z2z2_composite}
\end{align} 
We note that even though the quantization of them are not independent, the periods of them, in particular those from attachment of particle, may be different as we can see below.

To summarize, for the SFP pattern  $e00$, the classification of SFL is $(\Z_2)^6$. The corresponding characterization of these SFL are given by the M3L statistical phases above (see Table~\ref{table_z2_z2z2_sfl}).
  \subsubsection{Computing SFL when SFP is either $eC0$ or $e0C$}\label{section_Z2_Z2Z2_computing_SFL_for_SFP_eC0_e0C}

Now we  discuss  the case with SFP being $eC0$, namely the gauge charge $e$ carries half integer symmetry charge of the first $\Z_2$ and integer charge of the second $\Z_2$ subgroup. In this case, $Q_{11}=1, Q_{21}=0$ and then $\hat Q_{11}=1, \hat Q_{21}=2$. Then (\ref{eqn_main_znzk1zk2_multiple_1}) reduce to
\begin{align}
M_{12}
 =&4
\label{eqn_main_znzk1zk2_invariant_1aa_2}.
  \end{align}
while (\ref{eqn_main_znzk1zk2_multiple_2}) remains in the same form. Then the M3L invariants above reduce to
\begin{subequations}
\label{eq_main_invariants_expr_znzkzk_eC0}
\begin{align}
&\mathcal{I}_{\alpha aa}=-\frac{ k Q_{12}}{2}\label{eq_main_znzkzk_invriant_1_eC0}\\
&\mathcal{I}_{\beta aa}=-\frac{k Q_{22}}{2}\label{eq_main_znzkzk_invriant_2_eC0}\\
&\mathcal{I}_{a\alpha\alpha}=\frac{ k Q_{12}}{4}-\frac{\bar k Q_{12}Q_{12}}{2}-\frac{\widetilde k Q_{12}Q_{13}}{2}\label{eq_main_znzkzk_invriant_3_eC0}\\
&\mathcal{I}_{a\beta\beta}=-\frac{\bar k Q_{22}Q_{22}}{2}-\frac{\widetilde k Q_{22}Q_{23}}{2}\label{eq_main_znzkzk_invriant_4_eC0}\\
&\mathcal{I}_{\alpha a \beta}=\frac{\bar k Q_{12}Q_{22}}{2}+\frac{\widetilde k Q_{12}Q_{23}}{2}\label{eq_main_znzkzk_invriant_5_eC0}\\
&\mathcal{I}_{\beta a \alpha}=-\frac{k Q_{22}}{4}+\frac{\bar k Q_{22}Q_{12}}{2}+\frac{\widetilde k Q_{22}Q_{13}}{2} \label{eq_main_znzkzk_invriant_6_eC0}\\
&\mathcal{I}_{\alpha \beta a}=\frac{k Q_{22}}{4}+\frac{k' M_{23} (Q_{13}Q_{22}-Q_{12}Q_{23})}{8}\label{eq_main_znzkzk_invriant_7_eC0}\end{align}
\end{subequations}

As below we will show that all the first six invariants (\ref{eq_main_znzkzk_invriant_1_eC0})-(\ref{eq_main_znzkzk_invriant_6_eC0}) result in only trivial M3L statistical phases while the last one (\ref{eq_main_znzkzk_invriant_7_eC0}) can give rise to nontrivial M3L statistical phase.

First of all, we discuss the invariant $\mathcal{I}_{\alpha \beta a}$ that can contribute to nontrivial classification and now becomes
\begin{align}
\mathcal{I}_{\alpha \beta a}=\frac{k Q_{22}}{4} +\frac{k' (\bar p_{23}Q_{13}-p_{23}Q_{23})}{4}\label{eq_main_znzkzk_invriant_7_1}
\end{align}
where we have used (\ref{eq_main_znzkzk_aux_5}) and (\ref{eq_main_znzkzk_aux_6}) and assumed that $Q_{22}$ and $Q_{12}$ are nonzero (Otherwise the corresponding term(s) would be vanishing.).
We  divide it into two parts for convenience
\begin{align}
&\mathcal{I}_{\alpha \beta a}^{(1)}=\frac{k' (\bar p_{23}Q_{13}-p_{23}Q_{23})}{4}\\
&\mathcal{I}_{\alpha \beta a}^{(2)}=\frac{k Q_{22}}{4}
\end{align}
Then the two statistical phases $\theta_{\Sigma,\sigma^2;\sigma^1}$ and $\theta_{\Sigma,\sigma^1;\sigma^2}$ can be divided into two parts. We first discuss the part one determined by $\mathcal{I}_{\alpha \beta a}^{(1)}$, that is,
\begin{align}
&\theta_{\Sigma,\sigma^2;\sigma^1}^{(1)}=2\pi \frac{k' (\bar p_{23}Q_{13}-p_{23}Q_{23})}{4}\\
&\theta_{\Sigma,\sigma^1;\sigma^2}^{(1)}=-2\pi \frac{k' (\bar p_{23}Q_{13}-p_{23}Q_{23})}{4}
\end{align}
When $Q_{22}=Q_{13}=1$(with $Q_{23}=0$), $\bar p_{23}=1$ according to (\ref{eq_main_znzkzk_aux_6}) and (\ref{eqn_main_znzk1zk2_invariant_2aa}), which means $\theta_{\Sigma,\sigma^2;\sigma^1}^{(1)}$ and $\theta_{\Sigma,\sigma^1;\sigma^2}^{(1)}$ can take the most general quantized values, that is, $\theta_{\Sigma,\sigma^2;\sigma^1}^{(1)}=\frac{2\pi k'}{4}$ and $\theta_{\Sigma,\sigma^1;\sigma^2}^{(1)}=-\frac{2\pi k'}{4}$. 
From $\mathcal{I}_{\alpha \beta a}^{(2)}$, we have
$\theta_{\Sigma,\sigma^2;\sigma^1}^{(2)}=2\pi \frac{k Q_{22}}{4} $ and
$\theta_{\Sigma,\sigma^1;\sigma^2}^{(2)}=-2\pi \frac{k Q_{22}}{4} $
which are always multiple of $\frac{\pi}{2}$, the most general value from $\mathcal{I}_{\alpha \beta a}^{(1)}$.

We then consider the periods of the two statistical phases that can come from two aspects. First, $\theta_{\Sigma,\sigma^2;\sigma^1}$ and $\theta_{\Sigma,\sigma^1;\sigma^2}$ can shift by $\pi$ and $\frac{\pi}{2}$ respectively from attaching $e$ particle to $\Sigma$ since $e$ particle carries integer charge of the second $\Z_2$ and half charge of the first $\Z_2$.  Secondly, the periods of these statistical phases can also come from twisted coefficients, which, however, are always multiple of those from attaching particle, as follows. From (\ref{eqn:twisted_coff_general_1}) and (\ref{eqn:twisted_coff_general_3}), the period of $k$ and $k'$ are given by 
\begin{align}
\Gamma=\text{lcm}[\frac{2}{\gcd(2, \hat Q_{12})},  \frac{2}{\gcd(2, \hat Q_{22})}]
\end{align}
and 
\begin{align}
\Gamma'=\gcd\{ \text{lcm}[&\frac{2}{\gcd(2, p_{23} \hat Q_{13})}, \frac{2}{\gcd(2, p_{23} \hat Q_{23})}, \nonumber \\
&\frac{2}{\gcd(2, \bar p_{23} \hat Q_{13})}, \frac{2}{\gcd(2, \bar p_{23} \hat Q_{23})}], \nonumber \\
\text{lcm}[&\frac{4}{\gcd(4, p_{23})},\frac{4}{\gcd(4, \bar p_{23})}, \nonumber \\
&\frac{4}{\gcd(4, p_{32} )},\frac{4}{\gcd(4, \bar p_{32} )}
]\}
\end{align}
respectively. As  $\Gamma Q_{22}$ is  even, then $\frac{2\pi \Gamma Q_{22}}{4}$ is always multiple of $\pi$. Meanwhile, we can see that both $\Gamma' \bar p_{23} Q_{13}$ and $\Gamma' p_{23} Q_{23}$ are also even, then  $\frac{2\pi \Gamma'\bar p_{23} Q_{13}}{4}$ and $\frac{2\pi \Gamma' p_{23} Q_{2j}}{4}$ are also multiple of $\pi$. 

We note that the statistical phases $\theta_{\Sigma,\sigma^2;\sigma^1}$ can also determined by invariant $\mathcal{I}_{\alpha a \beta}$,  as given by  (\ref{eqn_main_z2_z2z2_statistics_ec0_1}) below, which however are always multiple of the trivial value $\pi$.

Therefore, $\theta_{\Sigma,\sigma^1;\sigma^2}$ are always equivalent to being trivial while $\theta_{\Sigma,\sigma^2;\sigma^1}$ can take two inequivalent values, that is, $\frac{\pi p_7}{2}$ with $p_7\in \Z_2$. Therefore, the $\Z_2$ inequivalent statistical phases $\theta_{\Sigma,\sigma^2;\sigma^1}$ contributes to one $\Z_2$ classification of SFL (see Table~\ref{table_z2_z2z2_sfl}).

Below we discuss the other invariants in (\ref{eq_main_invariants_expr_znzkzk_eC0}) can only result in trivial values.
We first discuss the invariant $\mathcal{I}_{\alpha aa}$.
If $Q_{12}$ is zero, then $\mathcal{I}_{\alpha aa}$ vanishes. So we assume $Q_{12}$ is nonzero when discussing the  invariant. Then the M3L statistical phases take
{\small{
\begin{align}
&\theta_{\Sigma;\sigma^1}=\frac{-2\pi  k Q_{12}}{2}\label{eq_app_z2zkzk_eC0_i11_phase1}\\
&  \theta_{\sigma^1,\Sigma;\Sigma}=\frac{2\pi  k Q_{12} }{2}\label{eq_app_z2zkzk_eC0_i11_phase2}
\end{align}}}
when $Q_{12}=1$, they can take the most general quantized values, that is, $\theta_{\Sigma;\sigma^1}=-\pi  k $ and $\theta_{\sigma^1,\Sigma;\Sigma}=\pi  k $.
However, these quantized values of the two statistical phases are equivalent to being trivial since by attaching $e$ particle to $\Sigma$ or $\sigma^1$, $\theta_{\Sigma;\sigma^1}$  and $\theta_{\sigma^1,\Sigma;\Sigma}$ can shift by $\pi$ and $\frac{\pi}{2}$.

Similarly, we discuss the invariant $\mathcal{I}_{\beta aa}$, which now becomes (\ref{eq_main_znzkzk_invriant_2_eC0}).
Then the two correspondingly statistical phases $\theta_{\Sigma;\sigma^2}$ and $\theta_{\sigma^2,\Sigma;\Sigma}$ are also multiple of $\pi$, which however are also equivalent to being trivial since attaching particle $e$ to $\Sigma$ can shift the phases by  $\pi$.

We now discuss the invariant $\mathcal{I}_{a\alpha \alpha}$, which now becomes (\ref{eq_main_znzkzk_invriant_3_eC0}).
Then the two statistical phases can take the quantized values as
\begin{align}
&\theta_{\sigma^1;\Sigma}=2\pi \bigg[\frac{k Q_{12}}{4}-\frac{\bar k Q_{12}Q_{12}}{2}-\frac{\widetilde k Q_{12}Q_{13}}{2}\bigg]\\
&\theta_{\Sigma,\sigma^1;\sigma^1}=-2\pi \bigg[\frac{k Q_{12}}{4}-\frac{\bar k Q_{12}Q_{12}}{2}-\frac{\widetilde k  Q_{12}Q_{13}}{2}\bigg]
\end{align}
When $Q_{12}=Q_{13}=1$, they can take the most general quantized values, \textit{i.e.}, $\theta_{\sigma^1;\Sigma}=\frac{2\pi l}{4}$ and $\theta_{\Sigma,\sigma^1;\Sigma}=-\frac{2\pi l}{4}$ with $l$ being integral. However, these quantized values of the statistical phases are trivial due to the following reason: as $e$ carries half integer charge of the first $\Z_2$ symmetry,  attaching $e$ to  $\sigma^1$ for $\theta_{\sigma^1;\Sigma}$ and  to $\Sigma$ for $\theta_{\Sigma,\sigma^1;\sigma^1}$ can both shift them  by $\frac{\pi}{2}$ phase shift.

We turn to discuss the invariant $\mathcal{I}_{a\beta \beta}$, which now becomes  (\ref{eq_main_znzkzk_invriant_4_eC0}).
Then the two statistical phases can take the quantized values as
\begin{align}
&\theta_{\sigma^2;\Sigma}=2\pi \bigg[-\frac{\bar k Q_{22}Q_{22}}{2}-\frac{\widetilde k Q_{22}Q_{23}}{2}\bigg]\\
&\theta_{\Sigma,\sigma^2;\sigma^2}=-2\pi \bigg[-\frac{\bar k Q_{22}Q_{22}}{2}-\frac{\widetilde k Q_{22}Q_{23}}{2}\bigg]
\end{align}
These quantized values of the statistical phases are always multiple of $\pi$ and then trivial due to the following reason: as $e$ carries  integer charge of the second $\Z_K$ symmetry,  attaching $e$ to  $\sigma^2$ for $\theta_{\sigma^2;\Sigma}$(i.e., one $e$ to each $\sigma^2$ in the exchanging braiding) or  to $\Sigma$ for $\theta_{\Sigma,\sigma^2;\sigma^2}$ can both shift them  by $\pi$ phase shift.

Now we discuss the invariant $\mathcal{I}_{\alpha a \beta}$, which now becomes (\ref{eq_main_znzkzk_invriant_5_eC0}).
Then the two statistical phases now can take the quantized values 
\begin{align}
&\theta_{\Sigma,\sigma^2;\sigma^1}=2\pi\bigg[\frac{\bar k Q_{12}Q_{22}}{2}+\frac{\widetilde k Q_{12}Q_{23}}{2}\bigg]
\label{eqn_main_z2_z2z2_statistics_ec0_1}\\
&\theta_{\sigma^2,\sigma^1;\Sigma}=-2\pi\bigg[\frac{\bar k Q_{12}Q_{22}}{2}+\frac{\widetilde k Q_{12}Q_{23}}{2}\bigg]
\end{align}
These quantized values of the statistical phases are always multiple of $\pi$, and then trivial due to the following reason: as $e$ carries half integer charge of the first $\Z_2$ symmetry and integer charge of the second $\Z_2$ symmetry,  attaching $e$ to  $\Sigma$ for $\theta_{\Sigma,\sigma^2;\sigma^1}$ and  to $\sigma^2$ for $\theta_{\sigma^2,\sigma^1;\Sigma}$ can  shift them  by  $\pi$ and $\frac{\pi}{2}$ respectively.

We turn to consider  the invariant $\mathcal{I}_{\beta a \alpha }$, which now becomes (\ref{eq_main_znzkzk_invriant_6_eC0}).
Then the two statistical phases now can take the quantized values as
\begin{align}
&\theta_{\Sigma,\sigma^1;\sigma^2}=2\pi\bigg[-\frac{ k Q_{22}}{4}+\frac{\widetilde k Q_{22}Q_{12}}{2}+\frac{\widetilde k Q_{22}Q_{13}}{2}\bigg]\\
&\theta_{\sigma^1,\sigma^2;\Sigma}=2\pi\bigg[\frac{ k Q_{22}}{4}-\frac{\widetilde k Q_{22}Q_{12}}{2}-\frac{\widetilde k Q_{22}Q_{13}}{2}\bigg]
\end{align}
The quantized values of the statistial phases are always multiple of $\frac{\pi}{2}$ and then trivial due to the following reason: as the $e$ particle carries half  charge of the first $\Z_K$ symmetry, attaching $e$ partcle to $\Sigma$ for $\theta_{\Sigma,\sigma^1;\sigma^2}$ and to $\sigma^2$ for $\theta_{\sigma^1,\sigma^2;\Sigma}$ can shift them by $\frac{\pi}{2}$ phase factor.

Now we discuss the M3L statistical phases $\theta_{\Sigma, \sigma^{12};\sigma^i}$. From  the relation (\ref{eqn_main_z2_z2z2_composite}) and the above discussions, we can see that  the quantization values of $\theta_{\Sigma, \sigma^{12};\sigma^i}$ are always multiple of $\frac{\pi}{2}$, which however is trivial as attaching $e$ particle to $\Sigma$ can result in phases shift of $\frac{\pi}{2}$.

To summaize, for SFP $eC0$, the classification for SFL is $\Z_2$, which is  characterized by $\theta_{\Sigma,\sigma^2;\sigma^1}=\frac{\pi p_7}{2}$ with $p_7\in \Z_2$.\\

 Similarly to the SFP $eC0$, the discussion for SFP with $e0C$ for the classification for SFL can be easily obtained by just exchanging the two $\Z_2$ symmetries. Therefore, we can  obtain the conclusion: for SFP $e0C$, the  classification for SFL is $\Z_2$, giving by the two inequivalent values of the statistical phases $\theta_{\Sigma,\sigma^1;\sigma^2}=\frac{\pi p_8}{2}$ where $p_8\in \Z_2$(see Table~\ref{table_z2_z2z2_sfl}).

  \subsubsection{Computing SFL when SFP is   $eCC$}\label{section_Z2_Z2Z2_computing_SFL_for_SFP_eCC}

This case can also be connected to by rearranging the symmetry group $\Z_2\times \Z_2$ generated by $g,h$ to  another isomorphic one $\Z_2\times \widetilde{\Z}_2$ whose generators are $g, gh$. In this case, $e$ particle still carries half charge of the first $\Z_2$ symmetry, but integer charge of the second $\widetilde{\Z}_{2}$ symmetry now. In other words, the SFP $eCC$ now becomes $eC\widetilde 0$ where the superscript reminds of the second $\widetilde{\Z}_2$ is generated by $gh$.  Then we can apply the above discussion to this case and we can draw the conclusion: for SFP $eCC$ (or equivalently $eC\widetilde 0$), the classification for SFL is $\Z_2$, characterizing by the two inequivalent statistical phases $\theta_{\Sigma, \sigma^{12};\sigma^1}=\frac{\pi p_9}{2}$
 with $p_9\in \Z_2$ where $\sigma^{12}$ is the fusion of $\sigma^1$ and $\sigma^2$(see Table~\ref{table_z2_z2z2_sfl}).
Similarly argument also show that $\theta_{\Sigma, \sigma^{12};\sigma^2}=\frac{\pi p_9}{2}$.  Therefore, for $eCC$, the classification of SFL is $\Z_2$ characterized by $\theta_{\Sigma,\sigma^{12};\sigma^1}=\theta_{\Sigma,\sigma^{12};\sigma^2}=0, \frac{\pi}{2}$ mod $\pi$.

\section{Summary and outlook}\label{section_summary_outlook}

In summary, we have systematically developed a field-theoretical framework to completely    characterize and classify symmetry fractionalization of topological excitations (especially loop excitations) in 3D symmetry enriched topological phases.
  The procedure is illustrated via concrete examples, as listed in Tables~\ref{znset}, \ref{table_z4z4sfl}, \ref{table_z2z2_z2_sfl}, and \ref{table_z2_z2z2_sfl}. Especially, the three examples  {studied} in the main text   {respectively stand for three typical   combinations of gauge fields and symmetries, namely, untwisted gauge theory with cyclic symmetry group, twisted gauge theory with cyclic symmetry group, and untwisted gauge theory with product symmetry group.}  In addition, anomalous symmetry fractionalization and the underlying algebraic structure of  the classification in our field-theoretical  framework have been discussed. 

%
%

Some interesting questions   remain open, which are left for future study:
      \begin{enumerate}[label=\textbf{(\roman*)}]
\item In this paper, SFP is characterized by fractionalized symmetry charges, corresponding to one-dimensional   projective representations of  {the} symmetry group,  {which}  indicates that the gauged group $G_g^*$ is always Abelian; SFL is characterized by the M3L statistical phases  that are also Abelian.  We have not yet considered  ``{non-Abelian fractionalization of Abelian symmetry}''. The SETs that involve non-Abelian properties may come from two aspects:   (i) the particles carry   two- or higher-dimensional  irreducible projective representations of symmetry group, which upon gauging results in a non-Abelian   {gauge theory};  (ii) the non-Abelian MML invariants (related to M4L  or non-Abelian M3L).  
 
\item We have not yet considered  permutation of topological superselection sectors under symmetry transformations. It is interesting to add this layer of complexity into 3D SET classification. Several initial attempts to  {treat} untwisted topological orders were made in Ref.~\cite{ye16a}.
 
\item Another interesting  {issue} is to   {introduce extra topological terms in the $BF$ field theory which is beyond the Dijkgraaf-Witten twisted gauge fields,}  such as $bb$ type \cite{bti2,Kapustin2014,QRW2019prb} or $aab$ type \cite{ye17b,PhysRevResearch.3.023132}. With these topological terms, it is very interesting to discuss the symmetry fractionalization.

\item  {Generalizing our theory to} fermionic SETs via spin-TQFT  {is interesting and challenging},  {where the} periods and quantization rules    {are} in general  substantially modified when transparent particles are fermionic.   And it is also possible to  implement spatial symmetry into topological orders and study generalized Wen-Zee terms \cite{PhysRevB.99.205120}.

\item  {The} underlying algebraic structure  {of our specific field-theoretical data of symmetry fractionalization} is not completely  {understood}. Some aspects of algebraic structure are  preliminarily  analyzed in Sec.~\ref{classification_principle}, inspired by the present field-theoretical framework. It is fundamentally important  but definitely   challenging to find a more coherent and more complete algebraic description.

\item Recently, field theories with exotic higher-rank symmetry and higher-moment conservation, e.g., dipole conservation were motivated in part from the rapid development of fracton topological order, have gained intensive attentions from both condensed matter physics and high energy physics~\cite{fracton_qft1,fracton_qft2,fracton_qft3,fracton_qft4,fracton_qft5,fracton_qft6,fracton_qft7,fracton_qft8,fracton_qft9,fracton_qft10,fracton_qft11,fracton_qft12,fracton_qft13,fracton_qft14,fracton_qft15,fracton_qft16,fracton_qft20,fracton_qft21,fracton_qft22,fracton_qft23,fracton_qft24,fracton_qft17,fracton_qft18}. It will be exciting to further implement global symmetry in these unconventional field theories and see how symmetry is fractionalized.

    \end{enumerate}
  
  \acknowledgements
We sincerely thank many colleagues, especially Meng Cheng and Qing-Rui Wang  for their discussions since we initiated this work in 2016. Most parts of this work done by S.Q.N. and P.Y. were conducted in Guangzhou South Campus of  Sun Yat-sen University (SYSU), with the support of   SYSU Talent Plan Startup Grant, Guangdong Basic and Applied Basic Research Foundation under Grant No.~2020B1515120100, and National Natural Science Foundation of China (NSFC)
Grant (No.~11847608 \& No.~12074438).   The work (2016-2018) done by P.Y. at the University of Illinois was supported in part by National Science Foundation grant DMR 1408713 and grant of the Gordon and Betty Moore Foundation EPiQS Initiative through Grant No. GBMF4305.  S.Q.N. and Z.X.L. were also supported by the NSF of China (No. 11974421, No. 12134020).

\appendix

 \section{Frequently used abbreviations and concepts}\label{appendix_abbr_list}
 
 \begin{itemize}
 
  \item AB: Aharonov-Bohm
  \item I3L: intrinsic 3-loop
\item MML: mixed multi-loop
 \item M3L: mixed 3-loop
 \item M4L: mixed 4-loop
\item SPT: symmetry-protected topological phases
 \item SET: symmetry-enriched topological phases
 \item SF: symmetry fractionalization
 \item SFP: symmetry fractionalization on particles
 \item SFL: symmetry fractionalization on loops
 
 \item TQFT: topological quantum field theory
 \item Charge matrix $Q_{ij}$ is defined in Sec.~\ref{sec_adding_symmetry}
 \item Reduced charge matrix  $\mathcal{Q}_{ij}$ is defined in Sec.~\ref{SF}
 \item SFP matrix  $\mathcal{C}_{ij}$ is defined in Sec.~\ref{SF}
 \item M3L statistical phases $\theta_{..;.}$ and $\theta_{.;.}$ are defined in Sec.~\ref{section_mmt_general_discussion}
 \item M3L invariant  $\mathcal{I}_{...}$ is defined in Sec.~\ref{mmt_general_discussion_more_General}
 
 \end{itemize}
 
  \section{Review on 3D bosonic topological order: particle excitations, loop excitations,  and  twisted gauge theories}
 \label{sec_prel_1}

 3D  Abelian  topological orders we consider {here}  are described by TQFT of Dijkgraaf-Witten type, \textit{i.e.}, twisted $BF$ gauge theories  with gauge group $G_g=\prod_i \Z_{N_i}$  \cite{PhysRevLett.114.031601, jian_qi_14,wang_levin1,string5,YeGu2015, ye17b, 2016arXiv161008645Y,3loop_ryu, 2016arXiv161209298P,Tiwari:2016aa, string6, bti2, Kapustin2014, PhysRevB.97.085147}. In such topological orders,  particle excitations are   Abelian and bosonic. Since they  carry gauge charge (representation of gauge group), we can define $n$ species of elementary particle excitations denoted by ${e}_i$ ($i=1,2,...,n$ denote $n $ different $\Z_{N_i}$ gauge groups) while   other excitations can be realized by fusing these elementary ones. 
One of basic fusion rules is the fact that $N_i$  ${e}^i$'s fuse to the trivial particle (\textit{i.e.}, vacuum), ${e}_i\times {e}_i \times ...\times {e}_i=1$, or symbolically  
  \begin{align}
  ({e}_i)^{N_i} =1\,.
  \label{eqn_chage_fusion_rule}
  \end{align}
  It is convenient to introduce integer-valued vectors to denote particle excitations.  For example, ${e}_i=(0,...,1,...,0)$  where only the $i$-th component is nonzero. All other particles can be geometrically denoted by lattice sites of $n$-dimensional hypercubic lattice.  But there is an explicit redundancy in this labeling system: the fusion rules in Eq.~(\ref{eqn_chage_fusion_rule}) indicate that the $i$-th  component of vectors is defined modulo $N_i$.
   
  As for loop excitations carrying gauge fluxes, there are also elementary loop excitations $\Sigma^i$ whose gauge fluxes, denoted as ${\phi}_i$ ($i=1,2,...,n$), 
  satisfy fusion rules  that $N_i$ different copies of ${m}_i$ gauge flux fuse to the trivial flux (\textit{i.e.}, $2\pi$), ${\phi}_i\times {\phi}_i \times ...\times {\phi}_i=1$, or symbolically 
  \begin{align}
  {\phi}_i^{N_i} =1\,.
  \label{eqn_flux_fusion_rule}
  \end{align}
  Likewise,  we may denote all loop excitations by using $n$-dimensional vectors. For example,   ${\phi}_i=(0,...,\frac{2\pi }{N_i}, ..., 0)$  where only the $i$-th component is nonzero. All other loops can be denoted by superposition of all these basis vectors with integer-valued coefficients.  
 The gauge flux of a generic  loop denoted as $\prod_i (\Sigma^i)^{k_i}$   can be labeled by $\phi=(\phi_1, \phi_2,...\phi_n)$ where $\phi_i=\frac{2\pi k_i}{N_i}$ with $k_i=0,1,..., N_i-1$ being the gauge fluxes corresponding to $\Z_{N_i}$ gauge subgroup.  

   One of   defining properties of the $\prod_i \Z_{N_i}$ twisted gauge theory is that the AB phase accumulated in the process that braids a particle with gauge charge $q=(q_1, q_2, ..., q_n)$ around a loop excitation with gauge flux $\phi=(\phi_1, \phi_2,...\phi_n)$ is given by
    \begin{align}
    \theta_{q,\phi}=\sum_{i} q_i\phi_i 
    \label{eqn_AB}
    \end{align}
    We also call such braiding process as charge-loop braiding.
      
  Another interesting braiding statistics of the Abelian topological order is  three loop braiding statistics.
 In the process of three loop braiding, one loop excitation is adiabatically braided around another while both are linked to the third one. While the fusion rules (\ref{eqn_chage_fusion_rule}), (\ref{eqn_flux_fusion_rule}) and the AB phases (\ref{eqn_AB}) are defined  the same  for all the bosonic topological orders with the same gauge group $\prod_i \Z_{N_i}$, the three loop braiding statistics can be different in different topological orders. The three loop braiding statistics are related to topological invariant in four dimensional manifold, and in principle can be detected numeraically and experimentally.

 The effective action of the 3D Abelian topological order we consider here is  
 \begin{align}
S=  \int \sum_i^n\frac{N_i}{2\pi} b^i d a^i+S^0_{int}\,,
\label{eq:action_of_pure_gauge}
 \end{align}
 where $S^0_{int}$ includes twisted terms\footnote{The terminology ``twisted term'' in this paper is always like $aada$. All symmetry-fractionalization types and topological order we consider in this paper are Abelian, so the twisted terms   $aaaa$ that induce non-Abelian four-loop braiding are not considered throughout this paper.\label{footnote_explain_twistedaada}}: 
 \begin{align}
 S^0_{int}=\int \sum_{i,j,k}\frac{q_{ijk}}{4\pi^2}  a^i a^j d a^k\,.\label{eq_twisted_pure_term}
 \end{align}
$\{a^i\}$ and $\{b^i\}$  are  respectively one- and  two-form gauge fields. All wedge products are implicit.
 $q_{ijk}=0$ for untwisted theories. The first term is a set of $BF$ terms that  describe   charge-loop braiding processes   while the second term is a set of  twisted terms that describe the  three-loop braiding processes. $N_i$ is the level of the $i$-th $BF$ term, corresponding to the $\Z_{N_i}$ gauge subgroup  while $q_{ijk}$ is  the coefficient  (alternative names are coupling constant,  level and so on) of the  twisted 
 term ${a^ia^jda^k}$. In the absence of twisted 
terms  (\textit{i.e.}, $q_{ijk}=0$), the gauge transformation is just $a^i\rightarrow a^i+d\chi^i$, $b^i\rightarrow b^i +dV^i$ where $\{\chi^i\}$ and $\{V^i\}$ are $0$-form and $1$-form gauge parameters respectively. Nevertheless, in the presence of  twisted terms, the gauge transformation of $b^i$ should be modified properly  by adding an additional term with   $q_{ijk}$ dependence  \cite{YeGu2015}.   In fact, several consistency conditions induced by twisted 
terms make   $q_{ijk}$  to be quantized and compact  (\textit{i.e.}, periodically identified), as shown in (\ref{eqn_eqi_rela^2}), \textit{i.e.},
\begin{align}
q_{ijk}=r\cdot \frac{ N_iN_j }{N_{ij}}, r\in \Z_{N_{ijk}}
\label{eqn_eqi_rela^2_app}
\end{align}
  As a result, all topologically  inequivalent values of $q_{ijk}$   determine inequivalent  three-loop braiding statistical phases. All these different gauge theories are classified\footnote{The full expression of $\Gamma_g$ should contain $\prod_{i<j<k<l}\Z_{N_{ijkl}}$ which comes from twisted terms $aaaa$. But in this paper, we do not consider this type of twisted terms.} by  
\begin{align} \Gamma_{G_g}=\prod_{i<j} (\mathbb{Z}_{N_{ij}})^2\!\times\!\prod_{i<j<k} (\mathbb{Z}_{N_{ijk}})^2,\label{equation_gamma_gg}
\end{align}  
From (\ref{equation_gamma_gg}),  it is   apparent that  in   the action  (\ref{eq:action_of_pure_gauge})   we are allowed to  arbitrarily  add      $BF$ terms  (\textit{i.e.}, ``trivial layers”) of  level-$1$  and   add  twisted terms containing   level-$1$ $a^i$ gauge fields. One can prove that addition and deletion of level-$1$ $BF$ terms doesn't affect  braiding data at all; in other words, $\Gamma_{G_g}$ is unchanged\footnote{It is   obvious that greatest common divisor of $1$ and any other integer is always $1$.}. More intuitively, adding $n$ ``trivial layers” is equivalent to adding $n$ species of bosons that are in trivial gapped phases. The coefficients of those twisted terms containing at least one level-$1$ $a^i$ gauge fields  are all equivalent to zero  {in the sense of bulk topological order, \textit{i.e.},  braiding data}.  Nevertheless, once   global symmetry is imposed, these auxiliary gauge fields will be useful and necessary in approaching complete classification of SET orders.

  \section{General properties of charge matrices}
 \label{subsection_syfra_particle_general}

%

Here we discuss the SFP for general charge matrices for system with gauge group  $\prod_{i=1}^n \Z_{N_i}$ and symmetry group $\prod_{i=1}^m \Z_{K_i}$. We begin with the following action:
\begin{align}
  S=\frac{1}{2\pi}\int \sum_{i=1}^{n+n'} N_i b^i da^i+\frac{1}{2\pi} \int \sum_{i=1}^m \sum_{j=1}^{n+n'} A^i Q_{ij} db^j+S_{int},
  \label{couplingaction}
  \end{align}
  where the first term is a set of $BF$ terms.  $N_i>1$ for $ 1\leq  i\leq n$ and $N_i=1$ for $n+1 \leq i  \leq n+n'$. The $Q$ matrix dependent term is a minimal coupling term that defines how global symmetry is implemented. The coefficients $Q_{ij}$ form the $m\times (n+n')$ charge matrix $Q$. $S_{int}$ includes all possible twisted terms (see footnote \ref{footnote_explain_twistedaada}) of $1$-form gauge fields $a^i$ with index $i=1,\cdots, n'$. One may further split $S_{int}$ into two terms:
  \begin{align}
  S_{int}=S^{0}_{int}+S^{1}_{int}\,	.\label{equation_twisted_term_split}
  \end{align}
The two terms are defined as follows. Each twisted term in $S^0_{int}$ is composed by $1$-form gauge gauge fields $a^i$ with  $i=1,\cdots, n$ only. In other words, $BF$ terms together with $S^{0}_{int}$  completely determine  bulk topological order, as shown in Eq.~(\ref{eq_twisted_pure_term}). But $S^0_{int}$, as a non-quadratic term, doesn't participate the determination of SFP at all.  On the other hand, each twisted term in $S^{1}_{int}$ at least contains a  gauge field $a^i$ from $i=n+1,\cdots, n+n'$. These gauge fields are called ``level-$1$ gauge fields'', see footnote \ref{footnote_level1}. Inclusion of $S^{1}_{int}$ doesn't introduce new twisted gauge theories  outside $\Gamma_{G_g}$ defined in Eq.~(\ref{equation_gamma_gg}), just like the inclusion of $BF$ terms with $N_i=1$.  But $S^1_{int}$ is expected to introduce nontrivial SFL, which is discussed  in Sec.~\ref{section_mmt_general_discussion}.  
\textit{For all above reasons, in the following discussions on SFP, we will temporarily ignore both  $S^0_{int}$ and  $S^1_{int}$.  }

    Now we insert a set of particle currents $\{j^i\}$ and a set of  loop currents $\{\Sigma^i\}$ with proper gauge\footnote{Loop currents are usually denoted by an antisymmetric tensor $\Sigma^i$, which, in Euclidean spacetime, form a two-dimensional closed world-sheet. A recent calculation related to this loop current in condensed matter can be found in Ref.~\cite{ye17b,bti2}.}, which minimally {couple to} the one-form and two-form gauge fields respectively:
  \begin{align}
   S'=&\frac{1}{2\pi}\int \sum_{i=1}^{n+n'} N_i b^i da^i+\frac{1}{2\pi} \int \sum_{i=1}^m \sum_{j=1}^{n+n'} A^i Q_{ij} db^j\nonumber\\
   &+\sum_{i=1}^{n+n'} q_{i} a^i *j^i+ \sum_{i=1}^{n+n'} q_{i}' b^i *\Sigma^i\,. \label{equation_sprime1}
   \end{align}
  Integrating out  all the two-form gauge fields $b^i$, we obtain a constraint   
   \begin{align}
a^i=-\frac{2\pi q_i'}{N_i} *d^{-1} \Sigma^i-\sum_{j=1}^m\frac{Q_{ji}}{N_i} A^j.
\label{Eqnmotion}
\end{align}
Substituting these constraints back into Eq.~(\ref{equation_sprime1}), we end up with the following effective functional that is fully determined by the configurations of external gauge fields $A^i$,  particle currents, and loop currents:
 \begin{align}
S[A,\!j,\!\Sigma]\!\!=\!\!-\!\!\sum_{i=1}^{n+n'} \!\!\!\int \!\frac{2\pi q_i q_i'}{N_i} j^id^{-1} \Sigma^i-\!\sum_{i=1}^{m}\!\!\sum_{j=1}^{n+n'} \! \!\!\int \!\!\frac{q_jQ_{ij}}{N_j}\! A^i \!*j^j .
\label{particle_SF0}
\end{align}
The first term in  Eq.~(\ref{particle_SF0}) describes  the braiding process in which a particle excitation   carrying $q_i$  unit gauge charge in $\Z_{N_i}$ gauge subgroup (i.e., $e_i^{q_i}$)  braids around a loop excitation carrying $q_i'$ unit gauge flux in $\Z_{N_i}$ gauge subgroup  generates a statistical phase $ {2\pi  q_i q_i' }/{N_i}$.   The second term indicates that the particle excitation carrying $q_j$  gauge charge in $\Z_{N_j}$ gauge subgroup (i.e., $e_j^{q_j}$) carries $ {q_jQ_{ij}}/{N_j}$ symmetry charge of global symmetry subgroup $\Z_{K_i}$. 
Since this symmetry charge is always integer-valued (\textit{i.e.}, ``non-fractionalized'') once $N_j=1$ for the labels $j=n+1,\cdots,n+n'$, it is sufficient to consider $j=1,\cdots,n$ during the following calculation of SFP.   Again, it is also sufficient to consider particles with unit gauge charge $q_j=1$\footnote{In order to simplify descriptive words, here we have already implicitly assumed that all other $q$'s vanish. By ``a particle with unit gauge charge in a certain gauge subgroup'', we have already assumed that the particle doesn't carry   nontrivial gauge charges in all other gauge subgroups, unless otherwise specified. These particles are also called ``elementary particles or elementary particle excitations'', denoted as $e_i$,  in this paper. Likewise, we have similar definitions for ``elementary gauge fluxes / loop excitations''.\label{footnote_definition_elementary_particle_loop}}. Since if  $ {q_jQ_{ij}}/{N_j}$ is fractional, ${Q_{ij}}/{N_j}$ is also fractional. The inverse is not always true. We conclude that, the complete answer to SFP has been encoded in  those  particles carrying unit gauge charge. For gauge group $G_g$, there are $n$ such elementary particles, denoted as $e_i$, each of which carries unit gauge charge of a specific gauge subgroup.

Next, we demonstrate that there exist equivalence relations such that infinite choices of symmetry charges $ {Q_{ij}}/{N_j}$ can be reduced to a finite set (here $Q_{ij}$ may take arbitary integer.).    For this purpose, let us examine the property of the minimal coupling term ``$\frac{1}{2\pi} \int \sum_{i=1}^m \sum_{j=1}^{n+n'} A^i Q_{ij} db^j$''. Since the external gauge field $A^i$ probes symmetry charge in symmetry subgroup $\Z_{K_i}$, the coefficient $Q_{ij}$ should be equivalent to $Q_{ij}+K_{i}$. Furthermore, one may perform the integral-by-part and obtain  ``$\frac{1}{2\pi} \int \sum_{i=1}^m \sum_{j=1}^{n+n'} b^j Q_{ij} dA^i$'', which indicates that the external symmetry flux is also charged in the gauge subgroup $\Z_{N_i}$. Therefore, the coefficient $Q_{ij}$ should be equivalent to $Q_{ij}+N_{j}$\footnote{Such equivalence relation of $Q_{ij}$ can also be understood as by attaching trivial particles to elementary  particle excitation $e_i$.}.  Combining two facts together, we see that for characterizing SFP,  $Q_{ij}$ is equivalent to $  Q_{ij}+\alpha_{ij} N_j +\beta_{ij}K_i$, \textit{i.e.},
\begin{align}
Q_{ij}\sim Q_{ij}+\alpha_{ij} N_j +\beta_{ij}K_i
\end{align} where $\alpha_{ij}$ and $\beta_{ij} $ are arbitrary integers.  From the B\'ezout's lemma, $Q$ can be decomposed in terms of  
\begin{align}
Q_{ij}=\alpha_{ij} N_j +\beta_{ij}K_i+\mathcal{Q}_{ij}\,,
\label{reducedQ}
\end{align}
  where the integer $|\mathcal{Q}_{ij}|$ is less than the greatest common divisor of $K_i$ and $N_j$ denoted by $\text{gcd}(K_i,N_j)$. Alternatively, one can generally express $Q_{ij}$ as:
 $  Q_{ij}=n_{ij}\text{gcd}(K_i,N_j)+\mathcal{Q}_{ij}$,  where the integer $n_{ij}$ is determined by $\alpha_{ij}$ and $\beta_{ij}$. $\mathcal{Q}$ is referred to as ``\textit{reduced charge matrix}'' whose element $\mathcal{Q}_{ij}$ takes values in a finite set:\footnote{One can always  redefine the set as $\{1,\cdots,   \text{gcd}(K_i,N_j)  \}$ since $0$ is equivalent to $ {\text{gcd}(K_i,N_j)}$.}
  \begin{align}
\mathcal{Q}_{ij}\in\{0,1,2,\cdots,   \text{gcd}(K_i,N_j)-1\}\,.\label{eq_reduced_element}
\end{align}
We use the SFP matrix $\mathcal{C}_{ij}$ to denote  the corresponding $\Z_{K_i}$ symmetry charges carried by particles $e_j$ with unit $\Z_{N_j}$ gauge charge: $\mathcal{C}_{ij}=\frac{\mathcal{Q}_{ij}}{N_j}\text{ mod }\frac{\text{gcd}(K_i,N_j)}{N_j}\,.  $ 
It has $\text{gcd}(K_i,N_j)$ inequivalent values,  indicating that there are $\text{gcd}(K_i,N_j)$ types of SFP of $e_j$ \footnote{For $j=n+1,\cdots, n+n'$, we have trivial $BF$ levels, \textit{i.e.}, $N_j=1$, therefore, $\mathcal{C}_{ij}$ is always $0\text{ mod }1$, \textit{i.e.}, non-fractionalized. As such, for the purpose of computing SFP, it is sufficient to consider reduced charge matrices $\mathcal{Q}$ of a reduced size $m\times n$. \label{footnote_reduced_column}}.  Considering all different subsymmetries for $i=1,...,m$, there are in total $\prod_{i}^m \text{gcd}(K_i,N_j)$ types of SFP of $e_j$. Furthermore, considering all elementary particles $e_j$ with $j=1,...,n$,   we obtain the total types of SFP should be: $\prod_{i=1}^m \prod_{j=1}^n \text{gcd} (K_i, N_j)$. 

In summary, we have successfully reduced infinite number of charge matrices $Q$  to finite number of ``reduced charge matrices'' $\mathcal{Q}$ whose elements are given by Eq.~(\ref{eq_reduced_element}). As a result, we end up with $\prod_{i=1}^m \prod_{j=1}^n \text{gcd} (K_i, N_j)$ different $\mathcal{Q}$s. Given a $\mathcal{Q}$, the $\Z_{K_i}$ symmetry charge $(i=1,2,\cdots,m)$ carried by $e_j$ (i.e., particles with unit $\Z_{N_j}$ gauge charge) ($j=1,2,\cdots,n$) is uniquely fixed and exactly provided by  $\mathcal{C}_{ij}$ in Eq.~(\ref{eq_mathcalc}). If we use $\nu_i$ to label SF type with $i=1,2,\cdots, N_v$ and
 $N_v=\prod_{i=1}^m \prod_{j=1}^n \text{gcd} (K_i, N_j)\,$,  each type of SFP is one-to-one correspondence to $\mathcal{Q}$ but there is a caveat about the anomalous SFP of twisted topological order, as mentioned in Sec.\ref{SF}. We finally note that  $\nu_1$ denotes the trivial SFP in which symmetry charge is integral.

 \section{Quantization and periods of coefficients of twisted topological terms} 
  \label{appendix_SEG_quantization}
   
 In this Appendix, we provide detailed calculations on quantization rules (i.e., levels and periods) of coefficients of twisted topological terms, with and without global symmetry.

We  consider  a gauge theory  with gauge group   $\prod_{i=1}^{n+n'}\Z_{N_i} $ and symmetry group   $\prod_{i=1}^m\Z_{K_i}$ that is twisted by one twisted term $q$, i.e., with general action 
\begin{eqnarray}
S&=&S_0+S_c \nonumber \\
S_0&=&\sum_{i=1}^{n+n'} \int \frac{N_i}{2\pi} a^i   db^i +\frac{q}{4\pi^2} a^1 a^2 da^3  \nonumber \\
S_c&=&\sum_{i=1}^m\sum_{j=1}^{n+n'} \int \frac{Q_{ij}}{2\pi} A^i  db^j.  
\label{action:review2}
\end{eqnarray}
In (\ref{action:review2}), besides the level-$N_{1,2,3}$ gauge fields that is included in the twisted term $q$, there are also additional $n$ different untwisted gauge fields, which might be level-1 or not and can also be coupled to external probe fields. As it is shown below, these untwisted gauge fields do not affect the quantization and period of the twisted coefficient $q$.

We first claim   that the quantization levels and period of $q$ are
\begin{align}
q=k  M, k\in\Z_\Gamma
\end{align}
with
\begin{widetext}
\small
\begin{align}
&M=\text{lcm}\bigg[N_1,N_2,\frac{N_1K_1}{\gcd(\hat Q_{11}, N_1K_1)},\frac{N_2K_1}{\gcd(\hat Q_{12},N_2K_1)},  
...,\frac{N_1K_m}{\gcd(\hat Q_{m1}, N_1K_m)},\frac{N_2K_m}{\gcd(\hat Q_{m2},N_2K_m)}\bigg]
\label{eqn:append_5} \\
&\Gamma=\gcd\bigg\{\text{lcm}\bigg[\frac{N_1N_2K_j}{\gcd(N_1N_2K_j,  \hat Q_{j2}M)}, \frac{N_1N_2K_i K_j}{\gcd({N_1N_2K_i K_j}, \hat Q_{i1} \hat Q_{j2} M)}, \forall i,j=1,...,m\bigg], \nonumber \\
&\qquad \qquad\, \,\text{lcm}\bigg[\frac{N_1N_2K_j}{\gcd(N_1N_2K_j,  \hat Q_{j1}M)},
\frac{N_1N_2K_i K_j}{\gcd({N_1N_2K_i K_j}, \hat Q_{i1} \hat Q_{j2} M)}, \forall i,j=1,...,m\bigg],\nonumber \\
&\qquad \qquad\, \,\text{lcm}\bigg[\frac{N_1N_3K_j}{\gcd(N_1N_3K_j,  \hat Q_{j1}M)},\,\frac{N_2N_3K_j}{\gcd(N_2N_3K_j,  \hat Q_{j2}M)},\nonumber \\
&\qquad \qquad\quad\, \,\,\,\,\,\,\,\frac{N_1N_3K_i K_j}{\gcd({N_1N_3K_i K_j}, \hat Q_{i3} \hat Q_{j1} M)}, \frac{N_2N_3K_i K_j}{\gcd({N_2N_3K_i K_j}, \hat Q_{i3} \hat Q_{j2} M)}, \forall i,j=1,...,m\bigg]
\bigg\}
\label{eqn:append_6}
\end{align}
\end{widetext}
where lcm and gcd mean the least common multiple and greatest common divisor respectively.  $Q_{ij}$ can take $0,1,...,K_i-1$ and we define $\hat Q_{ij}=Q_{ij}$ if $Q_{ij}$ is nonzero;  $\hat Q_{ij}=K_i$ if $Q_{ij}$ is zero.

Below we are going to prove  (\ref{eqn:append_5}) and (\ref{eqn:append_6}).
First of all, we consider the  gauge transformations, under which
the action $S_0$ is  invariant, 
\begin{align}
b^1 &\rightarrow b^1 +dV^1 -\frac{q}{2\pi N_1} \chi^2   da^3 \\
b^2 &\rightarrow b^2 +dV^2 +\frac{q}{2\pi N_2} \chi^1   da^3 \\
b^i &\rightarrow b^i +dV^i,
a^j\rightarrow a^j +d\chi^j 
\end{align}
where $i\ge 3$ and $j\ge 1$. The gauge transformations of two-form gauge fields $b^{1,2}$ are twisted due to the additional terms on r.h.s.. $V^{1}\,,V^{2}\,,\cdots$ are $1$-form gauge parameters; $\chi^{1}\,,\chi^{2}\,,\cdots$ are $0$-form gauge parameters.

The twisted gauge transformations should be  consistent  with    Dirac quantization conditions, namely $\frac{1}{2\pi}\int_{X^3} db^i\in\Z$ where $X^3$ is a 3D compact manifold. Thus,  $q$ is quantized as:
\begin{align}
\frac{q}{N_1}\in \Z,\,\frac{q}{N_2} \in \Z.\label{eqn:append_123_cond1_1}
\end{align}
by noting that $\frac{1}{2\pi}\int_{X^1} d\chi^i=\in\Z$ and $\frac{1}{2\pi}\int_{X^2}da^i\in\Z$ where $X^1$ and $X^2$ are respectively 1D and 2D compact manifolds. 

In addition,  $S_c$ is invariant up to multiple of $2\pi$ under the gauge transformations, which leads to another quantization constraints:
\begin{align}
\frac{qQ_{i1}}{N_1K_i}\in \Z,\,\frac{qQ_{i2}}{N_2K_i}\in \Z\,, i=1,...,m \,. \label{eqn:append_123_cond2_1}
\end{align}
Combining all these constraints together, we have the quantization of $q$ as 
\begin{equation}
q=k  M\,, k\in\Z\,,
\label{quantization:q123_1}
\end{equation}
where $M$ takes (\ref{eqn:append_5}).

Secondly, we consider the
 following shift operation to remove the redundancy of $k$
\begin{eqnarray}
\frac{1}{2\pi} \int db^1 &&\rightarrow \frac{1}{2\pi} \int db^1- \frac{\widetilde{K}_1 M}{4\pi^2 N_1} \int a^2   da^3\nonumber \\
\frac{1}{2\pi} \int db^2 &&\rightarrow \frac{1}{2\pi} \int db^2+\frac{\widetilde{K}_2 M}{4\pi^2 N_2} \int a^1  da^3\nonumber \\
\frac{1}{2\pi} \int db^3 &&\rightarrow \frac{1}{2\pi} \int db^3-\frac{\widetilde{K}_3 M}{4\pi^2 N_3} \int d(a^1a^2)\nonumber \\
k && \rightarrow k+\widetilde{K}_1 +\widetilde{K}_2+\widetilde{K}_3
\end{eqnarray}
(while other $2$-form gauge fields $b^i$ remain unshifted) which keeps the action $S_0$ invariant but shifts the coefficient $k$.  If  two $k$ can be related to under this shift operation, they are equivalent.  

This shift operation will  lead to two constraints on the  $\widetilde{K}_i$ (and hence on the period of $k$).  The first one comes from  consistency with the Dirac quantization conditions such that $\widetilde K_i$ must satisfy 
\begin{align}
 &    \frac{\widetilde{K}_1M  \hat Q_{j2}}{N_1N_2K_j} \in  \Z,\,\frac{ \widetilde{K}_2M \hat Q_{j1}}{N_1N_2K_j} \in \Z,  \\
 &\frac{  \widetilde{K}_3 M \hat Q_{j1}}{N_1N_3K_j}\in \Z,\,\frac{  \widetilde{K}_3 M \hat Q_{j2}}{N_2N_3K_j}\in \Z
 \end{align}
 for all $ j=1,...,m$. To derive the above constraints, we have considered the equation of motion: $\int_{X^1} a^i=-\sum_j\frac{Q_{ji}}{N_i}\int_{X^1} A_j=\sum_j\frac{Q_{ji}}{N_2}\frac{2\pi}{K_j}\times {\text{ integer}}$ with $i=1,2$. 
 In addition, the coupling action $S_c$ should be invariant (up to multiple of $2\pi$) under this shift operation, which leads to anther set of  constraints on $\widetilde K_i$: 
 \begin{align}
 & \frac{   Q_{i1} \hat Q_{j2} \widetilde{K}_1M }{N_1N_2K_i K_j} \in \Z,  \,\,  \frac{ Q_{i2} \hat Q_{j1}\widetilde{K}_2M}{N_1N_2K_iK_j} \in \Z,  \\
 &\frac{  Q_{i3} \hat Q_{j1}\widetilde{K}_3 M}{N_1N_3K_iK_j}\in\Z,\,\, \frac{  Q_{i3} \hat Q_{j2}\widetilde{K}_3 M}{N_2N_3K_iK_j} \in \Z  
\end{align}
for all $ i,j=1,...,m$.

The above  two set of constraints require that
\begin{align}
&\widetilde K_1/ \Gamma_1\in \Z,\, \widetilde K_2/\Gamma_2\in \Z, \, \widetilde K_3/\Gamma_3\in\Z
\end{align}
where
{\footnotesize
\begin{align}
\Gamma_1=\text{lcm}\bigg[&\frac{N_1N_2K_j}{\gcd(N_1N_2K_j,  \hat Q_{j2}M)}, \nonumber \\
&\frac{N_1N_2K_i K_j}{\gcd({N_1N_2K_i K_j}, \hat Q_{i1} \hat Q_{j2} M)}, \forall i,j=1,...,m\bigg],\\
\Gamma_2=\text{lcm}\bigg[&\frac{N_1N_2K_j}{\gcd(N_1N_2K_j,  \hat Q_{j1}M)}, \nonumber \\
&\frac{N_1N_2K_i K_j}{\gcd({N_1N_2K_i K_j}, \hat Q_{i1} \hat Q_{j2} M)}, \forall i,j=1,...,m\bigg],\\
\Gamma_3=\text{lcm}\bigg[&\frac{N_1N_3K_j}{\gcd(N_1N_3K_j,  \hat Q_{j1}M)},\,\frac{N_2N_3K_j}{\gcd(N_2N_3K_j,  \hat Q_{j2}M)} \nonumber \\
&\frac{N_1N_3K_i K_j}{\gcd({N_1N_3K_i K_j}, \hat Q_{i3} \hat Q_{j1} M)}, \nonumber \\
&\frac{N_2N_3K_i K_j}{\gcd({N_2N_3K_i K_j}, \hat Q_{i3} \hat Q_{j2} M)}, \forall i,j=1,...,m\bigg].
\end{align}
}

 By choosing different multiple of $\Gamma_i$, the minimal value  $\Gamma$ of $\widetilde K_1+\widetilde K_2+\widetilde K_3$ is equal to $\gcd[\Gamma_1,\Gamma_2,\Gamma_3]$, which is just given by (\ref{eqn:append_6}). Therefore, the minimal period  of $k$  is given by $\Gamma$.

\section{M3L statistical phases from M3L invariants}\label{footnote_multiloopinv}
 In this Appendix, we present more details  of Eq.~(\ref{eq_mml_first}). In a more rigorous treatment of  braiding statistical phases, one should split $\sigma$ into two spatially separate world-sheets: $\sigma=\sigma^{[1]}+\sigma^{[2]}$ where $\sigma^{[1]}$ and $\sigma^{[2]}$ are two symmetry fluxes that respectively form two different world-sheets in the (3+1)d spacetime. As a result,  the M3L invariant is split into four terms:
 \begin{align}
 &\frac{Q_{12}^2q\pi}{32} \int  (d^{-1} *\Sigma^1)  (d^{-1} *(\sigma^{[1]}+\sigma^{[2]}))   (*\sigma^{[1]}+*\sigma^{[2]})\nonumber\\
 =&\frac{Q_{12}^2q\pi}{32} \int  (d^{-1} *\Sigma^1)  (d^{-1} *\sigma^{[1]})   *\sigma^{[1]}  \nonumber\\
 &  +\frac{Q_{12}^2q\pi}{32} \int  (d^{-1} *\Sigma^1)  (d^{-1} * \sigma^{[2]})   *\sigma^{[2]} \nonumber\\
 &   +\frac{Q_{12}^2q\pi}{32} \int  (d^{-1} * \Sigma^1)  (d^{-1} * \sigma^{[1]})    *\sigma^{[2]}  \nonumber\\
 & +\frac{Q_{12}^2q\pi}{32} \int  (*d^{-1} \Sigma^1)  (*d^{-1}  \sigma^{[2]})   *\sigma^{[1]} .
 \end{align} If we want to calculate $\theta_{\sigma,\sigma;\Sigma^1}$, \textit{i.e.}, the   statistical phase of braiding one symmetry flux around another symmetry flux by letting $\Sigma^1$ be the base loop, then, the third and fourth terms simultaneously contribute to the phase, resulting in $\theta_{\sigma,\sigma;\Sigma^1}=\frac{Q_{12}^2q\pi}{32}+\frac{Q_{12}^2q\pi}{32}=\frac{Q_{12}^2q\pi}{16}$.  If we want to calculate $\theta_{\sigma, \Sigma^1;\sigma}$, \textit{i.e.}, the statistical phase of braiding one symmetry flux around $\Sigma^1$ by letting another symmetry flux be the base loop, then,   either the third term (if the base loop is $\sigma^{[1]}$)or the fourth term (if the base loop is $\sigma^{[2]}$) contributes to the phase, resulting in $\theta_{\sigma,\Sigma^1;\sigma}=\frac{Q_{12}^2q\pi}{32}$. The first term     directly gives  the exchange statistical (half-braiding or $2\pi$ self-rotation of a single symmetry flux denoted by $\sigma^{[1]}$) phase $\theta_{\sigma;\Sigma^1}=\frac{Q_{12}^2q\pi}{32}$   after a proper point-splitting regularization. One can also perform this self-rotation with $\sigma^{[2]}$ by using the second term, resulting in the same answer. In the current calculation scheme, one can always have the relation: $\theta_{\sigma,\sigma;\Sigma^1}=2\theta_{\sigma;\Sigma^1}$.

 \section{Procedure of gauging}   
  \label{section_gauging_loops}
  Below, we mainly discuss generally
 how to  gauge the symmetry  with a certain charge matrix, specially show how to obtain the  gauged group  denoted as $G_g^*$.  
  
   Again, we consider the gauge and symmetry group      to be $G_g=\prod_{i=1}^n\Z_{N_i}$ and $G_s=\prod_{i=1}^m \Z_{K_i}$, and the minimal coupling constant as $Q_{ij}$.  We start with the   action (\ref{couplingaction}),  rewritten as
  \begin{align}
  S=&\frac{1}{2\pi}\int \sum_{i=1}^{n+n'} N_i b^i da^i+S_{c}+S_{int} \,,\\
  S_c=&\frac{1}{2\pi} \int \sum_{i=1}^m \sum_{j=1}^{n+n'} A^i Q_{ij} db^j\,,
  \end{align}
  where  $N_i$ is the corresponding level of gauge fields $b^i$ and $a^i$.  $N_{i}=1$ for $n+n' \ge i\ge n+1$.  
  $A^i$  denotes the external probe field related to the symmetry subgroup $\Z_{K_i}$ and $S_{int}$ denotes the collection of all twisted terms. 
     Note that the external probe fields here can only introduce   static  (\textit{i.e.}, non-dynamical) symmetry fluxes. To gauge the symmetry $G_s$ is nothing but  to give the dynamics to those static symmetry fluxes by technically adding  new $BF$ terms $\frac{1}{2\pi}\sum_i^m K_i B^idA^i$\footnote{Actually, we should also   add   Maxwell terms which gives the dynamics to the loop and particle excitations. Nevertheless, these terms is less relevant than the $BF$ terms, and hence we ignore them when we focus on topological aspects of   gapped phases.}. The latter enforces the charge-loop braiding statistics onto the corresponding new gauge charges and flux loops. Besides we also need to add the corresponding functional integration over $A^i$ and $B^i$ in the partition function. As a result, we obtain the following \textit{gauged} theory:  
   \begin{align}
  S_g&=\frac{1}{2\pi}\int \mathcal{B}^T \mathcal{W} d\mathcal{A} +S_{int} \,,
  \label{gauged action}
  \end{align}  
  where $\mathcal{W}$ is an integer matrix given by
  \begin{align}
 \mathcal{W} =&\left(\begin{matrix}\mathbf{K}& \mathbf{0}\\
\mathbf{Q}^T& \mathbf{N}\oplus \mathbf{\mathbb{I}} \end{matrix}\right)   
.
\label{W} 
  \end{align} 
The entries of $\mathbf{K}$ are given by: $K_i\delta_{ij}$ for $i,j=1,2,\cdots m$. The entries of $\mathbf{Q}^T$ are determined by the charge matrix:  $Q_{ji}$ for $i=1,2,\cdots, n+n'$ and $j=1,2,\cdots, m$. $\mathbf{N}$ is diagonal: $N_{ij}=N_i \delta_{ij}$ for $i,j=1,\cdots, n$. $\mathbf{\mathbb{I}}$ is an $n'$-dimensional identity. $\mathbf{0}$ denotes an $m\times (n+n')$-dimensional null matrix.

All $2$-form and $1$-form gauge fields are collected in $\mathcal{B}$ and $\mathcal{A}$, respectively: 
\begin{align}
&\!\!\!\mathcal{B}= \!(B^1, B^2\cdots,B^m,b^1,\cdots,b^n,b^{n+1},\cdots,b^{n+n'})^T , \\
 &\!\!\!\mathcal{A}= \!(A^1, A^2\cdots,A^m,a^1,\cdots,a^n,a^{n+1},\cdots,a^{n+n'})^T  \! .
\end{align}
 Note that two gauge theories with   $\mathcal{W}$ and $\mathcal{W}'$ respectively are in fact  topologically equivalent if they can be connected by a general linear transformation: $\mathcal{W}'=\mathcal{UWV}$ with $\mathcal{U, V}\in \mathbb{GL} (m+n+n',\Z)$.

Next, we attempt to diagonalize the matrix $\mathcal{W}$ in Eq.~(\ref{W}).  $Q_{ij}$ can be any integer in principle. Through some elementary column or (and) row transformation (s), we can shift $Q_{ij}$ to be less than the greatest common divisor of $K_i$ and $N_j$. More precisely, from the B\'ezout's identity, $Q$ can be decomposed in terms of Eq.~(\ref{reducedQ}). 
  Under this constraint, {we recall the charge matrix $\mathcal{Q}$  as \textit{reduced} charge matrix}. 
   Technically, we can always perform the following transformation to simplify   $\mathcal{W}$ to be 
     
  \begin{align}
\! \!\!\! \mathcal{W}_1=\mathcal{U}_0\mathcal{W}\mathcal{V}_0\!\!=\! \!\!\left (
\begin{smallmatrix}
 K_1 & $0$ & \cdots &$0$ &$0$& \cdots & $0$ &$0$ &\cdots &$0$  \\
  $0$&K_2  & \cdots &$0$ &$0$& \cdots & $0$ &$0$ &\cdots &$0$  \\
 \cdots & \cdots  & \cdots &\cdots & \cdots&\cdots & \cdots  &\cdots &\cdots &\cdots \\
 $0$&$0$&\cdots& K_m & $0$ & \cdots & $0$ &$0$ &\cdots &$0$  \\  
 \mathcal{Q}_{11}&  \mathcal{Q}_{21} &\cdots &  \mathcal{Q}_{m1} &N_1& \cdots & $0$  &$0$ &\cdots &$0$  \\
 \cdots &\cdots &  \cdots & \cdots & \cdots &\cdots  &\cdots &\cdots &\cdots &\cdots \\
   \mathcal{Q}_{1n} & \mathcal{Q}_{2n} &  \cdots & \mathcal{Q}_{mn}& $0$&\cdots & N_n&   $0$ &\cdots &$0$  \\
   $0$ &$0$ &  \cdots &$0$ & $0$&\cdots & $0$&   $1$ &\cdots &$0$  \\
   \cdots &\cdots & \cdots &  \cdots & \cdots &\cdots  &\cdots &\cdots &\cdots &\cdots \\
 $0$ &$0$ &  \cdots &$0$&$0$ &\cdots & $0$&   $0$ &\cdots &$1  $
 \end{smallmatrix}
\right)  \!.\!\!
\label{simplifiedW} 
\end{align}
Note that $\mathcal{V}_0$ depends on $\alpha_{ij}$ while $U_0$ depends on $\beta_{ij}$. Their explicit forms are given by:
\begin{align}
\!\!\!\!\!\mathcal{V}_0\!=\!\!
\left (
\begin{smallmatrix}
 $1$ & $0$ & \cdots &$0$ &$0$& \cdots & $0$ &$0$ &\cdots &$0$  \\
  $0$&$1$  & \cdots &$0$ &$0$& \cdots & $0$ &$0$ &\cdots &$0$  \\
 \cdots & \cdots  & \cdots &\cdots & \cdots&\cdots & \cdots  &\cdots &\cdots &\cdots \\
 $0$&$0$&\cdots& $1 $& $0$ & \cdots & $0$ &$0$ &\cdots &$0$  \\  
 -\alpha_{11}& -\alpha_{21} &\cdots & -\alpha_{m,1} &$1$& \cdots & $0$  &$0$ &\cdots &$0$  \\
 \cdots &\cdots &  \cdots & \cdots & \cdots &\cdots  &\cdots &\cdots &\cdots &\cdots \\
  -\alpha_{1,n} &-\alpha_{2,n} &  \cdots &-\alpha_{m,n}& $0$&\cdots & $1$&   $0$ &\cdots &$0$  \\
   -Q_{1,n+1} &-Q_{2,n+1} &  \cdots &-Q_{m,n+1} & $0$&\cdots & $0$&  $ 1$ &\cdots &$0$  \\
   \cdots &\cdots & \cdots &  \cdots & \cdots &\cdots  &\cdots &\cdots &\cdots &\cdots \\
 -Q_{1,n+n'} &-Q_{2,n+n'} &  \cdots &-Q_{k,n+n'}&$0$ &\cdots & $0$&   $0$ &\cdots &$1$  
 \end{smallmatrix}
\right)  \,,
  \end{align}
  and 
   \begin{align}
\mathcal{U}_0=
\left (
\begin{smallmatrix}
$ 1$ & $0$ & \cdots &$0$ &$0$& \cdots & $0$ &$0$ &\cdots &$0$  \\
  $0$&$1$  & \cdots &$0$ &$0$& \cdots & $0$ &$0$ &\cdots &$0$  \\
 \cdots & \cdots  & \cdots &\cdots & \cdots&\cdots & \cdots  &\cdots &\cdots &\cdots \\
 $0$&$0$&\cdots& $1$ & $0$ & \cdots & $0$ &$0$ &\cdots &$0$  \\  
 -\beta_{11}& -\beta_{21} &\cdots & -\beta_{m,1} &$1$& \cdots & $0$  &$0$ &\cdots &$0$  \\
 \cdots &\cdots &  \cdots & \cdots & \cdots &\cdots  &\cdots &\cdots &\cdots &\cdots \\
  -\beta_{1,n} &-\beta_{2,n} &  \cdots &-\beta_{m,n}& $0$&\cdots &$ 1$&   $0$ &\cdots &$0$  \\
   $0$ &$0$ &  \cdots &$0$ & $0$&\cdots & $0$&   $1 $&\cdots &$0$  \\
   \cdots &\cdots & \cdots &  \cdots & \cdots &\cdots  &\cdots &\cdots &\cdots &\cdots \\
 $0$ &$0$ &  \cdots &$0$&$0$ &\cdots & $0$&   $0$ &\cdots &$1 $ 
 \end{smallmatrix}
\right)\,.
  \end{align}

We can further use the Smith normal form to diagonalize   $\mathcal{W}_1$ via  $\mathcal{W}_d=\mathcal{U}\mathcal{W}_1\mathcal{V}$ although the specific procedures  usually  depend on the values of $ \mathcal{Q}_{ij}$. Finally, from  non-unit diagonal elements in the diagonalized matrix, one can directly obtain 
 $G_g^*$, that is formally denoted as $G_g^*=G_s\times_{\nu}G_g$ in Sec.~\ref{classification_principle}.
 
We note that after  diagonalizing $\mathcal{W}$ we can further perform the transformation on those $2$-form and $1$-form gauge fields: $\mathcal{B}= (UU_0)^T \widetilde{\mathcal{B}}$ and $\mathcal{A}=\mathcal{V}_0\mathcal{V}\widetilde{\mathcal{A}}$ and  substitute them back to the action, especially back to the twisted terms, so that we obtain a new (twisted) gauged theory.  Those new twisted terms which involve level-one new gauge fields can be abandoned since their coefficients are topologically equivalent to zero. 
 
\section{$\Z_N$ topological order enriched by $\Z_{K}$ symmetry} 
\label{sec_app_znzk}

In this appendix, we calculate the classification through some examples (See Table~\ref{table_z4z4sfl}) for $\Z_N$ topological order enriched by $\Z_K$ symmetry.  These examples are illustrating which can be generalized to general $N$ and $K$.

From (\ref{eq_mathcalc}), we see that there are  $\gcd(N,K)$ different patterns of symmetry fractionalization on particles.

Now we discuss the SFL.
   From Table~\ref{table_three_M3L_example_list}, there are two M3L invariants $\mathcal{I}_{\alpha aa}$ and $\mathcal{I}_{a\alpha\alpha}$ for $\Z_N$ topological order enriched by $\Z_K$ symmetry. 
  As shown in Sec.\ref{section_generality_SFL_action}, without loss of generality, we study the following  actions (\ref{eq_main_z4z4_SFL_action0}), rewritten as follow
\begin{align}
S=&S_0+S_{int}+S_c+S_{sr}\label{eq_main_z4z4_SFL_action}\,,\nonumber \\
 S_0= & \frac{1}{2\pi} \int N b^1   da^1 +  b^2   da^2 + b^3   da^3  \,,\nonumber\\
 S_{int}=& \frac{1}{4\pi^2}  \int q_{211}  a^2  a^1da^1+q_{122}  a^1  a^2da^2+q_{123}  a^1  a^2da^3
 \,, \nonumber\\
 S_c=&\frac{1}{2\pi}\int  Q_{11}  A   db^1 +   Q_{12}A  db^2+ Q_{13} A   db^3   \,, \nonumber\\
 S_{sr}=&   \int a^1   *j^1 +  a^2   *j^2+a^3*j^3+ b^1   * \Sigma \,.
 \end{align}
After integrating out $b^1, b^2 ,b^3$, we can obtain the effective action
\begin{align}
S_\text{eff}=-&\frac{2\pi q_{211} Q_{12}}{N^2K}\int (*d^{-1}\sigma)(*d^{-1}\Sigma)(*\Sigma)\nonumber\\
+&[\frac{2\pi q_{211} Q_{11}Q_{12}}{N^2K^2}-\frac{2\pi  Q_{12}(q_{122}Q_{12}+q_{123}Q_{13})}{NK^2}]\nonumber \\
&\int (*d^{-1}\Sigma)(*d^{-1}\sigma)(*\sigma)
+S_{AB}
\end{align}
where $S_{AB}$ collects the  terms describing the braiding between charges(particles) and fluxes(defects). Therefore, the coefficients $\mathcal{I}_{\alpha aa}$ and $\mathcal{I}_{a\alpha\alpha}$ take
\begin{subequations}
\label{eq_invraint_znzk_general1}
\begin{align}
\mathcal{I}_{\alpha aa}&=-\frac{ q_{i11} Q_{12}}{N^2K}\\
\mathcal{I}_{a \alpha \alpha}&=\frac{ q_{211} Q_{11}Q_{12}}{N^2K^2}-\frac{  Q_{12}(q_{122}Q_{12}+q_{123}Q_{13})}{NK^2}
\end{align}
\end{subequations}
If $Q_{12}$ and $Q_{13}$ are set to be one, then the two invariants reduce to the form (\ref{eqn_simplified_Ia_a_alpha}) and (\ref{eqn_simplified_Ialpha_a_a}). Here we keep $Q_{12}$ and $Q_{13}$ generally and show below that the final classification of SFL for case with general $Q_{12}$ and $Q_{13}$ is the same as that by setting $Q_{12}=Q_{13}=1$ at first. In fact, that is also why we  can  safely set $Q_{12}=Q_{13}=1$ in the main text in order to simply the notation there.

From these expression of M3L invariants,  we assume that $Q_{12}$ is nonzero, otherwise the invariants vanish,  while $Q_{13}$ can be zero or nonzero. The  quantization rules  of the twisted coefficients are given by :
\begin{subequations}
\label{eqn_app_zNzK_gquant}
\begin{align}
q_{211}&=kM,\\
 q_{122}&=\bar k M, \\
 q_{123}&=\tilde k M
\end{align}
\end{subequations}
 where $k,\bar k,\tilde k$ are integral and from (\ref{eqn:twisted_coff_general_1}) and (\ref{eqn:twisted_coff_general_11})
\begin{align}
M=\text{lcm}[N, \frac{K}{\gcd(K, \hat Q_{12})}, \frac{NK}{\gcd(NK, \hat Q_{11})}]
\label{eqn_app_znzk_multiple}
\end{align}
and according to (\ref{eqn:twisted_coff_general_2}) and (\ref{eqn:twisted_coff_general_3}), the periods $\Gamma$ of $k,\bar k$ and $\tilde \Gamma$  of $\tilde k$ are given by
\begin{align}
\Gamma &=\frac{NK^2}{\gcd(NK^2, \hat Q_{11}\hat Q_{12}M)} \label{eqn_app_znzk_period1}\\
\tilde \Gamma &= \gcd\bigg\{ \frac{NK^2}{\gcd(NK^2, \hat Q_{11} \hat Q_{12} M)}, \nonumber \\
&\quad \,\,\, \text{lcm}[\frac{NK^2}{\gcd(NK^2, \hat Q_{11} \hat Q_{13}M)}, \frac{K^2}{\gcd(K^2, \hat Q_{12}\hat Q_{13}M)}]\bigg\}\label{eqn_app_znzk_period2}
\end{align}

As by assumption $Q_{12}$ nonzero,   $\hat Q_{12}=Q_{12}$.
Then the two M3L invariants can be simplified to be
\begin{align}
\mathcal{I}_{\alpha aa}&=-k\cdot \frac{ \text{lcm}[K,  NQ_{12}, \frac{NK  Q_{12}}{\gcd(NK, \hat Q_{11})}]}{N^2K} 
\label{eqn_app_z2zk_M3L_1aa}
\\
\mathcal{I}_{a \alpha \alpha}&=\mathcal{I}_{a \alpha \alpha}^{(1)}+\mathcal{I}_{a \alpha \alpha}^{(2)}\label{eqn_app_z2zk_M3L_a11} 
\end{align}
where
\begin{align}
\mathcal{I}_{a \alpha \alpha}^{(1)}&=k\cdot \frac{ \text{lcm}[KQ_{11},  NQ_{12}Q_{11}, NK  Q_{12}]|\text{Sign}({Q_{11}})|}{N^2K^2} \\
\mathcal{I}_{a \alpha \alpha}^{(2)}&=-(\bar k Q_{12}+\tilde k Q_{13})\frac{\text{lcm}[K,  NQ_{12}, \frac{NK  Q_{12}}{\gcd(NK, \hat Q_{11})}]}{NK^2}
\end{align}
where the Sign function encodes that when $Q_{11}=0$, i.e., Sign$(Q_{11})$=0, $\mathcal{I}^{(1)}_{a\alpha \alpha}=0$.

\subsection{$N=2$, $K=2n+1$}
\subsubsection{SFP}

Since $\gcd(2,2n+1)=1$, there is only trivial SFP. 
 Therefore, there is also only trivial SFP, no matter what values of $Q_{11}$ take.

\subsubsection{SFL}

Below we are going to show that the M3L statistical phases (see Table~\ref{table_three_M3L_example_list}) determined by the two M3L invariants $\mathcal{I}_{\alpha aa}$ and $\mathcal{I}_{a\alpha \alpha}$ are always equivalent to trivial values, namely, there is only trivial SFL.

First, we discuss $\mathcal{I}_{\alpha aa}$ in (\ref{eqn_app_z2zk_M3L_1aa}), which now can be simplified to be
\begin{align}
\mathcal{I}_{\alpha aa}=-\frac{k\text{lcm}[K,2Q_{12},\frac{2KQ_{12}}{\gcd(2K,\hat{Q}_{11})}]}{4K}
\end{align}

Then the two M3L statistical phases are given by
\begin{align}
\theta_{\Sigma,\sigma;\Sigma}&=-2\pi k \frac{\text{lcm}[K,2Q_{12},\frac{2KQ_{12}}{\gcd(2K,\hat{Q}_{11})}]}{4K}\\
\theta_{\Sigma;\sigma}&=2\pi k \frac{\text{lcm}[K,2Q_{12},\frac{2KQ_{12}}{\gcd(2K,\hat{Q}_{11})}]}{4K}.
\end{align}
Since $K$ is odd,   $\frac{\text{lcm}[K,2Q_{12},\frac{2KQ_{12}}{\gcd(2K,\hat{Q}_{11})}]}{2K} \in \Z$, and then we set  
\begin{align}
\text{lcm}[K,2Q_{12},\frac{2KQ_{12}}{\gcd(2K,\hat{Q}_{11})}]=2K p
\end{align}
where $p$ is integral and depends on $Q_{11}$ and $Q_{12}$.
Then the above two M3L statistical phases can be simplified to
\begin{align}
\theta_{\Sigma,\sigma;\Sigma}=- \pi  kp, \,\,\,\theta_{\Sigma;\sigma}= \pi kp.
\end{align}
It is easy to see that these two statistical values are trivial by checking the period from attaching particles with unit gauge charge to $\sigma$ or $\Sigma$.

Next, we discuss the M3L invariant $\mathcal{I}_{a\alpha \alpha}$. Since $\mathcal{I}_{a\alpha \alpha}^{(1)}$ and $\mathcal{I}_{a\alpha \alpha}^{(2)}$ have independent origins, we will discuss them separately and show that both of them determine trivial values of M3L statistical phases.  The two parts  $\mathcal{I}_{a\alpha \alpha}^{(1)}$ and $\mathcal{I}_{a\alpha \alpha}^{(2)}$ now take
\begin{align}
\mathcal{I}_{a \alpha \alpha}^{(1)}&=k\cdot \frac{ \text{lcm}[KQ_{11},  2Q_{12}Q_{11}, 2K  Q_{12}]|\text{Sign}({Q_{11}})|}{4K^2} \\
\mathcal{I}_{a \alpha \alpha}^{(2)}&=-(\bar k Q_{12}+\tilde k Q_{13})\frac{\text{lcm}[K,  2Q_{12}, \frac{2K  Q_{12}}{\gcd(2K, \hat Q_{11})}]}{2K^2}
\end{align}
Also due to odd $K$, $\text{lcm}[K,  2Q_{12}, \frac{2K  Q_{12}}{\gcd(2K, \hat Q_{11})}]/2K\in \Z$, and then we set 
 \begin{align}
 \text{lcm}[K,  2Q_{12}, \frac{2K  Q_{12}}{\gcd(2K, \hat Q_{11})}]=2Kp_1.
 \end{align}
Further, since $\text{lcm}[KQ_{11},  2Q_{12}Q_{11}, 2K  Q_{12}]|\text{Sign}({Q_{11}})|$  is equal to $Q_{11}\text{lcm}[K,  2Q_{12}, \frac{2K  Q_{12}}{\gcd(2K, \hat Q_{11})}]$, then we have
\begin{align}
\mathcal{I}_{a \alpha \alpha}^{(1)}&=\frac{kp_1Q_{11}}{2K}\\
\mathcal{I}_{a \alpha \alpha}^{(2)}&=-\frac{(\bar k Q_{12}+\tilde k Q_{13})p_1}{K}
\end{align}
Therefore, the correspondingly two M3L statistical phases are
\begin{align}
&\theta_{\Sigma,\sigma;\sigma}=-\frac{2\pi k p_1 Q_{11}}{2K}+\frac{2\pi (\bar k Q_{12}+\tilde k Q_{13}) p_1}{K} \label{eqn_app_z2_zoddk_statistics1}\\
&\theta_{\sigma;\Sigma}=\frac{2\pi k p_1 Q_{11}}{2K}-\frac{2\pi (\bar k Q_{12}+\tilde k Q_{13}) p_1}{K}\label{eqn_app_z2_zoddk_statistics2}
\end{align}
The trivialness of these M3L statistical phases can be seen by considering the attachment of particles to $\sigma$ or $\Sigma$. For $\theta_{\Sigma,\sigma;\sigma}$, one can attach particle with $t_1$ unit gauge charge  to $\sigma$  and particle with $t_2Q_{12}+t_3Q_{13}$ symmetry charge and $t_4$ unit gauge charge to $\Sigma$, which can cause a phase shift by $\frac{2\pi (t_1K+2t_2Q_{12}+2t_3Q_{13}+t_4Q_{11})}{2K}$. By choosing proper $t_i$, the minimal value of this phase shift is $\frac{2\pi \gcd(K, Q_{11}, Q_{12}, Q_{13})}{2K}$. So we can see that the values of $\theta_{\Sigma,\sigma;\sigma}$ in (\ref{eqn_app_z2_zoddk_statistics1}) are all multiple of this minimal value of the phase shift,  and then are  trivial.
As for $\theta_{\sigma;\Sigma}$, one can attach one  particle with $t_2Q_{12}+t_3Q_{13}$ symmetry charge and $t_2$ unit gauge charge to each $\sigma$ in the exchanging process, which results in a phase shift by $\frac{2\pi (2t_2Q_{12}+2t_3Q_{13}+t_4Q_{11})}{2K}$. The minimal value of the this shift is $\frac{2\pi\gcd(Q_{11},2Q_{12},2Q_{13})}{2K}$. Then we can see that the values of $\theta_{\sigma;\Sigma}$ in (\ref{eqn_app_z2_zoddk_statistics2}) are all multiple of this minimal value, and then are trivial.

\subsection{$N=2$, $K=2^n$}

\subsubsection{SFP}

Since $\gcd(2,2^n)=2$, there are two different patterns of SFP. Accordingly,  from (\ref{eq_mathcalc}), the elementary gauge charge can carry one-half or integer symmetry charge, corresponding to two reduced charge matrices: $\mathcal{Q}_{11}=0,1$. We denote these two patterns as $e0$ and $eC$.

\subsubsection{SFL}
Below we discuss the SFL for $e0$ and $eC$ case by case.\\

\textbf{Case one: $e0$}\\

We now discuss the case $e0$. Then $\mathcal{Q}_{11}=0$ and $Q_{11}=2m$, then $\hat Q_{11}=2\hat m$ (if $m=0$, $\hat m=2^{n-1}$; if $m\neq 0$, $\hat m=m$). We assume $Q_{12}\neq 0$(then $\hat Q_{12}=Q_{12}$), otherwise  all $\mathcal{I}_{\alpha aa}$ and $\mathcal{I}_{a\alpha \alpha}$ are automatically zero. The two M3L invariants $\mathcal{I}_{\alpha aa}$ and $\mathcal{I}_{a\alpha\alpha}$ become
\begin{align}
&\mathcal{I}_{\alpha aa}=-k\frac{\text{lcm}(K, 2Q_{12},\frac{K Q_{12}}{\gcd(K,\hat m)})}{4K}\\
&\mathcal{I}_{a\alpha \alpha}^{(1)}=k\frac{\text{lcm}(K Q_{11}, 2Q_{11} Q_{12},2K Q_{12} )|\text{Sign}({Q_{11}})|}{4K^2}\\
&\mathcal{I}_{a \alpha \alpha}^{(2)}=-(\bar k Q_{12}+\tilde k Q_{13})\frac{\text{lcm}[K, 2 Q_{12}, \frac{2K  Q_{12}}{\gcd(2K, \hat Q_{11})}]}{2K^2}
\end{align}
Then the  M3L statistical phases are given by
\begin{align}
&\theta_{\Sigma,\sigma;\Sigma}=2\pi k\frac{\text{lcm}(K, 2Q_{12},\frac{K Q_{12}}{\gcd(K,\hat m)})}{4K}\\
&\theta_{\Sigma;\sigma}=-2\pi k\frac{\text{lcm}(K, 2Q_{12},\frac{K Q_{12}}{\gcd(K,\hat m)})}{4K}\\
&\theta_{\Sigma,\sigma;\sigma}^{(1)}=-2\pi k \frac{\text{lcm}(K Q_{11}, 2Q_{11} Q_{12},2K Q_{12} )|\text{Sign}({Q_{11}})|}{4K^2}\\
&\theta_{\sigma;\Sigma}^{(1)}=2\pi k \frac{\text{lcm}(K Q_{11}, 2Q_{11} Q_{12},2K Q_{12} )|\text{Sign}({Q_{11}})|}{4K^2}\\
&\theta_{\Sigma,\sigma;\sigma}^{(2)}= 2\pi (\bar k Q_{12}+\tilde k Q_{13})\frac{\text{lcm}[K,  2Q_{12}, \frac{2K  Q_{12}}{\gcd(2K, \hat Q_{11})}]}{2K^2}\\
&\theta_{\sigma;\Sigma}^{(2)}=-2\pi (\bar k Q_{12}+\tilde k Q_{13})\frac{\text{lcm}[K,  2Q_{12}, \frac{2K  Q_{12}}{\gcd(2K, \hat Q_{11})}]}{2K^2}
\end{align}
We first discuss the $\theta_{\Sigma,\sigma;\Sigma}$ and $\theta_{\Sigma;\sigma}$. When $Q_{12}=1$, they can take the most general quantized values, \textit{i.e.},
\begin{align}
&\theta_{\Sigma,\sigma;\Sigma}=\frac{2\pi k}{4}\label{eqn_most_general_M3Lst_z2z2n_1}\\
&\theta_{\Sigma;\sigma}=-\frac{2\pi k}{4}.\label{eqn_most_general_M3Lst_z2z2n_2}
\end{align}

The periods of them can be discussed from two aspects.  First, from attachment of particles to $\sigma$ or $\Sigma$, $\theta_{\Sigma,\sigma;\Sigma}$ has a period to be $\frac{2\pi}{K}$ while $\theta_{\Sigma;\sigma}$ has $\pi$. Second, the periods of them can come from that of twisted coefficient. However, we can show that they are some multiple of those from attachment of particles. For this purpose, we note that from (\ref{eqn_app_znzk_period1}),  the period of $k$ is given by
\begin{align}
\Gamma=\frac{2K^2 }{\gcd[2K^2,\text{lcm}(\hat Q_{11}K, 2\hat Q_{11} Q_{12}, 2K Q_{12})]}
\label{eqn_app_z2z2n_period1}
\end{align}
where we have used the relation (\ref{eqn_app_znzk_multiple}) and  $\hat Q_{11} \hat Q_{12} M=\text{lcm}(\hat Q_{11}K, 2\hat Q_{11} Q_{12}, 2K Q_{12})$.
For convenience, 
we can set
 $\text{lcm}[K,  2Q_{12}, \frac{K  Q_{12}}{\gcd(K, \hat m)}]=pK$ with $p$ being  integral and  then we have 
$\text{lcm}(\hat Q_{11}K,2\hat Q_{11} Q_{12}, 2K Q_{12})=\hat Q_{11} pK$, which simplifies $\Gamma$  to be
\begin{align}
\Gamma=\frac{K}{\gcd[K, \hat m p]}.
\end{align}
The periods of $\theta_{\Sigma,\sigma;\Sigma}$ and $\theta_{\Sigma;\sigma}$ from twisted coefficient are both given by
\begin{align}
\frac{2\pi pK }{4\gcd[K, \hat m p]}
\end{align}
which must be some multiple of $\pi t$ and hence $\frac{2\pi}{K}$ since $K=2^n$ and $\hat m\le 2^{n-1}$. Therefore, the minimal periods of $\theta_{\Sigma,\sigma;\Sigma}$ and $\theta_{\Sigma;\sigma}$ are $\frac{2\pi}{K}$ and $\pi$ respectively.

 Therefore, from (\ref{eqn_most_general_M3Lst_z2z2n_1}) and (\ref{eqn_most_general_M3Lst_z2z2n_2}),  $k=0,1$ mod 2 are two inequivalent quantized M3L statistical phases, which contributes to a $\Z_2$ classification. When $K=2$, both  statistical phases are nontrivial when $k=1$. However, when $K=2^n$ with $n\ge 2$, $\theta_{\Sigma,\sigma;\Sigma}$ are trivial no matter what values of $k$ and $\theta_{\Sigma;\sigma}$ are still nontrivial with $k=1$.

Next we discuss $\theta_{\Sigma,\sigma;\sigma}$ and $\theta_{\sigma;\Sigma}$.  We first focus on $\theta_{\Sigma,\sigma;\sigma}^{(2)}$ and $\theta_{\sigma;\Sigma}^{(2)}$ since they originate from different twisted terms and  are independent of the other two statistical phases. When $Q_{12}=Q_{13}=1$, they can take the most general quantized values, \textit{i.e.},
\begin{align}
&\theta_{\Sigma,\sigma;\sigma}^{(2)}=\frac{2\pi (\bar k+\tilde k)}{2K}\label{eqn_most_general_M3Lst_z2z2n_3}\\
&\theta_{\sigma;\Sigma}^{(2)}=-\frac{2\pi (\bar k+\tilde k)}{2K}.\label{eqn_most_general_M3Lst_z2z2n_4}
\end{align}

From the attachment of particles, both $\theta_{\Sigma,\sigma;\sigma}$ and $\theta_{\sigma;\Sigma}$ have a period given by $\frac{2\pi}{K}$. The periods of twisted coefficients would also given rise to periods of these statistical phases, which however, as we can show, must be some multiple of $\frac{2\pi}{K}$. To show it, we note that the periods of $\bar k$ and $\tilde k$  are given by $\Gamma$ in (\ref{eqn_app_z2z2n_period1}) and
\begin{align}
\tilde \Gamma=\gcd[\frac{K}{\gcd(K,p\hat m)},\text{lcm}(\frac{K}{\gcd(K, p \hat Q_{13})},\frac{K}{\gcd(K,\bar p \hat Q_{13})})]
\end{align}
from (\ref{eqn_app_znzk_period2})
where $\bar p$ is defined through setting  $\text{lcm}(2K,2\hat Q_{11}, \frac{K\hat Q_{11}}{\gcd(K,\hat Q_{12})})=2\bar p K$. Since $\hat Q_{12},\hat Q_{13}\le K$,  then $\bar p\le K$. Then the periods of  $\theta_{\Sigma,\sigma;\sigma}$ and $\theta_{\sigma;\Sigma}$ from the periods of $\bar k$ and $\tilde k$ are given by
\begin{align}
\frac{2\pi pQ_{12}}{2\gcd(K,p\hat m)}
\label{eqn_M3Lst_z2z2n_period3}
\end{align}
and 
\begin{align}
\frac{2\pi  }{2K} \gcd[\frac{ Kp Q_{12}}{\gcd(K,p\hat m)},\text{lcm}(\frac{ K p Q_{12}}{\gcd(K, p \hat Q_{13})},\frac{ pQ_{13}K}{\gcd(K, \bar p \hat Q_{13})})]
\label{eqn_M3Lst_z2z2n_period4}
\end{align}
 If $p$ is even, it is easy to see both the above two periods are multiple of $\frac{2\pi}{K}$. Now we focus on odd $p$. We then have $\gcd(K,p\hat m)=\gcd(K,\hat m)$.
Since $\hat m\le K/2$,  $\frac{K}{\gcd(K,\hat m)}$ is even. Further, $\frac{p Q_{13} K}{\gcd(K, p \hat Q_{13})}=\frac{ p Q_{13} K}{\gcd(K,  \hat Q_{13})}$  is also even.  Then both the above two periods are multiple of $\frac{2\pi}{K}$. 
Therefore, the minimal periods of $\theta_{\Sigma,\sigma;\sigma}$ and $\theta_{\sigma;\Sigma}$  are both $\frac{2\pi}{K}$.

Now we turn to $\theta_{\Sigma,\sigma;\sigma}^{(1)}$ and $\theta_{\sigma;\Sigma}^{(1)}$. As $Q_{11}=2m$, the quantized values of them must be some multiple of $\frac{2\pi}{2K}$, so they do not contribute to any new values of $\theta_{\Sigma,\sigma;\sigma}$ and $\theta_{\sigma;\Sigma}$. Furthermore, one can show that they also do not contribute smaller periods of these two statistical phases.
Therefore, there is a $\Z_2$ classification for $\theta_{\Sigma,\sigma;\sigma}$ and $\theta_{\sigma;\Sigma}$ and the nontrivial values of them are given by $\theta_{\Sigma,\sigma;\sigma}=\theta_{\sigma;\Sigma}=\frac{2\pi}{2K}$ mod $\frac{2\pi}{K}$.

To summarize, for $e0$, the classification for SFL is given by $\Z_2\times \Z_2$.
\\
\\
\textbf{Case two: $eC$}\\

Now we turn to discuss the case with $eC$. For this case, $ Q_{11}=\hat Q_{11}=2m+1$. We assume $Q_{12}\neq 0$(then $\hat Q_{12}=Q_{12}$), otherwise both $\mathcal{I}_{\alpha aa}$ and $\mathcal{I}_{a\alpha \alpha}$ are  zero. In the case, the $\mathcal{I}_{\alpha aa}$ and $\mathcal{I}_{a\alpha \alpha}$ become
\begin{align}
\mathcal{I}_{\alpha aa}&=-k\cdot \frac{   Q_{12}}{2} \\
\mathcal{I}_{a \alpha \alpha}^{(1)}&=k\cdot \frac{ \tilde p}{2K} \\
\mathcal{I}_{a \alpha \alpha}^{(2)}&=-(\bar k Q_{12}+\tilde k Q_{13})\frac{  Q_{12}}{K}
\end{align}
where we have set $\text{lcm}[KQ_{11},  2Q_{12}Q_{11}, 2K  Q_{12}]=2\tilde pK$. Correspondingly, the M3L statistical phases are given by \begin{align}
&\theta_{\Sigma,\sigma;\Sigma}=2\pi k \frac{   Q_{12}}{2}\\
&\theta_{\Sigma;\sigma}=-2\pi k \frac{   Q_{12}}{2}\\
&\theta_{\Sigma,\sigma;\sigma}^{(1)}=-2\pi k \frac{ p}{2K}\\
&\theta_{\sigma;\Sigma}^{(1)}=2\pi k \frac{ p}{2K}\\
&\theta_{\Sigma,\sigma;\sigma}^{(2)}= 2\pi (\bar k Q_{12}+\tilde k Q_{13})\frac{  Q_{12}}{K}\\
&\theta_{\sigma;\Sigma}^{(2)}=-2\pi (\bar k Q_{12}+\tilde k Q_{13})\frac{  Q_{12}}{K}
\end{align}
We can  argue that all the values of the above M3L statistical phases are trivial simply by checking the periods from the attachment of particles: (1) for $\theta_{\Sigma,\sigma;\Sigma}$, $\theta_{\Sigma,\sigma;\sigma}$ and $\theta_{\sigma;\Sigma}$, by attaching particle with unit gauge charge, which now carries half symmetry charge, they can shift by $\frac{2\pi}{2K}$; (2) for $\theta_{\Sigma;\sigma}$, similar to the case, $e0$, by attaching one particle with unit gauge charge to each of $\Sigma$ in the exchanging process, it shifts by $\pi$. 

Therefore, there is only trivial SFL for $eC$.


\subsection{$N=4,K=4^n$}
\subsubsection{SFP}
Since $\gcd(4,4^n)=4$, there are four different patterns of SFP. Accordingly,  from (\ref{eq_mathcalc}), the elementary gauge charge can carry  integer, one half, plus and minus one fourth symmetry charge, corresponding to four reduced charge matrices: $\mathcal{Q}_{11}=0,2,1,3$ respectively. We denote these four patterns as $e0$, $eC$, and $eQ_{\pm}$.

\subsubsection{SFL}
Below we will discuss the SFL for different SFP case by case.\\
\\
\textbf{Case one: $e0$}\\
\\
We first discuss the case with SFP being $e0$. In this case,  $Q_{11}=4m$, and then $\hat Q_{11}=4\hat m$ with $\hat m=4^n$ if $m=0$; $\hat Q_{11}=4m$ if $m\neq 0$. Then (\ref{eqn_app_z2zk_M3L_1aa}) and (\ref{eqn_app_z2zk_M3L_a11}) reduce to
\begin{align}
\mathcal{I}_{\alpha aa}&=-k\cdot \frac{ \text{lcm}[4^n,  4Q_{12}, \frac{4^n  Q_{12}}{\gcd(4^n, \hat m)}]}{4^{n+2}} \\
\mathcal{I}_{a \alpha \alpha}^{(1)}&=k\cdot \frac{ \text{lcm}[4^n m ,  4Q_{12} m, 4^n  Q_{12}]|\text{Sign}({Q_{11}})|}{4^{2n+1}} \\
\mathcal{I}_{a \alpha \alpha}^{(2)}&=-(\bar k Q_{12}+\tilde k Q_{13})\frac{\text{lcm}[4^n, 4 Q_{12}, \frac{4^n  Q_{12}}{\gcd(4^n, \hat m)}]}{4^{2n+1}}
\end{align}
Then the M3L statistical phases can take
\begin{align}
&\theta_{\Sigma;\sigma}=-2\pi k\cdot \frac{ \text{lcm}[4^n,  4Q_{12}, \frac{4^n  Q_{12}}{\gcd(4^n, \hat m)}]}{4^{n+2}}\\
&\theta_{\Sigma,\sigma;\Sigma}=2\pi k\cdot \frac{ \text{lcm}[4^n,  4Q_{12}, \frac{4^n  Q_{12}}{\gcd(4^n, \hat m)}]}{4^{n+2}}\\
&\theta_{\sigma;\Sigma}^{(1)}=2\pi k\cdot \frac{ \text{lcm}[4^n m ,  4Q_{12} m, 4^n  Q_{12}]|\text{Sign}({Q_{11}})|}{4^{2n+1}}\\
&\theta_{\Sigma,\sigma;\sigma}^{(1)}=-2\pi k\cdot \frac{ \text{lcm}[4^n m ,  4Q_{12} m, 4^n  Q_{12}]|\text{Sign}({Q_{11}})|}{4^{2n+1}}\\
&\theta_{\sigma;\Sigma}^{(2)}=-2\pi(\bar k Q_{12}+\tilde k Q_{13})\frac{\text{lcm}[4^n,  4Q_{12}, \frac{4^n  Q_{12}}{\gcd(4^n, \hat m)}]}{4^{2n+1}}\\
&\theta_{\Sigma,\sigma;\sigma}^{(2)}=2\pi(\bar k Q_{12}+\tilde k Q_{13})\frac{\text{lcm}[4^n,  4Q_{12}, \frac{4^n  Q_{12}}{\gcd(4^n, \hat m)}]}{4^{2n+1}}
\end{align}

We first discuss $\theta_{\Sigma;\sigma}$ and $\theta_{\Sigma,\sigma;\Sigma}$. When $Q_{12}=1$, they can take the most general quantized values, \textit{i.e.},
\begin{align}
&\theta_{\Sigma;\sigma}=-\frac{2\pi k}{16}\label{eqn_most_general_M3Lst_z4z4n_1}\\
&\theta_{\Sigma,\sigma;\Sigma}=\frac{2\pi k}{16}.\label{eqn_most_general_M3Lst_z4z4n_2}
\end{align}
The periods of them can be discussed from two aspects.  First, from attachment of particles to $\sigma$ or $\Sigma$, $\theta_{\Sigma,\sigma;\Sigma}$ has a period to be $\frac{2\pi}{4^n}$ while $\theta_{\Sigma;\sigma}$ has $\frac{\pi}{2}$. 
Second, the periods of them can come from that of twisted coefficients. However, we will show that they are some multiple of those from attachment of particles. For this purpose, we note that from (\ref{eqn_app_znzk_period1}) the   period of $k$ is given by
\begin{align}
\Gamma=\frac{4^n}{\gcd[4^n, \hat m p]}.
\label{eqn_app_z4z4n_period1}
\end{align}
where we have set $\text{lcm}[4^n, 4 Q_{12}, \frac{4^n  Q_{12}}{\gcd(4^n, \hat m)}]=4^np$ with $p$ being  integral.
The periods of $\theta_{\Sigma,\sigma;\Sigma}$ and $\theta_{\Sigma;\sigma}$ from twisted coefficient are both given by
\begin{align}
\frac{2\pi 4^{n-2} p }{\gcd[4^n, \hat m p]}
\end{align}
which must be some multiple of $\frac{\pi}{2}$ and hence $\frac{2\pi}{K}$ since $K=4^n$ and $\hat m\le 4^{n-1}$. Therefore, the minimal periods of $\theta_{\Sigma,\sigma;\Sigma}$ and $\theta_{\Sigma;\sigma}$ are $\frac{2\pi}{K}$ and $\frac{\pi}{2}$ respectively.

 Therefore, from (\ref{eqn_most_general_M3Lst_z4z4n_1}) and (\ref{eqn_most_general_M3Lst_z4z4n_2}),  $k=0,1,2,3$ mod 4 are four inequivalent quantized M3L statistical phases, which contributes to a $\Z_4$ classification. When $K=4$, both  statistical phases are nontrivial when $k=1,2,3$. However, when $K=4^n$ with $n\ge 2$, $\theta_{\Sigma,\sigma;\Sigma}$ are trivial no matter what values of $k$ and $\theta_{\Sigma;\sigma}$ are still nontrivial with $k=1,2,3$.
 
 Next we discuss $\theta_{\Sigma,\sigma;\sigma}$ and $\theta_{\sigma;\Sigma}$.  We first focus on $\theta_{\Sigma,\sigma;\sigma}^{(2)}$ and $\theta_{\sigma;\Sigma}^{(2)}$ since they are independent of $k$, while the other two depend on. When $Q_{12}=Q_{13}=1$, they can take the most general quantized values, \textit{i.e.},
\begin{align}
&\theta_{\Sigma,\sigma;\sigma}^{(2)}=-\frac{2\pi (\bar k+\tilde k)}{4^{n+1}}\label{eqn_most_general_M3Lst_z4z4n_3}\\
&\theta_{\sigma;\Sigma}^{(2)}=\frac{2\pi (\bar k+\tilde k)}{4^{n+1}}.\label{eqn_most_general_M3Lst_z4z4n_4}
\end{align}
From the attachment of particles, both $\theta_{\Sigma,\sigma;\sigma}$ and $\theta_{\sigma;\Sigma}$ have a period given by $\frac{2\pi}{4^n}$. The periods of twisted coefficients would also give rise to periods of these statistical phases, which however, as we can show below, must be some multiple of $\frac{2\pi}{4^n}$. To show it, we note that the periods of $\bar k$ and $\tilde k$  are given by $\Gamma$ in (\ref{eqn_app_z4z4n_period1}) and
\begin{align}
\tilde \Gamma=\gcd[\frac{4^n}{\gcd(4^n,p\hat m)},\text{lcm}(\frac{4^n}{\gcd(4^n, p \hat Q_{13})},\frac{4^n}{\gcd(4^n,\bar p \hat Q_{13})})]
\end{align}
where $\bar p$ is defined through setting  $\text{lcm}(4^{n+1},4 \hat Q_{11}, \frac{4^n\hat Q_{11}}{\gcd(4^n,\hat Q_{12})})=4^{n+1}\bar p $.  Then the periods of  $\theta_{\Sigma,\sigma;\sigma}$ and $\theta_{\sigma;\Sigma}$ from the periods of $\bar k$ and $\tilde k$ are given by
\begin{align}
\frac{2\pi pQ_{12} }{4\gcd(4^n,p\hat m)}
\end{align}
and 
\begin{align}
\frac{2\pi Q_{13}  }{4^{n+1}}\gcd[\frac{4^np}{\gcd(4^n,p\hat m)},\text{lcm}(\frac{4^np}{\gcd(4^n, p \hat Q_{13})},\frac{4^np}{\gcd(4^n,\bar p \hat Q_{13})})]
\label{eqn_M3Lst_z2z2n_period4}
\end{align}
Since $\hat m\le 4^{n-1}$, both $\frac{4^n p}{\gcd(4^n,p\hat m)}$ and  $\frac{4^n p Q_{13} }{\gcd(4^n, p\hat Q_{13})}$ must be multiple of 4, so that both the above two periods are multiple of $\frac{2\pi}{4^n}$. 
Therefore, the minimal periods of $\theta_{\Sigma,\sigma;\sigma}$ and $\theta_{\sigma;\Sigma}$  are both $\frac{2\pi}{4^n}$.

Now we turn to $\theta_{\Sigma,\sigma;\sigma}^{(1)}$ and $\theta_{\sigma;\Sigma}^{(1)}$. As $Q_{11}=4m$, the quantized values of them must be some multiple of $\frac{2\pi}{4^{n+1}}$, so they do not contribute to any new values of $\theta_{\Sigma,\sigma;\sigma}$ and $\theta_{\sigma;\Sigma}$. Further, one can show that they do not contribute to smaller periods of $\theta_{\Sigma,\sigma;\sigma}$ and $\theta_{\sigma;\Sigma}$.
Therefore, there is a $\Z_4$ classification for $\theta_{\Sigma,\sigma;\sigma}$ and $\theta_{\sigma;\Sigma}$ and the nontrivial values of them are given by $\theta_{\Sigma,\sigma;\sigma}=\theta_{\sigma;\Sigma}=\frac{2\pi}{4^{n+1}}$ mod $\frac{2\pi}{4^n}$.

To summarize, for $e0$, the classification for SFL is given by $\Z_4\times \Z_4$.\\
\\
\textbf{Case two: $eC$}\\
\\
In this case, $Q_{11}=4m+2$ so $\hat Q_{11}=4m+2$. The two M3L invariants become
\begin{align}
\mathcal{I}_{\alpha aa}&=-k\frac{Q_{12}}{8} \\
\mathcal{I}_{a \alpha \alpha}^{(1)}&=k \frac{(2m+1)Q_{12}}{4^{n+1}} \\
\mathcal{I}_{a \alpha \alpha}^{(2)}&=-(\bar k Q_{12}+\tilde k Q_{13})\frac{Q_{12}}{2\times 4^n}
\end{align}
Then the M3L statistical phases quantize to
\begin{align}
&\theta_{\Sigma;\sigma}=-2\pi k \frac{Q_{12}}{8}\\
&\theta_{\Sigma,\sigma;\Sigma}=2\pi k\frac{Q_{12}}{8}\\
&\theta_{\sigma;\Sigma}^{(1)}=2\pi k\frac{(2m+1)Q_{12}}{4^{n+1}}\\
&\theta_{\sigma,\Sigma;\sigma}^{(1)}=-2\pi k\frac{(2m+1)Q_{12}}{4^{n+1}}\\
&\theta_{\sigma;\Sigma}^{(2)}=-2\pi(\bar k Q_{12}+\tilde k Q_{13})\frac{Q_{12}}{2\times 4^n}\\
&\theta_{\sigma,\Sigma;\sigma}^{(2)}=2\pi(\bar k Q_{12}+\tilde k Q_{13})\frac{Q_{12}}{2\times 4^n}
\end{align}
When $Q_{12}=Q_{13}=1$ and $m=0$, the above phases can take the most general values, \textit{i.e.},
\begin{align}
&\theta_{\Sigma;\sigma}=-2\pi  \frac{k}{8}\\
&\theta_{\Sigma,\sigma;\Sigma}=2\pi \frac{k}{8}\\
&\theta_{\sigma;\Sigma}^{(1)}=2\pi \frac{k}{4^{n+1}}\\
&\theta_{\sigma,\Sigma;\sigma}^{(1)}=-2\pi  \frac{k}{4^{n+1}}\\
&\theta_{\sigma;\Sigma}^{(2)}=-2\pi\frac{\bar k+\tilde k}{2\times 4^n}\\
&\theta_{\sigma,\Sigma;\sigma}^{(2)}=2\pi\frac{\bar k+\tilde k}{2\times 4^n}
\end{align}
The period of these statistical phases can come from two aspects. One the one hand, considering attachment of particles, 
$\theta_{\Sigma;\sigma}$ are ambiguous up to $\frac{2\pi}{4}$ while $\theta_{\Sigma,\sigma;\Sigma}$, $\theta_{\sigma;\Sigma}$ and $\theta_{\sigma,\Sigma;\sigma}$  up to $\frac{2\pi}{2\times 4^n}$. 
On the other hand, the periods of these statistical phases from those of twisted coefficients are always multiple of the ones from attachmeng particles, as we can show below.
So  $\theta_{\Sigma,\sigma;\Sigma}$, $\theta_{\sigma;\Sigma}^{(2)}$ and $\theta_{\sigma,\Sigma;\sigma}^{(2)}$ are equivalent to zero and $\theta_{\Sigma;\sigma}=\frac{2\pi k}{8}, \theta_{\sigma;\Sigma}^{(1)}=\frac{2\pi k}{4^{n+1}}$ and $\theta_{\sigma,\Sigma;\sigma}^{(1)}=-\frac{2\pi k}{4^{n+1}}$ can take two inequivalent values depending on $k=0,1$ mod 2. 

Now we return to prove that the periods of these statistical phases from those of twisted coefficients are always multiple of the ones from attachment of particles. From (\ref{eqn_app_znzk_period1}) and (\ref{eqn_app_znzk_period2}), the period of $k$ and $\bar k$ is 
\begin{align}
\Gamma=\frac{4^n}{\gcd(4^n,\hat Q_{12})}
\end{align}
and the one of $\tilde k$ is
\begin{align}
\tilde \Gamma=\gcd\{ \frac{4^n}{\gcd(4^n,\hat Q_{12})}, \text{lcm}[\frac{4^n}{\gcd(4^n,\hat Q_{13})}, \frac{ 4^{n-\frac{1}{2}}}{\gcd( 4^{n-\frac{1}{2}}, \hat Q_{12}\hat Q_{13})}]\}.
\end{align}
Then the corresponding period of $\theta_{\Sigma;\sigma}$ and $\theta_{\Sigma,\sigma;\Sigma}$ is 
\begin{align}
2\pi\frac{\text{lcm}(4^n,Q_{12})}{8}
\end{align}
which is apparently multiple of $\frac{\pi}{2}$ and also $\frac{2\pi }{2\times 4^n}$. And the corresponding period of $\theta_{\sigma;\Sigma}$ and $\theta_{\sigma,\Sigma;\sigma}$ due to $k$ is
\begin{align}
&2\pi \frac{(2m+1)Q_{12}}{4\gcd(4^n, Q_{12})} 
\end{align}
which  is multiple of $\frac{2\pi}{2\times 4^n}$; those due to
$\bar k$ and $\tilde k$ are
\begin{widetext}
\begin{align}
& 2\pi \frac{Q_{12}Q_{12}}{2\times 4^n}\gcd\{ \frac{4^n}{\gcd(4^n, Q_{12})}, \text{lcm}[\frac{4^n}{\gcd(4^n,\hat Q_{13})}, \frac{2\times 4^{n-1}}{\gcd(2\times 4^{n-1},  Q_{12}\hat Q_{13})}]\}\\
&2\pi \frac{Q_{12}Q_{13}}{2\times 4^n}\gcd\{ \frac{4^n}{\gcd(4^n, Q_{12})}, \text{lcm}[\frac{4^n}{\gcd(4^n, \hat  Q_{13})}, \frac{2\times 4^{n-1}}{\gcd(2\times 4^{n-1},  Q_{12}\hat Q_{13})}]\}
\end{align}
\end{widetext}
respectively, which is also apparently multiple of $\frac{2\pi}{2\times 4^n}$. 

Therefore, the classification for SFL for $eC$ is $\Z_2$.\\
\\
\textbf{Case three: $eQ_\pm$}\\
\\
In this case, $Q_{11}=4m\pm 1$, then $\hat Q_{11}=Q_{11}$. The two M3L invariants are simplified to
\begin{align}
&\mathcal{I}_{\alpha aa}=-k \frac{Q_{12}}{4}\\
&\mathcal{I}_{a\alpha \alpha}^{(1)}=k\frac{Q_{12} (4m\pm 1)}{4^{n+1}}\\
&\mathcal{I}_{a\alpha \alpha}^{(2)}=-(\bar kQ_{12}+\tilde kQ_{13}) \frac{Q_{12}}{4^n}
\end{align}
Then the M3L statistical phases quantize to
\begin{align}
&\theta_{\Sigma;\sigma}=-2\pi k \frac{Q_{12}}{4}\\
&\theta_{\Sigma,\sigma;\Sigma}=2\pi k\frac{Q_{12}}{4}\\
&\theta_{\sigma;\Sigma}^{(1)}=2\pi k\frac{(4m\pm1)Q_{12}}{4^{n+1}}\\
&\theta_{\sigma,\Sigma;\sigma}^{(1)}=-2\pi k\frac{(4m\pm 1)Q_{12}}{4^{n+1}}\\
&\theta_{\sigma;\Sigma}^{(2)}=-2\pi(\bar k Q_{12}+\tilde k Q_{13})\frac{Q_{12}}{ 4^n}\\
&\theta_{\sigma,\Sigma;\sigma}^{(2)}=2\pi(\bar k Q_{12}+\tilde k Q_{13})\frac{Q_{12}}{ 4^n}
\end{align}
 As for $\theta_{\Sigma;\sigma}$, it is ambiguous only up to $\frac{2\pi}{4}$ due to attaching particles, while as for $\theta_{\Sigma,\sigma;\Sigma}$, $\theta_{\sigma;\Sigma}$ and $\theta_{\sigma,\Sigma;\sigma}$,  they are ambiguous up to $\frac{2\pi}{4^{n+1}}$. So we see that all these values of statistical phases are equivalent to trivial values. 
 
 Therefore, for $eQ_\pm$, there is only trivial SFL.

\section{Details of substituting  Eq.~(\ref{eqn_formula_of_substituting3}) into the twisted terms}\label{appendix_substituting}
 For $a^1a^2da^2$ and $a^2a^1da^1$,  
  \begin{align}
 \frac{ q_{122}}{4\pi^2}a^1a^2da^2&=-2\pi q_{122}(\frac{*d^{-1}\Sigma^{1}}{N}+\frac{Q_{11}}{NK}*d^{-1}\sigma)\nonumber \\
&\quad  (\frac{*d^{-1}\Sigma^{2}}{N}+\frac{Q_{12}}{NK}*d^{-1}\sigma)
 (\frac{*\Sigma^{2}}{N}+\frac{Q_{12}}{NK}*\sigma)\nonumber \\
 &=-\frac{2\pi q_{122}}{N^3} (*d^{-1}\Sigma^{1})(*d^{-1}\Sigma^{2})(*\Sigma^{2})\nonumber \\
&\quad -\frac{2\pi q_{122}Q_{12}}{N^3K} (*d^{-1}\Sigma^{1})(*d^{-1}\Sigma^{2})(*\sigma)\nonumber \\
&\quad -\frac{2\pi q_{122}Q_{12}}{N^3K} (*d^{-1}\Sigma^{1})(*d^{-1}\sigma)(*\Sigma^{2})\nonumber \\
&\quad -\frac{2\pi q_{122}Q_{12}^2}{N^3K^2} (*d^{-1}\Sigma^{1})(*d^{-1}\sigma)(*\sigma)\nonumber \\
&\quad -\frac{2\pi q_{122}Q_{11}}{N^3K} (*d^{-1}\sigma)(*d^{-1}\Sigma^{2})(*\Sigma^{2})\nonumber \\
&\quad +\frac{2\pi q_{122}Q_{11}Q_{12}}{N^3K^2} (*d^{-1}\Sigma^{2})(*d^{-1}\sigma)(*\sigma)\nonumber
  \end{align}
  and 
   \begin{align}
 \frac{ q_{211}}{4\pi^2}a^2a^1da^1&=-2\pi q_{211}(\frac{*d^{-1}\Sigma^{2}}{N}+\frac{Q_{12}}{NK}d^{-1}\sigma)\nonumber \\
&\quad  (\frac{*d^{-1}\Sigma^{1}}{N}+\frac{Q_{11}}{NK}d^{-1}\sigma)
 (\frac{*\Sigma^{1}}{N}+\frac{Q_{11}}{NK}\sigma)\nonumber \\
 &=-\frac{2\pi q_{211}}{N^3} (*d^{-1}\Sigma^{2})(*d^{-1}\Sigma^{1})(*\Sigma^{1})\nonumber \\
&\quad -\frac{2\pi q_{211}Q_{11}}{N^3K} (*d^{-1}\Sigma^{2})(*d^{-1}\Sigma^{1})(*\sigma)\nonumber \\
&\quad -\frac{2\pi q_{211}Q_{11}}{N^3K} (*d^{-1}\Sigma^{2})(*d^{-1}\sigma)(*\Sigma^{1})\nonumber \\
&\quad -\frac{2\pi q_{211}Q_{11}^2}{N^3K^2} (*d^{-1}\Sigma^{2})(*d^{-1}\sigma)(*\sigma)\nonumber \\
&\quad -\frac{2\pi q_{211}Q_{12}}{N^3K} (*d^{-1}\sigma)(*d^{-1}\Sigma^{1})(*\Sigma^{1})\nonumber \\
&\quad +\frac{2\pi q_{211}Q_{11}Q_{12}}{N^3K^2} (*d^{-1}\Sigma^{1})(*d^{-1}\sigma)(*\sigma)\nonumber
  \end{align}
For more examples,  we have
\begin{align}
\frac{q_{\nu ii}}{4\pi^2} a^\nu a^ida^i&=-2\pi q_{\nu ii}  \frac{Q_{1i}^2}{NK^2} (*d^{-1}\Sigma^{\nu}) (*d^{-1}\sigma) (*\sigma)\nonumber \\
\frac{q_{\nu  ji}}{4\pi^2} a^\nu a^ida^j&=-2\pi q_{\nu ij}  \frac{Q_{1i}Q_{1j}}{NK^2} (*d^{-1}\Sigma^{\nu}) (*d^{-1}\sigma) (*\sigma) \nonumber \\
\frac{q_{\nu ij}}{4\pi^2} a^\nu a^jda^i&=-2\pi q_{\nu  ji}  \frac{Q_{1j}Q_{1i}}{NK^2} (*d^{-1}\Sigma^{\nu}) (*d^{-1}\sigma) (*\sigma) \nonumber \\
\frac{q_{ ij\nu}}{4\pi^2} a^i a^jda^\nu&=0 \nonumber \\
\frac{q_{i\nu \nu}}{4\pi^2} a^i a^\nu da^\nu
&= -\frac{2\pi q_{i\nu \nu}Q_{1i}}{N^2K} (*d^{-1}\sigma)(*d^{-1}\Sigma^{\nu})(*\Sigma^{\nu})\nonumber \\
&\quad +\frac{2\pi q_{i\nu \nu }Q_{1i}Q_{1\nu}}{N^2K^2} (*d^{-1}\Sigma^{\nu})(*d^{-1}\sigma)(*\sigma) \nonumber \\
\frac{q_{12i}}{4\pi^2} a^1a^2 da^i
 &=-\frac{2\pi q_{12i}Q_{1i}}{N^2K} (*d^{-1}\Sigma^{1})(*d^{-1}\Sigma^{2})(*\sigma)\nonumber \\
&\quad -\frac{2\pi q_{12i}Q_{12}Q_{1i}}{N^2K^2} (*d^{-1}\Sigma^{1})(*d^{-1}\sigma)(*\sigma)\nonumber \\
&\quad +\frac{2\pi q_{12i}Q_{11}Q_{1i}}{N^2K^2} (*d^{-1}\Sigma^{2})(*d^{-1}\sigma)(*\sigma)\nonumber\\
 \frac{ q_{i12}}{4\pi^2}a^ia^1da^2
 &= -\frac{2\pi q_{i12}Q_{1i}}{N^2K} (*d^{-1}\sigma) (*d^{-1}\Sigma^{1})(*\Sigma^{2})\nonumber \\
&\quad +\frac{2\pi q_{i12}Q_{12}Q_{1i}}{N^2K^2} (*d^{-1}\Sigma^{1})(*d^{-1}\sigma)(*\sigma)\nonumber 
\end{align}
\begin{align}
 \frac{ q_{i21}}{4\pi^2}a^ia^2da^1
 &= -\frac{2\pi q_{i21}Q_{1i}}{N^2K}(*d^{-1}\sigma)  (*d^{-1}\Sigma^{2})(*\Sigma^{1})\nonumber \\
&\quad +\frac{2\pi q_{i21}Q_{11}Q_{1i}}{N^2K^2} (*d^{-1}\Sigma^{2})(*d^{-1}\sigma)(*\sigma)\nonumber 
\end{align}

\section{$\Z_N\times \Z_N$ topological order enriched by $\Z_{K}$ symmetry} 
\label{sec_app_znznzk}
From (\ref{eq_mathcalc}), we see that there are  $(\gcd(N,K))^2$ different patterns of symmetry fractionalization on particles.

Now we discuss the SFL.  Generally, from the discussion in Sec.~\ref{mmt_general_discussion_more_General}, we  need to consider seven M3L invariants for this case
$ \mathcal{I}_{a\alpha \alpha}$, 
 $ \mathcal{I}_{\alpha aa}$, 
 $ \mathcal{I}_{b\alpha \alpha}$, 
 $ \mathcal{I}_{\alpha bb}$, 
 $ \mathcal{I}_{\alpha ab}$, 
 $ \mathcal{I}_{\alpha ba}$ and
$ \mathcal{I}_{ab\alpha }$
where we have used $a, b$ to represent the two elementary loop excitations $\Sigma^1$ and $\Sigma^2$ respectively and $\alpha$ to represent the symmetry flux $\sigma$.
 Correspondingly,  statistical phases determined by these invariants can be seen in Table~\ref{table_three_M3L_example_list}.
In fact, there are in total 15 different M3L statistical phases.  Furthermore, in the following we ignore the statistical phases $\theta_{\sigma, \sigma; \Sigma^i}$ and $\theta_{\Sigma^i, \Sigma^i; \sigma}$ since they are determined by $\theta_{\sigma; \Sigma^i}$ and $\theta_{\Sigma^i; \sigma}$.


From Sec.(\ref{SF_loop_z2z2_z2}), we  can start with considering the   action (\ref{eqn_action_z2z2_z20}), rewritten as follows 
\begin{align}
&S=S_0+S_{int}+S_c+S_{sr}\,,\\
 &S_0=  \frac{1}{2\pi}\int \sum_{v=1}^2 N b^v   da^v+ b^3   da^3 +b^4   da^4 \,, \\
  &S_c=\frac{1}{2\pi}\int  \sum_{v=1}^2 Q_{1v}  A   db^v +  Q_{13}A   db^3+Q_{14}A   db^4 \,,   \\
 &S_{sr}=   \int \sum_{v=1}^2 (a^v   *j^v +b^v   * \Sigma^v)+ a^3*j^3+ a^4   *j^4 \,,  \\
 &S_{int}= \frac{1}{4\pi^2}  \int  q_{122}  a^1  a^2da^2 +q_{211}  a^2  a^1da^1  \nonumber \\
&+ \sum_{v=1}^2 q_{3vv}a^3  a^vda^v+q_{v33}  a^v  a^3da^3+q_{v34}  a^v  a^3da^4 + \nonumber \\
&\quad \qquad q_{123}a^1a^2da^3+q_{312}a^3a^1da^2+q_{321}a^3a^2da^1\,.
 \end{align}
After integrating out $b^1,b^2, b^i ,b^j$, we can obtain the effective action
\begin{align}
S =& S_{AB}+S_{TL}+S_{M3L}
\label{eqn_z2z2_z2_M3L_coefficients}
\end{align}
where $S_{M3L}$ is given by (\ref{eqn_main_z2z2_z2_m3laction}), $S_{AB}$ collects the  terms describing the braiding between charge(particle) flux(defect) and $S_{TL}$ collects two terms that characterize three-loop braiding statistics, 
given by 
\begin{align}
S_{TL}= \int \frac{- 2\pi q_{122}}{N^3} (*d^{-1}\Sigma^1)(*d^{-1}\Sigma^2)(*\Sigma^2)+\nonumber \\
\frac{-2\pi q_{211}}{N^3} (*d^{-1}\Sigma^2)(*d^{-1}\Sigma^1)(*\Sigma^1).\,\label{eqn_z2z2_z3_TL_action}
\end{align}
The coefficients of the M3L  invariants  in (\ref{eqn_z2z2_z2_M3L_coefficients}) are given by (\ref{eqn_main_z2z2_z2_m3l_typical}), which are rewritten as follows
\begin{subequations}
  \begin{align}
\!&\mathcal{I}_{\alpha aa}\!\!=\! \!-\frac{ q_{211}Q_{12}}{N^3K}\!-\!\frac{ q_{311}Q_{13}}{N^2K}, \\
&\mathcal{I}_{\alpha bb}\!\!=\!\!-\frac{ q_{122}Q_{11}}{N^3K}\!-\! \!\!\frac{ q_{322}Q_{13}}{N^2K},\\
\!&\mathcal{I}_{\alpha ab}\!\!=\!\! \frac{ q_{122}Q_{12}}{N^3K}\!-\!  \frac{ q_{312}Q_{13}}{N^2K} , \\
&\mathcal{I}_{\alpha ba}\!\!=\!\!\frac{ q_{211}Q_{11}}{N^3K}\!-\!  \frac{ q_{321}Q_{13}}{N^2K} ,\\
\!&\mathcal{I}_{ab\alpha} \!\!=\!\! -\frac{ q_{122}Q_{12}}{N^3K}+\frac{ q_{211}Q_{11}}{N^3K}\!-\!\frac{ q_{123}Q_{13}}{N^2K} ,\\
&\mathcal{I}_{a\alpha \alpha}=\nonumber\\
& \frac{- q_{122}Q_{12}^2}{N^3K^2}+\frac{ q_{211}Q_{11}Q_{12}}{N^3K^2} +\frac{ q_{311}Q_{13}Q_{11}}{N^2K^2}- \frac{ q_{133} Q^2_{13}}{NK^2}\nonumber \\
&- \frac{ q_{134}Q_{13} Q_{14}}{NK^2} + \frac{ q_{312}Q_{13}Q_{12}}{N^2K^2}- \frac{ q_{123}Q_{12}Q_{13}}{N^2K^2}\,,\\
& \mathcal{I}_{b\alpha \alpha}=
 \nonumber\\
& \frac{ q_{122}Q_{11}Q_{12}}{N^3K^2}-\frac{ q_{211}Q_{11}^2}{N^3K^2}+ \frac{ q_{322}Q_{13}Q_{12}}{N^2K^2}- \frac{ q_{233}Q^2_{13} }{N K^2}\nonumber \\
&- \frac{ q_{234}Q_{13}Q_{14}}{N K^2} + \frac{ q_{321}Q_{13}Q_{11}}{N^2K^2}+ \frac{q_{123}Q_{11}Q_{13}}{N^2K^2} \,.
\end{align}
\end{subequations}

We first discuss the terms in (\ref{eqn_z2z2_z3_TL_action}) that determine the three loop braiding statistics.
 The two terms determine that the three loop braiding statistics $(\theta_{\Sigma^1,\Sigma^2;\Sigma^2},\theta_{\Sigma^2;\Sigma^1},\theta_{\Sigma^1,\Sigma^2;\Sigma^1},\theta_{\Sigma^1;\Sigma^2} )$ take $(\frac{2\pi q_{122}}{N^3}, \frac{-2\pi q_{122}}{N^3}, \frac{2\pi q_{211}}{N^3}, \frac{-2\pi q_{211}}{N^3})$ respectively. 
 
 In the absence of symmetry, $q_{122}=kN$, $q_{211}=\bar k N$ with $k,\bar k\in \Z_N$. Therefore $(\theta_{\Sigma^1,\Sigma^2;\Sigma^2},\theta_{\Sigma^2;\Sigma^1},\theta_{\Sigma^1,\Sigma^2;\Sigma^1},\theta_{\Sigma^1;\Sigma^2} )$ are further quantized to ($\frac{2\pi k}{N^2}$, $-\frac{2\pi k}{N^2}$, $\frac{2\pi \bar k}{N^2}$, $-\frac{2\pi \bar k}{N^2}$).  The periods of these statistical phases should also consider two  aspects as M3L invariants, that is the periods of twisted terms and attachment of particles. For the present cases, combined these two aspects, the periods of these statistical phases are all $\frac{2\pi}{N}$. Therefore, for pure $\Z_N\times \Z_N$ topological order, their four statistical phases are given by 
 \begin{align}
 &\theta_{\Sigma^1,\Sigma^2;\Sigma^2}=-\theta_{\Sigma^1;\Sigma^2}=\frac{2\pi k}{N^2} \text{ mod }\frac{2\pi}{N} \label{eqn_app_znznzk_TLB_1} \\
 &\theta_{\Sigma^1,\Sigma^2;\Sigma^1}=-\theta_{\Sigma^2;\Sigma^1}=\frac{2\pi \bar k}{N^2} \text{ mod }\frac{2\pi}{N} \label{eqn_app_znznzk_TLB_2} 
  \end{align}
 with   $k,\bar k\in \Z_N$. Correspondingly, the $N^2$ patterns of these three loop braiding statistics can characterize the  $\Z_N\times \Z_N$ topological orders.  In particular, if both $k, \bar k \sim 0$, then all the three loop braiding statistics are trivial. Topological order with trivial  three loop braiding statistics is  untwisted; otherwise twisted. In these  topological orders, no matter whether there is symmetry or not, these three loop braiding statistics are always there. However, the presence of symmetry may affect the quantization levels of $q_{122}, q_{211}$ so that some of three loop braiding statistics are not compatible with certain symmetry realization, as we discuss more details below.
 
 In the presence of symmetry, i.e., $Q_{11}, Q_{12}$ are not both zero, the twisted coefficients  $q_{122},q_{211}$ are quantized to
$ q_{122}=k M$, $q_{211}=\bar k M$ with 
\begin{align}
M=\text{lcm}[N, \frac{NK}{\gcd(\hat Q_{11}, NK)},\frac{NK}{\gcd(\hat Q_{12}, NK)}].
\label{eqn_z2z2_z2_00_qntz_1}
\end{align}
  When any of $Q_{11}$ and $Q_{12}$ take 1, then $M=NK$, so  $q_{122},q_{211}$ are both multiple of $NK$, and then the four loop braiding statistics $\theta_{\Sigma^1,\Sigma^2;\Sigma^2}$, $\theta_{\Sigma^1;\Sigma^2}$ and $\theta_{\Sigma^1,\Sigma^2;\Sigma^1}$, $\theta_{\Sigma^2;\Sigma^1}$ both take multiple of $\frac{2\pi K}{N}$, which are equivalent to be trivial. As $Q_{11}$ and $Q_{12}$  also determine the SFP, we see that  only the untwisted $\Z_N\times \Z_N$ topological order is compatible with these symmetry realizations with $Q_{11}=1$ or $Q_{12}=1$ (\textit{i.e.}, certain SFP), while the twisted ones  are not compatible with these SFP, hence are anomalous. The phenomenon might also happen for other charge matrix (\textit{i.e.}, the values of $Q_{11}$ and $Q_{12}$), which depends on both $N$ and $K$.  

From (\ref{eqn:twisted_coff_general_1}) and (\ref{eqn:twisted_coff_general_11}), the quantization levels of various twisted coefficients take the form as 
 $q_{133}=k_1M_{13}$, $q_{311}= k_2 M_{13} $, $q_{233}=k_3 M_{23} $, $q_{322}= k_4 M_{23}$ and $q_{134}=\bar k_1M_{13}$, $q_{234}= \bar k_3 M_{23}$, $q_{312}= k_5 M_{13}$, $q_{321}= k_6 M_{23}$,  and $q_{123}=k_7 M$ where $k_i,\bar k_i$ are all integral where
 \begin{align}
 M_{13}&=\text{lcm}[N, \frac{NK}{\gcd(NK, \hat Q_{11})}, \frac{K}{\gcd(K, \hat Q_{13})}]\label{eqn_app_znznzk_quantized_multiple1i}\\
  M_{23}&=\text{lcm}[N,\frac{NK}{\gcd(NK, \hat Q_{12})}, \frac{K}{\gcd(K, \hat Q_{13})}]\label{eqn_app_znznzk_quantized_multiple2i}
 \end{align}
Then the above invariants can be simplified to be
\begin{align}
&\mathcal{I}_{\alpha aa}=\frac{- \bar k M Q_{12}}{{N^3K}}+\frac{- k_2 M_{13} Q_{13}}{{N^2K}}\nonumber\\
&\mathcal{I}_{\alpha bb} =\frac{- k M Q_{11}}{{N^3K}}+\frac{- k_4 M_{23} Q_{13}}{{N^2K}}\nonumber\\
&\mathcal{I}_{\alpha ab}=\frac{ kMQ_{12}}{{N^3K}}+\frac{- k_5 M_{13} Q_{13}}{{N^2K}}\nonumber\\
&\mathcal{I}_{\alpha ba}=\frac{ \bar k M Q_{11}}{{N^3K}}+\frac{- k_6 M_{23}Q_{13}}{{N^2K}}\nonumber\\
&\mathcal{I}_{ab\alpha}=\frac{- k M Q_{12}}{{N^3K}}+\frac{\bar k M Q_{11}}{{N^3K}}+\frac{- k_7 M Q_{13}}{{N^2K}}\nonumber\\
&\mathcal{I}_{a\alpha \alpha}=[\frac{- k M Q_{12}^2}{{N^3K^2}}+\frac{ \bar k M Q_{11}Q_{12}}{{N^3K^2}} +\frac{ k_2 M_{13} Q_{13}Q_{11}}{{N^3K}}\nonumber \\
&\qquad +\frac{- Q_{13}(k_1 M_{13} Q_{13}+\bar k_1 M_{13}Q_{14})}{{NK^2}} \nonumber \\
&\qquad +\frac{ k_5 M_{13} Q_{13}Q_{12}}{{N^2K^2}}+\frac{- k_7 M Q_{12}Q_{13}}{{N^2K^2}}]\nonumber\\
& \mathcal{I}_{b\alpha \alpha}=[\frac{ k M Q_{11}Q_{12}}{{N^3K^2}}+\frac{- \bar k M Q_{11}^2}{{N^3K^2}}+\frac{ k_4 M_{23} Q_{13}Q_{12}}{{N^3K}}\nonumber \\
&\qquad +\frac{- Q_{13}(k_3 M_{23} Q_{13}+\bar k_3 M_{23}Q_{14})}{{NK^2}}\nonumber \\
&\qquad +\frac{ k_6 M_{23} Q_{13}Q_{11}}{{N^2K^2}}-\frac{k_7 M Q_{11}Q_{13}}{{N^2K^2}}].\nonumber
\end{align}

From (\ref{eqn:twisted_coff_general_2}) and (\ref{eqn:twisted_coff_general_3}),  the periods of  $k, \bar k$ are given by
\begin{align}
\Gamma=\gcd\bigg\{ \text{lcm}[\frac{N^2K}{\gcd(N^2K, \hat Q_{12} M)}, \frac{N^2K^2}{\gcd(N^2K^2, \hat Q_{11}\hat Q_{12} M)}], \nonumber \\
\text{lcm}[\frac{N^2K}{\gcd(N^2K, \hat Q_{11} M)}, \frac{N^2K^2}{\gcd(N^2K^2, \hat Q_{11}\hat Q_{12} M)}]\bigg\};
\end{align}
those of $k_1,k_2$ are given by
\begin{align}
\Gamma_1=\gcd\bigg\{ \text{lcm}[\frac{NK}{\gcd(NK, \hat Q_{13} M_{13})}, \frac{NK^2}{\gcd(NK^2, \hat Q_{11}\hat Q_{13} M_{13})}], \nonumber \\
\text{lcm}[\frac{NK}{\gcd(NK, \hat Q_{11} M_{13})}, \frac{NK^2}{\gcd(NK^2, \hat Q_{11}\hat Q_{13} M_{13})}]\bigg\};
\end{align}
those of $k_3,k_4$ are given by
\begin{align}
\Gamma_2=\gcd\bigg\{ \text{lcm}[\frac{NK}{\gcd(NK, \hat Q_{13} M_{23})}, \frac{NK^2}{\gcd(NK^2, \hat Q_{12}\hat Q_{13} M_{23})}], \nonumber \\
\text{lcm}[\frac{NK}{\gcd(NK, \hat Q_{12} M_{23})}, \frac{NK^2}{\gcd(NK^2, \hat Q_{12}\hat Q_{13} M_{23})}]\bigg\};
\end{align}
those of $\bar k_1, \bar k_3$ are given by
\begin{align}
\bar\Gamma_1=\gcd\bigg\{ &\text{lcm}[\frac{NK}{\gcd(NK, \hat Q_{13} M_{13})}, \frac{NK^2}{\gcd(NK^2, \hat Q_{11}\hat Q_{13} M_{13})}], \nonumber \\
&\text{lcm}[\frac{NK}{\gcd(NK, \hat Q_{11} M_{13})}, \frac{NK^2}{\gcd(NK^2, \hat Q_{11}\hat Q_{13} M_{13})}],\nonumber \\
&\text{lcm}[\frac{NK}{\gcd(NK, \hat Q_{11} M_{13})}, \frac{K}{\gcd(K,\hat Q_{13}M_{13})}, \nonumber \\
&\quad \,\,\,\,\frac{NK^2}{\gcd(NK^2,\hat Q_{11} \hat Q_{14})},\frac{K^2}{\gcd(K^2, \hat Q_{13}\hat Q_{14}M_{13})}]\bigg\};
\end{align}
and 
\begin{align}
\bar\Gamma_1=\gcd\bigg\{ &\text{lcm}[\frac{NK}{\gcd(NK, \hat Q_{13} M_{23})}, \frac{NK^2}{\gcd(NK^2, \hat Q_{12}\hat Q_{13} M_{23})}], \nonumber \\
&\text{lcm}[\frac{NK}{\gcd(NK, \hat Q_{12} M_{23})}, \frac{NK^2}{\gcd(NK^2, \hat Q_{12}\hat Q_{13} M_{23})}],\nonumber \\
&\text{lcm}[\frac{NK}{\gcd(NK, \hat Q_{12} M_{23})}, \frac{K}{\gcd(K,\hat Q_{13}M_{23})}, \nonumber \\
&\quad \,\,\,\,\frac{NK^2}{\gcd(NK^2,\hat Q_{12} \hat Q_{14})},\frac{K^2}{\gcd(K^2, \hat Q_{13}\hat Q_{14}M_{23})}]\bigg\};
\end{align}
those of $k_5, k_6$ are given by 
\begin{align}
\Gamma_5=\gcd\bigg\{ &\text{lcm}[\frac{NK}{\gcd(NK, \hat Q_{11} M_{13})}, \frac{NK^2}{\gcd(NK^2, \hat Q_{11}\hat Q_{13} M_{13})}], \nonumber \\
&\text{lcm}[\frac{NK}{\gcd(NK, \hat Q_{13} M_{13})}, \frac{NK^2}{\gcd(NK^2, \hat Q_{11}\hat Q_{13} M_{13})}],\nonumber \\
&\text{lcm}[\frac{NK}{\gcd(NK, \hat Q_{13} M_{13})}, \frac{N^2K}{\gcd(N^2K,\hat Q_{11}M_{13})}, \nonumber \\
&\quad \,\,\,\,\frac{NK^2}{\gcd(NK^2,\hat Q_{12} \hat Q_{13}M_{13})},\frac{N^2K^2}{\gcd(N^2K^2, \hat Q_{11}\hat Q_{12}M_{13})}]\bigg\};
\end{align}
and 
\begin{align}
\Gamma_6=\gcd\bigg\{ &\text{lcm}[\frac{NK}{\gcd(NK, \hat Q_{12} M_{23})}, \frac{NK^2}{\gcd(NK^2, \hat Q_{12}\hat Q_{13} M_{23})}], \nonumber \\
&\text{lcm}[\frac{NK}{\gcd(NK, \hat Q_{13} M_{23})}, \frac{NK^2}{\gcd(NK^2, \hat Q_{12}\hat Q_{13} M_{23})}],\nonumber \\
&\text{lcm}[\frac{NK}{\gcd(NK, \hat Q_{13} M_{23})}, \frac{N^2K}{\gcd(N^2K,\hat Q_{12}M_{23})}, \nonumber \\
&\quad \,\,\,\,\frac{NK^2}{\gcd(NK^2,\hat Q_{11} \hat Q_{13}M_{23})},\frac{N^2K^2}{\gcd(N^2K^2, \hat Q_{11}\hat Q_{12}M_{23})}]\bigg\};
\end{align}
and that of $k_7$ is given by
\begin{align}
\Gamma_7=\gcd\bigg\{ &\text{lcm}[\frac{N^2K}{\gcd(N^2K, \hat Q_{12} M_{12})}, \frac{N^2K^2}{\gcd(N^2K^2, \hat Q_{11}\hat Q_{12} M_{12})}], \nonumber \\
&\text{lcm}[\frac{NK}{\gcd(N^2K, \hat Q_{11} M_{12})}, \frac{N^2K^2}{\gcd(N^2K^2, \hat Q_{11}\hat Q_{12} M_{12})}],\nonumber \\
&\text{lcm}[\frac{NK}{\gcd(NK, \hat Q_{11} M_{12})}, \frac{NK}{\gcd(NK,\hat Q_{12}M_{12})}, \nonumber \\
&\quad \,\,\,\,\frac{NK^2}{\gcd(NK^2,\hat Q_{11} \hat Q_{13}M_{12})},\frac{NK^2}{\gcd(NK^2, \hat Q_{12}\hat Q_{13}M_{12})}]\bigg\};
\end{align}

Below we discuss a concrete example $\Z_2\times \Z_2$ topological order enriched by $\Z_{2^n}$ symmetry. The case of $n=1$  is also discussed in  Sec.\ref{SET:ZNZN} where to obtain the classification we have taken $Q_{1i}=1$ for $i\ge 3$ for simplicity. Here we prove that  the specific choice with $Q_{1i}=1$  can result in the same results as that  the general $Q_{1i}$ which is discussed here.

Due to the specific value $N=2, K=2^n$ and the fact $1\le \hat Q_{ij}\le K$, the above  periods $\Gamma$s can be further simplified as follows.
\begin{align}
\Gamma=&
 \frac{N^2K^2}{\gcd(N^2K^2, \hat Q_{11}\hat Q_{12} M)};\nonumber \\
\Gamma_1=& \frac{NK^2}{\gcd(NK^2, \hat Q_{11}\hat Q_{13} M_{13})};\nonumber \\
\Gamma_2=&\frac{NK^2}{\gcd(NK^2, \hat Q_{12}\hat Q_{13} M_{23})};\nonumber \\
\bar\Gamma_1=&\gcd\bigg[  \frac{NK^2}{\gcd(NK^2, \hat Q_{11}\hat Q_{13} M_{13})}, \nonumber \\
&\text{ lcm}[\frac{NK^2}{\gcd(NK^2,\hat Q_{11} \hat Q_{14}M_{13})},\frac{K^2}{\gcd(K^2, \hat Q_{13}\hat Q_{14}M_{13})}]\bigg];\nonumber \\
\bar\Gamma_2=&\gcd\bigg\{  \frac{NK^2}{\gcd(NK^2, \hat Q_{12}\hat Q_{13} M_{23})}, \nonumber \\
&\text{ lcm}[\frac{NK^2}{\gcd(NK^2,\hat Q_{12} \hat Q_{14}M_{23})},\frac{K^2}{\gcd(K^2, \hat Q_{13}\hat Q_{14}M_{23})}]\bigg];\nonumber \\
\Gamma_5=&\gcd\bigg[\frac{NK^2}{\gcd(NK^2, \hat Q_{11}\hat Q_{13} M_{13})},\nonumber \\
&\text{ lcm}[\frac{NK^2}{\gcd(NK^2,\hat Q_{12} \hat Q_{13}M_{13})},\frac{N^2K^2}{\gcd(N^2K^2, \hat Q_{11}\hat Q_{12}M_{13})}]\bigg];\nonumber \\
\Gamma_6=&\gcd\bigg[ \frac{NK^2}{\gcd(NK^2, \hat Q_{12}\hat Q_{13} M_{23})}],\nonumber \\
&\text{ lcm}[\frac{NK^2}{\gcd(NK^2,\hat Q_{11} \hat Q_{13}M_{23})},\frac{N^2K^2}{\gcd(N^2K^2, \hat Q_{11}\hat Q_{12}M_{23})}]\bigg];\nonumber \\
\Gamma_7=&\gcd\bigg[\frac{N^2K^2}{\gcd(N^2K^2, \hat Q_{11}\hat Q_{12} M)}],\nonumber \\
&\text{ lcm}[\frac{NK^2}{\gcd(NK^2,\hat Q_{11} \hat Q_{13}M)},\frac{NK^2}{\gcd(NK^2, \hat Q_{12}\hat Q_{13}M)}]\bigg];
\end{align}
To further simplify the expressions, we introduce  integers $p_{ij}$($\tilde p_{ij}$) by
\begin{subequations}
\label{eqn_app_znznzk_auxil_5}
\begin{align}
\hat Q_{11} M &= p_{11} NK\\
\hat Q_{12} M &= p_{12} NK\\
\hat Q_{13} M_{13} &= p_{13} K\\
\hat Q_{11} M_{13} &= \tilde p_{13} NK\\
\hat Q_{13} M_{23} &= p_{23} K\\
\hat Q_{12} M_{23} &= \tilde p_{23} NK
\end{align}
\end{subequations}
From the expressions of $M, M_{13}, M_{23}$ above, one can see that $p_{ij}$ and $\tilde p_{ij}$ are indeed integers. Using these relations, the periods $\Gamma$s can be further simplified as
\begin{align}
\Gamma=&
 \frac{NK}{\gcd(NK, p_{11}\hat Q_{12} )}=\frac{NK}{\gcd(NK, p_{12}\hat Q_{11})};\nonumber \\
\Gamma_1=& \frac{K}{\gcd(K, \tilde p_{13}\hat Q_{13} )}= \frac{NK}{\gcd(NK,  p_{13}\hat Q_{11} )};\nonumber \\
\Gamma_2=&\frac{K}{\gcd(K, \tilde p_{23}\hat Q_{13} )}=\frac{NK}{\gcd(NK,  p_{23}\hat Q_{12} )};\nonumber \\
\bar\Gamma_1=&\gcd\bigg[  \frac{K}{\gcd(K, \tilde p_{13}\hat Q_{13})}, \text{ lcm}[\frac{K}{\gcd(K,\tilde p_{13} \hat Q_{14})},\frac{K}{\gcd(K, p_{13}\hat Q_{14})}]\bigg]\nonumber \\
=&\gcd\bigg[  \frac{K}{\gcd(K,  p_{13}\hat Q_{11})}, \text{ lcm}[\frac{K}{\gcd(K,\tilde p_{13} \hat Q_{14})},\frac{K}{\gcd(K, p_{13}\hat Q_{14})}]\bigg];\nonumber \\
\bar\Gamma_2=&\gcd\bigg\{  \frac{K}{\gcd(K, \tilde p_{23}\hat Q_{13} )},\text{lcm}[\frac{K}{\gcd(K, \tilde p_{23}  \hat Q_{14})},\frac{K}{\gcd(K, p_{23}\hat Q_{14})}]\bigg]\nonumber \\
=&\gcd\bigg\{  \frac{K}{\gcd(K,  p_{23}\hat Q_{12} )},\text{lcm}[\frac{K}{\gcd(K, \tilde p_{23}  \hat Q_{14})},\frac{K}{\gcd(K, p_{23}\hat Q_{14})}]\bigg];\nonumber \\
\Gamma_5=&\gcd\bigg[\frac{K}{\gcd(K, \tilde p_{13}\hat Q_{13} )},\text{lcm}[\frac{K}{\gcd(K,\tilde p_{23}\hat Q_{13})},\frac{NK}{\gcd(NK, \tilde p_{13}\hat Q_{12})}]\bigg]\nonumber \\
=&\gcd\bigg[\frac{K}{\gcd(K,  p_{13}\hat Q_{11} )},\text{lcm}[\frac{K}{\gcd(K,\tilde p_{23}\hat Q_{13})},\frac{NK}{\gcd(NK, \tilde p_{13}\hat Q_{12})}]\bigg];\nonumber \\
\Gamma_6=&\gcd\bigg[ \frac{K}{\gcd(K, \tilde p_{23}\hat Q_{13})}],\text{lcm}[\frac{NK}{\gcd(NK,\hat Q_{11} p_{23})},\frac{NK}{\gcd(NK, \tilde p_{23}\hat Q_{11})}]\bigg]\nonumber \\
=&\gcd\bigg[ \frac{K}{\gcd(K,  p_{23}\hat Q_{12})}],\text{lcm}[\frac{NK}{\gcd(NK,\hat Q_{11} p_{23})},\frac{NK}{\gcd(NK, \tilde p_{23}\hat Q_{11})}]\bigg];\nonumber \\
\Gamma_7=&\gcd\bigg[\frac{NK}{\gcd(NK, p_{11} \hat Q_{12})}],\text{lcm}[\frac{K}{\gcd(K,p_{11} \hat Q_{13})},\frac{K}{\gcd(K, p_{12}\hat Q_{13})}]\bigg]\nonumber \\
=&\gcd\bigg[\frac{NK}{\gcd(NK, p_{12} \hat Q_{11})}],\text{lcm}[\frac{K}{\gcd(K,p_{11} \hat Q_{13})},\frac{K}{\gcd(K, p_{12}\hat Q_{13})}]\bigg];
\label{eqn_app_z2z2_zk_general_period}
\end{align}
With these expression of the quantization rules $M_{ij}$ and  periods $\Gamma$s, now we move to discuss the SFP and SFL of $\Z_2\times \Z_2$ topological order enriched by $\Z_{2^n}$ symmetry.

\subsection{$N=2,K=2^n$}

\subsubsection{SFP}

Since  $\gcd(2,2^n)\times \gcd(2,2^n)=4$, there are four different SFP. From (\ref{eq_mathcalc}), the  two elementary particle $e_1$ and $e_2$ can carry half and integer charge of $\Z_K$ symmetry. Accordingly, their reduced charge matrices are $\mathcal{Q}=(0,0), (0,1),(1,0), (1,1)$ respectively where $\mathcal{Q}=(\mathcal{Q}_{11},\mathcal{Q}_{12})$.  We denote these four patterns as $e_10e_20$, $e_10e_2C$, $e_1Ce_20$ and $e_1Ce_2C$ respectively.

\subsubsection{SFL}
Here we discuss the SFL for different SFP case by case. 
 \\
\\
\textbf{Case-1: $e_10e_20$}\\
\\ 
In this case, $Q_{11}=2m_1$ and $Q_{12}=2m_2$. We first discuss the two invariants in (\ref{eqn_z2z2_z3_TL_action}).   If $m_1$ or $m_2$ are zero,  then the three loop braiding statistics can take 
the values as those in (\ref{eqn_app_znznzk_TLB_1}) and (\ref{eqn_app_znznzk_TLB_2}), which means that all 
$\Z_2\times \Z_2$ topological order, including both untwisted and twisted ones, are compatible with this SFP.

Now we discuss the M3L statistical phases for all the $\Z_2\times \Z_2$ topological orders with this SFP. 
Now $M$ in (\ref{eqn_z2z2_z2_00_qntz_1}) becomes 
\begin{align}
M=\text{lcm}[\frac{K}{\gcd(\hat m_1, K)},\frac{K}{\gcd(\hat m_2, K)}].
\end{align}
where we have denoted $\hat Q_{11}=2\hat m_1$ and $\hat Q_{12}=2\hat m_2$ (If $ Q_{1\nu}=0, \hat m_\nu=K/2$, otherwise $\hat m_\nu=m_\nu$ with $\nu=1,2$.). Now (\ref{eqn_app_znznzk_quantized_multiple1i}) and (\ref{eqn_app_znznzk_quantized_multiple1i}) now take
\begin{align}
&M_{13}=\text{lcm}[\frac{K}{\gcd(\hat m_1, K)}, \frac{K}{\gcd(\hat Q_{13}, K)}]\\
&M_{23}=\text{lcm}[\frac{K}{\gcd(\hat m_2, K)}, \frac{K}{\gcd(\hat Q_{13}, K)}]
\end{align}
To further simplify the expression of M3L statistical phases, we introduce $\tau$ function such that that \begin{align}
m_\nu M&=\tau_{0\nu} p_\nu K, \label{eqn_app_znznzk_auxil_7}\\
  Q_{1 3} M_{1 3}&=\tau_{13} p_{13} K\label{eqn_app_znznzk_auxil_8}\\
     Q_{1 3} M_{2 3}&=\tau_{23} p_{23} K \label{eqn_app_znznzk_auxil_9}
     \end{align}
      with $\nu=1,2$ where $\tau_{ij}$ are defined as: $\tau_{0\nu}=0$ if $m_\nu=0$, otherwise $\tau_{0\nu}=1$; $\tau_{\nu 3}=0$ if $Q_{1 3}=0$, otherwise $\tau_{\nu 3}=1$.

The seven M3L invariants now become
\begin{align}
\mathcal{I}_{\alpha aa}=&\frac{- \bar k \tau_{02} p_2  }{{N^2}}+\frac{- k_2 \tau_{13} p_{13} }{{N^2}}\label{eqn_app_znznzk_case1_inv1}\\
\mathcal{I}_{\alpha bb} =&\frac{- k \tau_{01} p_1 }{{N^2}}+\frac{- k_4 \tau_{23} p_{23} }{{N^2}}\label{eqn_app_znznzk_case1_inv2}\\
\mathcal{I}_{\alpha ab}=&\frac{ k \tau_{02} p_2 }{{N^2}}+\frac{-k_5 \tau_{13} p_{13}  }{{N^2}}\label{eqn_app_znznzk_case1_inv3}\\
\mathcal{I}_{\alpha ba}=&\frac{\bar k \tau_{01} p_1 }{{N^2}}+\frac{- k_6 \tau_{23} p_{23}}{{N^2}}\label{eqn_app_znznzk_case1_inv4}\\
\mathcal{I}_{ab\alpha}=&\frac{ \bar k \tau_{01} p_1 }{{N^2}}-\frac{ k \tau_{02} p_2 }{{N^2}}-\frac{ k_7  \text{lcm}[\frac{Q_{13}K}{\gcd(\hat m_1, K)}, \frac{Q_{13}K}{\gcd(\hat m_2, K)}]}{{N^2K}}\label{eqn_app_znznzk_case1_inv5}\\
\mathcal{I}_{a\alpha \alpha}=&[\frac{- k \tau_{02} p_2 m_{2}}{{NK}}+\frac{ \bar k \tau_{01} p_1  m_{2}}{{NK}} +\frac{ k_2 \tau_{13} p_{13}  m_{1}}{{N^2}}\nonumber \\
&+\frac{-( Q_{13}k_1 \tau_{13} p_{13}  +\bar k_1 \tau_{13} p_{13}  Q_{14})}{{NK}}\nonumber\\ 
&+\frac{ k_5 \tau_{13} p_{13}  m_{2}}{{NK}}+\frac{- k_7 \tau_{23} p_2  Q_{13}}{{NK}}]\label{eqn_app_znznzk_case1_inv6}\\
 \mathcal{I}_{b\alpha \alpha}=&[\frac{ k \tau_{01}p_1 m_{2}}{{NK}}+\frac{-\bar k \tau_{01} p_{1} m_{1}}{{NK}}+\frac{ k_4 \tau_{23} p_{23}  m_{2}}{{N^2}}\nonumber \\
&+\frac{- (Q_{13}k_3\tau_{23} p_{23} +\bar k_3 \tau_{23} p_{23}  Q_{14})}{{NK}}\nonumber\\ 
&+\frac{ k_6 \tau_{23} p_{23}  m_{1}}{{NK}}-\frac{k_7 \tau_{01} p_1  Q_{13}}{{NK}}].\label{eqn_app_znznzk_case1_inv7}
\end{align}

When $Q_{13}=1$ and $m_1=m_2=0$, then $p_{13}=p_{23}=1$ and also $\tau_{13}=\tau_{23}=1$. Then all the M3L statistical phases can take the most general quantized values, that is,
\begin{align}
 & (\theta_{\Sigma^1, \sigma; \sigma}, \theta_{\sigma; \Sigma^1}) = (\frac{2\pi  (k_7+k_1-\bar k_1) }{NK}, \frac{-2\pi  (k_7+k_1-\bar k_1) }{NK})\label{eqn_app_znznzk_statisticalpha_mostgen1}\\
&(\theta_{\sigma, \Sigma^1; \Sigma^1}, \theta_{\Sigma^1; \sigma}) = (\frac{2\pi  k_2 }{N^2} , -\frac{2\pi  k_2 }{N^2})\label{eqn_app_znznzk_statisticalpha_mostgen2}\\
 & (\theta_{\Sigma^2, \sigma; \sigma}, \theta_{\sigma; \Sigma^2}) = (\frac{2\pi(k_7+k_3-\bar k_3)   }{NK}, \frac{-2\pi  (k_7+k_3-\bar k_3) }{NK})\label{eqn_app_znznzk_statisticalpha_mostge3}\\
& (\theta_{\sigma, \Sigma^2; \Sigma^2},\theta_{\Sigma^2; \sigma}) = ( \frac{2\pi  k_4 }{N^2}, \frac{-2\pi  k_4 }{N^2})\label{eqn_app_znznzk_statisticalpha_mostgen4}\\
& (\theta_{\sigma, \Sigma^2; \Sigma^1}, \theta_{\Sigma^2, \Sigma^1; \sigma}) = (\frac{2\pi  k_5 }{NK}, \frac{-2\pi  k_5 }{NK})\label{eqn_app_znznzk_statisticalpha_mostgen5}\\
 &(\theta_{\sigma, \Sigma^1; \Sigma^2}, \theta_{\Sigma^2, \Sigma^1; \sigma}) = (\frac{2\pi  k_6 }{NK}, \frac{-2\pi  k_6 }{NK})\label{eqn_app_znznzk_statisticalpha_mostgen6}\\
 & (\theta_{\sigma,\Sigma^1;\Sigma^2}, \theta_{\sigma,\Sigma^2;\Sigma^1}) = (\frac{2\pi  k_7 }{NK},  \frac{-2\pi  k_7 }{NK})\label{eqn_app_znznzk_statisticalpha_mostgen7}
 \end{align}
 
Now we discuss the periods of these M3L statistical phases. For this purpose, we need to consider two origins of the periods of the M3L statistical phases.  On the one hand, the periods of these M3L statistical phases can come from attaching particles. For $\theta_{\Sigma^1,\sigma;\sigma}$, $\theta_{\Sigma^2,\sigma;\sigma}$, $\theta_{\sigma; \Sigma^1}$,  $\theta_{\sigma; \Sigma^2}$, $\theta_{\sigma, \Sigma^1; \Sigma^1}$, $\theta_{\sigma, \Sigma^2; \Sigma^2}$, $\theta_{\sigma, \Sigma^1; \Sigma^2}$ and $\theta_{\sigma, \Sigma^2; \Sigma^1}$, the periods from attaching $e_1$ or $e_2$  or even trivial particle to $\Sigma^1$, $\Sigma^2$ or $\sigma$ are all $\frac{2\pi}{K}$ since $e_i$ or trivial particle can carry unit charge of $\Z_K$ symmetry which can induce  particle loop braiding statisitcs, and result in an amount of $\frac{2\pi}{K}$ phase factor.   For $\theta_{\Sigma^1; \sigma}$, $\theta_{\Sigma^2; \sigma}$ and $\theta_{\Sigma^1,\Sigma^2; \sigma}$, their periods due to attaching particles are all $\frac{2\pi}{N}$.

On the other hand, the periods of these M3L statistical phases can also come from those of twisted coefficients. However, as we can show below, the periods from twisted coefficients are always multiple of those from attaching particles. From (\ref{eqn_app_z2z2_zk_general_period}), the  periods of $k$ and $\bar k$   are 
\begin{align}
\Gamma=\frac{K}{\gcd(K,\hat m_1 p_{12})}=\frac{K}{\gcd(K,\hat m_2 p_{11})};
\end{align}
those of $k_1$ and $k_2$ are 
\begin{align}
\Gamma_1=\frac{K}{\gcd(K, \hat m_1 p_{13})};
\end{align}
 those of $k_3$ and $k_4$ are 
\begin{align}
\Gamma_2=\frac{K}{\gcd(K, \hat m_2 p_{23})};
\end{align}
those of $\bar k_1$ and $\bar k_3$ are  
\begin{align}
\overline{\Gamma}_1=\gcd\{\frac{K}{\gcd(K, \hat m_1 p_{13})}, \text{lcm}[&\frac{K}{\gcd(K,\hat Q_{14} \tilde p_{13})}, \nonumber \\
&\frac{K}{\gcd(K,\hat Q_{14} p_{13})}]\}\\
\overline{\Gamma}_2=\gcd\{\frac{K}{\gcd(K, \hat m_2 p_{23})}, \text{lcm}[&\frac{K}{\gcd(K,\hat Q_{14} \tilde p_{23})}, \nonumber \\
&\frac{K}{\gcd(K,\hat Q_{14} p_{23})}]\}
\end{align}
respectively;
those of $k_5,k_6, k_7$ are 
\begin{align}
\Gamma_5=\gcd\{\frac{K}{\gcd(K, \hat m_1 p_{13})}, \text{lcm}[&\frac{K}{\gcd(K,\hat m_2  p_{13})}, \nonumber \\
&\frac{K}{\gcd(K,\hat m_2 \tilde p_{13})}]\}\\
\Gamma_6=\gcd\{\frac{K}{\gcd(K, \hat Q_{13} \tilde p_{23})}, \text{lcm}[&\frac{K}{\gcd(K,\hat m_1  p_{23})}, \nonumber \\
&\frac{K}{\gcd(K,\hat m_1 \tilde p_{23})}]\}, \\
\Gamma_7=\gcd\{\frac{K}{\gcd(K, \hat m_{1}  p_{12})}, \text{lcm}[&\frac{K}{\gcd(K,  p_{11} \hat Q_{13})}, \nonumber \\
&\frac{K}{\gcd(K, \tilde p_{12} \hat Q_{13})}]\}
\end{align}
We then can evaluate the periods of the M3L statistical phases due to twisted coefficients. For $\theta_{\sigma, \Sigma^\nu; \Sigma^\nu}$ and $\theta_{\Sigma^\nu; \sigma}$ ($\nu=1,2$) which  come from invariants $\mathcal{I}_{\alpha aa}$ or $\mathcal{I}_{\alpha bb}$, since $\hat m_{1,2}\le n-1$, $p_2 \Gamma$,  $p_1\Gamma$, $p_{13}\Gamma_1$, and $p_{23}\Gamma_2$ must be multiple of $N=2$, the periods of these statistical phases from the periods of $k,\bar k$, $k_2$ and $k_4$ are all multiple of $\frac{2\pi}{N}$.  For $\theta_{\Sigma^\nu, \sigma; \sigma}$ and $\theta_{\sigma; \Sigma^\nu}$ ($\nu=1,2$) which  come from invariants $\mathcal{I}_{a\alpha \alpha}$ or $\mathcal{I}_{b\alpha\alpha}$, six types of different twisted terms are related  for each case (\textit{i.e.}, $\nu=1$ or $2$). As for $\nu=1$, the six types of twisted terms are represented by $k,\bar k, k_2, k_1,\bar k_1, k_5$ and $k_7$ which appear in the expression of $\mathcal{I}_{a\alpha \alpha}$ in (\ref{eqn_app_znznzk_case1_inv6}). As $p_2 \Gamma, p_1\Gamma$ must be even, then periods of  $\theta_{\Sigma^1, \sigma; \sigma}$ and $\theta_{\sigma; \Sigma^1}$ from $k$ and $\bar k$ must be $\frac{2\pi}{K}$; 
as $p_{13} m_1\Gamma_1 $ must be multiple of $K$, then the periods  of  $\theta_{\Sigma^1, \sigma; \sigma}$ and $\theta_{\sigma; \Sigma^1}$ from $k_2$ must be multiple of $\pi$; as $p_{13} \Gamma_1$ and $p_{13}Q_{14}\bar \Gamma_1$ must be even, then  the periods  of  $\theta_{\Sigma^1, \sigma; \sigma}$ and $\theta_{\sigma; \Sigma^1}$ from $k_1$ and $\bar k_1$ must be $\frac{2\pi}{K}$; as $p_{13}m_2 \Gamma_5$ and $p_2\Gamma_7$ must be even, the periods of $\theta_{\Sigma^1, \sigma; \sigma}$ and $\theta_{\sigma; \Sigma^1}$ from $k_5$ and $k_7$ must be $\frac{2\pi}{K}$ and $\frac{2\pi}{K}$ respectively. 
Therefore,  for $\theta_{\Sigma^1, \sigma; \sigma}$ and $\theta_{\sigma; \Sigma^1}$, their periods from twisted coefficients are always multiple of $\frac{2\pi}{K}$. Similarly and straightforwardly, we also can see that the periods of $\theta_{\Sigma^2, \sigma; \sigma}$ and $\theta_{\sigma; \Sigma^2}$ from twisted coefficients are always multiple of $\frac{2\pi}{K}$.  For $\theta_{\Sigma^1,\Sigma^2;\sigma}$ which can come from $\mathcal{I}_{\alpha ab}$ and $\mathcal{I}_{\alpha ba}$, four types of twisted terms are related, which  become $k, k_5$ in 
(\ref{eqn_app_znznzk_case1_inv3}) and $\bar k, k_6$ in (\ref{eqn_app_znznzk_case1_inv4}). 
As $p_2\Gamma, p_1\Gamma, p_{13} \Gamma_5$ and $p_{23}\Gamma_6$ must be even, then the periods of $\theta_{\Sigma^1,\Sigma^2;\sigma}$ from twisted coefficientsis are always multiple of $\frac{2\pi}{N}$. As for $\theta_{\sigma,\Sigma^1;\Sigma^2}$ which can come from $\mathcal{I}_{\alpha ba}$ and $\mathcal{I}_{ab\alpha }$, four types of twisted terms are related, which become $\bar k, k_6, k$  and $k_7$ in (\ref{eqn_app_znznzk_case1_inv4}) and (\ref{eqn_app_znznzk_case1_inv5}). 
As the period of $\theta_{\sigma,\Sigma^1;\Sigma^2}$ from attaching particles can be $\frac{2\pi}{K}$, those period (even the quantized values) of its due to $\bar k, k, k_6$ are trivially (\textit{i.e.}, always be multiple of $\frac{2\pi}{K}$).  We then only need to consider that from $k_7$. As $\Gamma_7 M$ is always multiple of $N^2=4$, the period of $\theta_{\sigma,\Sigma^1;\Sigma^2}$ from $k_7$ are always multiple of $\frac{2\pi}{K}$. Therefore, the periods of $\theta_{\sigma,\Sigma^1;\Sigma^2}$ from twisted coefficients are always multiple of $\frac{2\pi}{K}$. Similarly, the periods of $\theta_{\sigma,\Sigma^2;\Sigma^1}$ from twisted coefficients are always multiple of $\frac{2\pi}{K}$.

Therefore, the M3L statistical phases   in (\ref{eqn_app_znznzk_statisticalpha_mostgen1})-(\ref{eqn_app_znznzk_statisticalpha_mostgen4}) can be generated independently by different twisted terms and charge matrices. In other words, they can give to four different types of patterns of M3L statistical phases. More explicitly, these M3L statistical phases can take: (1)$(\theta_{\Sigma^1, \sigma; \sigma}, \theta_{\sigma; \Sigma^1})=(\frac{2\pi l_1}{NK},-\frac{2\pi l_1}{NK})$ with $l_1 \in\Z_N$; (2) $(\theta_{\sigma, \Sigma^1; \Sigma^1}, \theta_{\Sigma^1; \sigma})$=($0, \frac{2\pi l_2}{N^2}$) with $l_2\in \Z_N$; (3)$(\theta_{\Sigma^2, \sigma; \sigma}, \theta_{\sigma; \Sigma^1})=(\frac{2\pi l_4}{NK},-\frac{2\pi l_4}{NK})$ with $l_1 \in\Z_N$; (4) $(\theta_{\sigma, \Sigma^2; \Sigma^2}, \theta_{\Sigma^2; \sigma})$=($0, \frac{2\pi l_4}{N^2}$) with $l_4\in \Z_N$. As patterns of  M3L statistcal phases in (\ref{eqn_app_znznzk_statisticalpha_mostgen5})-(\ref{eqn_app_znznzk_statisticalpha_mostgen7}) are not independent, they in fact can give two independent patterns of M3L statistical phases, that is (5) $(\theta_{\sigma, \Sigma^2; \Sigma^1}, \theta_{\Sigma^2,\Sigma^1;\sigma} )=(\frac{2\pi k_5}{NK}, -\frac{2\pi k_5}{NK})$ with $l_5\in\Z_N$; (6) $(\theta_{\sigma,\Sigma^1;\Sigma^2}, \theta_{\Sigma^2,\Sigma^1;\sigma}) = (\frac{2\pi  l_6 }{NK},  \frac{-2\pi  l_6 }{NK})$ with $l_6\in\Z_N$. We note that the above discussion of patterns of  M3L statistical phases  are compatible with all $\Z_N\times \Z_N$ topological order. Therefore, the classification of SFL for the SFP $e_10e_20$ is $(\Z_N)^6$  all $\Z_N\times \Z_N$ topological order.\\
\\
\textbf{Case-2: $e_1Ce_20$}\\
\\ 
 In this case, $Q_{11}=2m_1+1$ and $Q_{12}=2m_2$. We first discuss the two invariants in (\ref{eqn_z2z2_z3_TL_action}).  Now $M$ in (\ref{eqn_z2z2_z2_00_qntz_1}) becomes 
\begin{align}
M=\text{lcm}[\frac{NK}{\gcd(2 m_1+1, NK)},\frac{K}{\gcd(\hat m_2, K)}]=NK.
\end{align}
Then  the statistical phases $\theta_{\Sigma^1,\Sigma^2;\Sigma^2},\theta_{\Sigma^2;\Sigma^1},\theta_{\Sigma^1,\Sigma^2;\Sigma^1}$ and $\theta_{\Sigma^1;\Sigma^2}$ are all quantized to $\frac{2\pi}{N}$. Therefore, only the untwisted $\Z_N\times \Z_N$ topological order are compatible with this SFP, while the twisted ones are not compatible with, namely they are anomalous.

Now we discuss the classification of SFL for the untwisted $\Z_N\times \Z_N$ topological order with SFP being $e_1Ce_20$. Now (\ref{eqn_app_znznzk_quantized_multiple1i}) and (\ref{eqn_app_znznzk_quantized_multiple2i}) become
\begin{align}
M_{13}&=NK\\
M_{23}&=\text{lcm}[N, \frac{K}{\gcd(\hat m_2, K)}, \frac{K}{\gcd(\hat Q_{13}, K)}].
\end{align}
With definition of $p_{2i}$  and $\tau_{2i}$ in (\ref{eqn_app_znznzk_auxil_5}) and (\ref{eqn_app_znznzk_auxil_8}),  the seven invariants now can be simplified to
\begin{subequations}
\label{eqn_app_znznzk_M3L_inv_general_2}
\begin{align}
&\mathcal{I}_{\alpha aa}=\frac{- \bar k  m_{2}}{{N}}+\frac{- k_2  Q_{13}}{{N}}\nonumber\\
&\mathcal{I}_{\alpha bb} =\frac{- k  Q_{11}}{{N^2}}+\frac{- k_4 \tau_{23} p_{23}}{{N^2}}\nonumber\\
&\mathcal{I}_{\alpha ab}=\frac{ km_{2}}{{N}}+\frac{- k_5  Q_{13}}{{N}}\nonumber\\
&\mathcal{I}_{\alpha ba}=\frac{ \bar k  Q_{11}}{{N^2}}+\frac{- k_6 \tau_{23} p_{23}}{{N^2}}\nonumber\\
&\mathcal{I}_{ab\alpha}=\frac{- k  m_{2}}{{N}}+\frac{\bar k  Q_{11}}{{N^2}}+\frac{- k_7  Q_{13}}{{N}}\nonumber\\
&\mathcal{I}_{a\alpha \alpha}=[\frac{- k  m_{2}^2}{{K}}+\frac{ \bar k  Q_{11}m_{2}}{{NK}} +\frac{ k_2  Q_{13}Q_{11}}{{N^2}}\nonumber \\
&\qquad +\frac{- Q_{13}(k_1  Q_{13}+\bar k_1  Q_{14})}{{K}} \nonumber \\
&\qquad +\frac{ k_5 Q_{13}m_{2}}{{K}}+\frac{- k_7  m_{2}Q_{13}}{{K}}]\nonumber\\
& \mathcal{I}_{b\alpha \alpha}=[\frac{ k  Q_{11}m_{2}}{{NK}}+\frac{- \bar k  Q_{11}^2}{{N^2K}}+\frac{ k_4 \tau_{23} p_{23}m_{2}}{{N^2}}\nonumber \\
&\qquad +\frac{- Q_{13}k_3 \tau_{23} p_{2i}-\bar k_3 \tau_{23} p_{23} Q_{1j}}{{NK}}\nonumber \\
&\qquad +\frac{ k_6 \tau_{23} p_{2i} Q_{11}}{{N^2K}}-\frac{k_7  Q_{11}Q_{13}}{{NK}}].\nonumber
\end{align}
\end{subequations}

To simplify the discussion, we first discuss the periods of the correspondingly M3L statistical phases (see Table \ref{table_three_M3L_example_list}) from attachment of particles. Now the $e_1$ particle carries one-half charge of the  $\Z_K$ symmetry, therefore, in certain M3L statistics, attaching particles to $\Sigma^i$ or $\sigma$ may lead to an extra process that $e_1$ braiding around $\sigma$, which causes a $\frac{2\pi}{NK}$ phase  shift.  Specifically, for $\theta_{\Sigma^1,\sigma;\sigma}$, $\theta_{\Sigma^1,\sigma;\sigma}$, $\theta_{\sigma; \Sigma^1}$,  $\theta_{\sigma; \Sigma^2}$, $\theta_{\sigma, \Sigma^1; \Sigma^1}$, $\theta_{\sigma, \Sigma^2; \Sigma^2}$, $\theta_{\sigma, \Sigma^1; \Sigma^2}$ and $\theta_{\sigma, \Sigma^2; \Sigma^1}$, the periods from attaching $e_1$ to $\Sigma^1$, $\Sigma^2$ or $\sigma$ can all be $\frac{2\pi}{NK}$.   For $\theta_{\Sigma^1; \sigma}$, $\theta_{\Sigma^2; \sigma}$ and $\theta_{\Sigma^1,\Sigma^2; \sigma}$, their periods from attaching particles are all $\frac{2\pi}{N}$. 
Then the invariants $\mathcal{I}_{\alpha aa}$, $\mathcal{I}_{\alpha ab}$,  $\mathcal{I}_{ab\alpha}$ and $\mathcal{I}_{a\alpha\alpha}$ in (\ref{eqn_app_znznzk_M3L_inv_general_2})  can be ignored since they only contribute to trivial values of $(\theta_{\sigma,\Sigma^1;\Sigma^1}, \theta_{\Sigma^1;\sigma})$, ($\theta_{\Sigma^2,\sigma;\Sigma^1}, \theta_{\Sigma^2,\Sigma^1;\sigma}$), ($\theta_{\sigma, \Sigma^2;\Sigma^1}, \theta_{\sigma, \Sigma^1;\Sigma^2}$) and ($\theta_{\sigma, \Sigma^1;\sigma},\theta_{\sigma;\Sigma^1}$) that  they determine, respectively.  For $\mathcal{I}_{b\alpha \alpha}$ in (\ref{eqn_app_znznzk_M3L_inv_general_2}), we can only need to consider the two parts with $\bar k$ and $k_6$ since its other parts  only contribute to trivial values of $(\theta_{\Sigma^2,\sigma;\sigma}, \theta_{\sigma;\Sigma^2})$.  

Therefore, among the above invariants, only $\mathcal{I}_{\alpha bb}$, $\mathcal{I}_{\alpha ba}$ and $\mathcal{I}_{b\alpha \alpha}$ can give rise to nontrivial values of M3L statistical phases, \textit{i.e.}, $\theta_{\Sigma^2, \sigma;\Sigma^2}$, $\theta_{\Sigma^2;\sigma}$, $\theta_{\Sigma^1,\Sigma^2;\sigma}$, $\theta_{\Sigma^1,\sigma;\Sigma^2}$, $\theta_{\Sigma^2,\sigma;\sigma}$ and $\theta_{\sigma;\Sigma^2}$.  When $Q_{13}=1$, $Q_{11}=1$(\textit{i.e.}, $M_{23}=K$, and $\tau_{23}=p_{23}=1$),   they can take the most general quantized values of M3L statistical phases, that is, $(\theta_{\Sigma^2;\sigma}, \theta_{\Sigma^1, \Sigma^2;\sigma}, \theta_{\sigma; \Sigma^2}, \theta_{\Sigma^2,\sigma;\sigma})$ = $(\frac{-2\pi (k+k_4)}{N^2}, \frac{2\pi(\bar k-k_6)}{N^2}, \frac{-2\pi (\bar k -k_6)}{N^2K},\frac{2\pi (\bar k -k_6)}{N^2K})$, while the other two $\theta_{\Sigma^2,\sigma;\Sigma^2}$ and $\theta_{\Sigma^1,\sigma;\Sigma^2}$ can only take trivial values as they have a period of $\frac{2\pi}{NK}$, as discussed above.

Below we discuss the periods of $(\theta_{\Sigma^2;\sigma}, \theta_{\Sigma^1, \Sigma^2;\sigma}, $ $ \theta_{\sigma; \Sigma^2}, \theta_{\Sigma^2,\sigma;\sigma})$ from those of twisted coefficients (as the periods from attachment of particles have already been discussed above).  From (\ref{eqn_app_z2z2_zk_general_period}), the  periods of $k$, $\bar k$, $k_4$ and $k_6$ are given by
\begin{align}
\Gamma\,\,\,&=\frac{K}{\gcd(K, \hat m_2)}\\
\Gamma_2&=\frac{K}{\gcd(K, \hat m_{2} p_{23})}\\
\Gamma_6&=\gcd\{ \frac{K}{\gcd(K,  \hat m_2 p_{23})}, \text{lcm}[\frac{NK}{\gcd(NK,  p_{23})}, \nonumber \\
&\qquad \qquad\qquad \qquad\qquad \qquad \frac{NK}{\gcd(NK,  \tilde p_{23})}]\}
\end{align}
As $\hat m_2\le n-1$, $\Gamma$, $p_{23} \Gamma_2$ and $p_{23} \Gamma_6$ must be even,  and then the periods of $\theta_{\Sigma^2;\sigma}$ and $\theta_{\Sigma^1, \Sigma^2;\sigma}$ from the periods of twisted coefficients are always multiple of $\frac{2\pi}{N}$ and those of $\theta_{\sigma; \Sigma^2}, \theta_{\Sigma^2,\sigma;\sigma}$  are always multiple of $\frac{2\pi}{NK}$. 

Combined with the periods from attaching particles(discussed above), the minimal periods of $\theta_{\Sigma^2;\sigma}$ and $\theta_{\Sigma^1, \Sigma^2;\sigma}$ are both $\frac{2\pi}{N}$ and those of $\theta_{\sigma; \Sigma^2}, \theta_{\Sigma^2,\sigma;\sigma}$  are both $\frac{2\pi}{NK}$.

Therefore, there are two independent types of patterns of M3L statistical phases, that is, 
$(1) \theta_{\Sigma^2;\sigma}=\frac{2\pi l_1}{N^2}$ with $l_1\in \Z_N$ and (2) $(\theta_{\Sigma^1, \Sigma^2;\sigma}, \theta_{\sigma; \Sigma^2}, \theta_{\Sigma^2,\sigma;\sigma})=(\frac{2\pi l_2}{N^2K}, \frac{-2\pi l_2}{N^2K},\frac{2\pi l_2}{N^2K}  )$ with $l_2\in \Z_N$. Therefore, the classification of SFL for $e_1Ce_20$ is $(\Z_N)^2$ for the untwisted $\Z_N\times \Z_N$ topological order while the twisted ones are anomalous with $e_1Ce_20$.
\\
\\
\textbf{Case-3: $e_10e_2C$}\\
\\ 
This case can be obtained similarly to the above case by just exchanging the two $\Z_N$ gauge group in the discussion. Therefore, there are two independent types of patterns of M3L statistical phases, that is, 
$(1) \theta_{\Sigma^1;\sigma}=\frac{2\pi l_1}{N^2}$ with $l_1\in \Z_N$ and (2) $(\theta_{\Sigma^1, \Sigma^2;\sigma}, \theta_{\sigma; \Sigma^1}, \theta_{\Sigma^1,\sigma;\sigma})=(\frac{2\pi l_2}{N^2K}, \frac{-2\pi l_2}{N^2K},\frac{2\pi l_2}{N^2K}  )$ with $l_2\in \Z_N$. Therefore, the classification of SFL for $e_10e_2C$ is $(\Z_N)^2$ for the untwisted $\Z_N\times \Z_N$ topological order while the twisted ones are anomalous with $e_1Ce_20$.
\\
\\
\textbf{Case-4: $e_1Ce_2C$}\\
\\ 
 In this case, $Q_{11}=2m_1+1$ and $Q_{12}=2m_2+1$. We first discuss the two invariants in (\ref{eqn_z2z2_z3_TL_action}).  Now $M$ in (\ref{eqn_z2z2_z2_00_qntz_1}) becomes 
\begin{align}
M=NK.
\end{align}
Then  the statistical phases $\theta_{\Sigma^1,\Sigma^2;\Sigma^2},\theta_{\Sigma^2;\Sigma^1},\theta_{\Sigma^1,\Sigma^2;\Sigma^1}$ and $\theta_{\Sigma^1;\Sigma^2}$ are all quantized to $\frac{2\pi}{N}$. Therefore, only the untwisted $\Z_N\times \Z_N$ topological order are compatible with this SFP, while the twisted ones are not compatible with, namely they are anomalous.

Now we discuss the classification of SFL for the untwisted $\Z_N\times \Z_N$ topological order with SFP being $e_1Ce_2C$. Now (\ref{eqn_app_znznzk_quantized_multiple1i}) and (\ref{eqn_app_znznzk_quantized_multiple2i}) become
\begin{align}
M_{13}&=NK\\
M_{23}&=NK.
\end{align}
The seven invariants now can be simplified to
\begin{subequations}
\label{eqn_app_znznzk_M3L_inv_general_4}
\begin{align}
&\mathcal{I}_{\alpha aa}=\frac{- \bar k  Q_{12}}{{N^2}}+\frac{- k_2 Q_{13}}{{N}}\\
&\mathcal{I}_{\alpha bb} =\frac{- k  Q_{11}}{{N^2}}+\frac{- k_4  Q_{13}}{{N}}\\
&\mathcal{I}_{\alpha ab}=\frac{ kQ_{12}}{{N^2}}+\frac{- k_5  Q_{13}}{{N}}\\
&\mathcal{I}_{\alpha ba}=\frac{ \bar k  Q_{11}}{{N^2}}+\frac{- k_6 Q_{13}}{{N}}\\
&\mathcal{I}_{ab\alpha}=\frac{- k  Q_{12}}{{N^2}}+\frac{\bar k  Q_{11}}{{N^2}}+\frac{- k_7  Q_{13}}{{N}}\\
&\mathcal{I}_{a\alpha \alpha}=[\frac{- k  Q_{12}^2}{{N^2K}}+\frac{ \bar k  Q_{11}Q_{12}}{{N^2K}} +\frac{ k_2  Q_{13}Q_{11}}{{N^2}} \\
&\qquad +\frac{- Q_{13}(k_1  Q_{13}+\bar k_1 Q_{14})}{{K}}  \\
&\qquad +\frac{ k_5  Q_{13}Q_{12}}{{NK}}+\frac{- k_7  Q_{12}Q_{13}}{{NK}}]\\
& \mathcal{I}_{b\alpha \alpha}=[\frac{ k  Q_{11}Q_{12}}{{N^2K}}+\frac{- \bar k  Q_{11}^2}{{N^2K}}+\frac{ k_4  Q_{13}Q_{12}}{{N^2}} \\
&\qquad +\frac{- Q_{13}(k_3  Q_{13}+\bar k_3 Q_{14})}{{K}} \\
&\qquad +\frac{ k_6 Q_{13}Q_{11}}{{NK}}-\frac{k_7  Q_{11}Q_{13}}{{NK}}].
\end{align}
\end{subequations}

To simplify the discussion, we first discuss the periods of the correspondingly M3L statistical phases (see Table \ref{table_three_M3L_example_list}) from attachment of particles. Now the $e_1$ and $e_2$ particles both carry one-half charge of the  $\Z_K$ symmetry, therefore, in certain M3L statistics, attaching particles to $\Sigma^i$ or $\sigma$ may lead to an extra process that $e_1$ or $e_2$ braiding around $\sigma$, which causes a $\frac{2\pi}{NK}$ phase  shift.  Specifically, for $\theta_{\Sigma^1,\sigma;\sigma}$, $\theta_{\Sigma^1,\sigma;\sigma}$, $\theta_{\sigma; \Sigma^1}$,  $\theta_{\sigma; \Sigma^2}$, $\theta_{\sigma, \Sigma^1; \Sigma^1}$, $\theta_{\sigma, \Sigma^2; \Sigma^2}$, $\theta_{\sigma, \Sigma^1; \Sigma^2}$ and $\theta_{\sigma, \Sigma^2; \Sigma^1}$, the periods from attaching $e_1$ to $\Sigma^1$, $\Sigma^2$ or $\sigma$ can all be $\frac{2\pi}{NK}$.   For $\theta_{\Sigma^1; \sigma}$, $\theta_{\Sigma^2; \sigma}$ and $\theta_{\Sigma^1,\Sigma^2; \sigma}$, their periods from attaching particles are all $\frac{2\pi}{N}$. 

Therefore, these invariants can determine the following possible nontrivial M3L statistical phases:
$\theta_{\Sigma^1;\sigma}=\frac{-2\pi \bar k Q_{12}}{N^2}$, $\theta_{\Sigma^2;\sigma}=\frac{-2\pi  k Q_{12}}{N^2}$,  $\theta_{\Sigma^1,\Sigma^2;\sigma}=\frac{2\pi (kQ_{11}+\bar k Q_{12})}{N^2}$, $\theta_{\Sigma^1,\sigma;\sigma}=\frac{2\pi (k Q_{12}^2-\bar k Q_{11}Q_{12})}{N^2K}$, $\theta_{\sigma;\Sigma^1}=\frac{-2\pi (k Q_{12}^2-\bar k Q_{11}Q_{12})}{N^2K}$, $\theta_{\Sigma^2,\sigma;\sigma}=\frac{2\pi (\bar k Q_{11}^2- k Q_{11}Q_{12})}{N^2K}$, $\theta_{\sigma;\Sigma^2}=\frac{-2\pi (\bar k Q_{11}^2- k Q_{11}Q_{12})}{N^2K}$ 
while others can only take trivial values due to the aforementioned periods. When $Q_{11}=Q_{12}=1$, they can take the most general values as $\theta_{\Sigma^1;\sigma}=\frac{-2\pi \bar k }{N^2}$, $\theta_{\Sigma^2;\sigma}=\frac{-2\pi  k }{N^2}$,  $\theta_{\Sigma^1,\Sigma^2;\sigma}=\frac{2\pi (k+\bar k )}{N^2}$, $\theta_{\Sigma^1,\sigma;\sigma} = -\theta_{\sigma;\Sigma^1} =-\theta_{\Sigma^2,\sigma;\sigma} = \theta_{\sigma;\Sigma^2}=\frac{2\pi (k -\bar k )}{N^2K}$.

Now we discuss another aspect of the periods of these M3L statistical phases that is from those of  twisted coefficients (as the ones from attachment of particles have already been discussed above).
From (\ref{eqn_app_z2z2_zk_general_period}), the  periods of $k$, $\bar k$ are both $\Gamma=NK.$ Therefore, it is easy to see that the periods from those of twisted coefficients are always multiple of those from attachment of particles. 

Therefore, there are $(\Z_N)^2$ different patterns of M3L statistical phases, that are given by
$(\theta_{\Sigma^1;\sigma},\theta_{\Sigma^2;\sigma}, \theta_{\Sigma^1,\Sigma^2;\sigma}, \theta_{\Sigma^1,\sigma;\sigma}, \theta_{\sigma;\Sigma^1}, \theta_{\Sigma^2,\sigma;\sigma},   \theta_{\sigma;\Sigma^2})$ = $(\frac{-2\pi \bar k}{N^2}, \frac{-2\pi k}{N^2}, \frac{2\pi (k+\bar k)}{N^2}, \frac{2\pi (k-\bar k)}{N^2}, \frac{-2\pi (k-\bar k)}{N^2}, \frac{-2\pi (k-\bar k)}{N^2}, \frac{2\pi (k-\bar k)}{N^2})$ with $k, \bar k\in \Z_N$. Therefore, the classification of SFL with  $e_1Ce_2C$ is $(\Z_N)^2$ for the untwisted $\Z_N\times \Z_N$ topological order while the twisted ones are anomalous with $e_1Ce_2C$.

\section{$\Z_2$ topological order enriched by $\Z_{2^n}\times \Z_{2^n}$ symmetry} 
\label{sec_app_znzkzk}

In this appendix, we discuss the symmetry fractionalization of $\Z_2$ topological order enriched by $\Z_{2^n}\times \Z_{2^n}$ symmetry. 
In Sec.\ref{sec_example_znzkzk}, we have discussed the case with $n=1$ and obtain the  quick classification using some trick by setting the trivial layers coupling constants to be one. Here we provide a more general   discussion that involve general charge matrix and justify that  setting is sufficient to obtain the classification of SFL.

From (\ref{eq_mathcalc}), we see that there are  $(\gcd(N,K))^2$ different patterns of symmetry fractionalization on particles.

Now we discuss the SFL.  Generally, from the discussion in Sec.~\ref{mmt_general_discussion_more_General}, we  need to consider seven M3L invariants  $\mathcal{I}_{a\alpha\alpha}$, $\mathcal{I}_{a\beta\beta}$, $\mathcal{I}_{\alpha aa}$, $\mathcal{I}_{\beta aa}$,  $\mathcal{I}_{ a\alpha \beta}$, $\mathcal{I}_{ a\beta \alpha}$ and $\mathcal{I}_{\alpha \beta a}$ and the corresponding statistical phases (see Table \ref{table_three_M3L_example_list}) where $\alpha, \beta$ are the defects with unit symmetry flux  of the two $\Z_K$ symmetry respectively, and $a$ is the loop excitation with unit gauge flux of $\Z_N$ where $N=2$ and $K=2^n$.
 
 From Sec.\ref{eqn_action_z2z2_z2_minimal_model}, it is sufficient to  consider the action (\ref{eq_main_z2_z2z2_SFL_action}), which is rewritten as
\begin{align}
&S=S_0+S_{int}+S_c+S_{sr}
 \\
 &S_0=  \frac{1}{2\pi}\int  N b^1   da^1+ b^2   da^2 +b^3   da^3  \\
  &S_c=\frac{1}{2\pi}\int  \sum_{i=1}^2 Q_{i1}  A^i   db^1 +Q_{i2}A^i   db^2+Q_{i3} A^i   db^3    \\
 &S_{sr}=   \int   a^1   *j^1 +b^1   * \Sigma+ a^2*j^2+ a^3   *j^3 \,,  \\
 &S_{int}= \frac{1}{4\pi^2}  \int q_{211}a^2 a^1da^1+q_{122}  a^1  a^2da^2+  \\
&\quad \qquad \qquad \quad\, q_{123}  a^1  a^2da^3 +q_{231}a^2a^3da^1.
 \end{align}
  
After integrating out $b^1, b^2 ,b^3$, we can obtain the effective action
\begin{align}
S=S_{{AB}}+S_{M3L}\\
\end{align} 
where $S_{AB}$ collects the  terms describing the braiding between charges(particles) and fluxes(defects) and $S_{M3L}$ contain the seven M3L invariants whose the  coefficient of the invariants are given by (\ref{eq_main_invariants_expr_znzkzk}), i.e.,
\begin{subequations}
\label{eq_app_invariants_expr_znzkzk}
\begin{align}
&\mathcal{I}_{\alpha aa}=-\frac{q_{211}Q_{12}}{NK^2}\\
&\mathcal{I}_{\beta aa}=-\frac{q_{211}Q_{22}}{NK^2}\\
&\mathcal{I}_{a\alpha\alpha}=\frac{q_{211} Q_{12}Q_{11}}{(NK)^2}-\frac{q_{122}Q_{12}Q_{12}}{NK^2}-\frac{q_{123}Q_{12}Q_{13}}{NK^2}\label{eq_app_znzkzk_invriant_3}\\
&\mathcal{I}_{a\beta\beta}=\frac{q_{211} Q_{22}Q_{21}}{(NK)^2}-\frac{q_{122}Q_{22}Q_{22}}{NK^2}-\frac{q_{123}Q_{22}Q_{23}}{NK^2}\label{eq_app_znzkzk_invriant_4}\\
&\mathcal{I}_{\alpha a \beta}=-\frac{q_{211}Q_{12}Q_{21}}{(NK)^2}+\frac{q_{122}Q_{12}Q_{22}}{NK^2}+\frac{q_{123}Q_{12}Q_{23}}{NK^2}\label{eq_app_znzkzk_invriant_5}\\
&\mathcal{I}_{\beta a \alpha}=-\frac{q_{211}Q_{22}Q_{11}}{(NK)^2}+\frac{q_{122}Q_{22}Q_{12}}{NK^2}+\frac{q_{123}Q_{22}Q_{13}}{NK^2} \label{eq_app_znzkzk_invriant_6}\\
&\mathcal{I}_{\alpha \beta a}=\frac{q_{211}(Q_{22}Q_{11}-Q_{12}Q_{21})}{(NK)^2}  +\frac{q_{231}(Q_{13}Q_{22}-Q_{12}Q_{23})}{NK^2}\label{eq_app_znzkzk_invriant_7}
\end{align}
\end{subequations}

We remark that if $Q_{21}=Q_{22}=Q_{23}=0$, the $\Z_{K_2}$ symmetry subgroup is not coupled to the system so that the system looks like only have $\Z_{K_1}$ symmetry. Then many of the above invariants become zero and only  the two $\mathcal{I}_{\alpha aa}$ and $\mathcal{I}_{a\alpha\alpha}$ remain
which is the same as those in (\ref{eq_invraint_znzk_general1}). 
If $Q_{11}=Q_{12}=Q_{13}=0$, the $\Z_{K_1}$ symmetry subgroup is not coupled to the system so that the system looks like only have $\Z_{K_2}$ symmetry and only the two invariants $\mathcal{I}_{\beta aa}$ and $\mathcal{I}_{a\beta\beta}$ remain.

From (\ref{eqn:twisted_coff_general_1}) and (\ref{eqn:twisted_coff_general_11}),  we have 
\begin{subequations}
\label{eqn_app_zNzK1zK2_gquant}
\begin{align}
q_{211}&=kM_{12},\\
 q_{122}&=\bar k M_{12}, \\
 q_{123}&=\widetilde k M_{12}\\
 q_{231}&= k' M_{23}
\end{align}
\end{subequations}
 where $k,\bar k,\widetilde k$ are integral and
\begin{align}
M_{12}=&\text{lcm}[N,\frac{NK}{\gcd(NK,\hat Q_{11})}, \frac{ K}{\gcd( K, \hat Q_{12})}, \nonumber \\
&\qquad   \frac{NK}{\gcd(NK,\hat Q_{21})}, \frac{ K}{\gcd( K,\hat Q_{22})}]\label{eqn_app_znzk1zk2_multiple_1}\\
M_{23}=&\text{lcm}[\frac{ K}{\gcd( K,\hat Q_{12})}, \frac{ K}{\gcd( K, \hat Q_{13})}, \nonumber \\
&\qquad   \frac{ K}{\gcd( K,\hat Q_{22})},\frac{  K}{\gcd( K,\hat Q_{23})}].
\label{eqn_app_znzk1zk2_multiple_2}
\end{align}
Plugging these results into the above M3L invariants, we can easily obtain quantization level of the corresponding M3L statistical phases. We can also further discuss the periods of these statistical phases from the periods of twisted coefficients and the symmetry fractionalization of ambient particles. In particular, we discuss them in details in the following example: $N=2,K=2^n$.

\subsection{SFP}
Since $\gcd(2,2^n)=2$, there are four different patterns of SFP. From (\ref{eq_mathcalc}), the elementary gauge charge, denoted as $e$, can carry one-half or integer  charge of the two $\Z_{K}$ subgroup. Correspondingly, their  reduced charge matrices are $\mathcal{Q}=(0,0)^T, (0,1)^T, (1,0)^T, (1,1)^T$ respectively where $\mathcal{Q}=(\mathcal{Q}_{11},\mathcal{Q}_{21})^T$ and $T$ denotes the matrix transposition. We denote these four patterns as $e00$ and $e0C$, $eC0$ and $eCC$, respectively.

\subsection{SFL}

Here we discuss the SFL for different SFP case by case.\\
\\
\textbf{Case-1: $e00$}\\
\\ 
We first discuss  the case with SFP being $e00$, namely the gauge charge carries both integer symmetry charge of the two $\Z_K$ subgroups. In this case, $Q_{11}=2m_1, Q_{21}=2 m_2$ and then $\hat Q_{11}=2\hat m_1, \hat Q_{21}=2\hat m_2$ with the conditions that $\hat m_i=2^{n-1}$ if the corresponding $ m_i=0$. Then (\ref{eqn_app_znzk1zk2_multiple_1}) and (\ref{eqn_app_znzk1zk2_multiple_2}) reduce to
\begin{align}
M_{12}
=&\text{lcm}[\frac{K}{\gcd(K,\hat m_1)}, \frac{ K}{\gcd( K, \hat Q_{12})}, \nonumber \\
&\qquad   \frac{K}{\gcd(K,\hat m_2)}, \frac{ K}{\gcd( K,\hat Q_{22})}]
\label{eqn_app_znzk1zk2_invariant_1aa}\\
M_{23}=&\text{lcm}[\frac{ K}{\gcd( K,\hat Q_{12})}, \frac{ K}{\gcd( K, \hat Q_{13})}, \nonumber \\
&\qquad   \frac{ K}{\gcd( K,\hat Q_{22})},\frac{  K}{\gcd( K,\hat Q_{23})}].
\label{eqn_app_znzk1zk2_invariant_2aa}
\end{align}

We first discuss the  invariant $\mathcal{I}_{\alpha aa}$. Using (\ref{eqn_app_znzk1zk2_invariant_1aa}) and (\ref{eqn_app_znzk1zk2_invariant_2aa}), the  invariant  becomes
\begin{align}
&\mathcal{I}_{\alpha aa}=-\frac{k Q_{12}}{KN^2}\text{lcm}[\frac{K}{\gcd(K,\hat m_1)}, \frac{ K}{\gcd( K,  \hat Q_{12})}, \nonumber \\
&\qquad\qquad \qquad  \qquad   \frac{K}{\gcd(K,\hat m_2)}, \frac{ K}{\gcd( K,\hat Q_{22})}]
\end{align} 
If $Q_{12}$ is zero, then $\mathcal{I}_{\alpha aa}$ vanishes. So we assume $Q_{12}$ is nonzero when discuss the  invariant. Under this assumption, $\hat Q_{12}=Q_{12}$. Then the invariant can simplified to
\begin{align}
&\mathcal{I}_{\alpha aa}=\frac{-k }{KN^2}\text{lcm}[K, \frac{KQ_{12}}{\gcd(K,\hat m_1)}, 
  \frac{KQ_{12}}{\gcd(K,\hat m_2)}, \frac{ KQ_{12}}{\gcd( K,\hat Q_{22})}]
\end{align} 
We note that $KQ_{12}/\gcd(K,\hat m_1)$ must be multiple of $Q_{12}$.
Then the M3L statistical phases take
{\small{
\begin{align}
&\theta_{\Sigma;\sigma^1}=\frac{-2\pi  k }{KN^2}\text{lcm}[K, \frac{KQ_{12}}{\gcd(K,\hat m_1)}, 
  \frac{KQ_{12}}{\gcd(K,\hat m_2)}, \frac{ KQ_{12}}{\gcd( K,\hat Q_{22})}]\label{eq_app_z2zkzk_e00_i11_phase1}\\
&  \theta_{\Sigma,\sigma^1;\Sigma}=\frac{2\pi  k }{KN^2}\text{lcm}[K, \frac{KQ_{12}}{\gcd(K,\hat m_1)}, 
  \frac{KQ_{12}}{\gcd(K,\hat m_2)}, \frac{ KQ_{12}}{\gcd( K,\hat Q_{22})}]\label{eq_app_z2zkzk_e00_i11_phase2}
\end{align}}}

When $Q_{12}=1$ and $Q_{22}=1$, these four statistical phases take the most general quantized values, that is, $\theta_{\Sigma;\sigma^1}=-\frac{2\pi k}{N^2}$ and  $\theta_{\Sigma,\sigma^1;\Sigma}=\frac{2\pi k}{N^2}$.

Now we discuss the periods of the two statistical phases. First of all, we consider those from attaching particles.  Since $e$ carries only integer symmetry charge, then by attaching particle, the statistical phases $\theta_{\Sigma;\sigma^1}$ can shift by $\frac{2\pi}{N}$, and  $\theta_{\Sigma,\sigma^1;\Sigma}$ can shift by $\frac{2\pi}{K}$ (since $K/N\in \Z$). Secondly, we can also consider the periods from those of twisted coefficients. However,  the periods from twisted coefficients are always multiple of the ones from attaching particles, as we show below.
From (\ref{eqn:twisted_coff_general_3}),  the period of $k$ is given by
\begin{align}
\Gamma=&\text{lcm}[\frac{K}{\gcd(K, \hat m_{1} p_{12})}, \frac{K}{\gcd(K, \hat m_{2} p_{12})}, \nonumber \\
&\qquad \frac{K}{\gcd(K, \hat m_{1} p_{22})}, \frac{K}{\gcd(K, \hat m_{2} p_{22})} ]\label{eq_app_z2zkzk_e00_i11_p1}
\end{align}
where we have defined $p_{ij}$ through setting 
\begin{align}
&  \hat Q_{12} M_{12}=p_{12}K, \label{eq_app_znzkzk_aux_1}\\
& \hat Q_{22} M_{12}=p_{22}K,\label{eq_app_znzkzk_aux_2} \\
& \hat Q_{11} M_{12}=p_{11}K, \label{eq_app_znzkzk_aux_3}\\
& \hat Q_{21} M_{12}=p_{21}K.\label{eq_app_znzkzk_aux_4}
\end{align} From (\ref{eqn_app_znzk1zk2_invariant_1aa}), we can easily see that all $p_{ij}$ are integral.  Plugging (\ref{eq_app_z2zkzk_e00_i11_p1}) into (\ref{eq_app_z2zkzk_e00_i11_phase1}) and (\ref{eq_app_z2zkzk_e00_i11_phase2}), the periods of these two statistical phases $\theta_{\Sigma;\sigma^1}$ and $\theta_{\Sigma,\sigma^1;\Sigma}$ are
\begin{align}
\frac{2\pi  p_{12}}{N^2}&\text{lcm}[\frac{K}{\gcd(K, \hat m_{1} p_{12})}, \frac{K}{\gcd(K, \hat m_{2} p_{12})}, \nonumber \\
&\qquad \frac{K}{\gcd(K, \hat m_{1} p_{22})}, \frac{K}{\gcd(K, \hat m_{2} p_{22})} ]
\end{align}
If $p_{12}$ is even, then this period of $\theta_{\Sigma;\sigma^1}$ and $\theta_{\Sigma,\sigma^1;\Sigma}$ is obviously multiple of $\pi$, the periods from attaching particles. If $p_{12}$ is odd, since $\hat m_{12}\le 2^{n-1}$, $\text{lcm}[\frac{K}{\gcd(K, \hat m_{1} p_{12})}, \frac{K}{\gcd(K, \hat m_{2} p_{12})}, \frac{K}{\gcd(K, \hat m_{1} p_{22})}, \frac{K}{\gcd(K, \hat m_{2} p_{22})} ]$ must also be even, so this period is also multiple of $\pi$.

Therefore, the statistical phase $\theta_{\Sigma;\sigma^1}$ can take  $N$ inequivalent values: $\frac{2\pi k}{N^2}$ with $k\in \Z_N$ while for the other  statistical phase $\theta_{\Sigma,\sigma^1;\Sigma}$, when $n\ge 2$, they can only take trivial value and when $n=1$(\textit{i.e.}, K=N), it can also take $N$ inequivalent values: $\frac{2\pi k}{N^2}$ with $k\in \Z_N$. In summary, these statistical phases contribute to one $\Z_N$ classification for SFL.

Similarly, the invariant $\mathcal{I}_{\beta aa}$ can also be discussed and the M3L statistical phases are quantized to be
{\small{
\begin{align}
&\theta_{\Sigma;\sigma^2}=\frac{-2\pi  k }{KN^2}\text{lcm}[K, \frac{KQ_{22}}{\gcd(K,\hat m_1)}, 
  \frac{KQ_{22}}{\gcd(K,\hat m_2)}, \frac{ KQ_{22}}{\gcd( K, Q_{12})}]\\
&  \theta_{\Sigma,\sigma^2;\Sigma}=\frac{2\pi  k }{KN^2}\text{lcm}[K, \frac{KQ_{22}}{\gcd(K,\hat m_1)}, 
  \frac{KQ_{22}}{\gcd(K,\hat m_2)}, \frac{ KQ_{22}}{\gcd( K,\hat Q_{12})}]
\end{align}}}
and they take the most general values
\begin{align}
&\theta_{\Sigma;\sigma^2}=-\frac{2\pi k}{N^2}\\
&\theta_{\Sigma,\sigma^2;\Sigma}=\frac{2\pi k}{N^2}
\end{align}
whose periods are  $\frac{2\pi}{N}$ and $\frac{2\pi}{K}$ respectively.  We note that as these statistical phases originate from $a_ia_1da_1$-like twisted terms, they can contribute to the classification  independently  by choosing different charge matrices. 
Therefore,  the statistical phases $\theta_{\Sigma;\sigma^2}$ and $\theta_{\Sigma,\sigma^2;\Sigma}$ can also contribute to another $\Z_N$ classification for SFL.

Now we consider the invariant $\mathcal{I}_{a\alpha \alpha}$. For convenience, we divide this invariant in (\ref{eq_app_znzkzk_invriant_3}) into two parts, \textit{i.e.},
\begin{align}
&\mathcal{I}^{(1)}_{a\alpha\alpha}=\frac{q_{211} Q_{12}Q_{11}}{K^2N^2}\\
&\mathcal{I}^{(2)}_{a\alpha\alpha}=-\frac{q_{122}Q_{12}Q_{12}}{K^2N}-\frac{q_{123}Q_{12}Q_{13}}{K^2N}
\end{align}
We  assume $Q_{12}$ nonzero, otherwise both parts vanish.    Then the two parts can be simplified to
\begin{align}
&\mathcal{I}^{(1)}_{a\alpha\alpha}=\frac{k  p_{12}  Q_{11}}{KN^2}\\
&\mathcal{I}^{(2)}_{a\alpha\alpha}=-\frac{\bar k p_{12} Q_{12}  }{KN}-\frac{\widetilde k p_{12}Q_{13} }{KN}
\end{align}
where we have used (\ref{eq_app_znzkzk_aux_1}). Then the M3L statistical phases $\theta_{\sigma^1;\Sigma}$ and $\theta_{\sigma^1,\Sigma;\sigma^1}$ from the two parts of $\mathcal{I}_{a\alpha \alpha}$ are quantized to 
\begin{align}
&\theta_{\sigma^1;\Sigma}^{(1)}=\frac{2\pi k  p_{12}  m_{1}}{KN}\label{eqn_app_znzkzk_e00_3}\\
&\theta_{\Sigma,\sigma^1;\sigma^1}^{(1)}=-\frac{2\pi k  p_{12}  m_{1}}{KN}\label{eqn_app_znzkzk_e00_4}\\
&\theta_{\sigma^1;\Sigma}^{(2)}=-\frac{2\pi (\bar k Q_{12}+ \widetilde k Q_{13} ) p_{12} }{KN}\label{eqn_app_znzkzk_e00_5}\\
&\theta_{\Sigma,\sigma^1;\sigma^1}^{(2)}=\frac{2\pi (\bar k Q_{12}+ \widetilde k Q_{13} ) p_{12} }{KN}\label{eqn_app_znzkzk_e00_6}
\end{align}
When $Q_{12}=Q_{13}=1$, from (\ref{eqn_app_znzk1zk2_invariant_1aa}) and (\ref{eq_app_znzkzk_aux_1}), we have $p_{12}=1$. Then the statistical phases  $\theta_{\sigma^1;\Sigma}^{(2)}$ and $\theta_{\Sigma,\sigma^1;\sigma^1}^{(2)} $ can take the most general values as
\begin{align}
&\theta_{\sigma^1;\Sigma}^{(2)}=-\frac{2\pi (\bar k + \widetilde k  )  }{KN}\\
&\theta_{\Sigma,\sigma^1;\sigma^1}^{(2)}=\frac{2\pi (\bar k + \widetilde k  )  }{KN}
\end{align}
with $\bar k$ and $\widetilde k$ being integral. We observe that $\theta_{\sigma^1;\Sigma}^{(1)}$ and $\theta_{\Sigma,\sigma^1;\sigma^1}^{(1)} $ are always multiple of $\frac{2\pi}{KN}$, hence they do not contribute any new quantized values. 
The periods of these statistical phases can come from attaching particles and periods of $\bar k$ and $\widetilde k$.  From attaching particles, both $\theta_{\sigma^1;\Sigma}$ and $\theta_{\Sigma,\sigma^1;\sigma^1}$ are ambiguous up to $\frac{2\pi}{K}$. On the other hand, the periods of the two statistical phases from the ones of $k, \bar k$ and $\widetilde k$ are always multiple of that from attaching particles, \textit{i.e.},  $\frac{2\pi}{K}$. 
 To see this point, using the period of $k$ given in (\ref{eq_app_z2zkzk_e00_i11_p1}), we can easily see that $\frac{2\pi p_{12}m_1}{KN}\Gamma$ is multiple of $\frac{2\pi}{N}$, hence also multiple of $\frac{2\pi}{K}$. 
 
 Below we show that the periods of the statistical phases $\theta_{\sigma^1;\Sigma}^{(2)}$ and $\theta_{\Sigma,\sigma^1;\sigma^1}^{(2)} $ from those of $\bar k$ and $\widetilde k$ are always multiple of $\frac{2\pi}{K}$.
The period of $\bar k$ is also given by (\ref{eq_app_z2zkzk_e00_i11_p1}). From (\ref{eqn:twisted_coff_general_2}), the period of $\widetilde k$ is given by
\begin{align}
\widetilde \Gamma
=\gcd\{ \text{lcm}[\frac{K}{\gcd(K, \hat m_{1} p_{12} )}, \frac{K}{\gcd(K, \hat m_{1} p_{22} )}, \nonumber\\\frac{K}{\gcd(K, \hat m_{2} p_{12})}, \frac{K}{\gcd(K, \hat m_{2}  p_{22})}],\nonumber \\
\text{lcm}[\frac{NK}{\gcd(NK, p_{11} \hat Q_{13})}, \frac{NK}{\gcd(NK, p_{11} \hat Q_{23})}, \nonumber \\
\frac{NK}{\gcd(NK, p_{21} \hat Q_{13})}, \frac{NK}{\gcd(NK, p_{21} \hat Q_{23})}, \nonumber \\
\frac{K}{\gcd(K, p_{12} \hat Q_{13})}, \frac{K}{\gcd(K, p_{12} \hat Q_{23})}, \nonumber \\
\frac{K }{\gcd(K, p_{22} \hat Q_{13})}, \frac{K}{\gcd(K,  p_{22} \hat Q_{23})}] \}
\label{eq_app_z2zkzk_e00_1ij_p1}
\end{align}
From (\ref{eqn_app_znzkzk_e00_5}) and (\ref{eqn_app_znzkzk_e00_6}),  since $\hat m_1\le 2^{n-1}$, then $\Gamma  p_{12}$ and $\widetilde \Gamma Q_{13} p_{12}$ must be even, so the periods of $\theta_{\sigma^1;\Sigma}$ and $\theta_{\Sigma,\sigma^1;\sigma^1}$ from  those of $\bar k$ and $\widetilde k$ must be multiple of $\frac{2\pi}{K}$.

Therefore, the statistical phases  ($\theta_{\sigma^1;\Sigma}$, $\theta_{\Sigma,\sigma^1;\sigma^1}$) can take $N$ inequivalent values, that is, ($\frac{2\pi k}{KN}$, $-\frac{2\pi k}{KN}$) with $k\in \Z_N$, which contributes to one $\Z_N$ classification for SFL. 

Similarly to $\mathcal{I}_{a\alpha \alpha}$, two statistical phases  ($\theta_{\sigma^2;\Sigma}$, $\theta_{\Sigma,\sigma^2;\sigma^2}$) determined by the invariant $\mathcal{I}_{a\beta \beta}$ also take $N$ inequivalent values,  that is, ($\frac{2\pi k}{KN}$, $-\frac{2\pi k}{KN}$) with $k\in \Z_N$, which contributes to another $\Z_N$ classification for SFL. 

Now we discuss the invariant $\mathcal{I}_{\alpha a \beta}$.  For convenience, we  also divide this invariant in (\ref{eq_app_znzkzk_invriant_5}) into two parts, \textit{i.e.},
\begin{align}
&\mathcal{I}_{\alpha a \beta}^{(1)}=-\frac{q_{211}Q_{12}Q_{21}}{K^2N^2}\\
&\mathcal{I}_{\alpha a \beta}^{(2)}=\frac{q_{122}Q_{12}Q_{22}}{K^2N}+\frac{q_{123}Q_{12}Q_{23}}{K^2N}
\end{align}
We  assume $Q_{12}$ nonzero, otherwise both parts vanish.    Then the two parts can be simplified to
\begin{align}
&\mathcal{I}^{(1)}_{\alpha a\beta}=-\frac{k  p_{12}  Q_{21}}{KN^2}\\
&\mathcal{I}^{(2)}_{\alpha a\beta}=\frac{\bar k p_{12} Q_{22}  }{KN}+\frac{\widetilde k p_{12}Q_{23} }{KN}
\end{align}
where we have used (\ref{eq_app_znzkzk_aux_1}). Then the M3L statistical phases $\theta_{\Sigma, \sigma^2;\sigma^1}$ and $\theta_{\sigma^2,\sigma^1;\Sigma}$ from the two parts of $\mathcal{I}_{\alpha  a\beta}$ are quantized to 
\begin{align}
&\theta_{\sigma^2,\Sigma;\sigma^1}^{(1)}=-\frac{2\pi k  p_{12}  m_{2}}{KN}\label{eqn_app_znzkzk_e00_7}\\
&\theta_{\sigma^2,\sigma^1;\Sigma}^{(1)}=\frac{2\pi k  p_{12}  m_{2}}{KN}\label{eqn_app_znzkzk_e00_8}\\
&\theta_{\sigma^2,\Sigma;\sigma^1}^{(2)}=\frac{2\pi (\bar k Q_{22}+ \widetilde k Q_{23} ) p_{12} }{KN}\label{eqn_app_znzkzk_e00_9}\\
&\theta_{\sigma^2,\sigma^1;\Sigma}^{(2)}=-\frac{2\pi (\bar k Q_{22}+ \widetilde k Q_{23} ) p_{12} }{KN}\label{eqn_app_znzkzk_e00_10}
\end{align}
When $Q_{12}=Q_{13}=1$, from (\ref{eqn_app_znzk1zk2_invariant_1aa}) and (\ref{eq_app_znzkzk_aux_1}), we have $p_{12}=1$. Together with the condition that $Q_{22}=Q_{23}=1$, the statistical phases  $\theta_{\sigma^1;\Sigma}^{(2)}$ and $\theta_{\sigma^1,\Sigma;\sigma^1}^{(2)} $ can take the most general values as
\begin{align}
&\theta_{\Sigma,\sigma^2;\sigma^1}^{(2)}=\frac{2\pi (\bar k + \widetilde k  )  }{KN}\label{eqn_app_znzkzk_e00_10_1}\\
&\theta_{\sigma^2,\sigma^1;\Sigma}^{(2)}=-\frac{2\pi (\bar k + \widetilde k  )  }{KN}\label{eqn_app_znzkzk_e00_10_2}
\end{align}
with $\bar k$ and $\widetilde k$ being integral. We observe that $\theta_{\Sigma,\sigma^2;\sigma^1}^{(1)}$ and $\theta_{\sigma^2,\sigma^1;\Sigma}^{(1)} $ are always multiple of $\frac{2\pi}{KN}$, hence they do not contribute any new quantized values. 

Now we consider the periods of  $\theta_{\Sigma,\sigma^2;\sigma^1}$ and $\theta_{\sigma^2,\sigma^1;\Sigma} $. The periods of these statistical phases can come from attaching particles and periods of $\bar k$ and $\widetilde k$.  From attaching particles, both $\theta_{\Sigma,\sigma^2;\sigma^1}$ and $\theta_{\sigma^2,\sigma^1;\Sigma}$ are ambiguous up to $\frac{2\pi}{K}$. On the other hand, the periods of the two statistical phases from the ones of $k, \bar k$ and $\widetilde k$ are always multiple of that from attaching particles, \textit{i.e.},  $\frac{2\pi}{K}$. 
First we consider $k$. Using the period of $k$ given in (\ref{eq_app_z2zkzk_e00_i11_p1}), we can easily see that $\frac{2\pi p_{12}m_2}{KN}\Gamma$ is multiple of $\frac{2\pi}{N}$, hence also multiple of $\frac{2\pi}{K}$ as $\frac{K}{N}\in \Z$.  The periods of $\bar k$ and $\widetilde k$ are also   $\Gamma$ and $\widetilde \Gamma$, given in (\ref{eq_app_z2zkzk_e00_i11_p1})  and (\ref{eq_app_z2zkzk_e00_1ij_p1}). From (\ref{eqn_app_znzkzk_e00_9}) and (\ref{eqn_app_znzkzk_e00_10}), since $\hat m_1 \le 2^{n-1}$, then $\Gamma p_{12}$ and $\widetilde \Gamma Q_{23} p_{12}$ must be even, so the periods of $\theta_{\Sigma,\sigma^2;\sigma^1}$ and $\theta_{\sigma^2,\sigma^1;\Sigma}$  from those of $\bar k$ and $\widetilde k$ must be multiple of $\frac{2\pi}{K}$.

Therefore, the statistical phases can take $N$ inequivalent values, \textit{i.e.}, $(\frac{2\pi k}{KN}, -\frac{2\pi k}{KN})$  with $k\in \Z_N$, which contributes to one $\Z_N$ classification for SFL.

Similarly, we can also discuss the invariant $\mathcal{I}_{\beta a \alpha }$. We note that one can observe that simply by exchanging the two $\Z_K$ subgroups, the above discussion can be carried over to the case of  $\mathcal{I}_{\beta a \alpha }$. Therefore, we can straightforwardly obtain the conclusion: from $\mathcal{I}_{\beta a \alpha }$, the statistical phases $\theta_{\Sigma,\sigma^1;\sigma^2}$ and $\theta_{\sigma^2,\sigma^1;\Sigma}$ can take $N$ inequivalent values that is, $(\frac{2\pi k}{KN}, -\frac{2\pi k}{KN})$  with $k\in \Z_N$, which contributes to one $\Z_N$ classification for SFL. We note that the statistical phase $\theta_{\sigma^2,\sigma^1;\Sigma}$ can also originate from the invariant  $\mathcal{I}_{\alpha a \beta }$, however its inequivalent values are compatible from these two different invariants.

Finally, we come to discuss the invariant $\mathcal{I}_{\alpha \beta a}$. We also divide the expression in (\ref{eq_app_znzkzk_invriant_7}) into four parts, \textit{i.e.},
\begin{align} 
&\mathcal{I}_{\alpha \beta a}^{(1)}=\frac{q_{231}Q_{13}Q_{22}}{K^2N}\\
&\mathcal{I}_{\alpha \beta a}^{(2)}=\frac{-q_{231}Q_{13}Q_{23}}{K^2N}\\
&\mathcal{I}_{\alpha \beta a}^{(3)}=\frac{q_{211}Q_{22}m_1}{K^2N}\\
&\mathcal{I}_{\alpha \beta a}^{(4)}=\frac{-q_{211}Q_{12}m_{2}}{K^2N}
\end{align}
where we have substituted   $Q_{11}=2m_1$ and $Q_{12}=2m_2$. For convenience, we define $p_{23}$ and $\bar p_{23}$ through
\begin{align}
\hat Q_{12} M_{23}=p_{23} K \label{eq_app_znzkzk_aux_5} \\
\hat Q_{22} M_{23}=\bar p_{23} K.  \label{eq_app_znzkzk_aux_6}\\
\hat Q_{13} M_{23}=p_{32} K \label{eq_app_znzkzk_aux_51} \\
\hat Q_{23} M_{23}=\bar p_{32} K.  \label{eq_app_znzkzk_aux_61}
\end{align}
We first discuss  $\mathcal{I}_{\alpha \beta a}^{(1)}$. If $Q_{22}=0$ or $Q_{13}=0$, it vanishes. So we focus on the case with nonzero $Q_{22}$ and nonzero $Q_{13}$. Then it can be simplified to
$\mathcal{I}_{\alpha \beta a}^{(1)}=\frac{k' Q_{13} \bar p_{23}}{KN}$,
which determines the statistical phases
\begin{align}
&\theta_{\Sigma, \sigma^1;\sigma^2}^{(1)}=-\frac{2\pi k’ Q_{13} \bar p_{23} }{KN}\label{eqn_app_znzkzk_e00_11}\\
&\theta_{\Sigma, \sigma^2;\sigma^1}^{(1)}=\frac{2\pi k' Q_{13} \bar p_{23} }{KN}\label{eqn_app_znzkzk_e00_12}
\end{align}
When $Q_{22}=1$, then $p_{23}=1$. Further if $Q_{13}=1$, the $\theta_{\Sigma, \sigma^1;\sigma^2}^{(1)}$ and $\theta_{\Sigma, \sigma^2;\sigma^1}^{(1)}$ can take the most general quantized values, that is,
$\theta_{\Sigma, \sigma^1;\sigma^2}^{(1)}=-\frac{2\pi k’ }{KN}
$, 
$\theta_{\Sigma, \sigma^2;\sigma^1}^{(1)}=\frac{2\pi k' }{KN}.
$

Similarly, we disucss the second part of the invariant $\mathcal{I}_{\alpha \beta a}^{(2)}$.  If $Q_{12}=0$ or $Q_{23}=0$, it vanishes. So we focus on the case with nonzero $Q_{12}$ and nonzero $Q_{23}$. Then it can be simplified to
$\mathcal{I}_{\alpha \beta a}^{(2)}=-\frac{k' Q_{23} p_{23}}{KN},
$
which determines the statistical phases
\begin{align}
&\theta_{\Sigma, \sigma^1;\sigma^2}^{(2)}=\frac{2\pi k’ Q_{23} p_{23} }{KN}\label{eqn_app_znzkzk_e00_15}\\
&\theta_{\Sigma, \sigma^2;\sigma^1}^{(2)}=-\frac{2\pi k' Q_{23} p_{23} }{KN}\label{eqn_app_znzkzk_e00_16}
\end{align}
When $Q_{12}=1$, then $p_{23}=1$. Further if $Q_{23}=1$, the $\theta_{\Sigma, \sigma^1;\sigma^2}^{(2)}$ and $\theta_{\Sigma, \sigma^2;\sigma^1}^{(2)}$ can take the most general quantized values, that is,
$\theta_{\Sigma, \sigma^1;\sigma^2}^{(2)}=\frac{2\pi k’ }{KN}$,
$\theta_{\Sigma, \sigma^2;\sigma^1}^{(2)}=-\frac{2\pi k' }{KN}$.

Next we consider $\mathcal{I}_{\alpha \beta a}^{(3)}$. If $Q_{22}$ is zero, it vanishes. So we focus on the case with nonzero $Q_{22}$.   Then it can be simplified to
$\mathcal{I}_{\alpha \beta a}^{(3)}=\frac{k p_{22} m_1}{KN}$
which determines the statistical phases
\begin{align}
&\theta_{\Sigma, \sigma^1;\sigma^2}^{(3)}=-\frac{2\pi k m_1 p_{22} }{KN}\label{eqn_app_znzkzk_e00_19}\\
&\theta_{\Sigma, \sigma^2;\sigma^1}^{(3)}=\frac{2\pi k m_1 p_{22} }{KN}.\label{eqn_app_znzkzk_e00_20}
\end{align}
When $Q_{22}=1$, then $p_{22}=1$. Further when $m_1=1$, they can take the most general values
$\theta_{\Sigma, \sigma^1;\sigma^2}^{(3)}=-\frac{2\pi k }{KN}$
$\theta_{\Sigma, \sigma^2;\sigma^1}^{(3)}=\frac{2\pi k }{KN}$. 

Now we discuss   $\mathcal{I}_{\alpha \beta a}^{(4)}$.  If $Q_{12}$ is zero, it vanishes. So we focus on the case with nonzero $Q_{12}$.   Then it can be simplified to
$\mathcal{I}_{\alpha \beta a}^{(4)}=-\frac{k p_{12} m_2}{KN}$
which determines the statistical phases
\begin{align}
&\theta_{\Sigma, \sigma^1;\sigma^2}^{(4)}=\frac{2\pi k m_2 p_{12} }{KN}\label{eqn_app_znzkzk_e00_23}\\
&\theta_{\Sigma, \sigma^2;\sigma^1}^{(4)}=-\frac{2\pi k m_2 p_{12} }{KN}.\label{eqn_app_znzkzk_e00_24}
\end{align}
When $Q_{12}=1$, then $p_{12}=1$. Further when $m_2=1$, they can take the most general values
$\theta_{\Sigma, \sigma^1;\sigma^2}^{(4)}=\frac{2\pi k }{KN}$,
$\theta_{\Sigma, \sigma^2;\sigma^1}^{(4)}=-\frac{2\pi k }{KN}$.

From  the above discussion, 
 we see the most general quantized values of the two statistical phases from $\mathcal{I}_{\alpha \beta a}$ are
\begin{align}
&\theta_{\Sigma, \sigma^1;\sigma^2}=\frac{2\pi l }{KN}\label{eqn_app_znzkzk_e00_27}\\
&\theta_{\Sigma, \sigma^2;\sigma^1}=-\frac{2\pi l }{KN}.\label{eqn_app_znzkzk_e00_28}
\end{align}
with $l$ being integral.

We recall that these two statistical phases $\theta_{\Sigma, \sigma^1;\sigma^2}$ and $\theta_{\Sigma, \sigma^2;\sigma^1}$ can also be determined by the invariants $\mathcal{I}_{\beta a \alpha}$ and $\mathcal{I}_{\alpha a \beta}$, respectively. As compared to (\ref{eqn_app_znzkzk_e00_15}) and (\ref{eqn_app_znzkzk_e00_10_1}),  the values of these two statistical phases from $\mathcal{I}_{\alpha \beta a}$ are not beyond those in (\ref{eqn_app_znzkzk_e00_15}) and (\ref{eqn_app_znzkzk_e00_10_1}).  And any pattern of the two statistical phases in (\ref{eqn_app_znzkzk_e00_27}) and (\ref{eqn_app_znzkzk_e00_28}) are not independent since they can be viewed as combination of those from invariants $\mathcal{I}_{\beta a \alpha}$ and $\mathcal{I}_{\alpha a \beta}$. While they do not contribute to new pattern of SFL,  $\mathcal{I}_{\alpha \beta a}$ might contribute to smaller periods of   these statistical phases.  However as we will show below it does not contribute any smaller period of these statistical phases, that is, they are always multiple of $\frac{2\pi}{K}$.

As we have discussed the periods of $\theta_{\Sigma, \sigma^1;\sigma^2}$ and $\theta_{\Sigma, \sigma^2;\sigma^1}$ from attaching particles above, here we only need to focus on the ones from twisted coefficients. We first discuss the period of $k'$, which from (\ref{eqn:twisted_coff_general_2}) takes
\begin{align}
\Gamma'=\gcd\{ \text{lcm}[\frac{K}{\gcd(K, p_{23} \hat Q_{13})}, \frac{K}{\gcd(K, p_{23} \hat Q_{23})}, \nonumber \\
\frac{K}{\gcd(K, \bar p_{23} \hat Q_{13})}, \frac{K}{\gcd(K, \bar p_{23} \hat Q_{23})}], \nonumber \\
\text{lcm}[\frac{K}{\gcd(K, p_{23} \hat m_{1})},\frac{K}{\gcd(K, \bar p_{23} \hat m_{1})}, \nonumber \\\frac{K}{\gcd(K, p_{23} \hat m_{2})}, \frac{K}{\gcd(K, \bar p_{23} \hat m_{2})}, \nonumber \\
\frac{K}{\gcd(K, p_{32} \hat m_{1})},\frac{K}{\gcd(K, \bar p_{32} \hat m_{1})}, \nonumber \\\frac{K}{\gcd(K, p_{32} \hat m_{2})}, \frac{K}{\gcd(K, \bar p_{32} \hat m_{2})}
]\}
\end{align}
Then $\Gamma' Q_{13} \bar p_{23}$ and $\Gamma' Q_{23} \bar p_{23}$ are even since $\hat m_1, \hat m_2\le 2^{n-1}$. So from (\ref{eqn_app_znzkzk_e00_11}), (\ref{eqn_app_znzkzk_e00_12}), (\ref{eqn_app_znzkzk_e00_15}), and (\ref{eqn_app_znzkzk_e00_16}),  we see that the periods of  $\theta_{\Sigma, \sigma^1;\sigma^2}$ and $\theta_{\Sigma, \sigma^2;\sigma^1}$ are both multiple of $\frac{2\pi}{K}$. 

As the period of $k$ in (\ref{eqn_app_znzkzk_e00_19})-(\ref{eqn_app_znzkzk_e00_20}) and also (\ref{eqn_app_znzkzk_e00_23})-(\ref{eqn_app_znzkzk_e00_24}) is given by $\Gamma$ in  (\ref{eq_app_z2zkzk_e00_i11_p1}), we easily see that both $\Gamma m_2 p_{12}$ and $\Gamma m_1 p_{22}$  are multiple of $K$. Therefore, the periods of $\theta_{\Sigma, \sigma^1;\sigma^2}$ and $\theta_{\Sigma, \sigma^2;\sigma^1}$ from (\ref{eqn_app_znzkzk_e00_19})-(\ref{eqn_app_znzkzk_e00_20}) and also (\ref{eqn_app_znzkzk_e00_23})-(\ref{eqn_app_znzkzk_e00_24}) are always multiple of $\frac{2\pi}{N}$ and then also multiple of $\frac{2\pi}{K}$.

To summarize, for the SFP pattern  $e00$, the classification of SFL is $(\Z_N)^6$.\\
\\
\textbf{Case-2: $eC0$}\\

Now we  discuss  the case with SFP being $eC0$, namely the gauge charge carries half integer symmetry charge of the first $\Z_K$ and integer charge of the second $\Z_K$ subgroup. In this case, $Q_{11}=2m_1+1, Q_{21}=2 m_2$ and then $\hat Q_{11}=2 m_1+1, \hat Q_{21}=2\hat m_2$ with the conditions that $\hat m_2=2^{n-1}$ if the corresponding $ m_2=0$. Then (\ref{eqn_app_znzk1zk2_multiple_1}) reduce to
\begin{align}
M_{12}
=&NK
\label{eqn_app_znzk1zk2_invariant_1aa_2}.
\end{align}
while (\ref{eqn_app_znzk1zk2_multiple_2}) remains in the same form.

We first discuss the invariant $\mathcal{I}_{\alpha aa}$. Using (\ref{eqn_app_znzk1zk2_invariant_1aa_2}), the  invariant  becomes
\begin{align}
&\mathcal{I}_{\alpha aa}=-\frac{k Q_{12}}{N}
\end{align} 
If $Q_{12}$ is zero, then $\mathcal{I}_{\alpha aa}$ vanishes. So we assume $Q_{12}$ is nonzero when discuss the  invariant. Then the M3L statistical phases take
{\small{
\begin{align}
&\theta_{\Sigma;\sigma^1}=\frac{-2\pi  k Q_{12}}{N}\label{eq_app_z2zkzk_eC0_i11_phase1}\\
&  \theta_{\Sigma,\sigma^1;\Sigma}=\frac{2\pi  k Q_{12} }{N}\label{eq_app_z2zkzk_eC0_i11_phase2}
\end{align}}}
when $Q_{12}=1$, they can take the most general quantized values, that is, $\theta_{\Sigma;\sigma^1}=\frac{-2\pi  k }{N}$ and $\theta_{\Sigma,\sigma^1;\Sigma}=\frac{2\pi  k  }{N}$.
However, these quantized values of the two statistical phases are equivalent to being trivial since attaching $e$ particle to $\Sigma$ or $\sigma^1$ can shift by at least a phase factor $\frac{2\pi}{N}$.

Similarly, we discuss the invariant $\mathcal{I}_{\beta aa}$, which now becomes
\begin{align}
\mathcal{I}_{\beta aa}=-\frac{k Q_{22}}{N}.
\end{align}
Then the two correspondingly statistical phases $\theta_{\Sigma;\sigma^2}$ and $\theta_{\Sigma,\sigma^2;\Sigma}$ are also multiple of $\frac{2\pi}{N}$, which however are also equivalent to being trivial since attaching particle $e$ to $\Sigma$ or $\sigma^2$ can shift the phases by at least $\frac{2\pi}{N}$.

We now discuss the invariant $\mathcal{I}_{a\alpha \alpha}$, which now becomes
\begin{align}
\mathcal{I}_{a\alpha \alpha}=\frac{k Q_{12}Q_{11}}{KN}-\frac{\bar k Q_{12}Q_{12}}{K}-\frac{\widetilde k Q_{12}Q_{13}}{K}
\end{align}
Then the two statistical phases can take the quantized values as
\begin{align}
&\theta_{\sigma^1;\Sigma}=2\pi \bigg[\frac{k Q_{12}Q_{11}}{KN}-\frac{\bar k Q_{12}Q_{12}}{K}-\frac{\widetilde k Q_{12}Q_{13}}{K}\bigg]\\
&\theta_{\Sigma,\sigma^1;\sigma^1}=-2\pi \bigg[\frac{k Q_{12}Q_{11}}{KN}-\frac{\bar k Q_{12}Q_{12}}{K}-\frac{\widetilde k  Q_{12}Q_{13}}{K}\bigg]
\end{align}
When $Q_{12}=Q_{11}=1$, they can take the most general quantized values, \textit{i.e.}, $\theta_{\sigma^1;\Sigma}=\frac{2\pi l}{KN}$ and $\theta_{\Sigma,\sigma^1;\Sigma}=-\frac{2\pi l}{KN}$ with $l$ being integral. However, these quantized values of the statistical phases are trivial due to the following reason: as $e$ carries half integer charge of the first $\Z_K$ symmetry,  attaching $e$ to  $\sigma^1$ for $\theta_{\sigma^1;\Sigma}$ or  to $\Sigma$ for $\theta_{\Sigma,\sigma^1;\sigma^1}$ can both shift them  by $\frac{2\pi}{NK}$ phase shift.

We turn to discuss the invariant $\mathcal{I}_{a\beta \beta}$, which now becomes 
\begin{align}
\mathcal{I}_{a\beta\beta}=\frac{k Q_{22}m_2}{K}-\frac{\bar k Q_{22}Q_{22}}{K}-\frac{\widetilde k Q_{22}Q_{23}}{K}\label{eq_app_znzkzk_invriant_4_1}
\end{align}
Then the two statistical phases can take the quantized values as
\begin{align}
&\theta_{\sigma^2;\Sigma}=2\pi \bigg[\frac{k Q_{22}m_2}{K}-\frac{\bar k Q_{22}Q_{22}}{K}-\frac{\widetilde k Q_{22}Q_{23}}{K}\bigg]\\
&\theta_{\Sigma,\sigma^2;\sigma^1}=-2\pi \bigg[\frac{k Q_{22}m_2}{K}-\frac{\bar k Q_{22}Q_{22}}{K}-\frac{\widetilde k Q_{22}Q_{23}}{K}\bigg]
\end{align}
These quantized values of the statistical phases are trivial due to the following reason: as $e$ carries  integer charge of the second $\Z_K$ symmetry,  attaching $e$ to  $\sigma^2$ for $\theta_{\sigma^2;\Sigma}$ or  to $\Sigma$ for $\theta_{\Sigma,\sigma^2;\sigma^2}$ can both shift them  by $\frac{2\pi}{K}$ phase shift.

Now we discuss the invariant $\mathcal{I}_{\alpha a \beta}$, which now becomes 
\begin{align}
\mathcal{I}_{\alpha a \beta}=-\frac{k Q_{12}m_{2}}{K}+\frac{\bar k Q_{12}Q_{22}}{K}+\frac{\widetilde k Q_{12}Q_{23}}{K}.\label{eq_app_znzkzk_invriant_5_1}
\end{align}
Then the two statistical phases now can take the quantized values 
\begin{align}
&\theta_{\Sigma,\sigma^2;\sigma^1}=-2\pi\bigg[\frac{k Q_{12}m_{2}}{K}-\frac{\bar k Q_{12}Q_{22}}{K}-\frac{\widetilde k Q_{12}Q_{23}}{K}\bigg]\\
&\theta_{\sigma^2,\sigma^1;\Sigma}=2\pi\bigg[\frac{k Q_{12}m_{2}}{K}-\frac{\bar k Q_{12}Q_{22}}{K}-\frac{\widetilde k Q_{12}Q_{23}}{K}\bigg]
\end{align}
These quantized values of the statistical phases are trivial due to the following reason: as $e$ carries half integer charge of the first $\Z_K$ symmetry and integer charge of the second $\Z_K$ symmetry,  attaching $e$ to  $\Sigma$ for $\theta_{\Sigma,\sigma^2;\sigma^1}$ or  to $\sigma^1$ for $\theta_{\sigma^2,\sigma^1;\Sigma}$ can both shift them  by at least  $\frac{2\pi}{K}$ phase shift.

We turn to consider the the invariant $\mathcal{I}_{\beta a \alpha }$, which now becomes 
\begin{align}
\mathcal{I}_{\beta a \alpha}=-\frac{ k Q_{22}(2m_1+1)}{KN}+\frac{\widetilde k Q_{22}Q_{12}}{K}+\frac{\widetilde k Q_{22}Q_{13}}{K}
\end{align} 
Then the two statistical phases now can take the quantized values as
\begin{align}
&\theta_{\sigma^1, \Sigma;\sigma^2}=-2\pi\bigg[\frac{ k Q_{22}(2m_1+1)}{KN}+\frac{\widetilde k Q_{22}Q_{12}}{K}+\frac{\widetilde k Q_{22}Q_{13}}{K}\bigg]\\
&\theta_{\sigma^1,\sigma^2;\Sigma}=2\pi\bigg[-\frac{ k Q_{22}(2m_1+1)}{KN}+\frac{\widetilde k Q_{22}Q_{12}}{K}+\frac{\widetilde k Q_{22}Q_{13}}{K}\bigg]
\end{align}
The quantized values of the statistical phases are trivial due to the following reason: as the $e$ particle carries half  charge of the first $\Z_K$ symmetry, attaching $e$ particle to $\Sigma$ for $\theta_{\sigma^1, \Sigma;\sigma^2}$ or to $\sigma^2$ for $\theta_{\sigma^1,\sigma^2;\Sigma}$ can shift them by $\frac{2\pi}{2K}$ phase factor.

Finally, we discuss the invariant $\mathcal{I}_{\alpha \beta a}$, which now becomes
\begin{align}
\mathcal{I}_{\alpha \beta a}=\frac{k Q_{22}(2m_1+1)}{KN}-\frac{ kQ_{12}m_2}{K}  +\frac{k' (\bar p_{23}Q_{13}-p_{23}Q_{23})}{KN}\label{eq_app_znzkzk_invriant_7_1}
\end{align}
where we have used (\ref{eq_app_znzkzk_aux_5}) and (\ref{eq_app_znzkzk_aux_6}) and assumed that $Q_{22}$ and $Q_{12}$ are nonzero (otherwise the corresponding term(s) would  vanish).
We can divide it into two parts
\begin{align}
&\mathcal{I}_{\alpha \beta a}^{(1)}=\frac{k' (\bar p_{23}Q_{13}-p_{23}Q_{23})}{KN}\\
&\mathcal{I}_{\alpha \beta a}^{(2)}=\frac{k Q_{22}(2m_1+1)}{KN}-\frac{ kQ_{12}m_2}{K} 
\end{align}
Then the two statistical phases $\theta_{\Sigma,\sigma^2;\sigma^1}$ and $\theta_{\Sigma,\sigma^1;\sigma^2}$ can be divided into two parts. We first discuss the part one determined by $\mathcal{I}_{\alpha \beta a}^{(1)}$, that is,
\begin{align}
&\theta_{\Sigma,\sigma^2;\sigma^1}^{(1)}=2\pi \frac{k' (\bar p_{23}Q_{13}-p_{23}Q_{23})}{KN}\\
&\theta_{\Sigma,\sigma^1;\sigma^2}^{(1)}=-2\pi \frac{k' (\bar p_{23}Q_{13}-p_{23}Q_{23})}{KN}
\end{align}
When $Q_{22}=Q_{13}=1$(with $Q_{23}=0$), $\bar p_{23}=1$ and $\theta_{\Sigma,\sigma^2;\sigma^1}^{(1)}$ and $\theta_{\Sigma,\sigma^1;\sigma^2}^{(1)}$ can take the most general quantized values, that is, $\theta_{\Sigma,\sigma^2;\sigma^1}^{(1)}=\frac{2\pi k'}{KN}$ and $\theta_{\Sigma,\sigma^1;\sigma^2}^{(1)}=-\frac{2\pi k'}{KN}$. 
From $\mathcal{I}_{\alpha \beta a}^{(2)}$, we have
$\theta_{\Sigma,\sigma^2;\sigma^1}^{(2)}=2\pi [\frac{k Q_{22}(2m_1+1)}{KN}-\frac{ kQ_{12}m_2}{K} ]$ and
$\theta_{\Sigma,\sigma^1;\sigma^2}^{(2)}=-2\pi [\frac{k Q_{22}(2m_1+1)}{KN}-\frac{ kQ_{12}m_2}{K} ]$
which are always multiple of $\frac{2\pi}{KN}$,the most general value from $\mathcal{I}_{\alpha \beta a}^{(1)}$.

The periods of the two statistical phases can come from two aspects. First, $\theta_{\Sigma,\sigma^2;\sigma^1}$ and $\theta_{\Sigma,\sigma^1;\sigma^2}$ can shift by $\frac{2\pi}{K}$ and $\frac{2\pi}{2K}$ respectively from attaching $e$ particle to $\Sigma$ since $e$ particle carries integer charge of the second $\Z_K$ and half charge of the first $\Z_K$.  Secondly, the periods of these statistical phases can also come from twisted coefficients, which, however, are always multiple of those from attaching particle, as follow. From (\ref{eqn:twisted_coff_general_2}) and (\ref{eqn:twisted_coff_general_3})), the period of $k$ and $k'$ are given by 
\begin{align}
\Gamma=\text{lcm}[\frac{K}{\gcd(K, \hat Q_{12})},  \frac{K}{\gcd(K, \hat Q_{22})}]
\end{align}
and 
\begin{align}
\Gamma'=\gcd\{ \text{lcm}[\frac{K}{\gcd(K, p_{23} \hat Q_{13})}, \frac{K}{\gcd(K, p_{23} \hat Q_{23})}, \nonumber \\
\frac{K}{\gcd(K, \bar p_{23} \hat Q_{13})}, \frac{K}{\gcd(K, \bar p_{23} \hat Q_{23})}], \nonumber \\
\text{lcm}[\frac{NK}{\gcd(NK, p_{23})},\frac{K}{\gcd(K,  p_{23} \hat m_{2})}, \nonumber \\\frac{NK}{\gcd(NK, \bar p_{23})}, \frac{K}{\gcd(K, \bar p_{23} \hat m_{2})}, \nonumber \\
\frac{NK}{\gcd(NK, p_{32} )},\frac{K}{\gcd(K,  p_{32} \hat m_{2})}, \nonumber \\\,\,\frac{NK}{\gcd(NK, \bar p_{32} )}, \frac{K}{\gcd(K, \bar p_{32} \hat m_{2})}
]\}
\end{align}
respectively. As both $\Gamma Q_{12}$ and $\Gamma Q_{22}$ are both multiple of $K$, then $\frac{2\pi \Gamma Q_{22}(2m_1+1)}{NK}$ and $\frac{2\pi \Gamma Q_{12}(2m_2)}{NK}$ are both multiple of $\frac{2\pi}{N}$ and hence $\frac{2\pi}{K}$. Meanwhile, we can see that both $\Gamma' \bar p_{23} Q_{13}$ and $\Gamma' p_{23} Q_{23}$ are also multiple of $K$, then  $\frac{2\pi \Gamma'\bar p_{23} Q_{13}}{NK}$ and $\frac{2\pi \Gamma' p_{23} Q_{23}}{NK}$ are also multiple of $\frac{2\pi}{N}$ and hence multiple of $\frac{2\pi}{K}$. 

Therefore, $\theta_{\Sigma,\sigma^1;\sigma^2}$ are always equivalent to being trivial while $\theta_{\Sigma,\sigma^2;\sigma^1}$ can take $N$ inequivalent values, that is, $\frac{2\pi l}{NK}$ with $l\in \Z_N$. Therefore, the $\Z_N$ inequivalent statistical phases $\theta_{\Sigma,\sigma^2;\sigma^1}$ contributes to one $\Z_N$ classification of SFL.

To summaize, for SFP $eC0$, the classification for SFL is $\Z_N$.\\
\\
\textbf{Case-3: $e0C$}\\

Similarly to the Case-2 with SFP $eC0$, the discussion for the classification for SFL can be easily obtained by just exchanging the two $\Z_K$ symmetry. Therefore, we can  obtain the conclusion: for SFP $eC0$, the  classification for SFL is $\Z_N$, giving by the $N$ inequivalent values of the statistical phases $\theta_{\Sigma,\sigma^1;\sigma^2}=\frac{2\pi l}{NK}$ where $l\in \Z_N$.\\
\\
\textbf{Case-4: $eCC$}\\

This case can also be connected to by rearranging the symmetry group $\Z_K\times \Z_K$ generated by $g,h$ to become another isomorphic one $\Z_K\times \widetilde{\Z}_K$ whose generators are $g, gh$. In this case, $e$ particle still carries half charge of the first $\Z_K$ symmetry, but integer charge of the second $\widetilde{\Z}_{K}$ symmetry now. In other words, the SFP $eCC$ now becomes $eC\widetilde 0$ where the superscript reminds of the second $\widetilde{\Z}_K$ is generated by $gh$.  Then we can apply the above discussion to this case and we can draw the conclusion: for SFP $eCC$ (or equivalently $eC\widetilde 0$), the classification for SFL is $\Z_N$, characterizing by the $N$ inequivalent statistical phases $\theta_{\Sigma, \sigma^{12};\sigma^1}=\frac{2\pi l}{NK}$
 with $l\in \Z_N$ where $\sigma^{12}$ is the fusion of $\sigma^1$ and $\sigma^2$.

%
 
 

\end{document}